\documentclass[a4paper,twoside,11pt,headings=optiontohead]{scrbook}

\usepackage[toc,page]{appendix}
\usepackage[utf8]{inputenc}
\usepackage{lmodern}
\usepackage{epstopdf}

\title{Topological States of Matter in Frustrated Quantum Magnetism}
\author{Alexander Wietek}

\usepackage{braket}
\usepackage{amsmath}
\usepackage{amssymb}
\usepackage{amsthm}
\usepackage{chemformula}
\usepackage{bm}

\newtheorem*{thm*}{Theorem}

\newtheorem*{defn*}{Definition}

\newtheorem*{exmpl*}{Example}
\usepackage{makeidx}
\makeindex

\numberwithin{equation}{section}

 \def\@chapter[#1]#2{\ifnum \c@secnumdepth >\m@ne
   \if@mainmatter
   \refstepcounter{chapter}%
   \typeout{\@chapapp\space\thechapter.}%
   \addcontentsline{toc}{chapter}%
   {\protect\numberline{\thechapter}#2}%
   \else
   \addcontentsline{toc}{chapter}{#2}%
   \fi
   \else
   \addcontentsline{toc}{chapter}{#2}%
   \fi
   \chaptermark{#1}%
   \addtocontents{lof}{\protect\addvspace{10\p@}}%
   \addtocontents{lot}{\protect\addvspace{10\p@}}%
   \if@twocolumn
   \@topnewpage[\@makechapterhead{#2}]%
   \else
   \@makechapterhead{#2}%
   \@afterheading
   \fi}

\usepackage[Bjornstrup]{fncychap}
\usepackage{xcolor}

\colorlet{partbgcolor}{gray!30} % shaded background color for parts
\colorlet{partnumcolor}{gray}% color for numbers in parts
\colorlet{chapbgcolor}{gray!30}% shaded background color for chapters
\colorlet{chapnumcolor}{gray}% color for numbers in chapters

\renewcommand*\partformat{%
  \fontsize{76}{80}\usefont{T1}{pzc}{m}{n}\selectfont%
  \hfill\textcolor{partnumcolor}{\thepart}}

\makeatletter
\renewcommand*{\@part}{}
\def\@part[#1]#2{%
  \ifnum \c@secnumdepth >-2\relax
    \refstepcounter{part}%
    \@maybeautodot\thepart%
    \addparttocentry{\thepart}{#1}%
  \else
    \addparttocentry{}{#1}%
  \fi
  \begingroup
    \setparsizes{\z@}{\z@}{\z@\@plus 1fil}\par@updaterelative
    \raggedpart
    \interlinepenalty \@M
    \normalfont\sectfont\nobreak
    \setlength\fboxsep{0pt}
    
    \colorbox{partbgcolor}{\rule{0pt}{40pt}%
      \makebox[\linewidth]{%
    \begin{minipage}{\dimexpr\linewidth+20pt\relax}
      \ifnum \c@secnumdepth >-2\relax
        \vskip-25pt
        \size@partnumber{\partformat}%
      \fi      %
      \vskip\baselineskip
      \hspace*{\dimexpr\myhi+10pt\relax}%
      \parbox{\dimexpr\linewidth-2\myhi-20pt\relax}{\raggedleft\LARGE#2\strut}%
      \hspace*{\myhi}\par\medskip%
    \end{minipage}%
      }%
    }%
    \partmark{#1}\par
  \endgroup
  \@endpart
}

\renewcommand\DOCH{%
  \settowidth{\py}{\CNoV\thechapter}
  \addtolength{\py}{-10pt}
  \fboxsep=0pt%
  \colorbox{chapbgcolor}{\rule{0pt}{40pt}\parbox[b]{\textwidth}{\hfill}}%
  \kern-\py\raise20pt%
  \hbox{\color{chapnumcolor}\CNoV\thechapter}\\%
}

 \makeatother

\DeclareMathOperator{\spn}{span}
\DeclareMathOperator{\integer}{int}

\DeclareMathOperator{\aut}{Aut}
\DeclareMathOperator{\tr}{Tr}
\newcommand\norm[1]{\left\lVert#1\right\rVert}
\newcommand{\argmin}{\operatornamewithlimits{argmin}}

\newcommand{\Mod}[1]{\ (\mathrm{mod}\ #1)}

\makeatletter
\def\moverlay{\mathpalette\mov@rlay}
\def\mov@rlay#1#2{\leavevmode\vtop{%
   \baselineskip\z@skip \lineskiplimit-\maxdimen
   \ialign{\hfil$\m@th#1##$\hfil\cr#2\crcr}}}
\newcommand{\charfusion}[3][\mathord]{
    #1{\ifx#1\mathop\vphantom{#2}\fi
        \mathpalette\mov@rlay{#2\cr#3}
      }
    \ifx#1\mathop\expandafter\displaylimits\fi}
\makeatother

\newcommand{\bigcupdot}{\charfusion[\mathop]{\bigcup}{\cdot}}

\usepackage{caption}
\usepackage{subcaption}

\usepackage{hyperref}
\usepackage{cleveref}
\crefname{equation}{Eq.}{Eqs.}
\Crefname{equation}{Eq.}{Eqs.}
\crefname{figure}{Fig.}{Figs.}
\Crefname{figure}{Fig.}{Figs.}
\crefname{part}{part}{parts}
\Crefname{part}{Part}{Parts}
\crefname{chapter}{chapter}{chapters}
\Crefname{chapter}{Chapter}{Chapters}
\crefname{section}{section}{sections}
\Crefname{section}{Section}{Sections}
\crefname{subsection}{section}{sections}
\Crefname{subsection}{Section}{Sections}
\crefname{appendix}{appendix}{appendices}
\Crefname{appendix}{Appendix}{Appendices}

\usepackage{algorithm}
\usepackage[noend]{algpseudocode}
\algblock[ForEach]{ForEach}{EndForEach}
\algblockdefx[ForEach]{ForEach}{EndForEach}[1]{\textbf{for each } #1 \textbf{:}}{}
\makeatletter
\ifthenelse{\equal{\ALG@noend}{t}}{\algtext*{EndForEach}}{}
\makeatother
\usepackage[
backend=biber,
style=numeric,
doi=true,
url=false,
sorting=none,
citestyle=numeric-comp,
maxcitenames=10]{biblatex}
\AtEveryBibitem{%
  \ifentrytype{article}{%
    \clearfield{pages}%
    \clearfield{issn}%
    \clearfield{Issn}%
    \clearfield{month}%
    \clearfield{Month}%
    \clearfield{language}%
  }{%
  }%
}
\AtEveryBibitem{%
  \ifentrytype{inbook}{%
    \clearfield{Isbn}%
    \clearfield{isbn}%
  }{%
  }%
}

\usepackage{float}
\newfloat{inlinecitation}{h}{ext}

\usepackage{graphicx}
\graphicspath{
  {./chapters/introduction/pics/},
  {./chapters/exact_diagonalization/pics/},
  {./chapters/exact_diagonalization/pics/blocks/},
  {./chapters/largescaleed/pics/},  
  {./chapters/largescaleed/pics/sublattice_coding/},
  {./chapters/largescaleed/pics/parallel/},
  {./chapters/largescaleed/pics/benchmarks/},
  {./chapters/variational_mc/pics/},
  {./chapters/tower_of_states/pics/},
  {./chapters/chiral_kagome/pics/},
  {./chapters/chiral_triangular/pics/},
  {./chapters/spin_liquids/pics/},
  {./chapters/spin_liquids/pics/bandstructures/},
  {./chapters/sun-csl-paper/pics/},
  {./chapters/sun_multicolumn/pics/}
  {./pics/}
}

\usepackage[final]{pdfpages}

\DeclareUnicodeCharacter{2215}{/}
\DeclareUnicodeCharacter{2212}{-}

\usepackage{epigraph}
\usepackage{calc}

\newcommand{\mytextformat}{\itshape\epigraphsize}

\let\originalepigraph\epigraph 
\renewcommand\epigraph[2]%
   {\setlength{\epigraphwidth}{\widthof{\mytextformat#1}}\originalepigraph{#1}{#2}}

\usepackage{adjustbox}

\begin{document}
\frontmatter

%auto-ignore
\begin{titlepage}
  \begin{center}

    { \Huge{Topological states of matter in frustrated quantum
      magnetism} } \bigskip \vspace{1cm}
    \includegraphics[width=10cm]{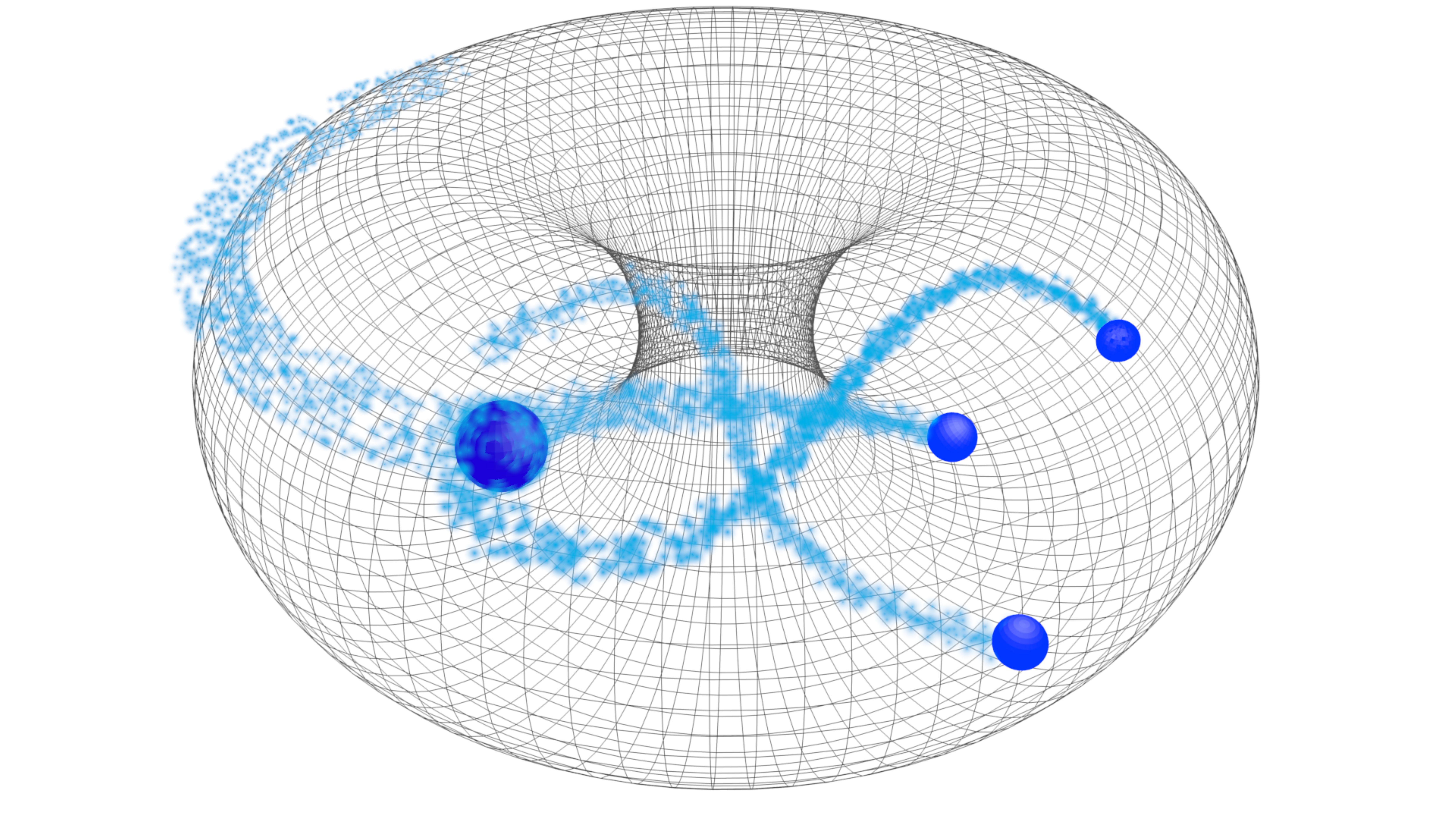}\\[3mm]
    \bigskip

    \huge{Dissertation}\\[5mm]
    
    \normalsize eingereicht von \\[5mm]

    {\Large
      Alexander Wietek, M.Sc. M.Sc. \\[5mm]
    }
    \normalsize
    zur Erlangung des akademischen Grades \\
    ``Doctor of Philosophy (Ph.D.)''\\[5mm]
    
    Leopold-Franzens-Universit\"at Innsbruck\\
    Fakult{\"a}t f{\"u}r Mathematik, Informatik und Physik\\[5mm]

    \includegraphics[height=3cm]{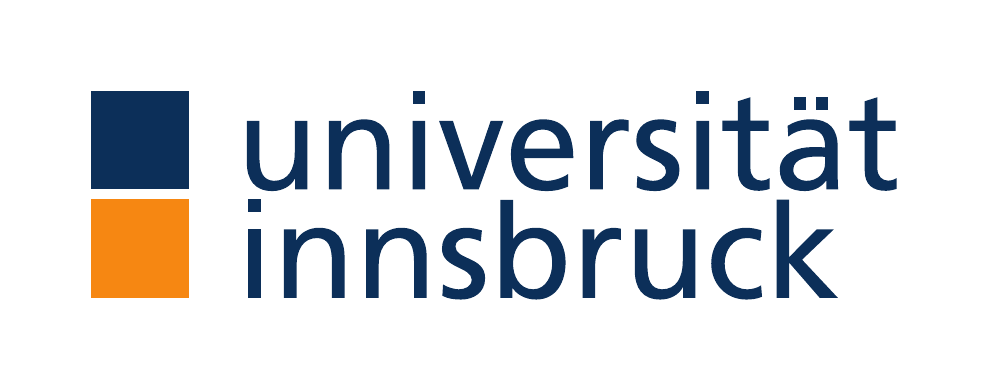}
    \\[5mm]

    Betreut von: Univ.-Prof. Dr. Andreas M. Läuchli \medskip
    
    Oktober 2017
  \end{center}

\end{titlepage}

%%% Local Variables:
%%% mode: latex
%%% TeX-master: "../thesis"
%%% End:

%auto-ignore
\chapter*{\centering Abstract / Zusammenfassung}
\label{sec:abstract}
Frustrated quantum magnets may exhibit fascinating collective
phenomena. The main goal of this dissertation is to provide conclusive
evidence for the emergence of novel phases of matter like quantum spin
liquids in local quantum spin models. After a general introduction to
frustrated magnetism, spin liquids and the numerical methods employed
in \cref{part:preliminaries} comprising
\cref{sec:introduction,sec:methods} we present the main results of the
thesis in \cref{part:research}.

We develop novel algorithms for large-scale Exact Diagonalization
computations in~\cref{sec:largescaleed}. So-called \textit{sublattice
  coding methods} for efficient use of lattice symmetries in the
procedure of diagonalizing the Hamiltonian matrix are
proposed. Furthermore, we suggest a randomized distributed memory
parallelization strategy. Benchmarks of computations on various
supercomputers with system size up to $50$ spin-$1/2$ particles have
been performed.

Results concerning the emergence of a chiral spin liquid in a
frustrated kagome Heisenberg antiferromagnet are presented
in~\cref{sec:paperkagome}. We confirm previous findings obtained via
DMRG calculations using Exact Diagonalization and propose that the
chiral spin liquid phase in this model is well described by
Gutzwiller-projected wave functions. Also, the stability and extent of
this phase are discussed.

In an extended Heisenberg model on the triangular lattice, we
establish another chiral spin liquid phase
in~\cref{sec:papertriangular} amongst several magnetically ordered
phases. We discuss the special case of the Heisenberg $J_1$-$J_2$
model with nearest and next-nearest neighbor interactions and present
a scenario where the critical point of phase transition from the
$120^\circ$ N{\'e}el to a putative $\mathbb{Z}_2$ spin liquid is
described by a Dirac spin liquid.

A generalization of the SU($2$) Heisenberg model with SU($N$) degrees
of freedom on the triangular lattice with an additional ring-exchange
term is discussed in~\cref{sec:sunchiral}. We present our contribution
to the project and the final results that suggest a series of chiral
spin liquid phases in an extended parameter range for $N=3,\ldots,10$.

Finally, we present preliminary data from a Quantum Monte Carlo study
of an SU($N$) version of the $J$-$Q$ model on a square lattice
in~\cref{sec:sunmulticolumn}. We study this model for $N=2,\ldots,10$
and multi-column representations of SU($N$) and establish the phase
boundary between the N{\'e}el ordered phase and the disordered phases
for $J,\;Q \geq 0$. The disordered phase in the four-column
representation is expected to be a two-dimensional analog of the
Haldane phase for the spin-$1$ Heisenberg chain.

%%% Local Variables:
%%% mode: latex
%%% TeX-master: "../../thesis"
%%% End:

\addcontentsline{toc}{chapter}{Abstract / Zusammenfassung}
%auto-ignore
% \chapter*{\centering Zusammenfassung}
% \label{sec:abstract_german}

\subsubsection{Deutsche Zusammenfassung}
Frustrierte Quantenmagnete können faszinierende kollektive Phänomene
aufweisen. Das Ziel dieser Dissertation ist es schlüssigen Nachweis
für die Emergenz neuer Materiezustände wie Quantenspinflüssigkeiten in
lokalen Quantenspinmodellen zu erbringen. Nach einer allgemeinen
Einleitung in frustrierten Magnetismus, Spinflüssigkeiten und die
verwendeten numerischen Methoden in Teil~\ref{part:preliminaries},
bestehend aus Kapitel~\ref{sec:introduction}~und~\ref{sec:methods}
stellen wir die wichtigsten Ergebnisse dieser Arbeit in
Teil~\ref{part:research} vor.

Wir entwickeln neue Algorithmen für skalierbare Exakte
Diagonalisierung in Kapitel~\ref{sec:largescaleed}. So genannte
\textit{Untergitterkodierungsmethoden} zur effizienten Nutzung von
Gittersymmetrien im Vorgang der Diagonalisierung der Hamiltonmatrix
werden vorgeschlagen. Desweiteren stellen wir eine randomisierte
Parellelisierungsstrategie für verteilte Speichersysteme vor.
Benchmarks mit Systemgrößen bis zu $50$ Spin-$1/2$ Teilchen wurden auf
mehreren Supercomputern durchgeführt.

Ergebnisse zur Emergenz von chiralen Spinflüssigkeiten in einem
erweiterten Heisenberg Antiferromagneten auf dem Kagome-Gitter
werden in Kapitel~\ref{sec:paperkagome} vorgestellt. Wir bestätigen
vorangegangene DMRG Studien mithilfe der Exakten Diagonalisierung und
zeigen auf, dass diese chirale Spinflüssigkeit gut mithilfe
Gutzwiller-projizierter Wellenfunktionen beschrieben werden kann.
Desweiteren werden die Stabilität und die Ausdehnung dieser Phase
behandelt.

In einem erweiterten Heisenbergmodell auf dem Dreiecksgitter bringen
wir den Nachweis einer weiteren chiralen Spinflüssigkeitsphase
zwischen mehreren magnetisch geordneten Phasen in
Kapitel~\ref{sec:papertriangular}. Wir besprechen den Spezialfall des
Heisenberg $J_1$-$J_2$ Modells mit nächster- und übernächster-Nachbar
Wechselwirkung und stellen ein Szenario vor, bei welchem der kritische
Punkt des Phasenübergangs zwischen der $120^\circ$ N{\'e}el geordneten
und der mutmaßlichen $\mathbb{Z}_2$ Spinflüssigkeit durch eine Dirac
Spinflüssigkeit beschrieben wird.

Eine Verallgemeinerung des SU($2$) Heisenberg Modells mit SU($N$)
Freiheitsgraden auf dem Dreiecksgitter mit zusätzlichem
Ringaustauschterm wird in Kapitel~\ref{sec:sunchiral} diskutiert.  Wir
stellen unseren Beitrag zum Projekt und die endgültigen Ergebnisse
vor, welche eine Reihe von chiralen Spinflüssigkeitsphasen in einer
ausgedehnten Parameterregion für $N=3,\ldots,10$.

Zuletzt stellen wir vorläufige Daten einer Quanten Monte Carlo Studie
zu einer SU($N$) Version des $J$-$Q$ Modells auf dem Quadratgitter
vor.  Wir untersuchen dieses Modell für $N=2,\ldots,10$ und
Darstellungen mit mehreren Spalten von SU($N$) und stellen den
Phasenübergangspunkt von der N{\'e}el geordneten zur ungeordeten Phase
für $J,\;Q \geq 0$ fest. Es wird erwartet, dass die ungeordnete Phase in
der vier-Spalten Darstellung ein zweidimensionales Analogon zur
Haldane Phase der Spin-$1$ Heisenberg Kette ist.

%%% Local Variables:
%%% mode: latex
%%% TeX-master: "../../thesis"
%%% End:

% \addcontentsline{toc}{chapter}{Zusammenfassung}

%auto-ignore
\chapter*{\centering Acknowledgements}
First things first. During my doctoral studies, I learned that
scientific work and writing a Ph.D. thesis is quite an
undertaking. Luckily, I received support from many people who I would
like to explicitly mention here and to whom I want to express my
sincere gratitude.

As my supervisor, Andreas Läuchli introduced me to the exciting world
of physics research. I am deeply grateful for his scientific advice
and the interesting projects he pointed out to me. I greatly enjoyed
the frequent scientific discussions that have always been very
inspiring for me, his friendly and forthcoming attitude and his
ongoing support. Moreover, I would like to thank Andreas for giving me
the opportunity to visit many conferences and workshops. Thank you
very much!

I am very happy to have had such amazing working colleagues, which
lightened up the everyday life at the Institute for Theoretical
Physics. It has been a pleasure to share an office with Michael
Schuler all the time. Discussing, working and developing software
together has had very positive impact on this thesis. My work
benefited a lot all the discussions with Antoine Sterdyniak,
Louis-Paul Henry, and Thomas Lang, to whom I could always turn to for
scientific questions. I would also like to thank Carlo Krimphoff, Lars
Bonnes, Elke Stenico, Andreas Parteli, Christian Romen, Michael Rader
and Clemens Ganahl for making the time at the institute so enjoyable.

I had the great opportunity to visit Synge Todo at the University of
Tokyo for six months, which has been nothing short of an amazing
experience. I would like to thank Synge Todo for accepting me as an
exchange student, for introducing me to the Quantum Monte Carlo
technique and for his scientific advice. Working in his group
broadened my horizon both scientifically as well as personally. I
benefited a lot from discussions with Hidemaro Suwa and would like to
thank also the other members of the lab, especially Kai Shimagaki,
Toshiki Horita and Fumihiro Ishikawa for showing me a good time in
Tokyo. Furthermore, I would once again like to thank Andreas Läuchli
for his financial support during that time and acknowledge the
financial support from the Marietta Blau Stipendium.

The good times I shared with my friends and family mean a lot to me
and I would like to sincerely thank all of them. I feel very lucky to
have received so much help and understanding, especially by my
parents.  Most of all I would like to thank my wonderful wife,
Christina Kurzthaler, who always had an open ear for my smaller and
sometimes bigger troubles. Her ongoing support and encouragement have
made everything, be it research or everyday life, so much easier and
so much more delightful.

%%% Local Variables:
%%% mode: latex
%%% TeX-master: "../thesis"
%%% End:

\addcontentsline{toc}{chapter}{Acknowledgements}

\newpage
\tableofcontents
\newpage

\mainmatter
\part{Preliminaries}\label{part:preliminaries}
\chapter{Introduction}
\label{sec:introduction}
\epigraph{ \parbox{6.8cm}{\flushright A beginning is the time for taking the most delicate care that the balances are correct.} }{Frank Herbert, \textit{Dune}}

%auto-ignore
% \section{Introduction}
% The whole is greater than the sum of its parts.
Microscopic particles
like electrons, neutrons, and protons build up the matter that
surrounds us in everyday life. While the fundamental physical laws
that describe the interaction between electrons and the atomic nuclei
are as of today well understood, predicting the behavior of many
particles interacting with each other remains a great challenge. The
behavior of a macroscopic amount of non-interacting particles can be
deduced from averaging over the behavior of single particles. Often,
also weakly interacting particles behave as if they are essentially
non-interacting. Strong interactions, on the other hand, may lead to
exciting collective phenomena. Understanding the emergence of
macroscopic behavior from strongly interacting microscopic
constituents is a challenging, yet fascinating, task.

The ratio between kinetic and interaction energy of microscopic
particles is set by temperature. Low temperatures decrease the kinetic
energy of particles, thus making interaction effects more pronounced,
as can prominently be observed in ferromagnetic materials. While the
spin degrees of freedom strongly fluctuate at high temperatures, they
collectively align in the same direction once cooled down below the
Curie temperature. In particular, the tendency to align in the same
direction is due to the exchange interaction of the spin degrees of
freedom. This order-to-disorder transition in ferromagnets can be
explained by Landau's theory of symmetry breaking. The disordered
state is symmetric under spin rotations, while the ordered state
chooses a preferred direction and the symmetry is not respected by the
state of the system. The principle of symmetry breaking explains most
known phase transitions and is a major cornerstone in the theory of
condensed matter physics.

\begin{figure}[t!]
  \centering
  \begin{minipage}[c]{0.4\textwidth}
    \centering
    \includegraphics[width=\textwidth]{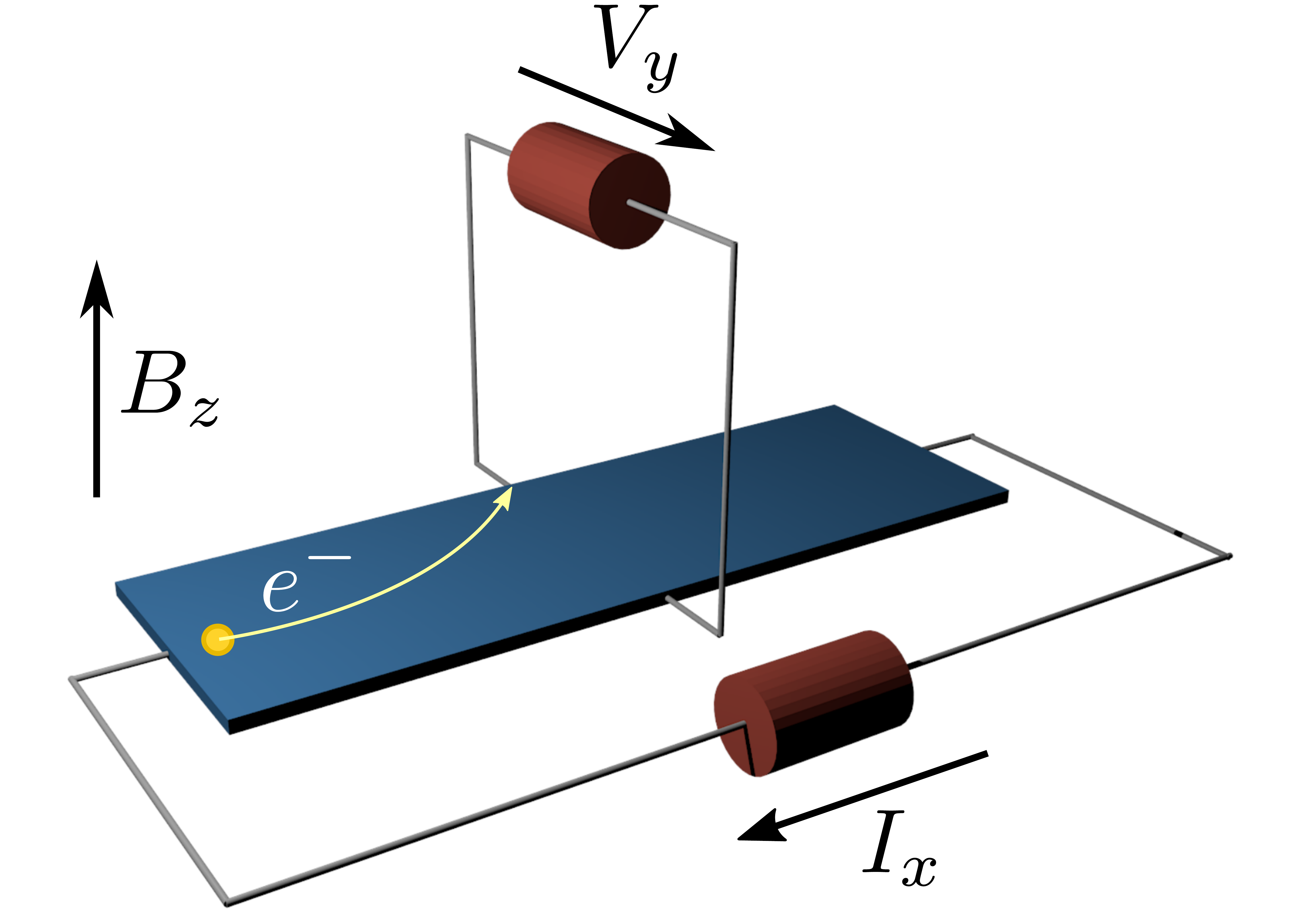}
  \end{minipage}%
  \quad
  \begin{minipage}[c]{0.5\textwidth}
    \includegraphics[width=\textwidth]{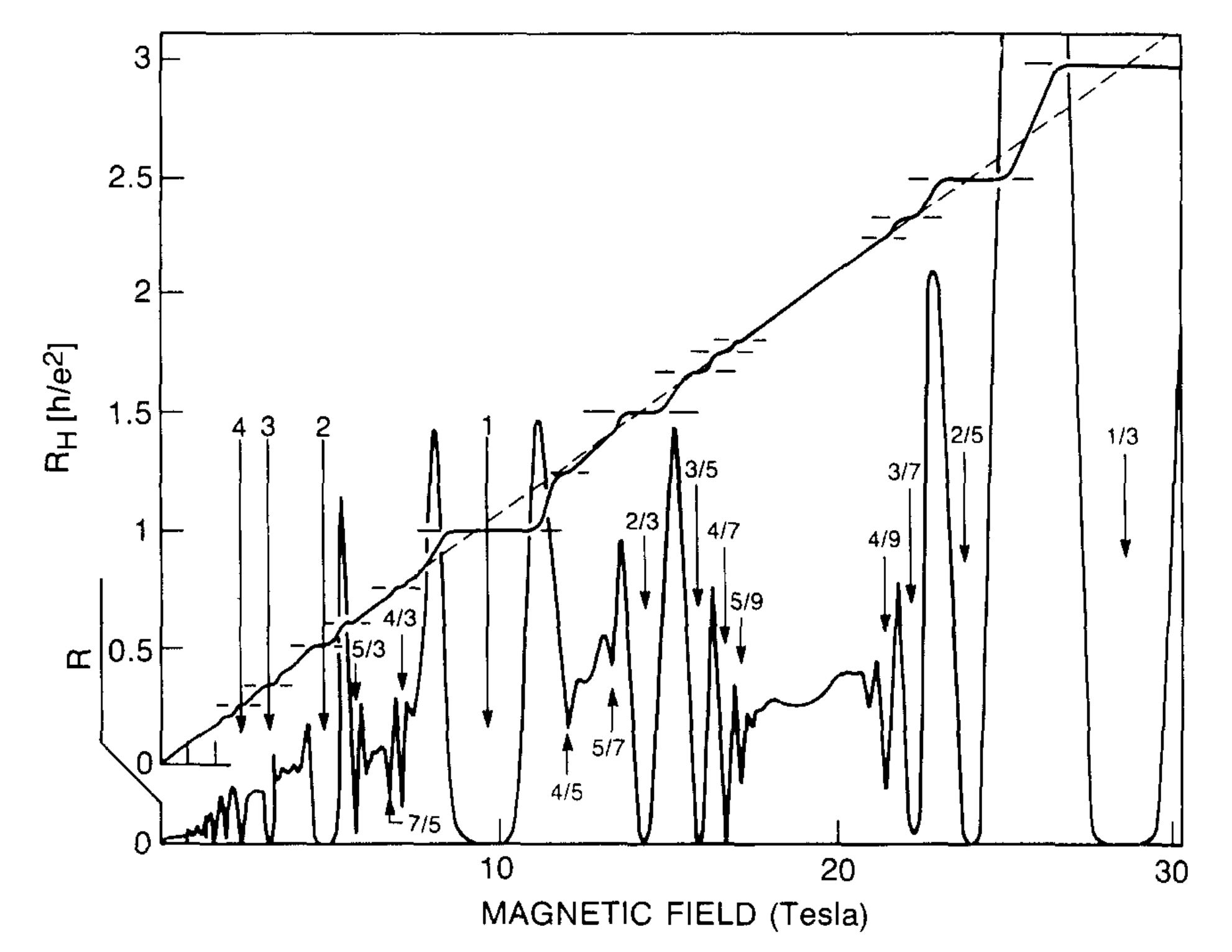}
  \end{minipage} \\
  \begin{minipage}[c]{0.4\textwidth}
    \subcaption{\label{fig:fqhe:hallprobe}}
  \end{minipage}%
  \quad
  \begin{minipage}[c]{0.5\textwidth}
    \subcaption{\label{fig:fqhe:resistivity}}
  \end{minipage}
  \caption{The Fractional Quantum Hall effect. a) Sketch of the
    experimental setup for measuring Hall effects, $B_z$ denotes the
    magnetic field, $I_x$ the applied current and $V_y$ the measured
    Hall voltage b) Measurements of the low-temperature Hall
    resistance $R_H (= R_{xy})$ and the diagonal resistivity $R$
    from~\cite{Stormer1992}. Plateaux appear at fractional values of
    the filling fraction $\nu$.}
  \label{fig:fqhe}
\end{figure}

However, there are phases of matter whose theoretical description is
beyond Landau's theory. The fractional quantum Hall effect, short
FQHE, discovered experimentally by Tsui, Stormer and Gossard in
1982~\cite{Tsui1982} is understood to exhibit a novel kind of
ordering. The FQHE is a variant of the Hall effect, where electrons
are confined to two dimensions at low temperatures and high magnetic
fields, as can be realized in \ch{GaAs}-\ch{AlGaAs} heterojunctions
\cite{Tsui1982,Stormer1983}. The Hall resistance is defined by
\begin{equation}
  \label{eq:hallresistance}
  R_{xy} = \frac{V_y}{I_x},
\end{equation}
where $V_y$ is the Hall voltage and $I_x$ the applied current. It
exhibits plateaux at certain quantized values
\begin{equation}
  \label{eq:fqhe}
  R_{xy} = \frac{1}{\nu}\frac{2\pi\hbar}{e^2},
\end{equation}
which depends on fundamental physical constants as the elementary
charge $e$ and Plank's constant $\hbar$ , cf. \cref{fig:fqhe}.
$\nu$ is called the filling fraction and sets the plateau of the Hall
resistance. Integer filling fractions $\nu = 1,2,3,\ldots$ constitute
the integer quantum Hall effect and fractional filling fractions
$\nu = \frac{1}{3},\frac{1}{5},\frac{2}{3},\ldots$ constitute the
fractional quantum Hall effect, respectively. Phase transitions
between different filling fractions are observed when tuning the
applied magnetic field. The electrons are strongly interacting via the
Coulomb force and form collective many-body quantum states with
intriguing properties in FQHE plateaux. Firstly, different FQHE
plateaux share the same symmetry properties. A phase transition
between these plateaux cannot be explained by Landau's theory of phase
transitions, instead, the novel concept of topological order was
introduced~\cite{Wen1995}. Different phases in the FQHE are understood
to have different topological orders which cannot be adiabatically
connected without observing a phase transition.

Topologically ordered phases exhibit a wide variety of exciting
phenomena. One of which is the fractionalization of quasiparticles.
Elementary excitations of FQHE states may have a charge which is a
fraction of the elementary charge of an electron. Thus, while the
microscopic constituents are electrons with exactly one elementary
charge, emergent quasiparticles may have, e.g. $1/3$ of the elementary
charge. This charge fractionalization has already been observed in
experiments \cite{Clark1988}. Another theoretically predicted property
of the quasiparticles is non-trivial statistical behavior. If two
bosons or fermions are interchanged their many-body wave function
attains a phase of $0$ or $\pi$ respectively due to
(anti-)symmetry. Emergent quasiparticles of topologically ordered
phases in two dimensions may exhibit a different behavior. The phase
for interchanging may be different from $0$ or $\pi$. In this case,
the quasiparticles are called \textit{anyons}\index{anyons}. More
generally, the process of interchanging or braiding topological
quasiparticles may act nontrivially on the space of degenerate states
with a given number of quasiparticles in which case the quasiparticles
are called non-Abelian anyons.

The discovery of novel phases of matter has in the past often given
rise to new technologies. The impact of, for example, semiconductors
or liquid crystal devices is hard to overestimate. Since topological
order is a completely new paradigm of phases of matter, naturally the
question arises what technologies such states can be used
for. Although technological breakthroughs are hard to predict, it has
been suggested that topological states of matter can be used to
implement fault-tolerant quantum computation or quantum memory
devices~\cite{Nayak2008}. Many approaches to quantum computation, like
trapped ions~\cite{Cirac1995,Haffner2008} suffer the problem that
small local disturbances to the physical system cause decoherence
since information is stored locally in those systems. The approaches
to topological quantum computation circumvent this problem by encoding
the information globally over the entire system. Changing the state of
the system with local perturbations is exponentially suppressed in the
system size, thus information can be stored robustly. The
implementation of quantum gates for performing computations can be
achieved by using the statistical properties of non-Abelian anyons.

Apart from the FQHE also different physical systems are expected to
exhibit phase transitions that are beyond Landau's theory. In the
context of high-temperature superconductivity of
cuprates~\cite{Bednorz1986} like \ch{La_2CuO_4} it was found that the
system can approximately be described by the Hubbard
model~\cite{Anderson1959}
\begin{equation}
  \label{eq:sunhubbard}
  H = - t\sum \limits_{\substack{\left< i, j\right>\\
      \alpha=\uparrow, \downarrow}}
  ( c_{i\alpha}^\dagger c_{j\alpha} +  \text{H.c.} )  + U \sum\limits_{i}  n_{i,\uparrow} n_{i,\downarrow},      
\end{equation}
where $c_{i\alpha}^\dagger$, $c_{i\alpha}$ are electron creation and
annihilation operators,
$n_{i,\alpha} = c_{i,\alpha}^\dagger c_{i,\alpha}$,
$\langle i,j \rangle$ denotes the sum over nearest neighbour pairs and
$t$ and $U$ are the hopping and on-site repulsion strength. Compounds
like \ch{La_2CuO_4} are expected to be close to half-filling, i.e. the
number of up or down spin fermions is exactly half the number of
lattice sites. In the strong coupling, or Mott-insulating limit
$t \ll U$ the effective low energy model is given by a quantum
Heisenberg model of the form
\begin{equation}
  H = J\sum\limits_{\langle i,j \rangle}\bm{S}_i\cdot\bm{S}_j,
  \label{eq:heisenbergmodellattice}
\end{equation}
The coupling constant is given by $J=t^2/U$ and
$\bm{S}_i = (S^x_i, S^y_i , S^z_i)^T$ denotes spin operators at
position $i$. Anderson proposed in 1987 that the ground state of the
Heisenberg model~\cref{eq:heisenbergmodellattice} could be a so-called
\textit{resonating valence bond}, short RVB state under certain
circumstances which would yield an explanation for high temperature
superconductivity~\cite{Anderson1987}. The RVB state is an example of
a so-called \textit{quantum spin liquid} state which is, apart from
the FQHE states, another class of states that may exhibit topological
order. We will discuss these states in more detail
in~\cref{sec:spinliquids}.

The Heisenberg model~\cref{eq:heisenbergmodellattice} and extensions
thereof have been studied in a plethora of situations. The physics of
the system~\cref{eq:heisenbergmodellattice} depends strongly on the
sign of the coupling constant $J$. For $J<0$ the model is called
\textit{ferromagnetic}\index{ferromagnetic} and it becomes
energetically favorable for neighboring spins to align in the same
direction. Similarly, the model is called \textit{antiferromagnetic}
\index{antiferromagnetic} for $J>0$ and neighboring spins are
energetically favored to align in the opposite direction.

A famous exact analytical solution by Hans Bethe~\cite{Bethe1931} is
known for the case of local spin-$1/2$ on a chain lattice. In most
cases, an analytical solution is not known. Until today an analytical
solution of the two-dimensional quantum spin-$1/2$ Heisenberg
antiferromagnet on a square lattice has not been found.  Yet,
analytical approximations like linear spin-wave theory~\cite{Huse1988}
or numerical techniques like Exact Diagonalization\cite{Oitmaa1978} or
Quantum Monte Carlo~\cite{Reger1988} have by now firmly established
that the ground state of this system is magnetically ordered. We will
discuss magnetic ordering in more detail in~\cref{sec:magneticorder}.

\begin{figure}[t!]
  \centering
  \begin{minipage}[c]{0.65\textwidth}
    \centering
    \includegraphics[width=0.9\textwidth]{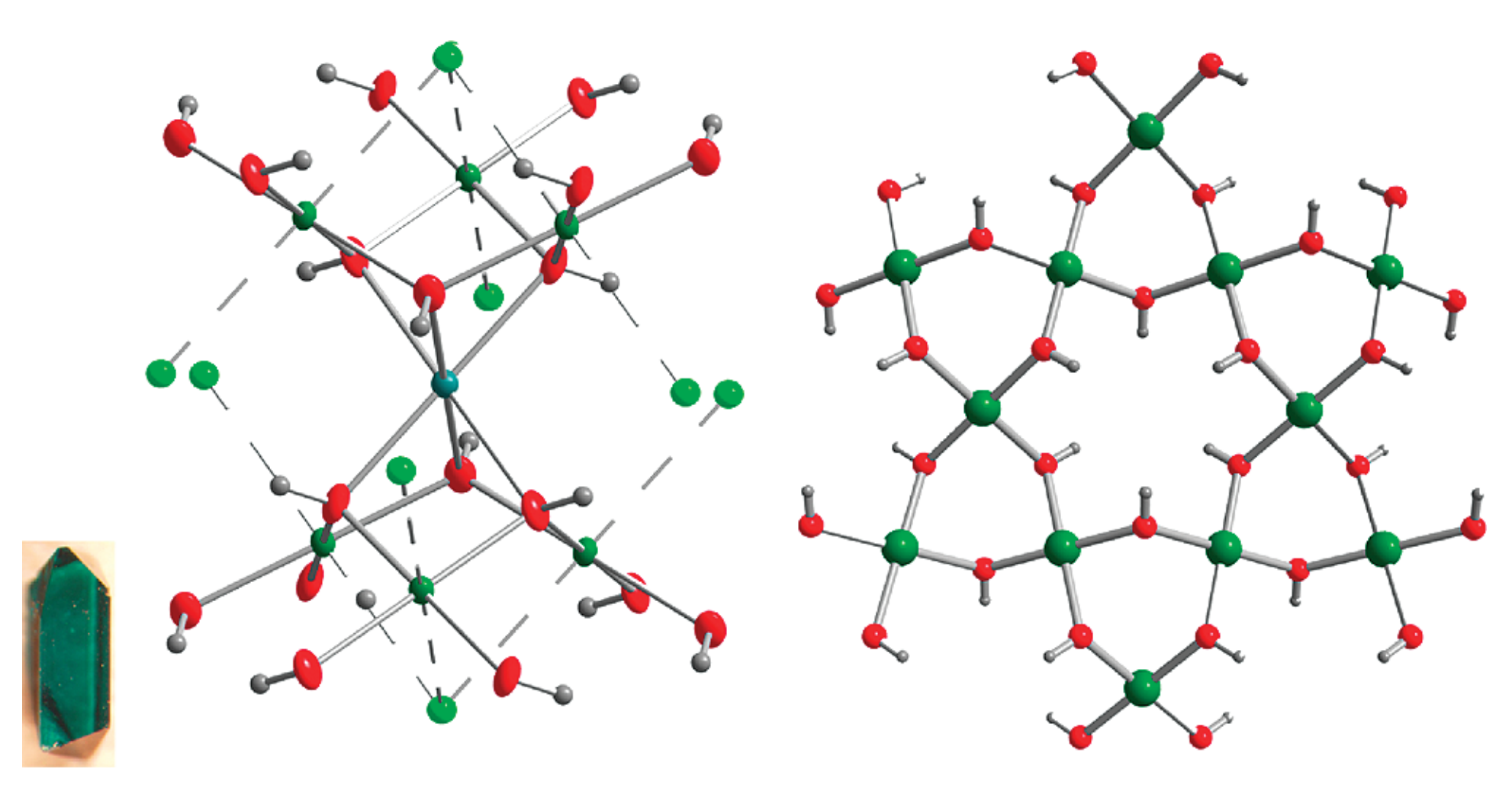}
  \end{minipage}%
  \quad
  \begin{minipage}[c]{0.25\textwidth}
    \includegraphics[width=0.9\textwidth]{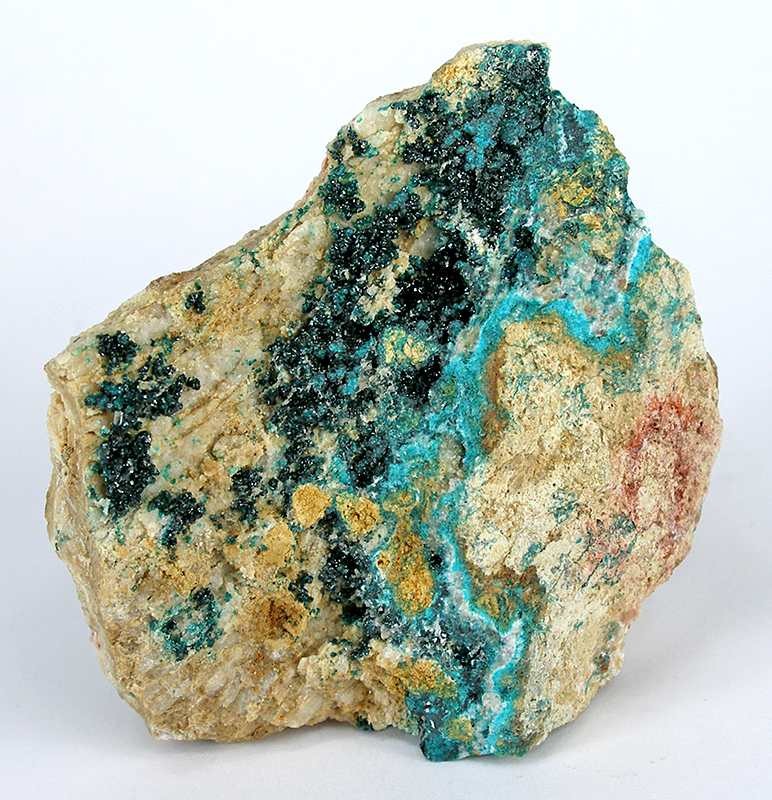}
  \end{minipage} \\
  \begin{minipage}[c]{0.65\textwidth}
    \subcaption{\label{fig:herbertsmithite:crystalstructure}}
  \end{minipage}%
  \quad
  \begin{minipage}[c]{0.25\textwidth}
    \subcaption{\label{fig:herbertsmithite:picture}}
  \end{minipage}
  \caption{Herbertsmithite \ch{ZnCu_3(OH)_6Cl_2}. a) schematic crystal
    structure. Reprinted with permission from~\cite{Freedman2010}. Copyright
    2017 American Chemical Society. Blue, green, bright green,
    red, and gray spheres represent \ch{Zn}, \ch{Cu}, \ch{Cl}, \ch{O},
    and \ch{H} atoms, respectively~\cite{Freedman2010}. The magnetic
    \ch{Cu^{2+}} ions form two-dimensional kagome planes which are
    weakly coupled amongst each other. b) Naturally ocurring
    Herbertsmithite (blue crystal) in a rock compund (from Wikimedia
    Commons, photograph by Rob Lavinsky, iRocks.com – CC-BY-SA-3.0)}
  \label{fig:herbertsmithite}
\end{figure}

In many situations, magnetic materials which are described by models
like~\cref{eq:heisenbergmodellattice} are disordered at high
temperatures and ordered at low temperatures. This need not
necessarily be the case, as certain mechanisms may prevent a system to
develop magnetic order at low temperatures or even in the ground
state. Famous materials not ordering at even lowest experimental
accessible temperatures include Herbertsmithite
\ch{ZnCu_3(OH)_6Cl_2}~\cite{Han2012} or organic Mott insulators like
% $\mathrm{EtMe_3Sb[Pd(dmit)_2]_2}$
\ch{EtMe_3Sb[Pd(dmit)_2]_2} \cite{Yamashita2010,Itou2008} or
\ch{$\kappa$-(BEDT-TTF)_2Cu_2(CN)_3} \cite{Kurosaki2005,
  Shimizu2003}. The main ingredients preventing magnetic ordering are
a low spin quantum number of the magnetic ions, reduced dimension of
the effective interactions at low temperatures, and geometrical
frustration. We will describe the effect of these mechanisms in detail
in \cref{sec:disorder}. The material Herbertsmithite
\Cref{fig:herbertsmithite}, for example, fulfills these criteria very
well: the local magnetic \ch{Cu^{2+}} ions with spin-$1/2$ are aligned
in two-dimensional layers which form a highly frustrated kagome
geometry. Indeed neutron scattering experiments show absence of
magnetic ordering at lowest experimentally attainable
temperatures~\cite{Han2012}.

The nature of disordered states in low-dimensional frustrated quantum
magnets is a highly interesting topic. In many cases, these disordered
states are expected to differ drastically from trivial paramagnetic
phases with a low degree of quantum entanglement. Indeed, some of
these states may exhibit topological order similar to the FQHE. Such
states are called \textit{Quantum Spin Liquids}\index{Quantum spin
  liquid}.  Many of them have been proposed as the ground state of
frustrated quantum antiferromagnets. Prominently, Anderson's RVB
liquid has been proposed as the ground state of the triangular lattice
spin $1/2$ Heisenberg antiferromagnet~\cite{Anderson1973}. Similarly,
so-called \textit{chiral spin liquids} have been envisioned as the
ground state for certain frustrated models~\cite{Wen1989a}. Chiral
spin liquids are a direct analog of FQHE wave functions for spin
systems~\cite{Kalmeyer1987}. Whereas, the ground state of the
triangular lattice spin-$1/2$ Heisenberg antiferromagnet is by now
understood to be magnetically ordered~\cite{Bernu1992}, the nature
ground state of the model on the kagome lattice is still highly
debated and suggested to be a quantum spin
liquid~\cite{Hastings2000,Ran2007,Hermele2008,Iqbal2013,Sachdev1992,Moessner2001,Balents2002,Misguich2002,Wang2006,Lu2011,Iqbal2011,Marston1991,Messio2012,Messio2013,Yang1993}. We
will discuss quantum spin liquids in detail in \cref{sec:spinliquids}.

There are several exactly solvable models which can be proven to
stabilize a quantum spin liquid ground state, famously Kitaev's Toric
Code~\cite{Kitaev2003a}, Kitaev's honeycomb model~\cite{Kitaev2006a},
or quantum dimer models~\cite{Rokhsar1988,Moessner2001}. Also recently
exact parent Hamiltonians for chiral spin liquids have been
found~\cite{Thomale2009,Schroeter2007,Nielsen2013}. The interactions
between the particles in those models are nevertheless often not
realistic to be realized in an experimental setup. It is thus an
important question whether simpler models might give rise to the
emergence of quantum spin liquids.

In absence of exact analytical solutions, one has to employ other
methods for studying quantum spin systems. Frustrated quantum magnets
are strongly correlated phases of matter where the interactions
between particles dominate the physics of the system. Therefore, many
standard analytical techniques like perturbation theory from a
non-interacting limit or mean-field approaches often fail to predict
the correct behavior. Numerical methods, on the other hand, have
proven as valuable tools to gain insight to strongly correlated
electron systems. Amongst the most prominent numerical methods for
studying quantum spin systems are Exact Diagonalization, Quantum Monte
Carlo, Variational Monte Carlo~\cite{Ceperley1977}, density matrix
renormalization group, short DMRG~\cite{White1992}, or tensor network
algorithms~\cite{Orus2014}. Each of the methods comes with its
advantages and disadvantages and some are particularly suited for
certain problems. In the course of this thesis the Exact
Diagonalization, Variational Monte Carlo and Quantum Monte Carlo
methods have been developed and applied. We give a short description
of these methods in \cref{sec:ed}, \cref{sec:vmc}.

%%% Local Variables:
%%% mode: latex
%%% TeX-master: "../../thesis"
%%% End:

\newpage
\section{Frustrated Magnetism}
\epigraph{ \parbox{6.4cm}{
    \flushright That's what the world is, after all: \\
     an endless battle of contrasting memories.}}{Haruki Murakami, \textit{1Q84}}
\label{sec:frustratedmagnetism}
%auto-ignore
% \section{Frustrated Magnetism}
Frustrated quantum magnets are currently studied intensely by both
experimentalists and theoreticians. On the experimental side, many
materials are studied for their intriguing properties. A detailed
survey of those and the experimental methods for investigating their
physical properties can be found in Ref.~\cite{Lacroix2011}. Here we
give a short introduction to the theoretical aspects. We describe how
the principle of spontaneous symmetry breaking manifests itself in
quantum magnets as magnetic order in
\cref{sec:magneticorder}. Interesting phenomena such as the emergence
of quantum spin liquids may occur when the system is disordered even
at lowest temperatures. There are several mechanisms preventing the
system from ordering when cooled down which we explain in
\cref{sec:disorder}. Often, these mechanisms lead to a massive
degeneracy of ground states. The order-by-disorder principle discussed
in \cref{sec:orderbydisorder} may favor certain submanifold of these
states due to thermal or quantum fluctuations and therefore give rise
to unexpected emergent states of matter.

%%% Local Variables:
%%% mode: latex
%%% TeX-master: "../../thesis"
%%% End:

%auto-ignore
\subsection{Magnetic order}
\label{sec:magneticorder}
Patterns of magnetic ordering are energetically favorable in many spin
systems. Consider the ferromagnetic Ising model
\begin{equation}
  \label{eq:isingferro}
  H_{\text{Ising}} = J \sum\limits_{\langle i,j\rangle}\sigma_{i}\cdot\sigma_{j}
  + h \sum\limits_i \sigma_i,
\end{equation}
where $\sigma_{i} = \pm 1$ are classical Ising spins with $J<0$ and
$h$ denotes an external magnetic field. For $h=0$ the ferromagnetic
states $\ket{\uparrow\cdots\uparrow}$ and
$\ket{\downarrow\cdots\downarrow}$ with all spins aligned in the same
direction are the ground states. The system in the ground state at
$h=0$ is magnetically ordered. Thermal fluctuations favor entropy and
the system becomes a disordered paramagnetic state at high
temperatures.

Magnetic order is an instance of spontaneous symmetry breaking. The
Ising model \cref{eq:isingferro} at $h=0$ is invariant under global
spin flips, $\sigma_i \rightarrow -\sigma_i$, thus possesses a
discrete $\mathbb{Z}_2$ symmetry. In two or more dimensions the model
exhibits an order-to-disorder transition at finite
temperature~\cite{Onsager1944}. In the ordered phase this symmetry is
not respected by the equilibrium ground states and the magnetization
order parameter
\begin{equation}
  \label{eq:isingmag}
  \langle m \rangle= \frac{1}{N}\sum\limits_j \langle \sigma_{j} \rangle, 
\end{equation}
attains a finite value in the thermodynamic limit in the sense that
\begin{equation}
  \label{eq:spontaneousferromagnetization}
  \lim\limits_{h\rightarrow 0}\lim\limits_{N\rightarrow \infty}
  \left< m\right> \neq 0.
\end{equation}
Here, $\langle \ldots \rangle$ denotes a classical ensemble average
and $N$ the number of lattice sites. The degeneracy of the ground
state is twofold, thus discrete in the thermodynamic limit.

Another fundamental model describing magnetic materials is the quantum
Heisenberg model
\begin{equation}
  \label{eq:heisenberggeneric}
  H = \sum\limits_{i,j}J_{ij}\bm{S}_i\cdot\bm{S}_j +
  h \sum\limits_j e^{-i\bm{Q}\cdot\bm{r}_j} S^z_{j} + \text{H.c.},
\end{equation}
where $\bm{S}_i = (S^x_i, S^y_i , S^z_i)^T$ are quantum mechanical
SU($2$) spin operators and $J_{ij}$ are coupling constants. $\bm{r}_j$
denotes the position of the $i$-th spin and $\bm{Q}$ defines an
ordering vector. The Heisenberg model \cref{eq:heisenberggeneric} is
invariant under continuous global SU($2$) spin rotations for vanishing
external field $h=0$, i.e.
\begin{equation}
  \label{eq:spinrotation}
  \left[H,\; \sum\limits_i S_i^\alpha\right] = 0 \quad \text{at} \quad h=0,
\end{equation}
for $\alpha = x,y,z$. The ordered magnetization with ordering vector
$\bm{Q}$ is defined by
\begin{equation}
  \label{eq:orderedmagnetization}
  \langle m^{\bm{Q}} \rangle = \frac{1}{N}\sum\limits_j e^{-i\bm{Q}\cdot\bm{r}_j} \langle S^z_{j} \rangle, 
\end{equation}
where
$\langle \ldots \rangle = \frac{1}{\mathcal{Z}}\tr(e^{-\beta H}
\ldots)$ denotes the quantum statistical average at inverse
temperature $\beta$ and partition function
$\mathcal{Z} =\tr(e^{-\beta H})$. The ordered magnetization
$\langle m^{\bm{Q}} \rangle$ is not invariant under the SU($2$)
symmetry group of spin rotations and can, therefore, be used as an
order parameter to detect spontaneous symmetry breaking,
\begin{equation}
  \label{eq:spontaneousmagnetization2}
  \lim\limits_{h\rightarrow 0}\lim\limits_{N\rightarrow \infty}
  \langle m^{\bm{Q}}\rangle \neq 0.
\end{equation}
The phenomenology of spontaneous symmetry breaking of a continuous
symmetry is more diverse than in the discrete case. In the
thermodynamic limit, the ground state becomes infinitely degenerate
due to the continuous nature of the symmetry
group~\cite{Anderson1952,Azaria1993}.

There are certain characteristic phenomena when approaching the
thermodynamic limit in systems breaking a continuous symmetry.
Ferromagnetic Heisenberg models with $J_{ij} < 0$ for all $i$ and $j$
always admit the fully polarized ground state
\begin{equation}
  \label{eq:ferrogroundstate}
  \ket{\text{ferro}} = \ket{\uparrow\uparrow\ldots \uparrow},
\end{equation}
with all spins aligned in one direction. Yet, the ground state is at
least $(2SN + 1)$ degenerate where $S$ denotes the local value of the
spin. This is because the multiplet with total spin $SN(SN + 1)$ can
be generated by rotating the fully polarized state in
\cref{eq:ferrogroundstate}.

\begin{figure}[ht!]
  \centering
  \includegraphics[width=0.7\textwidth]{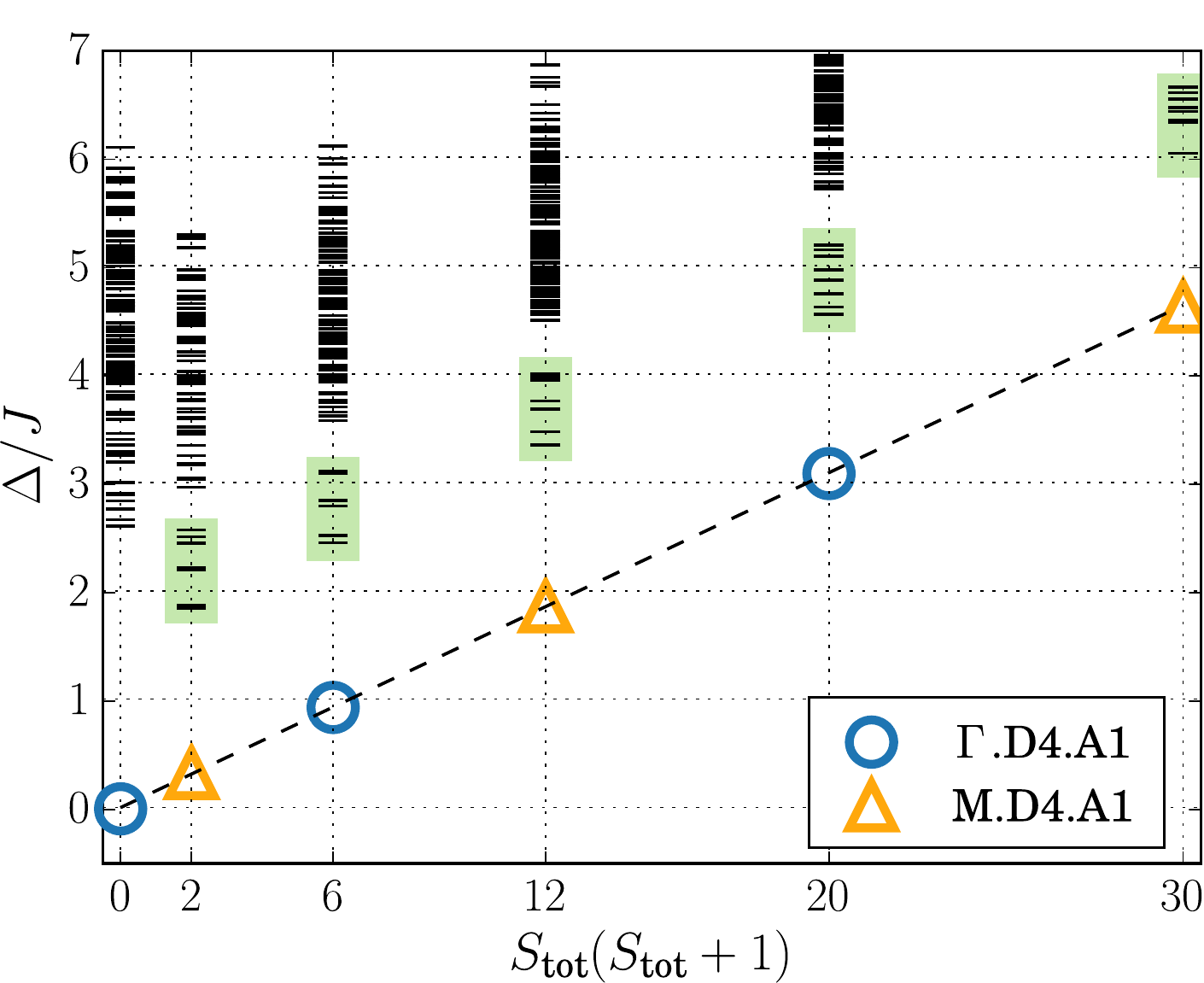}
  \caption{Many-body energy spectrum of the 32 site spin-$1/2$
    Heisenberg antiferromagnet on a square lattice geometry. States
    are differentiated by their total spin quantum number
    $S(S+1)$. The quasimomentum of the states in the Anderson tower is
    given by $\Gamma$ and $M$ with $D4$ point group representation
    $A1$. For details on the representation theory and quantum numbers
    of space groups see \cref{sec:spacegroupreptheory}. The levels
    with the green background are the magnon (Goldstone mode) levels
    and exhibit a linear dispersion relation.}
  \label{fig:andersontowersquare32}
\end{figure}

The situation is quite different for antiferromagnetic models
$J_{ij} > 0$. The simple analog of the fully polarized state in
\cref{eq:ferrogroundstate} would be the classical \textit{N{\'e}el
  state}\index{N{\'e}el state} \cite{Neel1948}
\begin{equation}
  \label{eq:neelstate}
  \ket{\text{N{\'e}el}} = \ket{\uparrow\downarrow\uparrow\downarrow\ldots}.
\end{equation}
Even for an antiferromagnetic Ising model this state is only the
ground state if the lattice is bipartite and antiferromagnetic
interactions only connect sites between the two disjoint
sublattices. This is not the case for frustrated geometries,
cf. \cref{sec:disorder}. Moreover, this state is not an eigenstate of
the antiferromagnetic Heisenberg model. For a two-site spin-$1/2$
Heisenberg interaction the eigenstates and eigenvalues are given by
\begin{alignat}{5}
  \label{eq:twositeheisenbergeigenstates}
  &\bm{S}_1\cdot\bm{S}_2\ket{S=0, m=0} & &= -\frac{3}{4}\ket{S=0, m=0}
  & &=
  -\frac{3}{4}(\ket{\uparrow\downarrow} - \ket{\downarrow\uparrow}) \\
  &\bm{S}_1\cdot\bm{S}_2
  \begin{cases}
    \ket{S=1, m=1} \\
    \ket{S=1, m=0} \\
    \ket{S=1, m=-1}
  \end{cases}
  & &= +\frac{1}{4}
  \begin{cases}
    \ket{S=1, m=1} \\
    \ket{S=1, m=0} \\
    \ket{S=1, m=-1}
  \end{cases}
  & &= +\frac{1}{4}
  \begin{cases}
    \ket{\uparrow\uparrow} \\
    \ket{\uparrow\downarrow} + \ket{\downarrow\uparrow} \\
    \ket{\downarrow\downarrow}
  \end{cases}
\end{alignat}
The ground state with energy $-\frac{3}{4}$ is a rotationally
invariant spin singlet. The fact that the ground state has total spin
zero is not only true for the two site interaction but for generic
bipartite antiferromagnetic Heisenberg models on a finite
lattice. This is known as Marshall's theorem and has been
mathematically proven in Refs.~\cite{Marshall1955a,Lieb1962a}. This
result only holds for finite size geometries. In order to break the
SU($2$) symmetry, a set of higher spin-$S$ states approaches the
ground state energy. This set of states is called the \textit{Anderson
  tower of states}\index{tower of states} and is characteristic for
spontaneous breaking of a continuous symmetry. The excitation energy
of these states is proportional to $S(S+1)/N$, thus collapses linearly
to the ground state energy for $N \rightarrow \infty$. A typical
finite size spectrum obtained from Exact Diagonalization of the
Heisenberg spin-$1/2$ square lattice antiferromagnet on $32$ lattice
sites is shown in \cref{fig:andersontowersquare32}.
Quantum numbers such as total spin, quasimomentum and point group
representations of the states occurring in the tower of states can
actually be predicted for a given magnetic order and give a strong
evidence that a certain order is realized by the system. The method of
extracting these quantum numbers is explained in
\cref{sec:towerofstates} and more detailed in Ref.~\cite{Wietek2017}.

% Spin correlation functions between two spins at positions $\bm{r}_i$
% and $\bm{r}_j$ are defined by
% \begin{equation}
%   \label{eq:spincorrelations}
%   C(\bm{r}_i,\bm{r}_j) = \langle \bm{S}_i\cdot\bm{S}_j\rangle
% \end{equation}
Spontaneous symmetry breaking in the sense of
\cref{eq:spontaneousmagnetization2} implies \textit{long-range
  order}\index{long-range order} of the spin correlation functions
\begin{equation}
  \label{eq:longrangeorder}
  \lim\limits_{|\bm{r}_i - \bm{r}_j|\rightarrow \infty}\langle \bm{S}_i\cdot\bm{S}_j\rangle  \neq 0.
\end{equation}
For translationally invariant systems, the static magnetic structure
factor
\begin{equation}
  \label{eq:structurefactor}
  \mathcal{S}(\bm{k}) = \sum_{j}
  e^{i\bm{k}\cdot(\bm{r}_j-\bm{r}_0)}\langle \bm{S}_j\cdot\bm{S}_0 \rangle,
\end{equation}
is the Fourier-transform of the spin correlation function. Long-range
order implies the divergence of the structure factor at an ordering
wave vector $\bm{Q}$.

\begin{figure}[ht!]
  \centering
  \begin{minipage}[c]{0.4\textwidth}
    \centering \includegraphics[height=5.5cm]{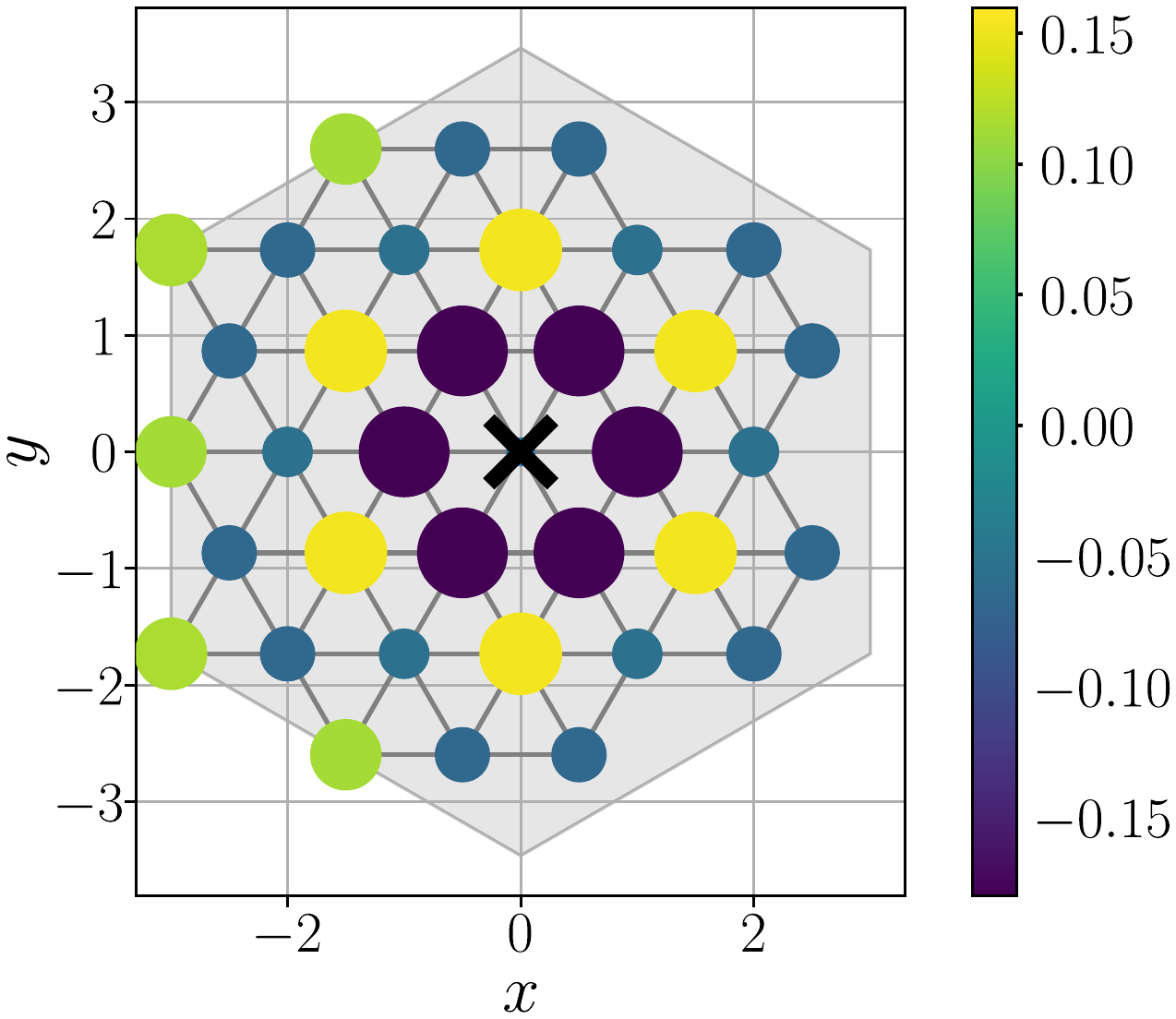}
  \end{minipage}%
  \quad
  \begin{minipage}[c]{0.5\textwidth}
    \includegraphics[height=5.5cm]{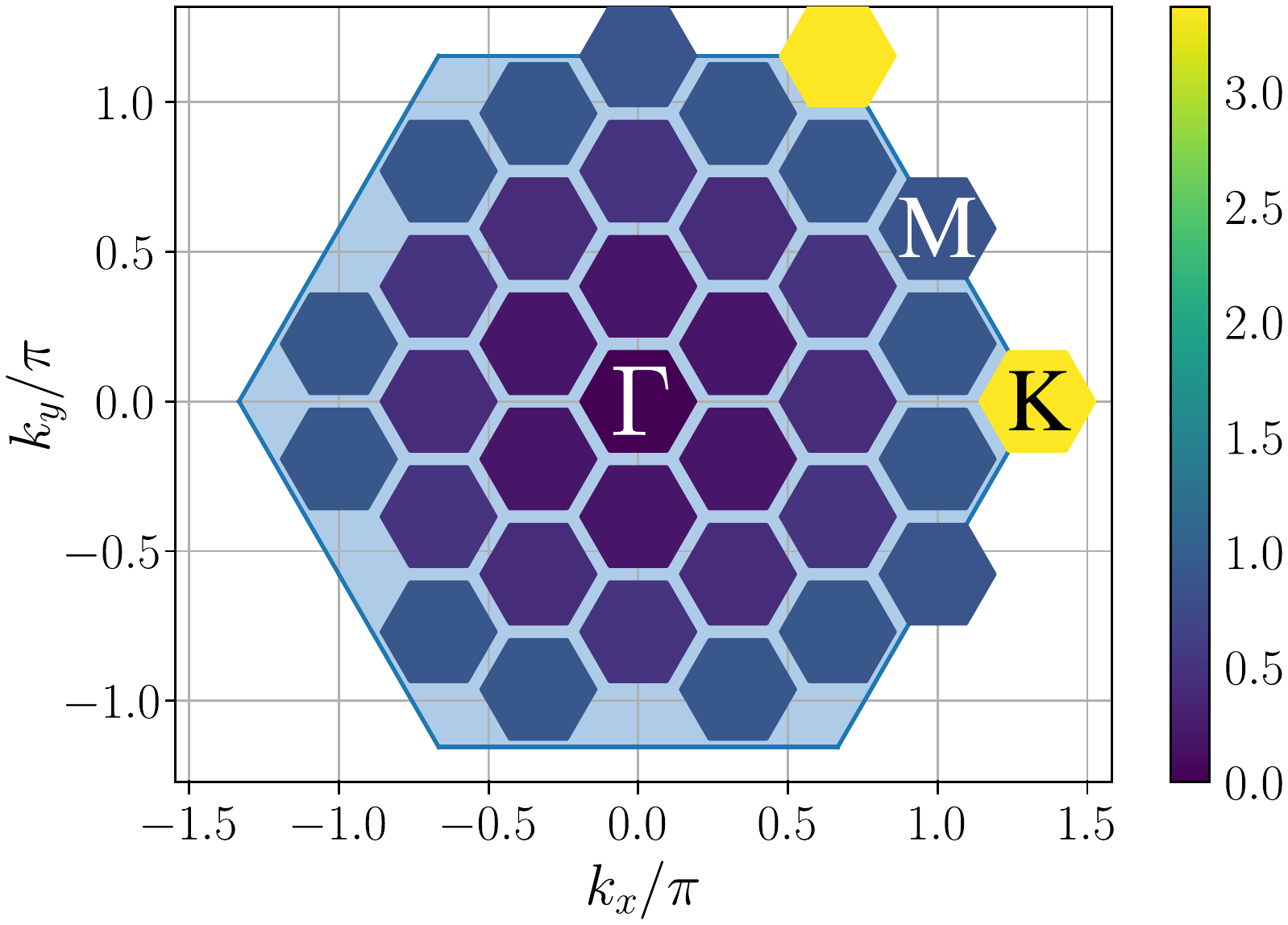}
  \end{minipage}
  \\
  \begin{minipage}[t]{0.44\linewidth}
    \subcaption{\label{fig:sdotscorr}}
  \end{minipage}
  \quad
  \begin{minipage}[t]{0.5\linewidth}
    \subcaption{\label{fig:structurefactor}}
  \end{minipage}

  \caption{Spin correlations $\langle \bm{S}_i\cdot\bm{S}_0\rangle$
    (a) and structure factor $\mathcal{S}(\bm{k})$ (b) for the ground
    state of the antiferromagnetic spin-$1/2$ Heisenberg nearest
    neighbour model on a 36 site triangular lattice. The black cross
    in (a) marks the spin $\bm{S}_0$. The ground state is $120^\circ$
    N{\'e}el ordered~\cite{Bernu1992} and shows a peak in the
    structure factor at the $K$ point in the Brillouin zone.}
  \label{fig:spincorrstruc}
\end{figure}

Every pattern of magnetic ordering on regular lattice has one or
several characteristic ordering vectors $\bm{Q}$. Ferromagnetic order
is peaked at $\bm{Q} = \Gamma \equiv \mathbf{0}$ whereas a square lattice
N{\'e}el antiferromagnetic order is peaked at
$\bm{Q} = M \equiv (\pi, \pi)$. The ordering vector $\bm{Q}$ does not
uniquely define an ordering pattern. Thus, there can be multiple
ordering patterns with the same ordering
vector. Ref.~\cite{Messio2011} gives a classification of possible
patterns of magnetic orderings and presents their prominent features,
including the peaks of the structure factor in reciprocal
space. \cref{fig:spincorrstruc} shows the spin correlations of the
$120^\circ$ N{\'e}el ordered ground state of the triangular lattice
spin-$1/2$ Heisenberg antiferromagnet on a $36$ site lattice and its
corresponding static spin structure factor in reciprocal space.

A hallmark feature in systems breaking a continuous symmetry is the
occurrence of gapless low-energy modes. These modes are called
Goldstone modes \cite{Goldstone1961, Goldstone1962} and are a general
phenomenon occurring in many branches of physics. For spin systems,
they are called \textit{magnons}\index{magnon} or \textit{spin
  waves}\index{spin wave}. The Goldstone theorem states that these
gapless modes necessarily need to exist once a continuous symmetry is
spontaneously broken. We state the theorem in the form presented in
Ref.~\cite{Auerbach} where also an elementary proof is given.

\begin{thm*}[Goldstone's theorem]
  Consider the generic Heisenberg Hamiltonian
  \begin{equation}
    \label{eq:heisenberggeneric}
    H = \sum \limits_{ij} J_{ij} \bm{S}_i \cdot \bm{S}_j,
  \end{equation}
  satisfying the following locality condition
  \begin{equation}
    \label{eq:goldstonelocalitycondition}
    \frac{1}{N}| J_{ij} | |\bm{r}_i - \bm{r}_j | < \infty.
  \end{equation}
  If the spin structure factor \cref{eq:structurefactor} diverges at
  some finite wave vector $\bm{Q}$
  \begin{equation}
    \label{eq:structurefactordivergence}
    \lim\limits_{\bm{k} \rightarrow \bm{Q}} \mathcal{S}(\bm{k}) \rightarrow \infty,
  \end{equation}
  then there exists an eigenstate with momentum $\bm{k}$, whose energy
  $E(\bm{k})$ vanishes at $\bm{Q}$
  \begin{equation}
    \label{eq:goldstoonevanishenergy}
    \lim\limits_{\bm{k} \rightarrow \bm{Q}} E(\bm{k}) = 0. 
  \end{equation}
\end{thm*}
The magnon modes can already be detected in finite size spectra,
cf. \cref{fig:andersontowersquare32}. Their behavior can be
approximated by linear spin wave theory (see
e.g. Refs.~\cite{Anderson1952,Kubo1952,Toth2015a}) and allow for the
prediction of many physical properties of magnetically ordered states.

%%% Local Variables:
%%% mode: latex
%%% TeX-master: "../../thesis"
%%% End:

\newpage
\subsection{Mechanisms of Disorder}
\label{sec:disorder}
%auto-ignore

% \subsubsection{Temperature}
The equilibrium behavior of a physical system in statistical mechanics
system is determined by its free energy functional 
\begin{equation}
  \label{eq:freeenergy_tmp}
  F = E - TS,
\end{equation}
where $E$ is the internal energy, $T$ the temperature and
$S$ the entropy of the system. In equilibrium, the state minimizing
the free energy is realized. At higher temperatures, states with
higher entropy are favored, at lower temperatures states with lower
energy are favored. Ordered states are often minimizing energy
constraints and are consequently often realized in a ground
state. Temperature tends to introduce disorder to the system. In the
infinite temperature limit, the equilibrium state becomes 
equidistributed in phase space. For spin systems, this state is a
featureless paramagnetic state.

Temperature is not the only physical mechanism that induces
fluctuations in spin systems. There are several other mechanisms that
increase the entropy of states even at low or zero temperature. These
mechanisms may give rise to the emergence of fascinating new phenomena
such as quantum spin liquids. We will now discuss the most important
disorder mechanisms for frustrated quantum spin systems.

%%% Local Variables:
%%% mode: latex
%%% TeX-master: "../../thesis"
%%% End:

%auto-ignore
\subsubsection{Quantum fluctuations}
\label{sec:quantumfluctuations}

Disorder can be introduced by going from classical spin models to
quantum mechanical spin models. Such a transition can be observed in
the spin-$1/2$ anisotropic Heisenberg model
\begin{equation}
  \label{eq:anisotropicheisenberg}
  H =  J_z\sum\limits_{\langle i, j \rangle} S^z_iS^z_j + \frac{J_{xy}}{2}\sum\limits_{\langle i, j \rangle}S_i^+S_j^- + S_i^-S_j^+.   
\end{equation}
Consider a classical N{\'e}el state
\begin{equation}
  \label{eq:neelstate2}
  \ket{\text{N{\'e}el}} = \ket{\uparrow\downarrow\uparrow\downarrow\ldots}.
\end{equation}
In the classical Ising limit $J_{xy}= 0$ on a bipartite lattice this
state is the ground state of \cref{eq:anisotropicheisenberg} for
$J_z > 0$. When turning on the quantum mechanical exchange interaction
for $J_{xy} > 0$ this state is not an eigenstate anymore, since the
exchange term introduces spin flips like
\begin{equation}
  \label{eq:exchangespinflip}
  (S_i^+S_j^- + S_i^-S_j^+)\ket{\ldots\uparrow_i\ldots\downarrow_j\ldots}
  = \ket{\ldots\downarrow_i\ldots\uparrow_j\ldots}.
\end{equation}
The true ground state in \cref{eq:anisotropicheisenberg} must, therefore,
be a superposition of multiple spin configurations with locally
fluctuating spins.

This effect is most pronounced in the $S=1/2$ case. An exchange term
turns a fully polarized state $\ket{\uparrow\downarrow}$ to the
oppositely polarized state $\ket{\downarrow\uparrow}$. For higher spin,
the effect of the exchange term is less pronounced since applying the
off-diagonal term to a fully polarized state
\begin{equation}
  \label{eq:exchangespinfliphighers}
  (S_i^+S_j^- + S_i^-S_j^+)\ket{\ldots(S)_i\ldots(-S)_j\ldots}
  \propto \ket{\ldots(S-1)_i\ldots(-S+1)_j\ldots},
\end{equation}
does not fully invert the local polarization. Processes reverting the
local polarization only occur at $2S$-th order in perturbation theory.
Hence, higher spin $S$ yields less fluctuations in the local spin
polarization. Indeed, the large-$S$ limit of the quantum Heisenberg
model can be proven to correspond to the classical Heisenberg model in
a mathematically rigorous way~\cite{Lieb1973}.

For the square anisotropic spin-$1/2$ Heisenberg model in the
classical Ising limit $J_{xy}= 0$ we know that the model exhibits
long-range order at low but finite temperatures due to Onsager's
solution \cite{Onsager1944}. Yet, in the isotropic Heisenberg case,
the system is disordered at any finite temperature
\cite{Mermin1966}. So perturbing a classical system with terms
introducing quantum fluctuations may result in an order-to-disorder
transition. In one dimension the Ising model at exactly $T=0$ is
long-range ordered whereas the Heisenberg case exhibits algebraically
decaying correlation functions according to the exact Bethe ansatz
solution~\cite{Bethe1931}.

%%% Local Variables:
%%% mode: latex
%%% TeX-master: "../../thesis"
%%% End:

%auto-ignore
\subsubsection{Geometric Frustration}
\label{sec:frustration}

\begin{figure}[t!]
  \centering
  \begin{minipage}[c]{0.3\textwidth}
    \centering
    \includegraphics[width=0.7\textwidth]{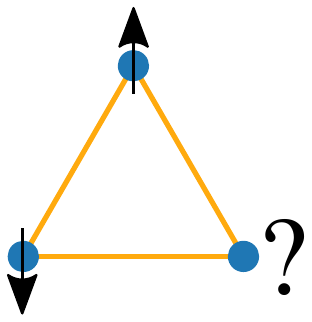}
  \end{minipage}%
  \quad
  \begin{minipage}[c]{0.2\textwidth}
    \includegraphics[width=\textwidth]{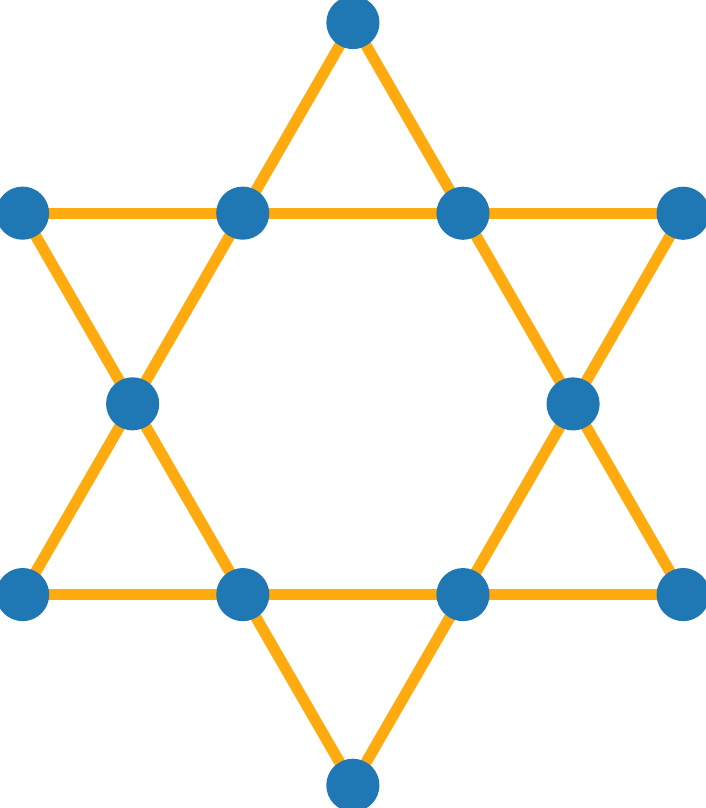}
  \end{minipage}
  \quad
  \begin{minipage}[c]{0.3\textwidth}
    \quad \includegraphics[width=0.9\textwidth]{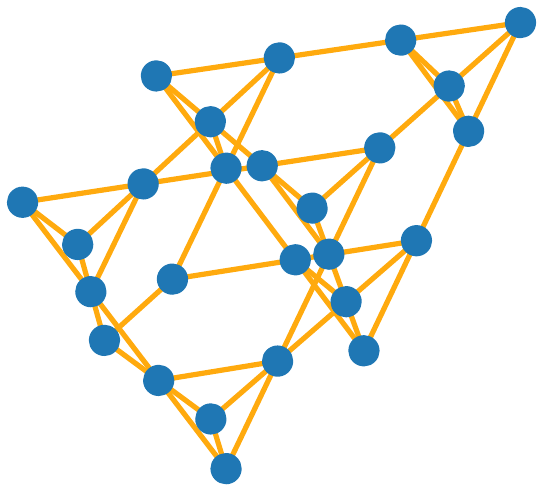}
  \end{minipage}
  \\
  \begin{minipage}[t]{0.3\linewidth}
    \subcaption{\label{fig:frustration:triangle}}
  \end{minipage}
  \quad
  \begin{minipage}[t]{0.2\linewidth}
    \subcaption{\label{fig:frustration:kagome}}
  \end{minipage}
  \quad
  \begin{minipage}[t]{0.3\linewidth}
    \subcaption{\label{fig:frustration:pyrochlore}}
  \end{minipage}

  \caption{Geometrically frustrated lattice geometries. The triangle
    (a) shows the basic dilemma leading to frustration in
    antiferromagnetic models. Local interactions cannot be
    simultaneously be minimized. The kagome (b) and the pyrochlore
    (c) lattice both give rise to high degrees of frustration.}
  \label{fig:frustration}
\end{figure}

\begin{figure}[t!]
  \centering \includegraphics[width=0.4\textwidth]{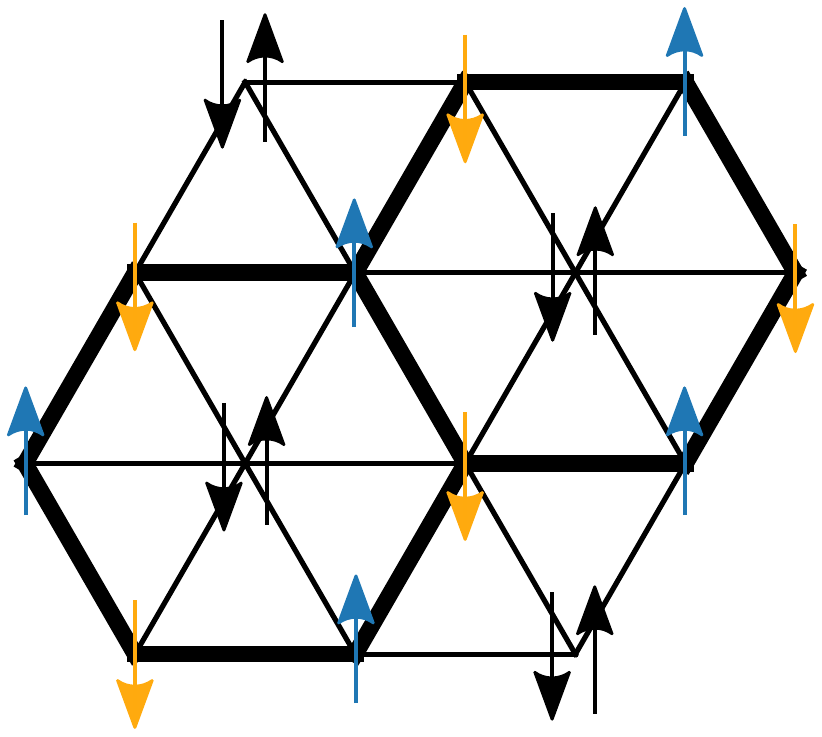}
  \caption{A subset of ground states of the triangular lattice Ising
    antiferromagnet. The direction of the spins in the middle of the
    hexagons can be freely chosen without altering the energy of the
    state. Redrawn freely from \cite{Mila2015}.}
  \label{fig:triangularisingdegenerate}
\end{figure}

As in real life, frustration occurs in physical systems if too many
constraints cannot be simultaneously satisfied. The most basic example
in physics is the antiferromagnetic Ising model
\begin{equation}
  \label{eq:isingmodel}
  H = J \sum\limits_{\left< i,j\right>}\sigma_i\cdot \sigma_j \text{, }\quad J > 0,\quad \sigma_i = \pm 1,
\end{equation}
on the triangular lattice. \cref{fig:frustration:triangle} illustrates
the dilemma.
Out of the $2^3$ possible spin configurations on a single triangle,
six states are of energy $-J$ and two are of energy $3J$. The ground
state is thus sixfold degenerate. This is a typical effect if not all
local energy constraints can be simultaneously minimized. For spin
systems, triangular geometries may give rise to frustration. Besides
the triangular lattice geometry, prominent lattice geometries with a
high degree of geometric frustration are the two-dimensional
kagome and the three-dimensional pyrochlore lattice shown in
\cref{fig:frustration}. The ground state degeneracy can even become an
extensive thermodynamic quantity. On an extended triangular lattice
consider a state where on a hexagonal sublattice all energy
constraints are minimized as in
\cref{fig:triangularisingdegenerate}. Then the direction of each third
spin in the middle of the hexagon can be freely chosen. This gives
rise to a degeneracy of $D= 2^{N/3}$ states, where $N$ is the number
of sites. A measure of the ground state degeneracy is given by the
entropy at zero temperature $S_0$, the \textit{residual
  entropy}\index{residual entropy}. Since the states in
\cref{fig:triangularisingdegenerate} are a subset of all degenerate
ground states we have a simple estimation for the residual entropy of
the triangular lattice Ising antiferromagnet,
\begin{equation}
  \label{eq:triangularisingresidualestimation}
  S_0/ N > \frac{1}{3}\log{2} \approx 0.231049.
\end{equation}
The ground state degeneracy thus becomes an extensive thermodynamic
quantity. We see that geometric frustration induces large fluctuations
even at zero temperature. Similar counting arguments have also been
performed for Heisenberg models \cite{Moessner1998}.

For the square lattice Ising model, an exact analytical solution was
found by Onsager \cite{Onsager1944}. Following his work, several other
authors found solutions to Ising models on the honeycomb, triangular
and kagome lattice
\cite{Wannier1950,Houtappel1950,Syozi1951,Kano1953}. All unfrustrated
cases, i.e. the (anti-)ferromagnetic square and honeycomb geometry and
the ferromagnetic triangular and kagome geometries have been
proven to exhibit an order-disorder transition at a finite critical
temperature $T_c > 0$. The frustrated cases, on the other hand, do not
order even at zero temperature. The residual entropies have been
computed exactly. For the triangular lattice \cite{Wannier1950} is
given by
\begin{equation}
  \label{eq:triangularisingresidual}
  S_0/ N = \frac{2}{\pi}\int\limits_{0}^{\pi/3}\log(2\cos(\omega))
  d\omega \approx 0.3383 ,
\end{equation}
and for the kagome lattice \cite{Kano1953} by
\begin{equation}
  \label{eq:kagomeisingresidual}
  S_0/ N = \frac{1}{24\pi^2} \int\limits_{0}^{2\pi}
  \int\limits_{0}^{2\pi}\log\left[21 - 4(\cos(\omega_1) + \cos(\omega_2)
    + \cos(\omega_1 + \omega_2))\right] d\omega_1 d\omega_2
  \approx 0.50183.
\end{equation}
The kagome lattice may, therefore, be regarded as being more
frustrated than the triangular lattice geometry.

%%% Local Variables:
%%% mode: latex
%%% TeX-master: "../../thesis"
%%% End:

%auto-ignore
\subsubsection{Dimensionality}
\label{sec:dimensionality}

Lower dimensionality may introduce additional fluctuations to a
system. Systems exhibiting a global continuous symmetry like SU($2$)
spin rotational symmetry in Heisenberg models
\cref{eq:heisenberggeneric} become disordered at dimensions
$d \leq 2$. A nice heuristic scaling argument for this has been given
by John Cardy \cite{Cardy1996}. Consider an ordered state like a
ferromagnet with a domain of linear length $l$ where the local
magnetic moment is rotated $180^\circ$ in the center with respect to
the ferromagnetic state. Since we have continuous degrees of freedom
we can perform this rotation gradually. The energy cost between two
neighboring spins is of the order $\mathcal{O}(l^{-2})$. In $d$
dimensions, a domain of length $l$ has $\mathcal{O}(l^{d})$ local
interactions. Thus the energy cost of such a domain is of the order
$\mathcal{O}(l^{d-2})$. On the other hand allowing for domain walls
increases the entropy. For minimizing the free energy,
\begin{equation}
  \label{eq:freeenergy_dim}
  F = E - TS,
\end{equation}
it is therefore favorable to have states with several domain walls
which yield a higher entropy in $d \leq 2$, whereas such states become
energetically unfavorable in $d > 2$.

This heuristic argument can be made very precise in the sense of a
rigorous mathematical proof for several systems. First works
\cite{Bogoliubov1962,Mermin1966,Hohenberg1967} showed the absence of
long-range order in bosonic superfluids, fermionic superconductors,
and spin systems in $d \leq 2$. For Heisenberg spin systems with mild
locality assumptions on the range of interactions, the precise
statement is given by the Mermin-Wagner theorem \cite{Mermin1966}. We
state the theorem from Ref.~\cite{Auerbach}.
\begin{thm*}[Mermin-Wagner theorem]
  For the quantum Heisenberg model
  \begin{equation}
    \label{eq:merminwagnerheisenberg}
    H = \sum\limits_{i,j}J_{ij}\bm{S}_i\cdot\bm{S}_j 
    - h\sum\limits_j e^{-i\bm{Q}\cdot\bm{r}_j} S^z_{j} + \text{H.c.},
  \end{equation}
  satisfying the locality condition
  \cref{eq:goldstonelocalitycondition} there can be no spontaneous
  symmetry breaking in one dimension for $T\geq 0$ or in two
  dimensions at finite temperature $T>0$,
  \begin{equation}
    \label{eq:merminwagnersymbreak}
    \lim\limits_{h\rightarrow 0}\lim\limits_{N\rightarrow \infty}
    \langle m^{\bm{Q}}\rangle = 0.
  \end{equation}

\end{thm*}
The theorem has been generalized to various different systems. A
review on these generalizations is given by Ref.~\cite{Gelfert2001}.

From a field theoretical perspective, spontaneous symmetry breaking of
a continuous symmetry implies the emergence of massless Goldstone
bosons. It has been shown \cite{Coleman1973a} that in two space time
dimensions the correlation functions of these fields are infrared
divergent and thus the theory is ill-defined. This can be seen as the
generic underlying principle that forbids the emergence of long-range
order in lower dimensions.

The Mermin-Wagner theorem does not make statements about the
two-dimensional case at zero temperature and the three-dimensional
case. The question about the behavior of the two-dimensional
spin-$1/2$ Heisenberg antiferromagnet on a square lattice at exactly
zero temperature thus remains unanswered by the Mermin-Wagner
theorem. Several analytical and numerical studies have investigated
this question. A first numerical Exact Diagonalization study
\cite{Oitmaa1978} on lattices up to $18$ sites indicated that the
ground state is indeed long-range ordered. This has been supported by
spin-wave calculations \cite{Huse1988} who predicted the behavior of
the spin correlation function
\begin{equation}
  \label{eq:heisenbergsquarecorr}
  \left | \langle \bm{S}_i\cdot \bm{S}_j \rangle \right| - (\tilde{m})^2\sim\frac{1}{|\bm{r}_i - \bm{r}_j|},
\end{equation}
where $\tilde{m} \equiv \langle m^{(\pi,\pi)} \rangle$ is the
staggered magnetization with wave vector $\bm{Q} = (\pi,\pi)$,
cf. \cref{eq:orderedmagnetization}. This was finally confirmed by
statistically exact Quantum Monte Carlo simulations \cite{Reger1988}
which found a value of
\begin{equation}
  \label{eq:heisenbergsquarestag}
  \tilde{m} = 0.30 \pm 0.02.
\end{equation}
This corresponds to $60 \%$ of the staggered magnetization of the
classical ground state. A review of the physics of the two-dimensional
spin $1/2$ Heisenberg model is given by \cite{Manousakis1991}. For
spin $S \geq 1$ a mathematical proof showed that at $T=0$ long-range
order is established \cite{Kubo1988}.

In three dimensions the spin-$1/2$ Heisenberg antiferromagnetic model
on the cubic lattice exhibits a finite temperature phase transition
from an ordered to a disordered phase. This has been demonstrated by a
QMC study \cite{Sandvik1998} finding a critical temperature
\begin{equation}
  \label{eq:heisenbergcubictcrit}
  T_c/J = 0.946 \pm 0.001.
\end{equation}

%%% Local Variables:
%%% mode: latex
%%% TeX-master: "../../thesis"
%%% End:

% \input{chapters/introduction/symmetries.tex}
%auto-ignore
\subsection{Order-by-Disorder principle}
\label{sec:orderbydisorder}
Quantum or thermal fluctuations usually introduce disorder to a
system. Under special circumstances, they can also have the opposite
effect and order the system. Although several types of states may be
degenerate ground states in the classical regime, the fluctuations
around different types can be unequal in behavior. Introducing thermal
or quantum fluctuations then selects the type of states whose
fluctuations provide the largest entropy or lowest zero-point
energy. This somehow paradoxical effect is called
\textit{order-by-disorder}\index{order-by-disorder} mechanism.

\begin{figure}[t!]
  \centering
  \includegraphics[width=0.8\textwidth]{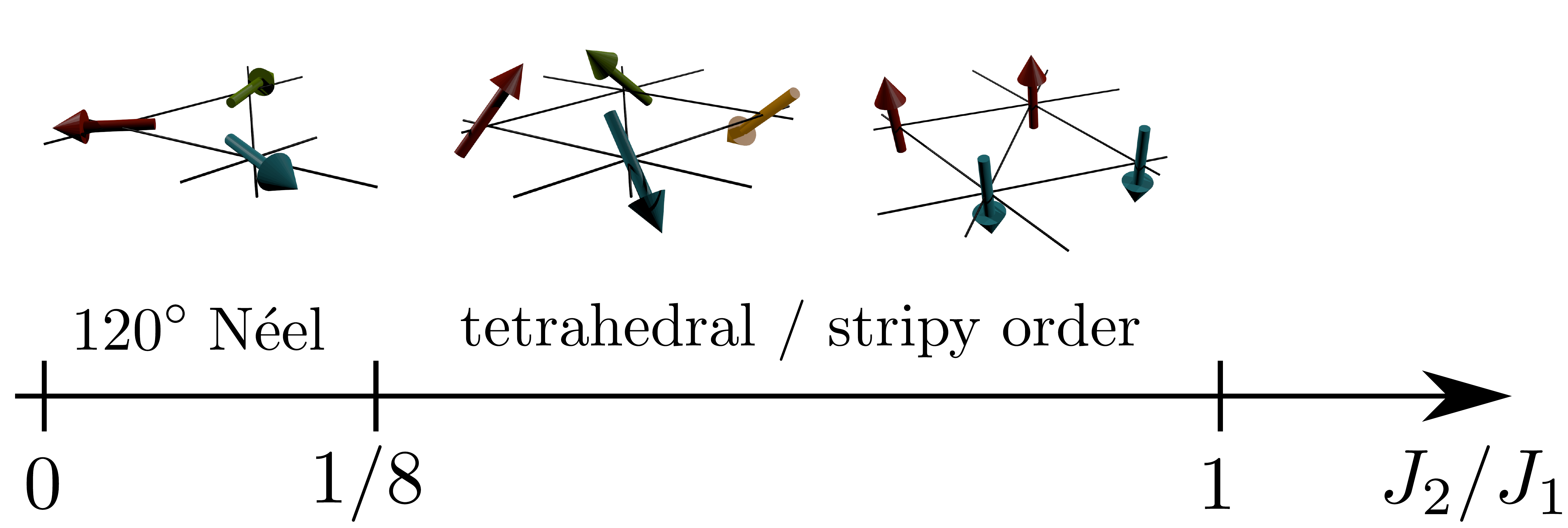}
  \caption{Ground states of the classical triangular lattice
    antiferromagnetic Heisenberg model with additional next-nearest
    neighbor interactions \cref{eq:triangularhbnn}. For
    $1/8 < J_2/J_1 < 1$ the tetrahedral and stripy states are
    degenerate.}
  \label{fig:orderbydisorderphase}
\end{figure}

A typical example of quantum fluctuations ordering a system can be
found in the frustrated antiferromagnetic triangular lattice
Heisenberg model with nearest and next-nearest neighbor interactions
~\cite{Jolicoeur1990}
\begin{equation}
  \label{eq:triangularhbnn}
  \mathcal{H} \,=\, J_1\sum\limits_{\left< i,j\right>}\bm{S}_i\cdot\bm{S}_j + 
  J_2\sum\limits_{\left<\left<  i,j\right>\right>}\bm{S}_i\cdot\bm{S}_j .
\end{equation}
For $0 < J_2/J_1 < 1/8$ the classical ground state is $120^\circ$
N{\'e}el ordered whereas for $1/8 < J_2/J_1 < 1$ two types of states
are degenerate: a collinear state where states are ordered
ferromagnetically in one direction of the triangular lattice and
antiferromagnetically along the other two directions and a tetrahedral
state where four spins are aligned in a way that they form a regular
tetrahedron, cf. \cref{fig:orderbydisorderphase}.
Quantum fluctuations can now be taken into account via spin-wave
theory around the classically ordered states. The correction to the
ground state energy yields a lower energy for the stripy ordered state
than the tetrahedral state. By numerically evaluating ground state
properties of the model one can show that indeed stripy order is
realized for
$J_2/J_1 \gtrsim 0.18$~\cite{Chubukov1992,Lecheminant1995}. We will
discuss an extended version of this model in detail in
\cref{sec:papertriangular}.

Several other examples of quantum and thermal fluctuations leading to
the realization of a subset of degenerate configurations have been
discussed in the
literature~\cite{Chalker1992,Chubukov1993,Bramwell1994,Villain1980,Shender1982a,Moessner1998}. The
order-by-disorder mechanism can be understood as a generic guiding
principle how specific states are selected amongst a highly degenerate
ground state manifold and therefore as an unconventional mechanism for
the emergence of order.

%%% Local Variables:
%%% mode: latex
%%% TeX-master: "../../thesis"
%%% End:

\newpage
\section{Quantum Spin Liquids}
\label{sec:spinliquids}
\epigraph{ \parbox{6.4cm}{\flushright
What if someone said \\
Promise lies ahead \\
Hopes are high in certain scientific circles } }{Dream Theater,
\textit{The Great Debate}}
%auto-ignore
% \subsection{Intro}

Discovering and understanding novel phases of matter is one of the
main objectives of condensed matter physics. Often, experimental
discovery precedes the theoretical understanding. The FQHE, for
example, was first discovered experimentally by Tsui, Stormer and
Gossard in 1982~\cite{Tsui1982} and important theoretical ideas such
as the Laughlin wave function~\cite{Laughlin1983} have been proposed a
posteriori. \textit{Quantum Spin Liquid} phases, on the other hand,
are envisioned by theoreticians and direct experimental proof of their
existence is still missing. They are disordered states of matter and
may exhibit fascinating phenomena such as the emergence of
quasiparticles described by gauge theories. Thus, they substantially
differ from a featureless disordered paramagnetic state. There are
many experimental candidate systems in frustrated magnetism for which
quantum spin liquid behavior could be a plausible
explanation~\cite{balents2010spin}, although a satisfactory proof of
their existence in nature has as of today not been given. Here we want
to give a very short introduction to these novel phases which is
tailored to the needs of this thesis. More comprehensive reviews have
been given by Refs.~\cite{balents2010spin,Savary2016a,Zhou2017}.

We discuss the implications of a gap in the excitation spectrum of a
system in \cref{sec:gaps}. We then introduce the resonating valence
bond state in \cref{sec:rvbstate}, which has been amongst the first
proposed spin liquid states. A general construction principle for spin
liquids generalizing the RVB state called the parton construction is
presented in \cref{sec:partonconstruction}.  We then discuss how this
construction is applied to yield two types of spin liquids: the chiral
spin liquid in \cref{sec:chiralspinliquids} which is closely related
to the FQHE and Dirac spin liquids in \cref{sec:diracspinliquids}
which are envisioned to be gapless states without long-range order.

%%% Local Variables:
%%% mode: latex
%%% TeX-master: "../../thesis"
%%% End:

%auto-ignore
\subsection{Gapless and gapped phases}
\label{sec:gaps}
Long-range order in systems with a continuous symmetry like the
Heisenberg model \cref{eq:heisenberggeneric} implies gapless
excitations according to Goldstone's theorem,
cf. \cref{sec:magneticorder}. The converse is not necessarily
true. Quantum critical states at continuous phase transitions usually
exhibit algebraically decaying correlations
\begin{equation}
  \label{eq:algebraicdecay}
  \langle \bm{S}_i\cdot \bm{S}_j \rangle \sim |\bm{r}_i -
  \bm{r}_{j}|^{-\eta},
\end{equation}
while having a gapless excitation spectrum. Also extended phases may 
show this kind of behavior. The most prominent example thereof is the
spin-$1/2$ Heisenberg chain, whose spin correlation function also
decays algebraically over distance according to the Bethe ansatz
solution~\cite{Bethe1931}. Also in higher dimensions, such states have
been proposed. One example is the so-called Dirac or algebraic spin
liquid~\cite{Affleck1988,Marston1989}, which we will discuss in more
detail in \cref{sec:diracspinliquids}.
\begin{figure}[t]
  \centering \centering
  \includegraphics[width=0.75\textwidth]{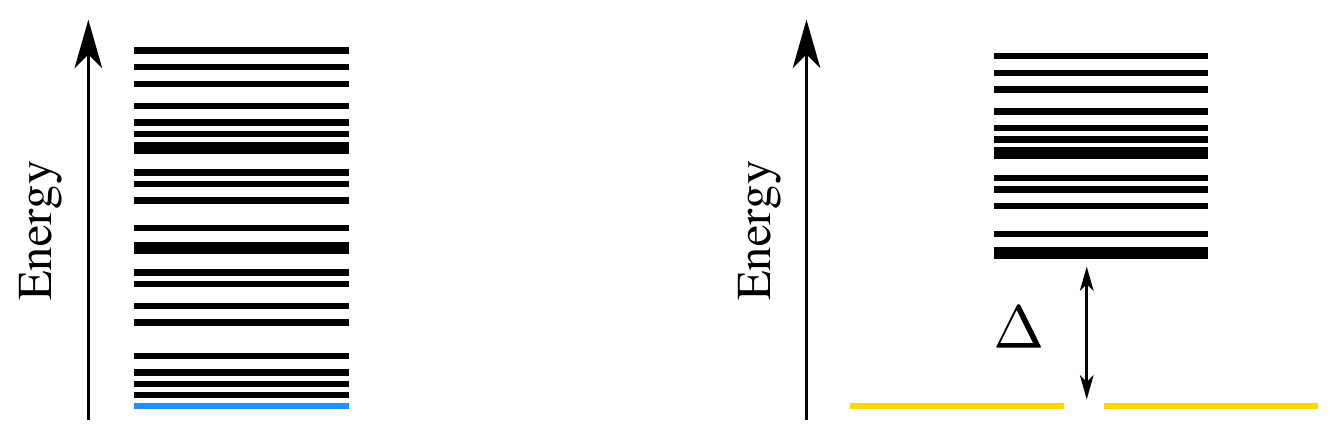}
  \\
  \begin{minipage}[t]{0.45\linewidth}
    \subcaption{\label{fig:gapless}}
  \end{minipage}
  \quad
  \begin{minipage}[t]{0.45\linewidth}
    \subcaption{\label{fig:gapped}}
  \end{minipage}
  \caption{Gapless (a) and gapped (b) excitation spectra. Long-range
    ordered phases are necessarily gapless. Other gapless states
    include critical states, the spin-$1/2$ Heisenberg
    chain~\cite{Bethe1931} and algebraic spin
    liquids~\cite{Affleck1988,Marston1989}. Gapped phases are
    necessarily short-ranged and for half-odd integer spin the ground
    state must be degenerate with periodic boundary conditions
    according to the Lieb-Schultz-Mathis-Hastings
    theorem~\cite{Lieb1961,Hastings2004}.}
  \label{fig:gaplessgapped}
\end{figure}
On the other hand, gapped phases necessarily exhibit exponentially
decaying correlation functions~\cite{lieb1972},
\begin{equation}
  \label{eq:exponentialdecay}
  \langle \bm{S}_i\cdot \bm{S}_j \rangle \sim
  \exp(-|\bm{r}_i -  \bm{r}_{j}|/\xi).
\end{equation}
The ground state of a gapped phase can be degenerate. This degeneracy
may be caused by symmetry breaking a discrete symmetry. An example of
such a phase would be valence bond solid states shown in
\cref{fig:vbs}.
\begin{figure}[t]
  \centering \includegraphics[width=0.7\textwidth]{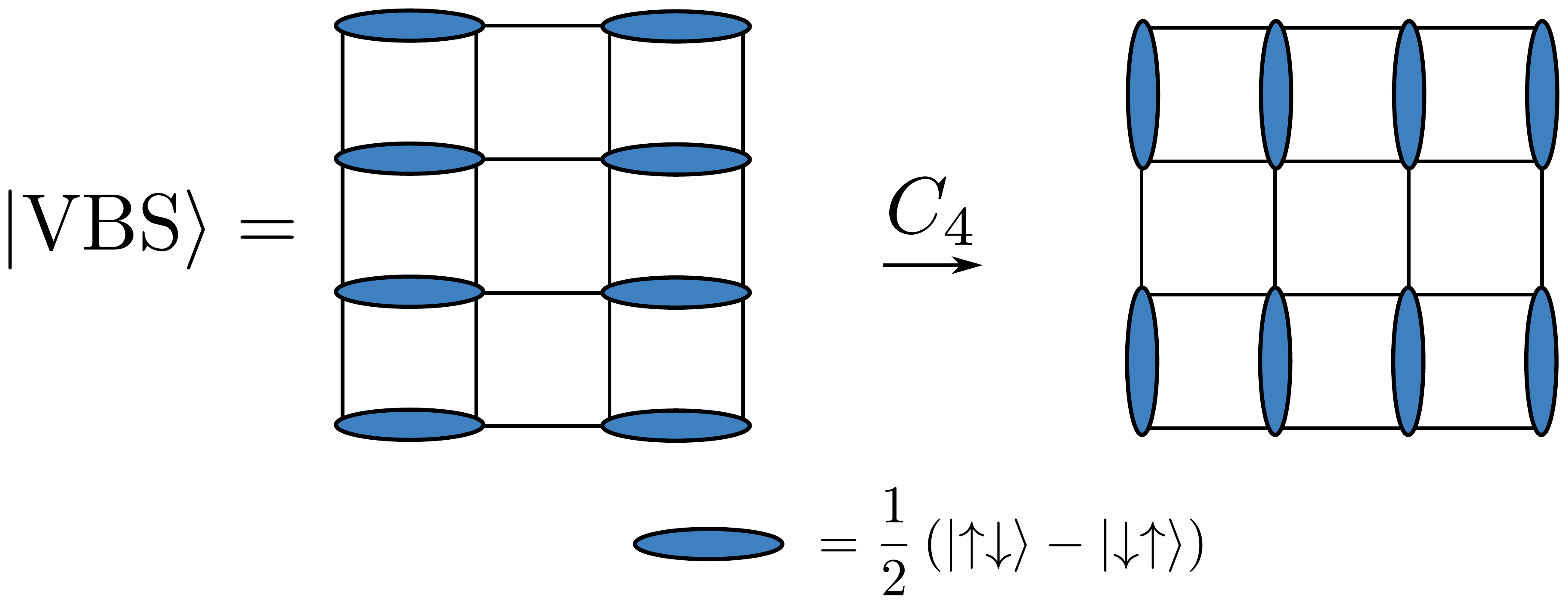}
  \caption{Valence Bond solids are regular coverings of a lattice with
    two spins paired up in a singlet. They are not invariant under
    lattice symmetries such as rotational or translational symmetry.}
  \label{fig:vbs}
\end{figure}
Valence bond solids are spin singlets, thus do not break rotational
SU($2$) symmetry. Yet, they break discrete lattice symmetries, like
translation or rotation symmetry. Such a state is for example realized
as a phase in the so-called J-Q models~\cite{Sandvik2007}.

\begin{figure}[t]
  \centering
  \begin{minipage}[t]{0.3\linewidth}
    \centering \includegraphics[width=0.8\textwidth]{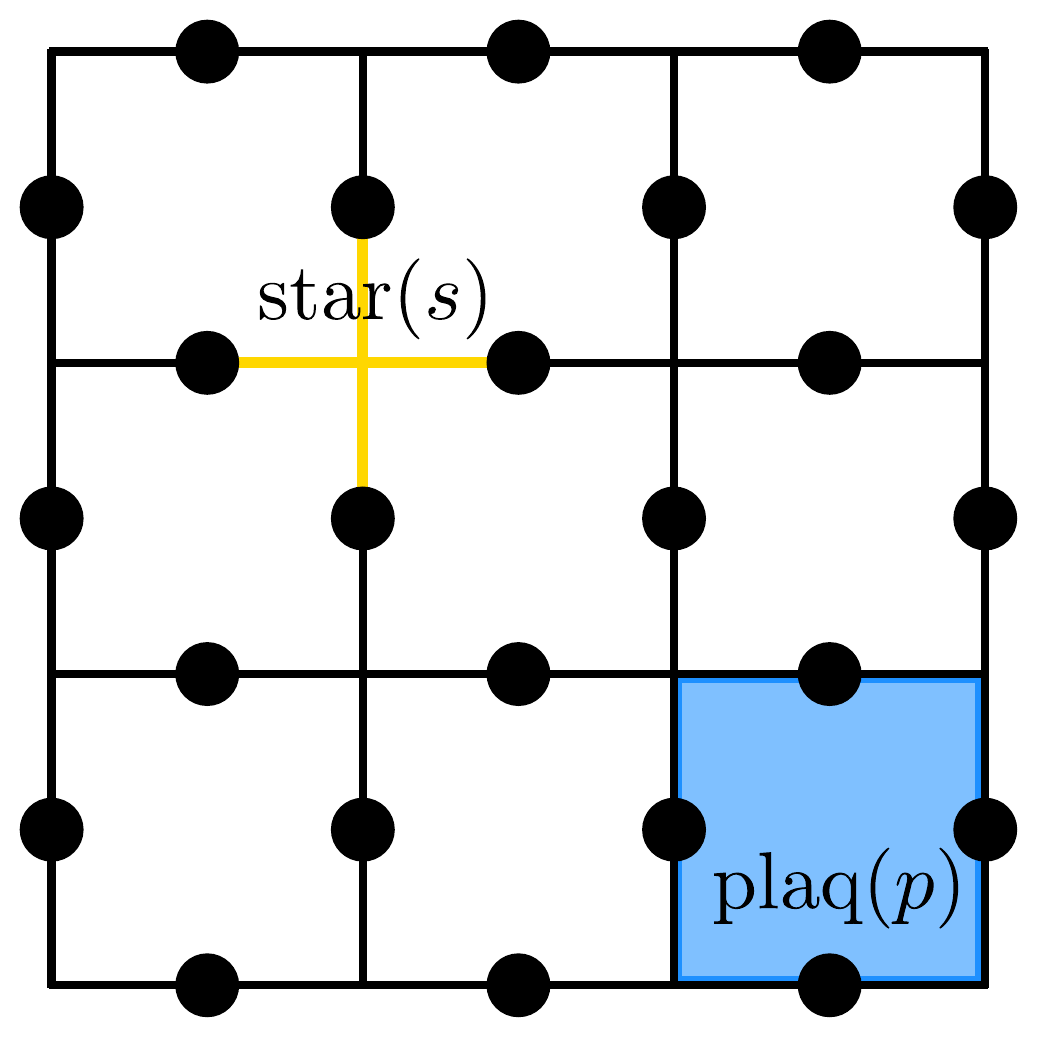}

  \end{minipage}
  \quad
  \begin{minipage}[t]{0.65\linewidth}
    \includegraphics[width=0.8\textwidth]{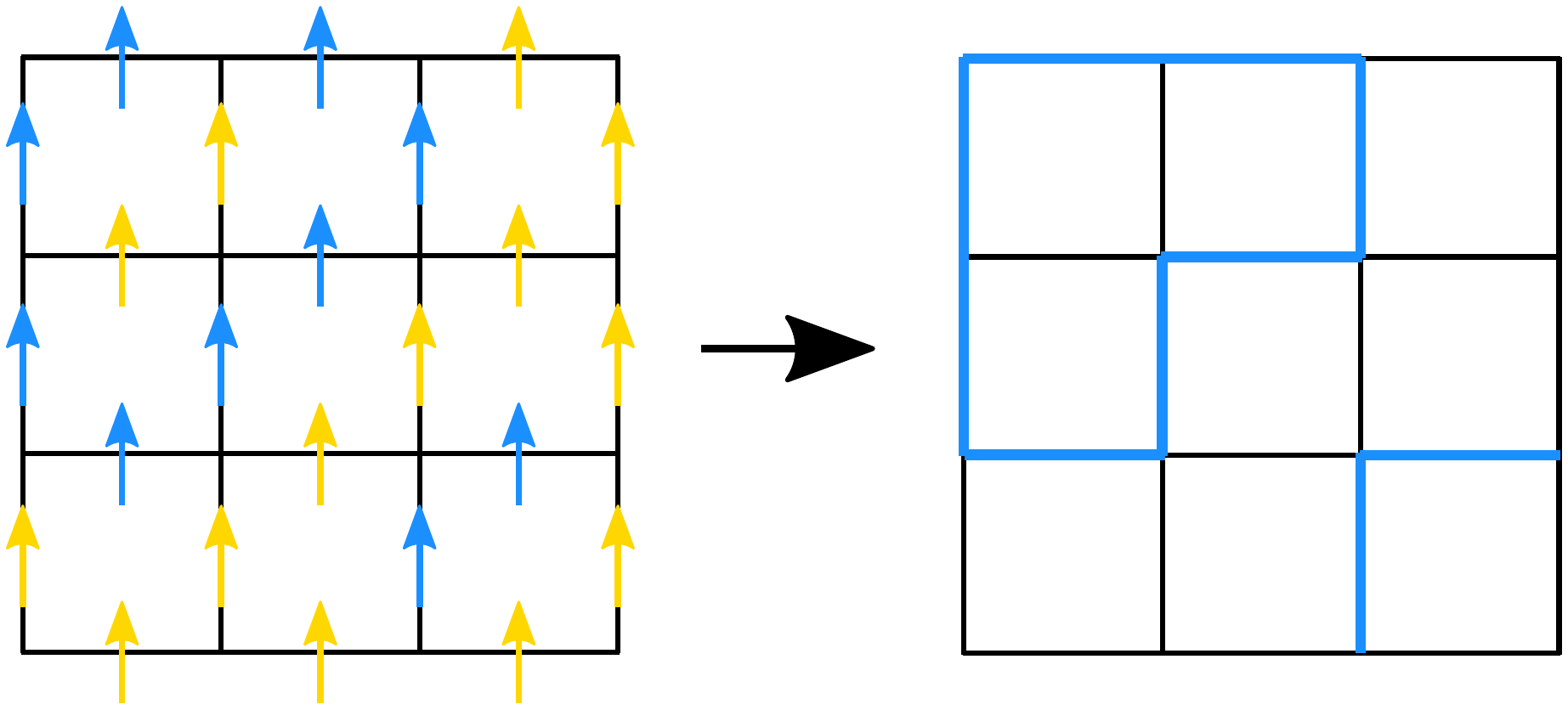}
    \centering
  \end{minipage}
  \\
  \begin{minipage}[t]{0.3\linewidth}
    \subcaption{\label{fig:toriccode:starplaq}}
  \end{minipage}
  \quad
  \begin{minipage}[t]{0.65\linewidth}
    \subcaption{\label{fig:toriccode:loops}}
  \end{minipage}
  \caption{The toric code model. The spins are placed on the links
    connecting the sites of a square lattice. The terms $A_s$ and
    $B_p$ as in \cref{eq:toriccodeoperators} are defined on the
    ``stars'' and ``plaquettes'' as shown in (a). The language of
    spins can be translated into the language of loops and strings as
    shown in (b). A string segment is present if the corresponding
    spin points up.  }
  \label{fig:toriccode}
\end{figure}

More excitingly, the degeneracy below the gap may be due to
topological ordering. A remarkable exactly solvable model that
illustrates this mechanism is given by Kitaev's toric code
model~\cite{Kitaev2003a},
\begin{equation}
  \label{eq:toriccode}
  H = - \sum \limits_s A_s - \sum\limits_p B_p,
\end{equation}
where
\begin{equation}
  \label{eq:toriccodeoperators}
  A_s = \prod\limits_{j \in \text{star}(s)}\sigma^x_j, \quad
  B_p = \prod\limits_{j \in \text{plaq}(s)}\sigma^z_j.
\end{equation}
Here, $\sigma^x_j$ and $\sigma^z_j$ denote the Pauli matrices.
$\text{star}(s)$ is the set of lattice sites in the ``star'' of
position $s$ and $\text{plaq}(s)$ is the set of lattice sites at the
boundary of ``plaquette'' $p$, cf. \cref{fig:toriccode}.  Although
this model is not realistic for actual materials in perfectly
illustrates many of the aspects of topological order. We will not
discuss the details of the solution here, but we refer the reader to
the original article~\cite{Kitaev2003a}. The gapped ground state is an
equal weight superposition of all possible loop configurations on the
square lattice (the language of loops can be directly translated to
spin configurations, cf. \cref{fig:toriccode:loops}), which is
sometimes called a \textit{loop soup}. Therefore, it does not break
any lattice symmetries. With periodic boundary conditions, there are 4
kinds of loop configurations that cannot be continuously deformed into
each other. They differ by having an even/odd number of incontractible
loops along the two periodic directions. These four kinds of loop
configurations are mutually orthogonal and thus yield a four-fold
degenerate ground state. The four-fold degeneracy for periodic
boundary conditions stems from the fact that a torus allows for
incontractible loops along both periodicity directions. 

The excitations of the toric code can also be exactly analyzed and the
phase for exchanging particles can be explicitly computed and it turns
out that these obey anyonic statistics. The toric code can be regarded
as a Hamiltonian implementation of a $\mathbb{Z}_2$ lattice gauge
theory~\cite{Wegner1971} which admits a deconfined phase in two
dimensions. Short-range correlations, ground state degeneracy
dependent on the topology (e.g. periodic boundary conditions) of the
lattice, anyonic braiding statistics and the relation to deconfined
phases of lattice gauge theories are typical hallmark features of
quantum spin liquids. The toric code exhibits what is called
$\mathbb{Z}_2$ topological order. There are also other $\mathbb{Z}_2$
topologically ordered states, as for example the nearest-neighbor RVB
states, discussed in \cref{sec:rvbstate} on a triangular
lattice~\cite{Moessner2001,Kivelson1989}. A different kind of
topological order is realized in chiral spin liquids which we will
discuss in \cref{sec:chiralspinliquids}.

Interestingly, there are certain restrictions on the nature of the
gapped, short-range ordered phases. The statement is known as the
\textit{Lieb-Schultz-Mattis-Hastings
  theorem}\index{Lieb-Schultz-Mattis-Hastings
  theorem}~\cite{Lieb1961,Hastings2004}. Assuming half-odd integer
spin per unit cell with periodic boundary conditions and certain
technical assumptions on the geometry of the finite lattice
(see~\cite{Lieb1961,Hastings2004} for the exact prerequisites) the
theorem states that a gapped system cannot have a unique ground state
in the thermodynamic limit. Instead, there must be degenerate ground
states below the gap. The prerequisites of the
Lieb-Schultz-Mattis-Hastings theorem are fulfilled by the extended
Heisenberg models we investigate in
\cref{sec:paperkagome,sec:papertriangular}. The toric code, on the
other hand, does not fulfill these conditions since its unit cell
consists out of two spin-$1/2$ degrees of freedom. Nevertheless, it is
an example of a gapped, topologically ordered phase.

\subsection{Resonating Valence Bond Liquids}
\label{sec:rvbstate}
The behavior of frustrated low dimensional quantum magnets may in many
ways differ from their classical unfrustrated counterparts.  Due to
the mechanisms discussed in \cref{sec:disorder} magnetic ordering may
be suppressed even at zero temperature. Instead of symmetry breaking
phases novel states of matter can emerge.

\begin{figure}[ht!]
  \centering \includegraphics[width=0.9\textwidth]{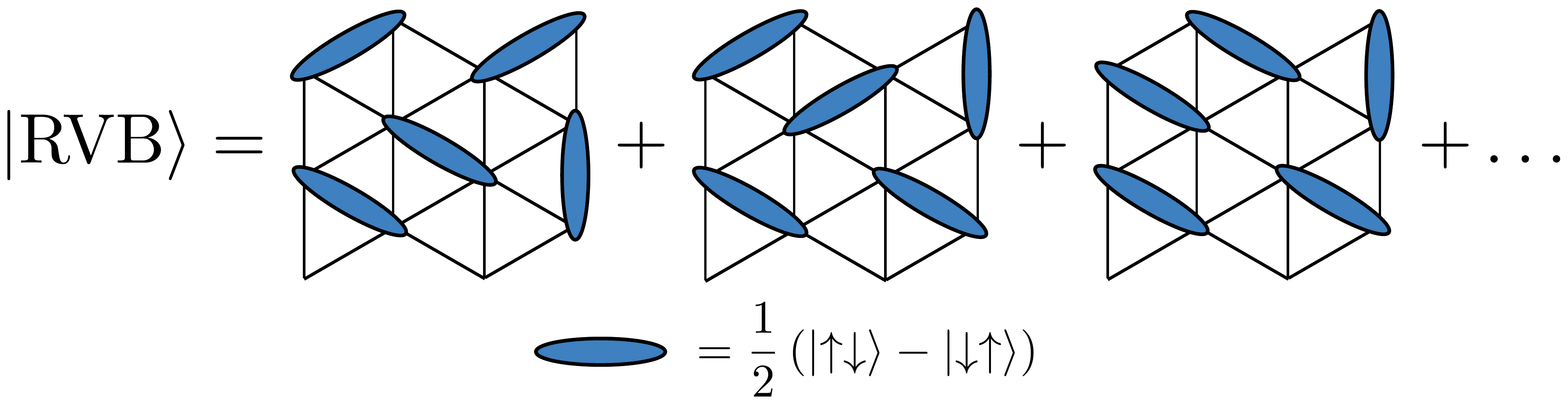}
  \caption{Valence bond configurations in an RVB state on the
    triangular lattice. The RVB state proposed by~\cite{Anderson1973}
    is an equal weight superposition of nearest-neighbor valence bond
    configurations. More general RVB states may include valence bond
    configurations with longer range singlet pairs and non-uniform
    weights~\cite{Anderson1987}.}
  \label{fig:rvbstate}
\end{figure}

A fundamental building block of quantum spin models in magnetism is
the Heisenberg bond between two neighboring spins
$J(\bm{S}_i\cdot\bm{S}_j)$. In the antiferromagnetic case $J>0$, its
energy is minimized by the singlet state in contrast to the N{\'e}el
state for classical interaction.  Given a disjoint covering of a
lattice with dimers (i.e. pairs of two sites) we can define a state
that is just a tensor product of singlets on these dimers,
cf.~\cref{fig:rvbstate}.
we will call such a state a \textit{valence bond configuration},
similar to the regular valence bond solid in
\cref{fig:vbs}. Heisenberg bonds on lattice sites connected by a dimer
are energetically minimized whereas other bonds are not
minimized. Hence, such states are reasonable variational wave functions
for quantum magnets. Importantly, these states are manifestly singlet
states and are thus spin rotational SU($2$) symmetric.

In a seminal paper~\cite{Anderson1973}, Anderson proposed that a
superposition of valence bond configurations yields an extremely low
variational energy for the triangular spin-$1/2$ Heisenberg model. He
considered the equal weight superposition of all dimer configurations
where dimers connect nearest neighbor bonds. This state is called a
\textit{resonating valence bond liquid}\index{Resonating Valence
  Bond}, short RVB liquid and schematically shown in
\cref{fig:rvbstate}. As a superposition of all possible
nearest-neighbor valence bond configurations, it is not only SU($2$)
symmetric but also symmetric under space group symmetries. The state
is a truly quantum mechanical state and does not possess a classical
equivalent. Anderson found that the RVB state had lower variational
energy than spin-wave estimations on the energy of the N{\'e}el state.
The energy estimate for the RVB state given by \cite{Anderson1973} is
\begin{equation}
  \label{eq:rvbenergyestimate}
  E_{\text{RVB}} \approx -(0.54 \pm 0.01) \; N J,
\end{equation}
where $N$ denotes the number of sites and $J$ the coupling constant.
We compare this energy to the actual ground state energy from
numerical Exact Diagonalization on a $48$ site cluster obtained in the
course of this thesis, cf. \cref{sec:edbanchmarks}. The precise energy
on this cluster is given in \cref{tab:edbenchmarks}
\begin{equation}
  \label{eq:triangulargroundstateenergyestimate}
  E_0 = -0.5586 \; N J.
\end{equation}
The variational energy of the RVB state is therefore extremely close
to the actual ground state energy. Nevertheless, numerical studies
later on provided conclusive evidence that the system breaks SU($2$)
symmetry and exhibits $120^\circ$ N{\'e}el
order~\cite{Jolicoeur1990,Bernu1994,Bernu1992}. Therefore the question
arises whether this state can be realized as the ground state of a
local spin Hamiltonian. Although a toy model exactly realizing the
above RVB state has for example been given by Rokhsar and
Kivelson~\cite{Rokhsar1988}, the question whether the RVB state is
realized in more realistic models remains open until today. It is now
understood that RVB states on non-bipartite two-dimensional lattices
may exhibit fractionalized excitations with $\mathbb{Z}_2$ topological
order~\cite{Moessner2001,Kivelson1989} similar to the toric code
model~\cref{eq:toriccode}~\cite{Kitaev2003a}.

The valence bond configurations (not necessarily nearest-neighbors)
span the space of singlet states on a lattice
\cite{Hulthen1938,Karbach1993} and thus, every singlet wave function
can be expressed as a linear combination of them. In order to extend
the idea of the RVB state Anderson~\cite{Anderson1987} proposed to
investigate a certain subset of possible superpositions of valence
bond configurations.
% Consider the Hubbard model
% \begin{equation}
%   \label{eq:hubbardmodel}
%   H = \sum \limits_{i,j,\sigma} ( t_{ij}c_{i,\sigma}^\dagger c_{j,\sigma} + \text{h.c.}) +
%   U\sum\limits_{i}n_{i,\uparrow}n_{i,\downarrow}
% \end{equation}
% where $n_{i,\sigma} = c^\dagger_{i,\sigma} c_{i,\sigma}$.  In the
% large $U$ limit double site occupancies are supressed by the on-site
% repulsion term.
For constructing these states we first consider a fermionic Hilbert
space of up and down electrons on the lattice. Thus a typical
configuration in this Hilbert space is given by
\begin{equation}
  \label{eq:fermionichilbert}
  \ket{\psi} = \ket{\uparrow\downarrow, \emptyset, \uparrow, \downarrow},
\end{equation}
where double site occupancies $\uparrow\downarrow$ and vacancies
$\emptyset$ are allowed in contrast to pure spin configurations. The
operator which sets the part of a many-body wave function with double
site occupancy to zero is given by
\begin{equation}
  \label{eq:gutzwillerprojectorfirst}
  P_{\text{GW}} = \prod\limits_i \left( 1 - n_{i,\uparrow}n_{i,\downarrow}\right),
\end{equation}
where $n_{i,\sigma} = c_{i\sigma}^\dagger c_{i\sigma}$ for fermionic
creation and annihilation operators $c_{i\sigma}^\dagger$ and
$c_{i\sigma}$, $\sigma = \uparrow\downarrow$.  The operator
$P_{\text{GW}}$ in \cref{eq:gutzwillerprojector} is called Gutzwiller
projection~\cite{Gutzwiller1963,Gros1989}. Anderson now proposed to
apply the Gutzwiller projection to BCS wave functions from the BCS
theory of superconductivity of the form
\begin{equation}
  \label{eq:bcsproductwavefunction}
  \ket{\psi_{\text{BCS}}} = \prod\limits_{{\bm{k}} \in \text{B.Z.}} \left(u_{\bm{k}} +
    v_{\bm{k}}c^\dagger_{{\bm{k}}\uparrow}c^\dagger_{-{\bm{k}}\downarrow}
  \right)\ket{0},
\end{equation}
where $u_{\bm{k}}$, $v_{\bm{k}}$ serve as variational parameters and
\begin{equation}
  \label{eq:fermifouriertrafo}
  c_{{\bm{k}}\sigma}^\dagger = \frac{1}{\sqrt{N}}\sum\limits_{i=1}^N
  e^{i{\bm{k}}\cdot\bm{r}_i} c_{i\sigma}^\dagger,
\end{equation}
are fermionic creation operators in reciprocal space. The following
computation shows how the Gutzwiller projected BCS state
$P_{\text{GW}}\ket{\psi_{\text{BCS}}}$ is related to the RVB state
\begin{align}
  P_{\text{GW}}  \ket{\psi_{\text{BCS}}}
  &\propto P_{\text{GW}}\prod\limits_{{\bm{k}} \in \text{B.Z.}}
    \left(1 + \frac{v_{\bm{k}}}{u_{\bm{k}}}c^\dagger_{{\bm{k}}\uparrow}c^\dagger_{-{\bm{k}}\downarrow}
    \right)\ket{0} \\
  &= P_{\text{GW}} \exp \left[\sum \limits_{\bm{k}} \frac{v_{\bm{k}}}{u_{\bm{k}}}c^\dagger_{{\bm{k}}\uparrow}c^\dagger_{-{\bm{k}}\downarrow}\right]\ket{0}\\
  &= \left(\sum \limits_{\bm{k}} \frac{v_{\bm{k}}}{u_{\bm{k}}}c^\dagger_{{\bm{k}}\uparrow}
    c^\dagger_{-{\bm{k}}\downarrow} \right)^{N/2}\ket{0}\\
  &= \left(\sum \limits_{i,j} a_{ij}c^\dagger_{i\uparrow}
    c^\dagger_{j\downarrow} \right)^{N/2}\ket{0},
    \label{eq:rvbexpanded}
\end{align}
where
\begin{equation}
  \label{eq:pairingbcs}
  a_{ij} = \sum\limits_{{\bm{k}} \in \text{B.Z.}}\frac{v_{\bm{k}}}{u_{\bm{k}}}e^{i{\bm{k}}(\bm{r}_i - \bm{r}_j)}.
\end{equation}
The second equality uses the fact that
$\exp(f^\dagger) = 1 + f^\dagger$ for fermionic operators $f^\dagger$,
the third equality projects to the correct particle number in the
Taylor expansion of the exponential function. If $a_{ij} = a_{ji}$ the
operator\begin{equation}
  \label{eq:singletcreation}
  a_{ij}(c^\dagger_{i\uparrow} c^\dagger_{j\downarrow} +
  c^\dagger_{j\uparrow} c^\dagger_{i\downarrow}) =  a_{ij}(c^\dagger_{i\uparrow} c^\dagger_{j\downarrow} -
  c^\dagger_{i\downarrow} c^\dagger_{j\uparrow}) 
\end{equation}
in \cref{eq:rvbexpanded} creates a singlet state on the sites $i$ and
$j$ with coefficient $a_{ij}$.  We see that the state in
\cref{eq:rvbexpanded} is again a superposition of dimer coverings on
the lattice, but now with different coefficients $a_{ij}$ defined by
$u_{\bm{k}}$ and $v_{\bm{k}}$ via \cref{eq:pairingbcs}. It may,
therefore, be regarded as a generalization of the initially proposed
state in~\cite{Anderson1973} and is thus also called an RVB state.

%%% Local Variables:
%%% mode: latex
%%% TeX-master: "../../thesis"
%%% End:

%auto-ignore
\subsection{Parton construction and mean-field theory}
\label{sec:partonconstruction}

% \subsubsection{Equivalence of the Heisenberg model to a fermionic
%   parton Hamiltonian}
The Gutzwiller projection operator \cref{eq:gutzwillerprojectorfirst}
transforms a fermionic wave function defined on a fermionic Hilbert
space into a spin wave function on the subspace with no
double-occupancies or vacancies. As in the case of the triangular
lattice spin-$1/2$ Heisenberg models, cf. \cref{sec:rvbstate}, these
Gutzwiller projected wave functions may yield low variational energies
and may even be realized as a ground state of pure spin systems
without charge degrees of freedom. A generic Heisenberg model 
\begin{equation}
  \label{eq:heisenberggeneric_sl}
  H =  \sum\limits_{i, j}J_{ij} \bm{S}_i\cdot\bm{S}_j , 
\end{equation}
may be considered as a low-energy effective Hamiltonian in the
Mott-insulating regime $t_{ij} \ll U$ of a Hubbard model of type
\begin{equation}
  \label{eq:hubbardmodelgeneric}
  H_{\text{Hubbard}} = \sum \limits_{i,j,\;\sigma} \left( t_{ij}c_{i\sigma}^\dagger c_{j\sigma} + \text{H.c.}\right) +
  U\sum\limits_{i}n_{i\uparrow}n_{i\downarrow}.
\end{equation}
The spin operators $\bm{S}_i$ on site $i$ are introduced by
\begin{equation}
  \label{eq:parton}
  \bm{S}_i = \frac{1}{2}\sum_{\alpha\beta}c_{i\alpha}^\dagger\bm{\sigma}_{\alpha\beta}c_{i\beta},
\end{equation}
where $\alpha, \beta = \downarrow, \uparrow$. The coupling constants
$J_{ij}$ are related to the original hopping $t_{ij}$ and on-site
repulsion $U$ terms via $J_{ij} = 4|t_{ij}|^2/U$. The effective
Heisenberg model can be defined on the smaller Hilbert space of spin
configurations with no double-occupancies or vacancies. It is thus a
simplification of the original Hubbard model.

Starting from a pure spin model like \cref{eq:heisenberggeneric_sl} we
can also take the opposite direction. We introduce fermionic operators
as in \cref{eq:parton} to the Heisenberg model
\cref{eq:heisenberggeneric_sl} and investigate the resulting fermionic
problem \cite{Wen2002}
\begin{equation}
  \label{eq:partonhamiltonianplusconst}
  H =  \sum\limits_{i, j, \alpha,\beta}-\frac{J_{ij}}{2}c_{i\alpha}^\dagger
  c_{j\alpha}c_{j\beta}^\dagger c_{i\beta} + \sum\limits_{i, j}
  \frac{J_{ij}}{2}  \left( n_i - \frac{1}{2}n_in_j \right),
\end{equation}
where
\begin{equation}
  \label{eq:numberoperator}
  n_i = c_{i\uparrow}^\dagger c_{i\uparrow} + c_{i\downarrow}^\dagger
c_{i\downarrow},
\end{equation}
is the number operator of fermions per site. The fermionic
operators $c_{i\alpha}^\dagger$ introduced via \cref{eq:parton} are
then called \textit{parton} or \textit{spinon} operators. Several
authors also use to call the parton operators Abrikosov fermions or
the decomposition \cref{eq:parton} slave-boson
approach~\cite{Wen2002}. Now, the Hilbert space is enlarged and we are
actually considering a different problem. In order to describe the
same model we have to enforce the constraint of single site occupancy
\begin{equation}
  \label{eq:singlesiteoccupancy}
   n_i = 1.
\end{equation}
The parton Hamiltonian \cref{eq:partonhamiltonianplusconst} together
with the single site occupation constraint
\cref{eq:singlesiteoccupancy} are now completely equivalent to the
original Heisenberg model \cref{eq:heisenberggeneric_sl}. The single site
occupancy constraint \cref{eq:singlesiteoccupancy} turns the second
term in \cref{eq:partonhamiltonianplusconst} into a constant. We are
thus left with the following Hamiltonian for the partons
\begin{equation}
  \label{eq:partonhamiltonian}
  H_{\text{parton}} =  \sum\limits_{i, j, \alpha,\beta}-\frac{ J_{ij}}{2}c_{i\alpha}^\dagger
  c_{j\alpha}c_{j\beta}^\dagger c_{i\beta} . 
\end{equation}
The Hamiltonian \cref{eq:partonhamiltonian} has an interesting
fundamental property: local U($1$) gauge
symmetry~\cite{Baskaran1988}. It is invariant under the local gauge
transformations
\begin{equation}
  \label{eq:partongaugesymmetry}
  c_{i\alpha}^\dagger \rightarrow e^{i \theta_i}  c_{i\alpha}^\dagger . 
\end{equation}
This symmetry is not present the original Hubbard model
\cref{eq:hubbardmodelgeneric} since the hopping terms
$c_{i\alpha}^\dagger c_{j\alpha}$ are only invariant under global
phase transformations. The local gauge symmetry
\cref{eq:partongaugesymmetry} is therefore an emergent property of the
system in the Mott insulating regime~\cite{Baskaran1988}.

\subsubsection{Generic ansatz wave functions from Gutzwiller
  projection}
The equivalence between the original spin model
\cref{eq:heisenberggeneric_sl} and the parton Hamiltonian
\cref{eq:partonhamiltonian} together with the constraint
\cref{eq:singlesiteoccupancy} gives rise to a principle of
constructing variational ansatz wave functions generalizing the RVB
approach. The problem in exactly solving \cref{eq:partonhamiltonian}
are the four-fermion operators. In a mean-field approach, we can
replace $c_{i\alpha}^\dagger c_{j\alpha}$ with its vacuum expectation
values
$\chi_{ij} \equiv \langle c_{i\alpha}^\dagger c_{j\alpha}\rangle$ or
the operators $c_{i\alpha}^\dagger c_{j\beta}^\dagger$ with
$\Delta_{ij} \equiv \langle c_{i\uparrow}^\dagger c_{j\downarrow}^\dagger
\rangle$. Furthermore, we replace the constraint
\cref{eq:singlesiteoccupancy} by a relaxing this constraint to be
valid only on average
\begin{equation}
  \label{eq:singlesiteoccupancyrelaxed}
  \langle n_i \rangle = 1.
\end{equation}
This condition can be realized by adding a chemical potential term of
the form
\begin{equation}
  \label{eq:chemicalpotential}
  \sum\limits_{i} \mu_i(n_i - 1)  
\end{equation}
to the parton Hamiltonian \cref{eq:partonhamiltonian} and adjusting
the chemical potential $\mu_i$. In summary, the Hamiltonian
\cref{eq:partonhamiltonian} can be approximated by a generic quadratic
mean-field Hamiltonian of the form \cite{Wen2002}
\begin{align}
  \label{eq:partonmeanfieldhamiltonian}
  \begin{split}
    H_{\text{mean}} = &\sum\limits_{i, j, \alpha}
    (\chi_{ij}c_{i\alpha}^\dagger c_{j\alpha} + \text{H.c.}) +
    \sum\limits_{i, j}(\Delta_{ij}^*c_{i\uparrow}^\dagger
    c_{j\downarrow}^\dagger + \text{H.c.}) + \sum\limits_{i} \mu_i(n_i
    - 1).
  \end{split}
\end{align}
The numbers $\chi_{ij}$, $\Delta_{ij}$ and $\mu_i$ define the hopping,
pairing and chemical potential amplitudes, respectively. Since the
mean-field Hamiltonian is quadratic in the parton operators it is
exactly solvable in the extended Hilbert space. In general the
non-interacting mean-field ground state $\ket{\psi_{0}^{\text{mf}}}$
will be given by a filled Fermi sea or a BCS type wave function,
depending on whether or not the pairing amplitudes $\Delta_{ij}$ are
non-zero. In order to create an ansatz wave function for the ground
state of the original spin model \cref{eq:heisenberggeneric_sl} we can
again set the coefficients of configurations with double site
occupancy or vacancies to zero via Gutzwiller projection
\cref{eq:gutzwillerprojectorfirst}
\begin{equation}
  \label{eq:gutzwillerprojwf}
  \ket{\psi_{\text{GPWF}}} = P_{\text{GW}}\ket{\psi_{0}^{\text{mf}}}.
\end{equation}
We will call such a state a \textit{Gutzwiller projected parton wave
  function}\index{Gutzwiller projected parton wave function}, short
GPWF.

The previously defined RVB states in \cref{eq:rvbexpanded} are a
special case of these GPWFs. Fourier transforming
\cref{eq:partonmeanfieldhamiltonian} with a translationally invariant
ansatz $\chi_{ij}$ and $\Delta_{ij}$ yields a BCS type mean-field
Hamiltonian of the form
\begin{equation}
  \label{eq:bcsmeanfield}
  H = \sum_{{\bm{k}},\sigma} \chi_{\bm{k}} c_{{\bm{k}}\sigma}^\dagger c_{{\bm{k}}\sigma} +
  \sum_{\bm{k}} (\Delta_{\bm{k}}^*  c_{{\bm{k}}\uparrow}^\dagger c_{-{\bm{k}}\downarrow}^\dagger + \text{H.c.}),
\end{equation}
whose solution after Bogoliubov transformation is given by the BCS
state in \cref{eq:bcsproductwavefunction}.

The coefficients of these ansatz wave functions in the parton
construction can be computed as a Slater determinant. We will discuss
the numerical evaluation of properties and coefficients of these wave
functions in \cref{sec:gpwfvmc}. It turns out that studying these wave
functions can be done numerically in a computationally efficient
way~\cite{Ceperley1977}.

The manipulations and approximations leading from
\cref{eq:heisenberggeneric_sl} to \cref{eq:partonmeanfieldhamiltonian}
are actually quite crude. First of all, the Hilbert space is enlarged and
a fermionic model \cref{eq:partonhamiltonianplusconst} is
introduced. From \cref{eq:partonhamiltonian} to
\cref{eq:partonmeanfieldhamiltonian} we replace the single occupancy
constraint \cref{eq:singlesiteoccupancy} by its relaxed average
version \cref{eq:singlesiteoccupancyrelaxed}.  The parameters
$\chi_{ij}$ and $\Delta_{ij}$ may then be chosen to satisfy
self-consistency equations,
\begin{equation}
  \label{eq:selfconsistencey}
  \langle c_{i\alpha}^\dagger c_{j\alpha}\rangle = \chi_{ij}, \quad
  \langle c_{i\uparrow}^\dagger c_{j\downarrow}^\dagger \rangle = \Delta_{ij},
\end{equation}
and the chemical potentials $\mu_i$ may be adjusted to satisfy the
average constraint \cref{eq:singlesiteoccupancyrelaxed}. The resulting
quadratic mean-field Hamiltonian can then be analyzed and parton
mean-field phase diagrams can be worked out. Yet, the predictive power
of such an approach is limited and necessarily needs to be
complemented with different approaches. Often, variational energies
are computed using Variational Monte Carlo, cf.~\cref{sec:gpwfvmc} or
overlaps with numerically precise ground states from Exact
Diagonalization can be calculated.

\subsubsection{Fluctuations around mean-field theory}
For taking into account fluctuations around the zeroth-order
mean-field theory of the parton Hamiltonian
\cref{eq:partonmeanfieldhamiltonian} we introduce the path integral
formulation. The partition function can be written as~\cite{Wen2002}
\begin{equation}
  \label{eq:partonfieldintegral}
  \mathcal{Z} = \int \mathcal{D}c_i \mathcal{D}\mu_i \mathcal{D} \chi_{ij}
  \exp\left\{i\int \text{d}t \;\mathcal{L} - \sum_i\mu_i(t)(n_i - 1) \right\},
\end{equation}
where the Lagrangian $\mathcal{L}$ is given by
\begin{equation}
  \label{eq:partonlagrangian}
  \mathcal{L} = \sum_i c_i^\dagger i \partial_tc_i - \sum\limits_{i, j, \sigma}-\frac{ J_{ij}}{2}\left[(c_{i\sigma}^\dagger
    c_{j\sigma}\chi_{ji} + \text{H.c.}) - |\chi_{ij}|^2\right],
\end{equation}
with the Hubbard-Stratonovich fields $\chi_{ij}$. The path integral
formulation also allows deriving effective theories for specific
ansätze for in \cref{sec:chiralspinliquids} and
\cref{sec:diracspinliquids}.

We emphasize that in \cref{eq:partonfieldintegral} the chemical
potentials $\mu_i(t)$ are now explicitly time-dependent. This
reproduces the original single occupancy constraint
\cref{eq:singlesiteoccupancy} since functional integration over the
$\mu_i$ fields yields
\begin{equation}
  \label{eq:funcintconst}
  \int\mathcal{D}\mu_i \exp\left\{i\int\text{d}t \; \mu_i
    (n_i - 1)\right\} =
  \delta\left(n_i - 1 \right).
\end{equation}
\cref{eq:partonfieldintegral} and \cref{eq:partonlagrangian} are
therefore again equivalent to the original spin system
\cref{eq:heisenberggeneric_sl}.

The fluctuating fields $\chi_{ij}$ can now be decomposed into
amplitude fluctuations $\bar{\chi}_{ij}$ and phase fluctuations
$a_{ij}$,
\begin{equation}
  \label{eq:hsampphase}
  \chi_{ij} = \bar{\chi}_{ij}e^{ia_{ij}}.
\end{equation}
According to~\cite{Wen2002} the amplitude fluctuations should be
gapped and $\bar{\chi}_{ij}$ may, therefore, be regarded as a
constant. Still, phase fluctuations $a_{ij}$ are assumed as dynamical
fields. The effective Hamiltonian is then given by
\begin{equation}
  \label{eq:partonfirstordermf}
  H = \sum\limits_{i, j, \sigma}
  (\bar{\chi}_{ij}c_{i\sigma}^\dagger c_{j\sigma}e^{ia_{ij}} + \text{H.c.}) +
  \sum_i \mu_i(n_i - 1).
\end{equation}
The system now exhibits a gauge structure. Gauge transformations of
the type
\begin{equation}
  \label{eq:partonmfgaugetrafo}
  c_{i\sigma}^\dagger \rightarrow e^{i\theta_i}c_{i\sigma}^\dagger, \quad
  a_{ij} \rightarrow a_{ij} + \theta_j - \theta_i,
\end{equation}
leave the Hamiltonian \cref{eq:partonfirstordermf} invariant.
\cref{eq:partonfirstordermf} now describes a lattice gauge theory with
U($1$) lattice gauge fields $a_{ij}$ and $\mu_i$.

Let us compare the two effective Hamiltonians
\cref{eq:partonmeanfieldhamiltonian} and
\cref{eq:partonfirstordermf}. A common feature for both is that we
have to make an ansatz for the coefficients $\chi_{ij}$ or
$\bar{\chi}_{ij}$ respectively. Apart from this
\cref{eq:partonmeanfieldhamiltonian} describes free partons which are
essentially uncorrelated. Consequently, this approximation cannot
correctly describe the strong correlations present in Heisenberg
systems. Correlations are taken into account via the gauge fields in
\cref{eq:partonfirstordermf}. The only approximation made when passing
from the Heisenberg model to \cref{eq:partonfirstordermf} is to assume
a constant amplitude in the Hubbard-Stratonovich fields $\chi_{ij}$,
because amplitude fluctuations are expected to be
gapped~\cite{Wen2002}. It is thus conjectured that the lattice gauge
theory \cref{eq:partonfirstordermf} may correctly describe strongly
correlated phases of Heisenberg spin systems. Phases of the lattice
gauge theory \cref{eq:partonfirstordermf} correspond to phases of the
original spin model.

A central question in the study of gauge theories is whether the
matter fields (in our case the partons) are confined or deconfined.
Particles are said to be confined if the energy cost for separating
them diverges at large distances. A typical example of this is quark
confinement in quantum chromodynamics. An example of deconfinement is
quantum electrodynamics in $3+1$ dimensions where the potential energy
between two electrons is given by the Coulomb interaction and two
electrons can be separated by an arbitrary distance with finite energy
cost. For spin systems, we may now ask the same question whether two
partons can be separated by an arbitrary distance at finite energy
cost. Deconfined phases of lattice gauge theories such as
\cref{eq:partonfirstordermf} are essentially quantum spin liquids.

%%% Local Variables:
%%% mode: latex
%%% TeX-master: "../../thesis"
%%% End:

%auto-ignore
\subsection{Chiral Spin Liquids}
\label{sec:chiralspinliquids}

One of the first proposed parton ansätze was the \textit{chiral spin
  liquid}\index{chiral spin liquid}, short CSL state~\cite{Wen1989a}.

The word "chiral" refers to the violation of parity symmetry
(i.e. spatial reflection symmetry) and time reversal symmetry. A
quantity that measures the breaking of these two symmetries is the
so-called \textit{scalar chirality} operator,
\begin{equation}
  \label{eq:scalarchirality}
  \bm{S}_i\cdot(\bm{S}_j\times \bm{S}_k).
\end{equation}
A reflection symmetry reverses the orientation of a path
$i \rightarrow j \rightarrow k$ whereas time reversal symmetry
transforms the spin operators as $\bm{S}_i \rightarrow -\bm{S}_i$,
thus changing the sign in \cref{eq:scalarchirality}. Certain magnetic
orderings may also not be parity or time reversal symmetry invariant
like the tetrahedral order in \cref{fig:orderbydisorderphase}.

\begin{figure}[t!]
  \centering
  \begin{minipage}[c]{0.45\textwidth}
    \centering
    \includegraphics[width=0.7\textwidth]{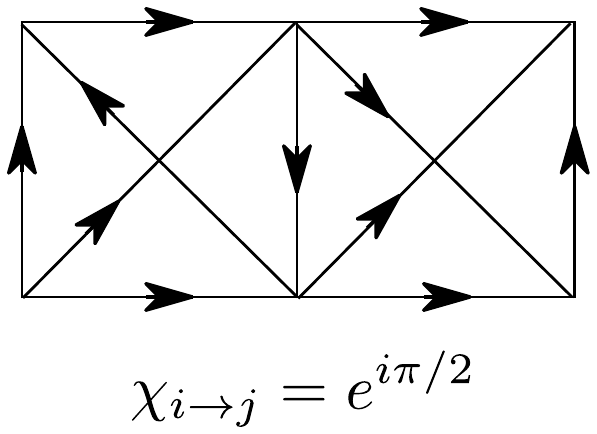}
  \end{minipage}%
  \quad
  \begin{minipage}[c]{0.45\textwidth}
    \centering
    \includegraphics[width=0.6\textwidth]{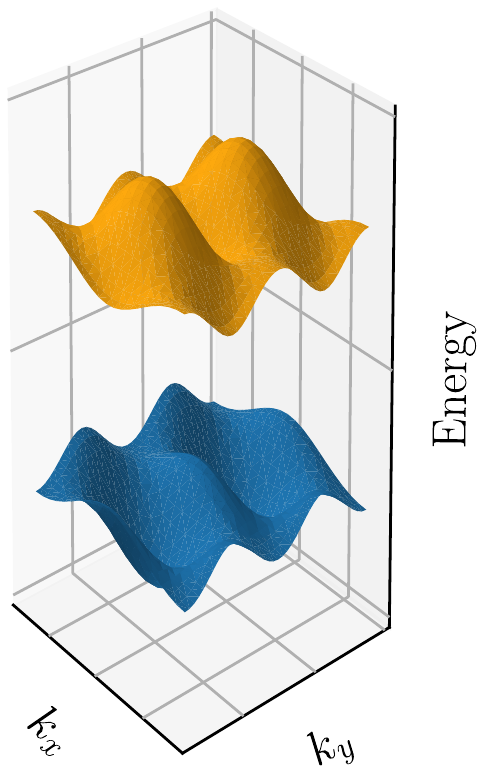}
  \end{minipage}
  \\
  \begin{minipage}[t]{0.45\linewidth}
    \subcaption{
      \label{fig:csl:ansatz}}
  \end{minipage}
  \quad
  \begin{minipage}[t]{0.45\linewidth}
    \subcaption{
      \label{fig:csl:bandstructure}}
  \end{minipage}

  \caption{Parton ansatz for chiral spin liquids as proposed
    by~\cite{Wen1989a}. The hopping amplitudes in (a) are chosen such
    that there is $\pi$ flux through the squares and $\pi/2$ flux
    through triangles. The band structure of this parton ansatz
    features two bands separated by a gap with Chern number $\pm 1$.
  }
  \label{fig:csl}
\end{figure}

In contrast to chiral magnetic orderings, a CSL is a chiral gapped
quantum state of matter whose low-energy degrees of freedom are
described by an effective Chern-Simons theory. We will explain the
latter statement in more detail below. One way of constructing this
phase is via a parton ansatz that breaks time-reversal symmetry as
proposed by Ref.~\cite{Wen1989a}. Consider complex hopping amplitudes
in the parton ansatz \cref{eq:partonmeanfieldhamiltonian}. The hopping
terms are then chosen such that we obtain a band structure where the
valence band has non-zero Hall conductivity $\sigma_{xy} \neq 0$,
\begin{equation}
  j_y = \sigma_{xy}E_x,
\end{equation}
where $j_y$ denotes the Hall current as a response to the applied
electric field $E_x$. The famous TKNN formula \cite{Thouless1982}
tells us that the Hall conductivity is quantized and proportional to
the integer Chern number of the occupied
bands. Ref.~\cite{Wen1989a}~proposed to choose a parton ansatz with a
gapped band structure where the bands have non-zero Chern number. The
specific ansatz proposed by Ref.~\cite{Wen1989a} is shown in
\cref{fig:csl}.
%%%%%%%%%%%%%%%%%%%%%%%%%%%%%%%%%%%%%%%%%%%%%%%%%%%%%%%%%%%%%%%%%%
%%%%%%%%%%%%%%%%%%%%%%%%%%%%%%%%%%%%%%%%%%%%%%%%%%%%%%%%%%%%%%%%%%
The unit cell of the square lattice is enlarged to two sites and the
fluxes of the $\chi_{ij}$ are chosen such that there is $\pi$ flux
through the squares and $\pi/2$ flux through triangles. The band
structure of this ansatz consists of two bands separated by a finite
gap with Chern number $\pm 1$. The ground state of the free parton
Hamiltonian is obtained by filling up the valence band with spin up
and spin down partons. The ansatz itself is not translationally
invariant. Nevertheless, there are still gauge degrees of freedom for
choosing a phase locally, i.e. the gauge transformation,
\begin{equation}
  \label{eq:partongaugesymmetry2}
  c_{i\alpha}^\dagger \rightarrow e^{i \theta_i}  c_{i\alpha}^\dagger ,
\end{equation}
leaves the ansatz invariant. A translation in the $x$-direction
followed by the gauge transformation,
\begin{equation}
  \label{eq:partongaugesymmetrychiral}
  c_{i\alpha}^\dagger \rightarrow (-1)^{i_x} c_{i\alpha}^\dagger ,
\end{equation}
restores the original ansatz. Thus, the projected wave function is
translationally invariant. Nevertheless, the flux through elementary
plaquettes or triangles cannot be altered by a gauge transformation
like \cref{eq:partongaugesymmetry2}. Hence, the state with
$\pi/2$-flux through the triangles is not gauge-equivalent to the
state with $-\pi/2$-flux. Time-reversal symmetry is thus explicitly
broken by this state.

Since the parton band structure is gapped, an effective theory of the
gauge fields $a_{ij}$ and $\mu_i$ in the parton mean-field Hamiltonian
\cref{eq:partonfirstordermf} can be obtained by integrating out the
parton fields in the path integral formulation
\cref{eq:partonfieldintegral}. The resulting theory is a pure gauge
theory. Taking the proper continuum limit it can be shown that the
effective action is given by~\cite{Wen1989a}
\begin{equation}
  \label{eq:csleffectivetheory}
  S = \int\text{d}^3x \frac{1}{2}\sigma_{xy}a_{\mu}\partial_\nu a_\lambda
  \epsilon_{\mu\nu\lambda} + \mathcal{O}(1/g^2),\quad \mu = 0,1,2.
\end{equation}
Here $a_0$ and $a_{1,2}$ denote the continuum limits of the lattice
gauge fields $\mu_i$ and $a_{ij}$ respectively and the neglected terms
are of order $1/g^2$, where $g$ is proportional to the gap of the
parton spectrum. The action \cref{eq:csleffectivetheory} is called a
Chern-Simons theory.

% The original ansatz proposed by~\cite{Wen1989a} for the chiral spin
% liquid state on a square lattice is shown in \cref{fig:csl}.  The
% pairing terms $\Delta_{ij}$ and $a_0^i$ in
% \cref{eq:partonmeanfieldhamiltonian} zero and only complex hopping
% terms $\chi_{ij}$ are considered.

% Another famous kind of spin liquid wave functions has been proposed
% by Kalmeyer and Laughlin~\cite{Kalmeyer1987,Kalmeyer1989} called
% \textit{chiral spin liquid}\index{chiral spin liquid}, short CSL.
% Similarly as the RVB wave functions, this state has been proposed as
% a candidate state for the ground state of the triangular lattice
% spin $1/2$ Heisenberg antiferromagnet.

Chern-Simons theory prominently occurs in the theory of the fractional
quantum Hall effect also as a low-energy effective field theory
describing the plateaux of the Hall resistivity.  Chiral spin liquids
can, therefore, be considered as a spin analog of FQHE wave
function. Response functions and quasiparticle statistics are encoded
in the effective Chern-Simons theory~\cref{eq:csleffectivetheory}. The
Chern-Simons theory for the ansatz in \cref{fig:csl} supports semionic
statistics, i.e. braiding two parton excitations yields a statistical
phase of $\pi/2$.  We will study analogous states for triangular and
kagome lattices in
\cref{sec:paperkagome,sec:papertriangular,sec:sunchiral}. Semionic
spinon statistics implies a twofold degenerate ground state for
periodic boundary conditions~\cite{Wen1989b}. This degeneracy can be
observed well in numerical Exact Diagonalization studies.

Historically, the first construction of the CSL phase by Kalmeyer and
Laughlin~\cite{Kalmeyer1987} is in close analogy to the Laughlin wave
functions of the fractional quantum Hall effect.  For a given spin
configuration $\ket{\sigma_1\ldots\sigma_{N}}$ with
$S^z_{\text{tot}} = 0$ let $(x_i, y_i)$ be the positions of the $N/2$
up spins $\ket{\uparrow}$. The chiral spin liquid wave function as
defined by Ref.~\cite{Kalmeyer1987} is given by its coefficients
$\braket{\sigma_1\ldots\sigma_{N}|\text{CSL}}$ in the $S^z$ basis,
\begin{equation}
  \label{eq:klcsl}
  \braket{\sigma_1\ldots\sigma_{N}|\text{CSL}} =
  \prod_{j<k} (z_j - z_k)^2\exp\left\{ -\frac{1}{4l_0^2}
    \sum\limits_{i=1}^{N/2}|z_i|^2\right\},
\end{equation}
where $z_i = x_i + i y_i$ is the complex coordinate of the $i$-th up
spin and $l_0$ is called the magnetic length. This corresponds to the
bosonic Laughlin state~\cite{Laughlin1983} at filling fraction
$\nu=1/2$ of the fractional Quantum Hall effect.

This CSL state \cref{eq:klcsl} can be proven to be a singlet
state~\cite{Kalmeyer1989}, which is not obvious at first sight. Spin
correlations,
\begin{equation}
  \label{eq:klcslspincorrs}
  \braket{\text{CSL}| \bm{S}_i\cdot\bm{S}_j| \text{CSL}},
\end{equation}
can be computed numerically and have been shown to decay exponentially
with distance. Quasihole wave functions can be regarded as excitations
of the CSL wave function. Their semionic statistics can then directly
be proven~\cite{Kalmeyer1989}. It is understood, that the
Kalmeyer-Laughlin construction~\cite{Kalmeyer1987} yields the same CSL
phase as the parton construction by Ref.~\cite{Wen1989a}.

%%% Local Variables:
%%% mode: latex
%%% TeX-master: "../../thesis"
%%% End:

%auto-ignore
\subsection{Dirac Spin Liquids}
\label{sec:diracspinliquids}

\begin{figure}[t!]
  \centering
  \begin{minipage}[c]{0.45\textwidth}
    \centering
    \includegraphics[width=0.7\textwidth]{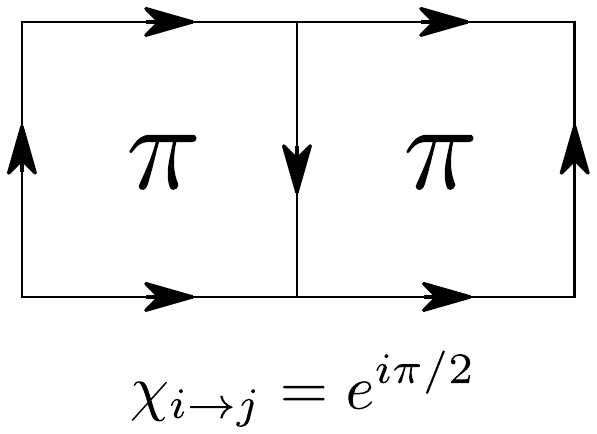}
  \end{minipage}%
  \quad
  \begin{minipage}[c]{0.45\textwidth}
    \centering
    \includegraphics[width=0.6\textwidth]{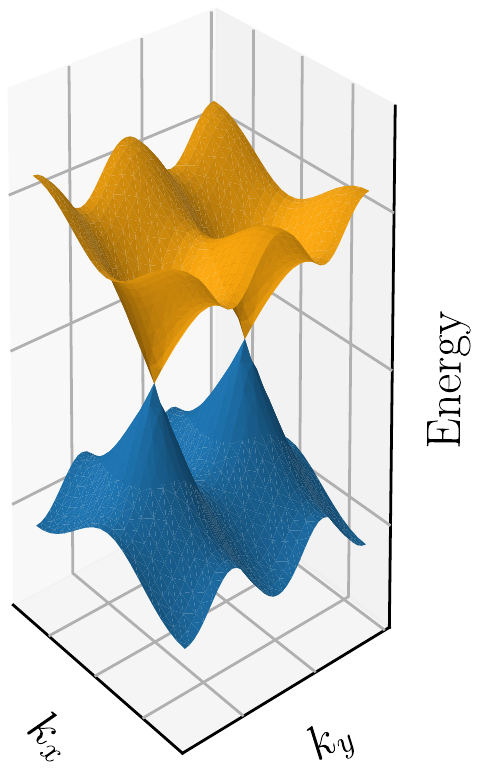}
  \end{minipage}
  \\
  \begin{minipage}[t]{0.45\linewidth}
    \subcaption{
      \label{fig:dsl:ansatz}}
  \end{minipage}
  \quad
  \begin{minipage}[t]{0.45\linewidth}
    \subcaption{
      \label{fig:dsl:bandstructure}}
  \end{minipage}

  \caption{Dirac spin liquids. (a) parton ansatz of the $\pi$-flux
    state in~\cite{Affleck1988,Marston1989}. The phases in the hopping
    parameters $\chi_{ij}$ are chosen that each plaquette has
    $\pi$-flux. (b) band structure of the $\pi$-flux parton
    ansatz. There are gapless excitations at the Fermi level with two
    distinct Dirac cones.}
  \label{fig:dsl}
\end{figure}

Another prominent type of mean field parton ansätze are Dirac spin
liquids, sometimes also called U($1$) or algebraic spin liquids. Their
parton band structure contains Dirac cones at the Fermi level, similar
to the graphene band structure. This kind of ansatz has first been
proposed by~\cite{Affleck1988,Marston1989}. The specific choice
in~\cite{Affleck1988,Marston1989} is called the $\pi$-flux state. Its
mean field parameters and band structure are shown in \cref{fig:dsl}.

In contrast to the chiral spin liquid ansatz in \cref{fig:csl} the
parton band structure is gapless. Consequently, the partons cannot
safely be integrated out as for the chiral spin liquids. The effective
low energy theory of the ansatz is described by $2\times2$ fermions,
two for spin up/down times two for the Dirac cones of the band
structure, which are coupled to the U($1$) gauge fields $a_{ij}$ and
$\mu_i$ in \cref{eq:partonfirstordermf}. The behavior of such gauge
theories is not fully understood as of today. It is therefore not
clear whether the gapless excitations are stable with respect to the
gauge fluctuations. It has has been shown that the Dirac spin liquid
indeed exhibits gapless excitations with algebraically decaying
correlation functions~\cite{Rantner2001}. It is thus an example of a
gapless state without long-range order. The Dirac spin liquid has been
proposed as a wave function able to describe deconfined
criticality~\cite{Hermele2005}, thus a wave function describing a
quantum critical point. Another proposal suggests that the Dirac spin
liquid may actually yield an extended phase in two
dimensions~\cite{Hermele2004}, at least if the number of Dirac cones
or the number of flavors (up/down) is increased. This can be viewed as
an analogy to the Heisenberg spin-$1/2$ chain in one dimension.

Similar states have been proposed on the kagome and the triangular
lattice~\cite{Iqbal2011,Iqbal2016} for describing spin liquid phases
whose nature is still debated.

%%% Local Variables:
%%% mode: latex
%%% TeX-master: "../../thesis"
%%% End:

\newpage
\chapter{Methods}
\epigraph{ \parbox{6.2cm}{Set the controls for the heart of the
    sun.}}{Pink Floyd}
  \label{sec:methods}
  \section{Exact Diagonalization}
  \label{sec:ed}
  %auto-ignore

% \section{asdf}
Solving the stationary Schr{\"o}dinger equation,
\begin{equation}
  \label{eq:stationaryschrodinger}
  H \ket{\psi} = E\ket{\psi},
\end{equation}
of nonrelativistic quantum mechanics amounts to computing all
eigenstates $\ket{\psi}$ and energies $E$ of the Hamiltonian $H$.  For
problems without an exact analytical solution, the diagonalization of
the Hamiltonian could be performed numerically. This method to
numerically compute properties of a quantum system is called
\textit{Exact Diagonalization} (ED). Since eigenvectors and
eigenvalues of matrices can be obtained up to high precision with
modern algorithms, ED is a powerful and reliable numerical
tool. Computing every single eigenvalue and eigenstate of the
Hamiltonian completely solves the stationary Schr{\"o}dinger equation,
such that all physical quantities can in principle be derived from
this solution. This numerical procedure is called \textit{full
  diagonalization}. In order to diagonalize the Hamiltonian this way,
a variety of numerical algorithms can be
employed~\cite{saad1992numerical}. Although there are many differences
between those, the computational effort of these methods is of the
order $\mathcal{O}(D^3)$, where $D$ is the linear dimension of the
matrix.  On present-day supercomputers matrices with linear dimension
up to approximately $N\approx 10^6$ can be fully diagonalized.

For quantum many-body systems the Hilbert spaces $\mathcal{H}$ are
typically fermionic or bosonic Fock spaces or tensor products of local
spin systems. The dimension of these kinds of Hilbert spaces increases
exponentially with the number of particles considered. Thus, full
diagonalizations become infeasible if the particle number is too
large. If one is only interested in the ground state or several
low-lying excitations iterative methods can be applied that yield
approximate eigenvalues at the boundary of the spectrum. The most
prominent iterative algorithm for computing extremal eigenvalues and
eigenvectors of Hermitian matrices is the Lanczos
algorithm~\cite{Lanczos1950}, which we will discuss in
\cref{sec:lanczosmethod}. For sparse matrices, the complexity of the
computation for achieving a fixed precision may effectively reduce to
$\mathcal{O}(D)$. This allows for computations with orders of
magnitude larger Hilbert spaces. For ED in condensed matter systems
Hilbert space dimensions of $D \approx 10^{11}$ can currently be
attained.

In this chapter, we introduce the basic principles of the ED
method. In \cref{sec:codinghilbertspace} we explain how to represent
Hilbert spaces and operators in a computer. We discuss basic
principles and some practical issues about the Lanczos algorithm in
\cref{sec:lanczosmethod}. Symmetries and the use of symmetry adapted
wave functions for block diagonalization of matrices are discussed in
\cref{sec:symsnadaptesdwaves}. We present novel algorithms for working
with symmetry adapted wave functions and efficient distributed memory
parallelization in \cref{sec:largescaleed}. There, we also present
benchmarks for our implementation of these ideas.

\subsection{Representing Hilbert spaces and operators}
\label{sec:codinghilbertspace}

The first step towards practical numerical computations is to choose a
basis of the Hilbert space. For spin systems with local dimension $d$
we may choose the canonical basis of tensor products of local
spins. Spin configurations are represented by an integer value via its
$d$-ary representation. For instance, the spin-$1/2$ state of four
particles $\ket{\uparrow \uparrow \downarrow \uparrow}$ is encoded as
\begin{align}
  \label{eq:exspinintencoding}
  \ket{\uparrow \uparrow \downarrow \uparrow}  \rightarrow  \ket{1101} \rightarrow (1101)_2 = (13)_{10}.
\end{align}
In general, we encode a spin configuration
$\ket{\bm{\sigma}} =\ket{\sigma_1, \dots, \sigma_N}$ on $N$ lattice
sites by
\begin{align}
  \label{eq:numberforspinconfig}
  \integer(\ket{\bm{\sigma}})\equiv \sum\limits_{k=1}^N \sigma_k d^{N - k},
\end{align}
where $d$ denotes the local dimension of the Hilbert space.
Bases of fermionic or bosonic Hilbert spaces may be encoded
correspondingly.

Given a basis $\ket{\bm{\sigma}_n}$, $n=1,\ldots,D$, of the Hilbert
space we can compute the matrix elements
$\braket{\bm{\sigma}_n|H|\bm{\sigma}_m}$ of the Hamiltonian. For
example, we can compute the matrix elements of a spin exchange term on
a spin-$1/2$ state
\begin{equation}
  \label{eq:exchangeapplyspin}
  \frac{1}{2}(S^+_2 S^-_3 + S^-_2 S^+_3)\ket{(13)_{10}} = 
  \frac{1}{2}(S^+_2 S^-_3 + S^-_2 S^+_3)
  \ket{\uparrow \uparrow \downarrow \uparrow} =
  \frac{1}{2}\ket{\uparrow \downarrow \uparrow \uparrow} =
  \frac{1}{2}\ket{(11)_{10}} .
\end{equation}
Thus, we have
\begin{equation}
  \label{eq:exchangespinmatelement}
  \braket{(11)_{10}| \frac{1}{2}(S^+_2 S^-_3 + S^-_2 S^+_3)| \ket{(13)_{10}} }
  = \frac{1}{2}.
\end{equation}
We can then store all the elements in a numerical $D \times D$ matrix
for full diagonalization. In case we want to apply iterative methods
we may also store the elements in a sparse-matrix format or compute
matrix-vector products on-the-fly without storing any matrix elements.

%%% Local Variables:
%%% mode: latex
%%% TeX-master: "../../thesis"
%%% End:

  %auto-ignore

\subsection{The Lanczos algorithm}
\label{sec:lanczosmethod}

Iterative methods drastically reduce the computational effort in
computing extremal eigenvalues and eigenvectors,
\begin{equation}
  \label{eq:eigenvalueeq}
  A\ket{v}=\lambda\ket{v},
\end{equation}
of large matrices $A \in \mathbb{C}^{D\times D}$, where $D$ is the
dimension of the matrix. Typically a series of matrix-vector
multiplications is performed to yield improving approximations to
eigenvalues and eigenvectors. Krylov subspace methods such as the
Lanczos algorithm~\cite{Lanczos1950} for Hermitian matrices are
powerful and efficient iterative methods.

We give a short rationale for this kind of algorithms where we
restrict ourselves to the case when $A$ is Hermitian, $A^\dagger =
A$. The Rayleigh coefficient of a matrix $A$ is defined as
\begin{equation}
  \label{eq:rayleighcoeff}
  r_A(\ket{v}) \equiv \frac{\braket{v|A|v}}{\braket{v|v}},
\end{equation}
for all vectors $\ket{v} \in V \equiv \mathbb{C}^D$.  The Rayleigh
coefficient evaluated at an eigenvector $\ket{\lambda}$ yields the
corresponding eigenvalue $\lambda$,
\begin{equation}
  \label{eq:rayleighcoefflambda}
  r_A(\ket{\lambda}) = \lambda.
\end{equation}
Minimizing the Rayleigh coefficient corresponds to finding a minimal
eigenvalue $\lambda_0$ of the matrix $A$
\begin{equation}
  \label{eq:minimizerayleigh}
  \lambda_0 = \min\limits_{v \in V} r_A(\ket{v}).
\end{equation}
The minimal eigenvalue can be approximated by minimizing the Rayleigh
coefficient in an $n$-dimensional subspace

\begin{equation}
  \label{eq:rayleighsubspace}
  V_n = \spn\{\ket{v_1}, \ldots, \ket{v_{n}}\} \subseteq V.
\end{equation}
To further improve the approximation we can apply the gradient descent
method. The direction of steepest descent for Hermitian matrices $A$
is given by
\begin{equation}
  \label{eq:rayleighgradient}
  -\nabla r_A(\ket{v}) = \frac{2}{\braket{v|v}}
  ( r_A(\ket{v}) \ket{v} - A\ket{v}).
\end{equation}
The Krylov space of order $n$ of a matrix $A$ and a starting vector
$\ket{V_1}$ is defined as
\begin{equation}
  \label{eq:krylowspace}
  \mathcal{K}_n(A, \ket{v_1}) \equiv \spn{\left\{\ket{v_1}, A\ket{v_1}, \ldots, A^n\ket{v_1} \right\}}.
\end{equation}
The direction of steepest descent \cref{eq:rayleighgradient} in the
$n$-th Krylov space $\mathcal{K}_n(A, \ket{v_1})$ is an element of the
next $(n+1)$-th Krylov space $\mathcal{K}_{n+1}(A, \ket{v_1})$. Thus,
minimizing the Rayleigh coefficient in higher order Krylov spaces
increases the accuracy of the approximation in
\cref{eq:minimizerayleigh} by means of the gradient descent method.

In general the vectors $\ket{v_1}, A\ket{v_1}, \ldots, A^n\ket{v_1}$
are not orthonormal. The Lanczos method \cite{Lanczos1950} iteratively
constructs an orthonormal basis of the Krylov spaces. Given an
orthonormal basis $\ket{v_1}, \ket{v_2}, \ldots, \ket{v_{n}}$ of the
$n$-th Krylov space we can use Gram-Schmidt orthogonalization to
construct an orthonormal basis of the $(n+1)$-th Krylov space. The
next $(n+1)$-th orthonormal basis vector is given by
\begin{equation}
  \label{eq:lanczosgramschmidt}
  \ket{v_{n+1}} = \frac{\ket{\hat{v}_{n+1}}}{\norm{\hat{v}_{n+1}}}, \text{ where }
  \ket{\hat{v}_{n+1}} = A \ket{v_{n}} - \sum\limits_{i=0}^{n}  \braket{v_{n} | A | v_i}\ket{v_i}.
\end{equation}
Using the orthogonality of the vectors $\ket{v_k}$ and the hermitecity
$A^\dagger = A$, this procedure simplifies to (see
e.g. \cite{Allaire2008})
\begin{equation}
  \label{eq:lanczosrecursion}
  \ket{v_{n+1}} = \frac{\ket{\hat{v}_{n+1}}}{\norm{\hat{v}_{n+1}}} \text{ with }
  \ket{\hat{v}_{n+1}} = A \ket{v_{n}} -
  \braket{v_{n} | A | v_{n}}\ket{v_{n}} - \norm{\hat{v}_{n}}\ket{v_{n-1}}.
\end{equation}
The prescription \cref{eq:lanczosrecursion} is called the
\textit{Lanczos recursion}\index{Lanczos recursion}. The vectors
$\ket{v_k}$ are called the \textit{Lanczos vectors}\index{Lanczos
  vectors}. We introduce the usual abbreviations found in literature,
\begin{equation}
  \label{eq:lanczosalphabeta}
  \alpha_k \equiv \braket{v_k | A | v_k} , \quad \quad
  \beta_k \equiv \norm{\hat{v}_{k}}.
\end{equation}
With these abbreviations the Lanczos recursion reads
\begin{equation}
  \label{eq:lanczosrecursionabbrev}
  \ket{v_{n+1}} = \frac{\ket{\hat{v}_{n+1}}}{\beta_{n+1}}, \text{ where }
  \ket{\hat{v}_{n+1}} = A \ket{v_{n}} -
  \alpha_n\ket{v_{n}} - \beta_n\ket{v_{n-1}}.
\end{equation}
By defining the matrix of Lanczos vectors
\begin{equation}
  \label{eq:lanczosmatrixdef}
  V_n = \left( v_1 | \cdots | v_{n} \right ),
\end{equation}
we can write the recursion \cref{eq:lanczosrecursionabbrev} as
\cite{Allaire2008}
\begin{equation}
  \label{eq:lanczoserror}
  A V_n = V_n T_n + \ket{\hat{v}_{n+1}}e_k^T,
\end{equation}
where $e_k$ is the $k$-th canonical basis vector of $\mathbb{R}^k$ and
the matrix $T_n$, called the $n$-th T-matrix, is a tridiagonal matrix
given by
\begin{equation}
  \label{eq:tmatrix}
  T_n =
  \begin{pmatrix}
    \alpha_1 & \beta_1  & 0        &       & \cdots & 0 \\
    \beta_1  & \alpha_2 & \beta_2  & 0      &   & \vdots  \\
    0        & \beta_2  & \ddots   &        &        & \\
    &          &          &        & \ddots & 0 \\
    \vdots   &          &          & \ddots & \alpha_{n-1} & \beta_{n-1}\\

    0   &  \cdots &   & 0       & \beta_{n-1}       & \alpha_n \\
                     
  \end{pmatrix}.
\end{equation}
\begin{figure}[t]
  \centering
  \begin{minipage}[t]{0.48\textwidth}
    \includegraphics[width=\textwidth]{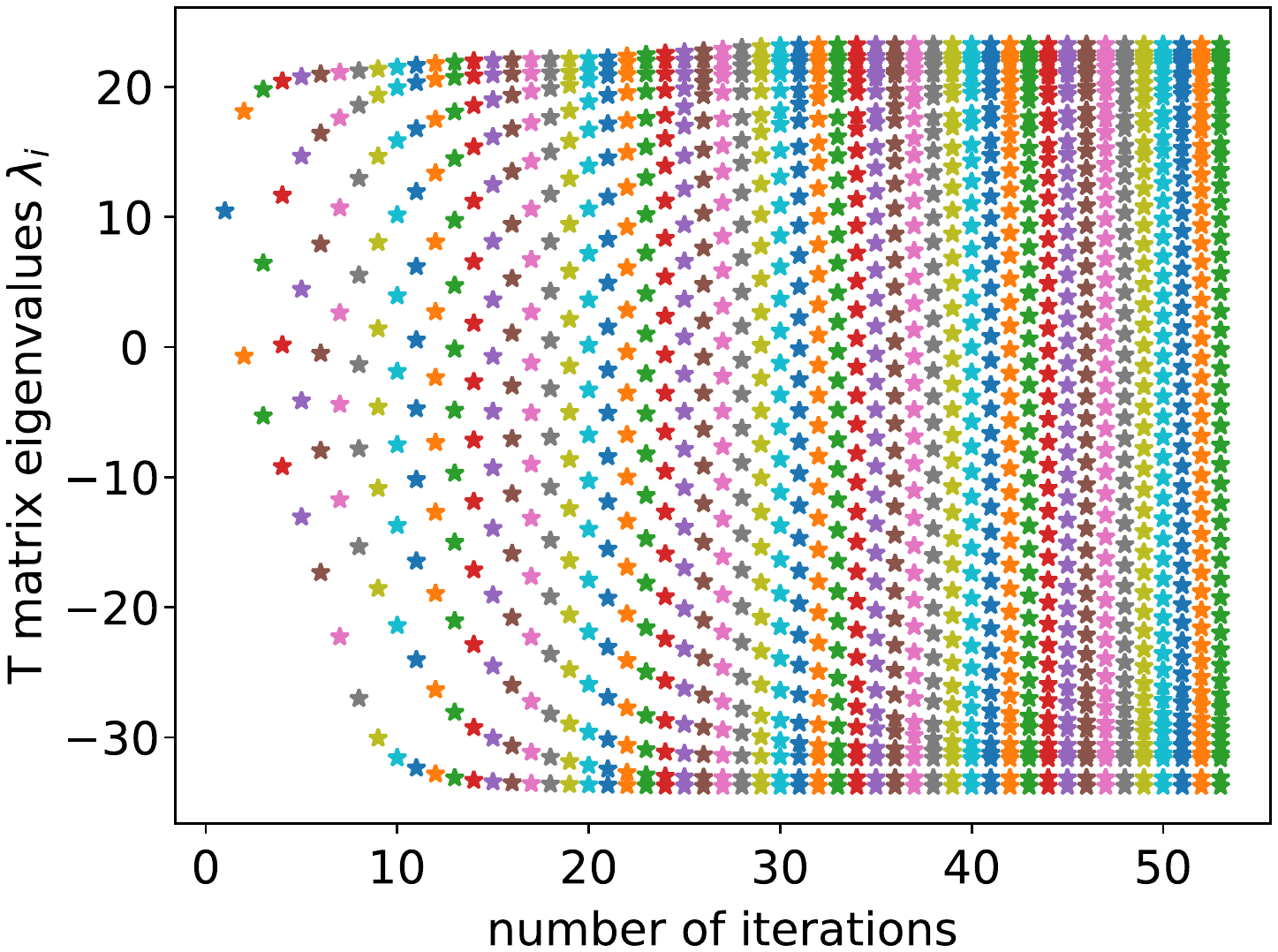}
  \end{minipage} \quad
  \begin{minipage}[t]{0.48\textwidth}
    \includegraphics[width=\textwidth]{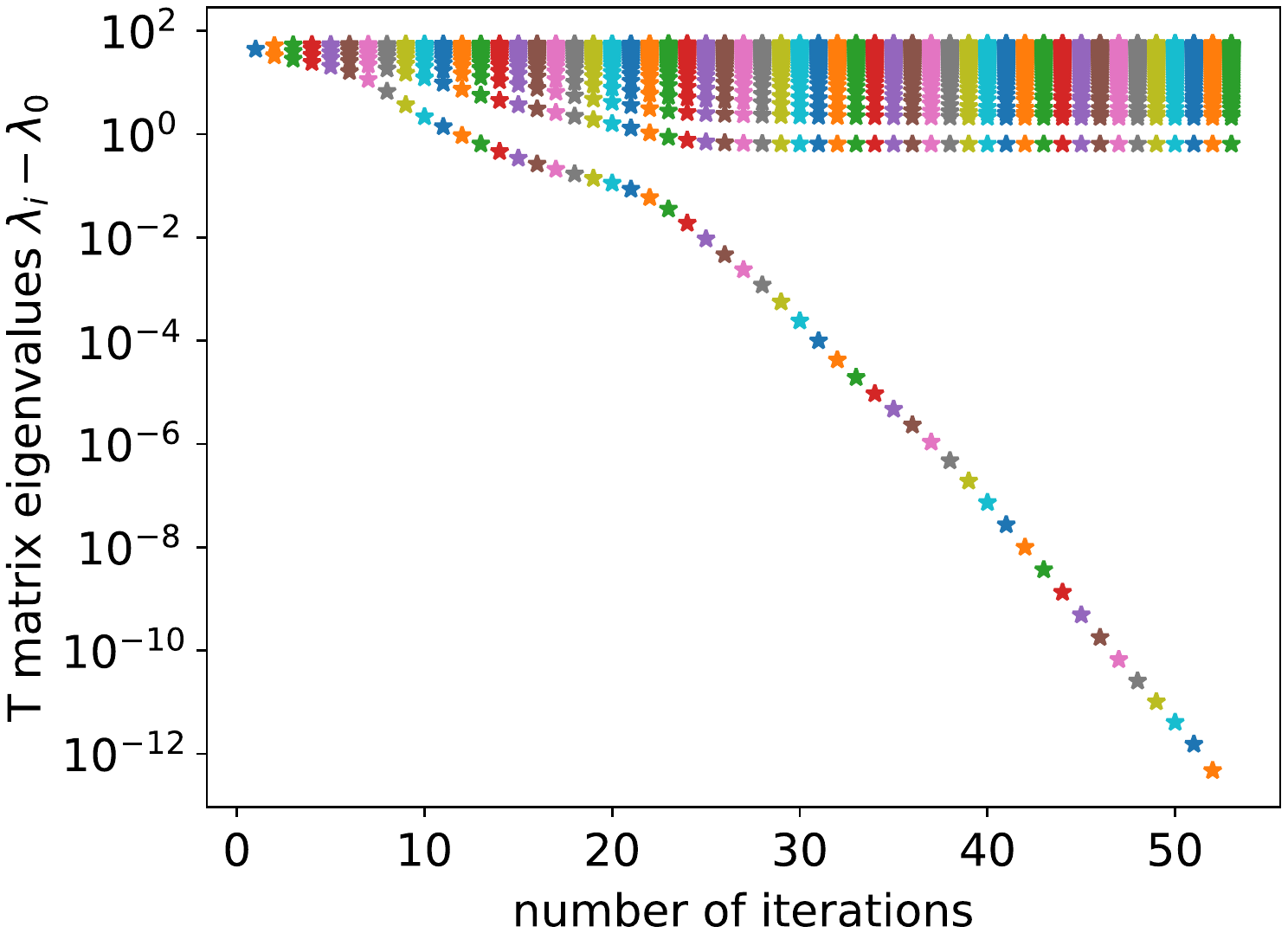}
  \end{minipage}\\
  \begin{minipage}[t]{0.48\textwidth}
    \subcaption{\label{fig:lanczosconvergence50:normal}}
  \end{minipage} \quad
  \begin{minipage}[t]{0.48\textwidth}
    \subcaption{\label{fig:lanczosconvergence50:log}}
  \end{minipage}
  \caption{(a) Convergence of the T-matrix eigenvalues. The matrix $A$
    diagonalized is the Hamiltonian of the spin-$1/2$ Heisenberg
    antiferromagnetic model on a $50$ site square lattice in the
    ground state symmetry sector. The dimension of this sector is
    $D\approx 3\cdot 10^{11}$. (b) Logarithmic plot of
    $\lambda_i - \lambda_0$. The speed of convergence of the ground
    state energy is exponential and a precision of $10^{-12}$ is
    reached after $53$ Lanczos iterations.  }
  \label{fig:lanczosconvergence50}
\end{figure}
Since $T_n = V_n^\dagger A V_n$, the T-matrix can be interpreted as
the projection of the matrix $A$ onto $\mathcal{K}_n(A,
\ket{v_1})$. The T-matrix $T_n$ thus approximates the matrix $A$ on
the $n$-th Krylov space. By $\lambda_{k,T_n}$ and
$\ket{\lambda_{k,T_n}}$ we denote the eigenvalues and eigenvectors of
the $n$-th T-matrix. The sequence $(\lambda_{k,T_n})_{n\in\mathbb{N}}$
converges to an eigenvalue $\lambda_{k}$ of $A$ and the sequence
$(V_n\ket{\lambda_{k,T_n}})_{n\in\mathbb{N}}$ converges to the
corresponding eigenvector $\ket{\lambda_k}$. The speed of convergence
is typically exponential. Thus, already a small number of iterations
yields a good approximation to the eigenvalues and eigenvectors of
$A$.  To converge for the minimal eigenvalue in condensed matter
systems in this thesis typically $50-200$ iterations suffice. The
T-matrix is then diagonalized by a full diagonalization
algorithm. Yet, a detailed analysis of the convergence behavior is
subtle and we refer the reader to
Refs.~\cite{Kuczynski1992,Simon1984}. A practical example of this
convergence is shown in \cref{fig:lanczosconvergence50} for a matrix
$A$ of dimension $D \approx 3\cdot 10^{11}$.

There are several practical issues concerning the Lanczos iteration.
First of all, there are several ways of how to actually compute a
single Lanczos iteration which are compared in
Refs. \cite{Paige1972,Paige1980,Cullum1985}. If the matrix $A$ is not
stored in memory, thus matrix-vector multiplications are computed
on-the-fly, the main memory requirement of the Lanczos algorithm is
the storage of the Lanczos vectors. We make use of two simple variants
which require storing either $2$ or $3$ vectors depending on whether
the matrix $A$ is real symmetric or complex Hermitian, cf.
\cref{alg:lanczosreal} and \cref{alg:lanczoscomplex}.

\begin{algorithm}
  \caption{Real-symmetric Lanczos step \cite{Paige1980}, only the
    current Lanczos vector $\ket{v_n}$ and a temporary vector
    $\ket{w}$ are stored. Applicable, if $A$ is real and symmetric. In
    the first step $\ket{w}$ is initialized as the zero vector. }
  \begin{algorithmic}[0]
    \If{$n = 1$} $\ket{w}\leftarrow 0, \beta_1 \leftarrow 0$ \EndIf
    \State $\ket{w} \leftarrow A\ket{v_n} - \beta_n\ket{w}$ \Comment
    r.h.s. $\ket{w} = \ket{v_{n-1}}$ \State
    $\alpha_n \leftarrow \braket{w|v_n}$ \State
    $\ket{w} \leftarrow \ket{w} - \alpha_n\ket{v_n}$ \State
    $\beta_{n+1} \leftarrow \sqrt{\braket{w|w}}$ \State Swap
    $\ket{v_n} \leftrightarrow \ket{w}$ \State
    $\ket{v_{n+1}} \leftarrow \frac{1}{\beta_{n+1}}\ket{v_{n}}$
  \end{algorithmic}
  \label{alg:lanczosreal}
\end{algorithm}
% \todo{why cant I use the ``real Lanczos'' for complex calculations?}
\begin{algorithm}
  \caption{Complex-Hermitian Lanczos step, storage of two Lanczos
    vectors $\ket{v_n}$ and $\ket{v_{n-1}}$ and a temporary vector
    $\ket{w}$ is required. }
  \begin{algorithmic}[0]
    \If{$n = 1$} $ \beta_1 \leftarrow 0$ \EndIf \State
    $\ket{w} \leftarrow A\ket{v_n}$ \State
    $\alpha_n \leftarrow \braket{w|v_n}$ \State
    $\ket{w} \leftarrow \ket{w} - \alpha_n\ket{v_n} -
    \beta_n\ket{v_{n-1}}$ \State
    $\ket{v_{n-1}} \leftarrow \ket{v_{n}}$ \State
    $\beta_{n+1} \leftarrow \sqrt{\braket{w|w}}$ \State
    $\ket{v_{n+1}} \leftarrow \frac{1}{\beta_{n+1}}\ket{w}$
  \end{algorithmic}
  \label{alg:lanczoscomplex}
\end{algorithm}

Another issue concerning the convergence of the Lanczos algorithm is
the loss of orthogonality of the Lanczos vectors due to finite
precision machine arithmetic. In several applications an explicit
reorthogonalization of the Lanczos vectors is necessary. A survey of
these methods is given by Ref. \cite{Cullum1985}. In practice, if one
is only interested in a ground state or first excited state of some
specific condensed matter system, it turns out that this
reorthogonalization is not necessary. Also, reorthogonalization
increases the memory requirements. Therefore, we always employ the
simple Lanczos recursion without reorthogonalization in this thesis.

The Lanczos iterations construct the T-matrix which yields approximate
eigenvalues $\lambda_{k,T_n}$. In case we are interested in
approximate eigenvectors of $A$ we have to compute
$V_n\ket{\lambda_{k,T_n}}$.  In order not to store all Lanczos vectors
in $V_n$ we first run the Lanczos algorithm to compute the T-matrix
eigenvectors $\ket{\lambda_{k,T_n}}$ and then, in a second run,
starting from the same initial vector $\ket{v_1}$, compute the linear
combination $V_n\ket{\lambda_{k,T_n}}$. This requires one additional
vector to be stored in memory.

%%% Local Variables:
%%% mode: latex
%%% TeX-master: "../../thesis"
%%% End:

  %auto-ignore

\subsection{Symmetries and symmetry-adapted wave functions}
\label{sec:symsnadaptesdwaves}

Apart from being most fundamental properties of a system, symmetries
can be employed to divide an ED calculation into smaller pieces by
block diagonalizing the Hamiltonian. In this chapter, we review the
notion of quantum numbers and symmetry-adapted wave functions in a
group theoretical setting. For basics on group representation theory
in quantum mechanics, we refer to~\cite{Landau1988}. We then explain
how, in principle, numerical calculations in a symmetrized basis can be
performed.

\subsubsection{Quantum numbers and degenerate eigenstates}
A symmetry of a Hamiltonian $H$ is an operator ${g}$ that commutes
with the Hamiltonian,
\begin{equation}
  \left [ H, {g}\right ] = 0.
\end{equation}
Denote by $\ket{n, \alpha}$ the eigenfunctions of $H$ satisfying
\begin{equation}
  H \ket{n, \alpha} = E_n \ket{n, \alpha},
\end{equation}
where the index $\alpha$ denotes the different degenerate eigenstates
with eigenvalue $E_n$.  Due to
\begin{equation}
  H {g}\ket{n, \alpha} =  {g} H\ket{n, \alpha} =
  E_n {g}\ket{n, \alpha},
\end{equation}
the symmetry ${g}$ leaves the eigenspaces for a given eigenvalue
$E_n$ invariant.  A group $\mathcal{G}$ of symmetries defines a
representation $\rho_n$ on the degenerate eigenspaces via
\begin{equation}
  \label{eq:basisadaptedrepresentationdef}
  \Gamma^n: g \mapsto  \Gamma(g),
\end{equation}
where
$\Gamma(g)_{\alpha \beta} \equiv \left(\braket{n,\alpha| {g} |
    n,\beta}\right)_{\alpha \beta}$ are matrices with dimension equal
to the degeneracy of the eigenvalue $E_n$. Thus, every degenerate set
of eigenvalues can be labeled by irreducible representations of the
symmetry group, its \textit{quantum numbers}. Common examples are
momentum quantum numbers for translational symmetry, total spin
quantum numbers for spin rotational SU($2$) symmetry for spin systems
or parity for inversion symmetry.

\begin{figure}[t]
  \centering \includegraphics[width=\textwidth]{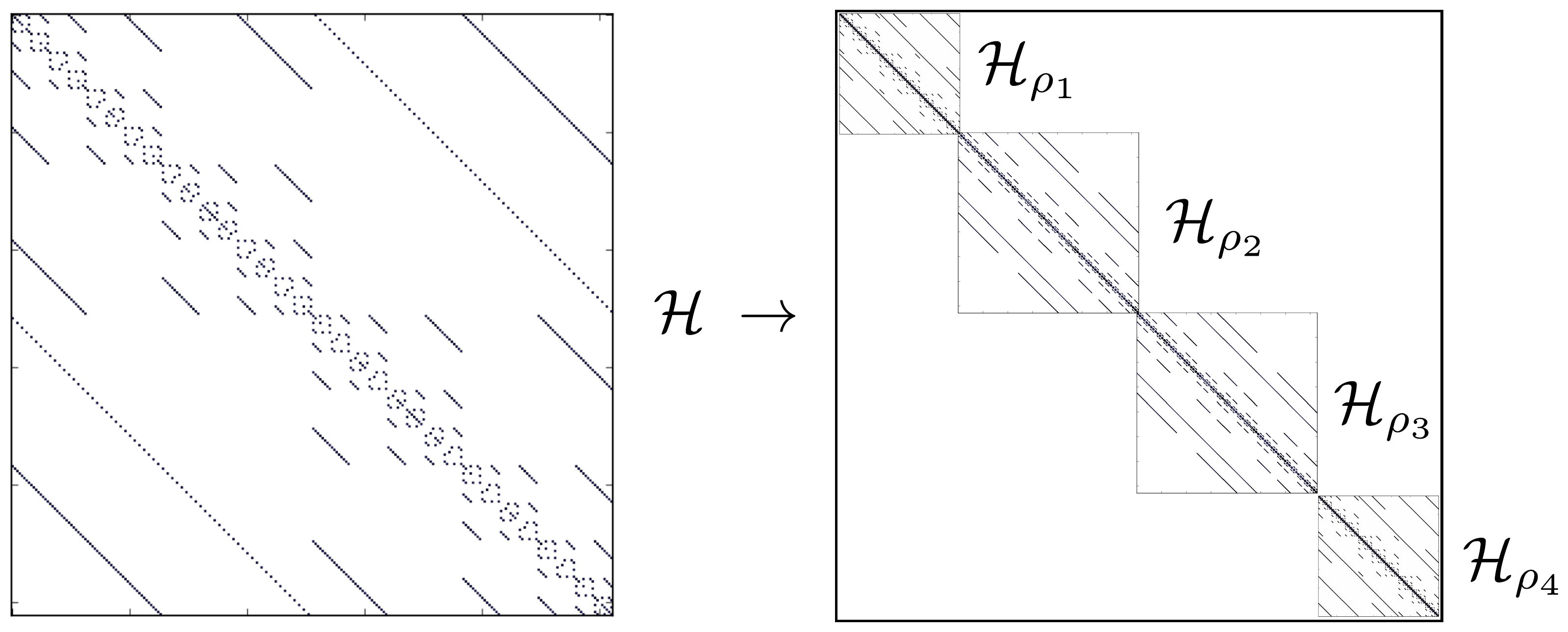}
  \caption{Block diagonalization of a Hamiltonian. The Hilbert space
    $\mathcal{H}$ is split into subspaces $\mathcal{H}_{\rho_i}$
    corresponding to an irreducible representation $\rho_i$ of the
    symmetry group. The Hamiltonian transformed to the basis of
    symmetry-adapted wave functions is block diagonal.}
  \label{fig:blockdiagonalization}
\end{figure}

A basis of the Hilbert space can be chosen such that the Hilbert space
decomposes into a direct sum of subspaces,
\begin{equation}
  \label{eq:hilbertspacedecomposition}
  \mathcal{H} = \bigoplus\limits_{\rho} \mathcal{H}_\rho,
\end{equation}
where the states in each subspace $\mathcal{H}_\rho$ transform
according to a given irreducible representation $\rho$ of the symmetry
group $\mathcal{G}$. We will work out how the exact form of these
\textit{symmetry-adapted wave functions}\index{symmetry-adapted wave
  function} and their transformation properties for different kinds of
symmetries in the following paragraphs. Importantly, the Hamiltonian
does not couple states in different irreducible representation spaces,
\begin{equation}
  \label{eq:irrepnocoupling}
  \rho \neq \sigma \Rightarrow
  \braket{\psi^\rho | H | \phi^\sigma} = 0 \text{ for } \ket{\psi^\rho} \in \mathcal{H}_\rho \text{ and } \ket{\phi^\sigma} \in \mathcal{H}_\sigma.
\end{equation}
Consequently, if we group the basis states of a given irreducible
representation together we will have block diagonalized the
Hamiltonian where every block is labeled by its irreducible
representation, as shown in \cref{fig:blockdiagonalization}. We will
now disuss the general block diagonalization principle for several
specific symmetries.

\subsubsection{Particle number conservation}
Particle number conservation is often encountered in spin systems. For
SU($2$) Heisenberg spin systems like
\begin{equation}
  \label{eq:heisenbergallover}
  H = \sum\limits_{i,j} J_{ij} \bm{S}_i\cdot\bm{S}_j,
\end{equation}
the total magnetization
\begin{equation}
  \label{eq:sztotal}
  S^z_{\text{tot}} = \sum_i S_i^z,
\end{equation}
is a conserved quantity.
% and therefore
% \begin{equation}
%   \label{eq:sznocoupling}
%   \frac{\braket{\psi|S^z_{\text{tot}}|\psi}}{\braket{\psi|\psi}} \neq
%   \frac{\braket{\phi|S^z_{\text{tot}}|\phi}}{\braket{\phi|\phi}}
%   \Rightarrow \braket{\psi | H | \phi} = 0
% \end{equation}
Grouping together states of the same $S^z_{\text{tot}}$ will therefore
block diagonalize the Hamiltonian. The total $S^z_{\text{tot}}$ labels
the irreducible representations of the continuous U($1$) symmetry,
\begin{equation}
  \label{eq:u1particlenumberconservation}
  {g}(\theta) = e^{i\theta S^z_{\text{tot}}},
\end{equation}
and the basis states transform according to
\begin{equation}
  \label{eq:u1particlenumberconservationtransformation}
  {g}(\theta)\ket{\psi} = e^{i\theta n}
  \ket{\psi}, \text{ if } S^z_{\text{tot}}\ket{\psi} = n \ket{\psi}.
\end{equation}
We see that eigenstates in the $S^z$ basis are already
symmetry-adapted wave functions. While
\cref{eq:u1particlenumberconservation} and
\cref{eq:u1particlenumberconservationtransformation} are not necessary
to understand and implement particle number conservation, their
analogs for space group symmetries are essential for constructing the
symmetry-adapted wave functions.

\subsubsection{Discrete symmetries}
Translational or point group symmetries imply (angular-)momentum
conservation. These symmetries form a finite discrete group on finite
size lattices. Following the general idea of decomposing the Hilbert
space into irreducible representations in
\cref{eq:hilbertspacedecomposition} we can use them to further
decompose out Hilbert space into momentum or point group
representation sectors. We review how to build the basis of
symmetry-adapted wave functions for a discrete symmetry group
$\mathcal{G}$. Detailed derivations can be found in
\cite{Landau1988}. Let
\begin{equation}
  \label{eq:basisadaptedrepresentationdef}
  \Gamma^\rho: g \mapsto  \Gamma^\rho(g),
\end{equation}
be an irreducible representation of the symmetry group $\mathcal{G}$
such that the representation matrices $\Gamma^\rho(g)$ are
unitary. Given an arbitrary state $\ket{\psi} \in H$ its projection
onto symmetry-adapted wave functions with irreducible representation
$\rho$ is given by~\cite{Landau1988}
\begin{align}
  \label{eq:symmetryadaptedwavefunction}
  \ket{\psi^\rho_n} \equiv \frac{1}{N_{\rho,n,\psi}}\sum\limits_{g \in \mathcal{G}}
  \left( \Gamma^\rho(g)_{nn} \right)^* {g}\ket{\psi},
\end{align}
where $N_{\rho,n,\psi}$ is a normalization constant and
$n \in \{1, \ldots, d_\rho\}$, where $d_\rho$ is the dimension of the
representation $\Gamma^\rho$. The symmetry-adapted wave function in
\cref{eq:symmetryadaptedwavefunction} now transforms according to the
$n$-th row of the representation $\Gamma^\rho$, i.e.
\begin{align}
  \label{eq:symmetryadaptedwavefunctiontrafo}
  {g}\ket{\psi^\rho_n} = \sum\limits_m \Gamma^\rho(g)_{nm} \ket{\psi^\rho_m}.
\end{align}
symmetry-adapted wave functions from different representations
$\Gamma^\rho$ are mutually orthonormal, i.e.
\begin{align}
  \label{eq:symmetryadaptedwavefunctionortho}
  \braket{\psi^\rho_n | \psi^\eta_m} = \delta_{\rho\eta}\delta_{nm}.
\end{align}
% \begin{proof} We first prove the transformation property in
%   \cref{eq:symmetryadaptedwave functiontrafo}
%   \begin{align}
%     \label{eq:symmetryadaptedwavefunctiontrafoproof}
%     {g}\ket{\psi^\rho_n} &= \frac{\todo{1}}{|\mathcal{G}|}\sum\limits_{h \in \mathcal{G}}
%                                \sum\limits_{k} \left( \Gamma^\rho(h)_{kn} \right)^*
%                                {g}{h}\ket{\psi} \\
%                              &= \frac{\todo{1}}{|\mathcal{G}|}\sum\limits_{h \in \mathcal{G}}
%                                \sum\limits_{k} \left( \Gamma^\rho(ghg^{-1})_{kn} \right)^*
%                                {g}{h}\ket{\psi} \\
%                              &= \frac{\todo{1}}{|\mathcal{G}|}\sum\limits_{h \in \mathcal{G}}
%                                \sum\limits_{m,k} \left( \Gamma^\rho(gh)_{km}\Gamma^\rho(g^{-1})_{mn} \right)^*
%                                {g}{h}\ket{\psi} \\
%                              &= \sum\limits_{m} \Gamma^\rho(g)_{nm}
%                                \frac{\todo{1}}{|\mathcal{G}|} \sum\limits_{h \in \mathcal{G}}
%                                \sum\limits_{k}\left( \Gamma^\rho(h)_{km}\right)^*
%                                {h}\ket{\psi} \\
%                              &= \sum\limits_m \Gamma^\rho(g)_{nm} \ket{\psi^\rho_m}
%   \end{align}
%   \todo{proof orthonormality and check prefactor}
% \end{proof}
In case we consider a one-dimensional representation of the symmetry
group \cref{eq:symmetryadaptedwavefunction} simplifies to
\begin{align}
  \label{eq:symmetryadaptedwavefunction1d}
  \ket{\psi^\rho} \equiv \frac{1}{N_{\rho,\psi}}\sum\limits_{g \in \mathcal{G}}
  \chi^\rho(g)^* {g}\ket{\psi},
\end{align}
with the transformation property
\begin{align}
  \label{eq:symmetryadaptedwavefunctiontrafocharacter}
  {g}\ket{\psi^\rho} = \chi^\rho(g) \ket{\psi^\rho},
\end{align}
where $\chi^\rho(g)$ denotes the character of the representation
$\Gamma^\rho$ and $N_{\rho,\psi}$ is a normalization constant.
\paragraph{Example (translational symmetries):}
We consider a spin-$1/2$ model on a $L \times L$ square lattice with
periodic boundary conditions and lattice constant $a=1$. The
translational symmetry group consists out of $L$ translations in the
$x$-direction times $L$ translations in the $y$-direction. We thus
consider the Abelian symmetry group
\begin{equation}
  \label{eq:squaresymmetrygroup}
  \mathcal{G} = \mathbb{Z}_L\times\mathbb{Z}_L.
\end{equation}
Its one-dimensional irreducible representations can be labeled by the
lattice momenta
\begin{equation}
  \label{eq:latticemomenta}
  \bm{k}_{mn} = (k_m, k_n) = (2\pi m /L, 2\pi n /L), \quad
  m,n = 0,\ldots, L-1 
\end{equation}
whose characters are given by the Bloch factors
\begin{equation}
  \label{eq:squarelatticemomenta}
  \chi_{\bm{k}_{mn}}(t_x^p \cdot t_y^q) = e^{i(k_m\cdot p + k_n\cdot
    q) },
\end{equation}
where $t_x^p \cdot t_y^q $ denotes a translation by $p$ lattice sites
in the $x$-direction and $q$ lattice sites in the $y$ direction. The
symmetry-adapted wave functions are then given by
\begin{equation}
  \label{eq:squcharform}
  \ket{\psi^{\bm{k}_{mn}}} = \frac{1}{N_{\bm{k}_{mn},\psi}}
  \sum\limits_{p,q = 1}^{L }e^{i(k_m\cdot p  + k_n\cdot q)}
  ({t}_x^p \cdot {t}_y^q) \ket{\psi}. 
\end{equation}

For a given finite discrete symmetry group all irreducible
representations $\Gamma^\rho(g)$ can in principle be worked out. From
the generic \cref{eq:symmetryadaptedwavefunction} and
\cref{eq:symmetryadaptedwavefunction1d} the symmetry-adapted
wave functions are then obtained. Discrete local symmetries like spin
flip symmetry for spin-$1/2$ systems,
\begin{equation}
  \label{eq:spinflipsymmetry}
  f = \prod_{i} \sigma^x_i,
\end{equation}
can also be considered within this framework. This symmetry defines a
$\mathbb{Z}_2$ symmetry group with an even and odd parity irreducible
representation.

Moreover, if the symmetry group is a direct product of two subgroups
$\mathcal{G} = \mathcal{A} \times \mathcal{B}$ the representations of
$\mathcal{G}$ are given by tensor products,
\begin{equation}
  \label{eq:tensorrep}
  \Gamma^{(\rho,\nu)}(a\cdot b) = \Gamma^{\rho}(a) \otimes \Gamma^{\nu}(b),
\end{equation}
of representations $\Gamma^{\rho}(a)$ and $\Gamma^{\nu}(b)$ of
$\mathcal{A}$ and $\mathcal{B}$. The characters are simply the product
of the characters of $\mathcal{A}$ and $\mathcal{B}$
\begin{equation}
  \label{eq:tensorchar}
  \chi^{(\rho,\nu)}(a\cdot b) = \chi^{\rho}(a) \cdot \chi^{\nu}(b).
\end{equation}
This can be applied to construct symmetry-adapted wave functions with a
space group $\mathcal{S}$ and a local symmetry group like
$\mathbb{Z}_2$ for spin flip symmetry. Since local symmetries like the
spin flip symmetry commute with space group symmetries the symmetry
group $\mathcal{G} = \mathcal{S}\times\mathbb{Z}_2$ is a direct
product. Hence, the representation matrices and characters of
$\mathcal{G}$ are given by \cref{eq:tensorrep} and
\cref{eq:tensorchar}.

For semi-direct products of groups, \cref{eq:tensorrep} and
\cref{eq:tensorchar} do not hold in general. Space groups are in
general only a semi-direct product of the translation group and the
point group. We briefly discuss the representation theory of
two-dimensional space groups in \cref{sec:spacegroupreptheory}.

% The basis of symmetry-adapted wave functions is particularly
% interesting since it can be used to block-diagonalize a
% Hamiltonian. Given two different irreducible representations $\rho$
% and $\eta$ of the symmetry group $\mathcal{G}$ of the Hamiltonian
% $H$ we have
% \begin{align}
%   \label{eq:symmetryadaptedwavefunctionortho}
%   \braket{\psi^\rho | H | \phi^\eta} = 0
% \end{align}
% if $\rho \neq \eta$.

%% Given the transformation to symmetry-adapted wave functions in
%% eq.~\eqref{eq:symmetryadaptedwavefunction1d} we can compute the
%% matrix elements of the Hamiltonian via
%% \begin{align}
%%   \label{eq:symmetryadaptedwavefunction1d}
%%   \braket{\psi^\rho | H | \phi^\rho} = \frac{1}{|\mathcal{G}|^2}\sum\limits_{g,h \in \mathcal{G}}
%%                            \chi^\rho(g)\chi^\rho(h)^*\braket{ {g} \psi|H| {h} \phi}
%% \end{align}

\subsubsection{Computations in the symmetrized basis}
From now we only consider one-dimensional representations of the
symmetry group. We change the basis from pure spin configurations,
\begin{equation}
  \label{eq:spinconf}
  \ket{\bm{\sigma}} = \ket{\sigma_1\ldots\sigma_{N}} \in \mathcal{H},
\end{equation}
to symmetry-adapted spin configurations,
\begin{equation}
  \label{eq:spinconfsym}
  \ket{\bm{\sigma}^\rho} \in \mathcal{H}_\rho,
\end{equation}
as defined in \cref{eq:symmetryadaptedwavefunction1d}.  For numerical
implementations the question arises how to encode the symmetrized
basis states $\ket{\bm{\sigma}^\rho}$ and how to explicitly compute
the matrix elements
\begin{align}
  \label{eq:symmetrizedmatrixelement}
  \braket{\bm{\tau}^\rho | H | \bm{\sigma}^\rho}.
\end{align}
The symmetry group $\mathcal{G}$ decomposes the space of pure spin
configurations into disjoint orbits,
\begin{align}
  \label{eq:hilbertspaceorbits}
  \text{Orbit}(\ket{\bm{\sigma}}) =
  \{ {g} \ket{\bm{\sigma}} | g \in \mathcal{G} \}.
\end{align}
For every irreducible representation an orbit corresponds to a
symmetrized state via \cref{eq:symmetryadaptedwavefunction1d}. To
encode such a symmetrized state we simply choose a single state
$\ket{\tilde{\bm{\sigma}}} \in \text{Orbit}(\ket{\bm{\sigma}})$ which
unambigously identifies the symmetrized state. We call the state
$\ket{\tilde{\bm{\sigma}}}$ the
\textit{representative}\index{representative} of
$\text{Orbit}(\ket{\bm{\sigma}})$.

% We introduce an ordering on the basis states via
% \begin{equation}
%   \ket{\sigma} < \ket{\bm{\tau}} :\Leftrightarrow \integer(\ket{\sigma}) < \integer(\ket{\bm{\tau}}).y
% \end{equation}
To choose a specific state in $\text{Orbit}(\ket{\bm{\sigma}})$ it is
canonical to choose the state with the smallest integer encoding,
\begin{align}
  \label{eq:representativecondition}
  \ket{\tilde{\bm{\sigma}}} = g_{\bm{\sigma}}\ket{\bm{\sigma}}, \quad
  \text{where} \quad g_{\bm{\sigma}} = \argmin_{g \in \mathcal{G}}
  \,\text{int}( g\ket{\bm{\sigma}}).
\end{align}
To compute the matrix elements of the Hamiltonian in
\cref{eq:symmetrizedmatrixelement} we decompose the Hamiltonian,
\begin{align}
  H = \sum \limits_k H_k,
\end{align}
as a sum of non-branching terms $H_k$, i.e. for every single spin
configurations $\ket{\bm{\sigma}}$ there is only one other spin
configuration $\ket{\bm{\tau}}$, such that
\begin{align}
  \label{eq:nonbranchingcondition}
  H_k \ket{\bm{\sigma}} = h_k\ket{\bm{\tau}},
\end{align}
where $h_k$ is in general just a complex number. For instance, a
simple spin-$1/2$ Heisenberg bond $\bm{S}_i\cdot \bm{S}_j$ does not
fulfill the non-branching condition in \cref{eq:nonbranchingcondition}
but it can be rewritten as a sum of a spin exchange bond
$\frac{1}{2}\left( S_i^+S_j^- + S_i^-S_j^+\right)$ and an Ising bond
$S_i^z S_j^z$ which do indeed fulfill the condition
\cref{eq:nonbranchingcondition} individually. Applying a non-branching
bond on a representative does in general not yield another
representative. Put differently, the state $\ket{\bm{\tau}}$ in
\begin{align}
  \label{eq:brangingnonrepresentative}
  \ket{\bm{\tau}} = H_k \ket{\tilde{\bm{\sigma}}}
\end{align}
is not necessarily minimal in $\text{Orbit}(\ket{\bm{\tau}})$. If
$g_{\bm{\tau}}$ is the element of the symmetry group that transforms
$\ket{\bm{\tau}}$ to its representative $\ket{\tilde{\bm{\tau}}}$,
i.e.
\begin{align}
  \ket{\tilde{\bm{\tau}}} = {g}_{\bm{\tau}}\ket{\bm{\tau}},
\end{align}
then the matrix element \cref{eq:symmetrizedmatrixelement} is given by
\begin{align}
  \label{eq:matrixelementrepresentatives}
  \braket{\tilde{\bm{\tau}}^\rho | H_k | \tilde{\bm{\sigma}}^\rho} =
  \chi^\rho(g_{\bm{\tau}}) \frac{N_{\rho,\bm{\tau}}}{N_{\rho,\bm{\sigma}}}
  \braket{\bm{\tau} | H_k | \tilde{\bm{\sigma}}}.
\end{align}
The right hand side of \cref{eq:matrixelementrepresentatives} contains
the matrix elements in the unsymmetrized basis, the character
$\chi^\rho(g_{\bm{\tau}})$ of the representation labeled by $\rho$
evaluated at $g_{\bm{\tau}}$ and the normalization constants
$N_{\rho,\bm{\tau}}$ and $N_{\rho,\bm{\sigma}}$. Once a complete set
of representatives is known, the Hamiltonian matrix in the symmetrized
basis can be constructed via
\cref{eq:matrixelementrepresentatives}. If the original Hamiltonian is
sparse, also the Hamiltonian in the symmetrized basis is sparse and a
Lanczos algorithm can be applied for diagonalization.

%%% Local Variables:
%%% mode: latex
%%% TeX-master: "../../thesis"
%%% End

  \section{Variational Monte Carlo}
  \label{sec:vmc}
  %auto-ignore
% \section{asdf}                  
According to science history \cite{Ulam1987}, Stanislav Ulam invented
the idea of Monte Carlo simulations while playing the card game
Solitaire. Since computing the winning probability is much more
interesting than actually playing the game, Ulam came up with an
inventive idea to compute this probability. Instead of working out all
possible permutations and ways of playing the cards a pretty good
guess of the winning probability can be attained by simply playing
several times and counting the number of wins. More generally, in case
we want to compute a stochastic expectation value of the form
\begin{equation}
  \label{eq:stochasticexpectation}
  \braket{f} = \sum\limits_{x} f(x) p(x),
\end{equation}
of a function $f(x)$ with respect to the probability measure $p(x)$.
Instead of summing over all microscopic configurations $x$ we could
compute the mean value,
\begin{equation}
  \label{eq:meanestimator}
  \overline{\braket{f}} = \frac{1}{n}\sum\limits_{i=1}^n f(X_i),
\end{equation}
of $n$ microscopic configurations $X_i$, which are chosen randomly
according to the probability distribution $p(x)$. This yields an
estimator $\overline{\braket{f}}$ for the true value $\braket{f}$,
according to the law of large numbers. In case the samples $X_i$ are
independently distributed an unbiased error estimator is given by
\begin{equation}
  \label{eq:errorestimatoriid}
  \sigma_{f}^2 = {\frac{1}{n-1}\sum\limits_{i=1}^n
    \left(f(X_i) - \overline{\braket{f}} \right)^2}.
\end{equation}
A refined method of stochastically estimating an expectation value
\cref{eq:stochasticexpectation} is given by the Markov Chain Monte
Carlo method, which we briefly review in \cref{sec:mcmc}. 
Expectation values of the form
\begin{equation}
  \label{eq:physexpvalfirst}
  \braket{\mathcal{O}} = \frac{\braket{\psi|\mathcal{O}|\psi}}{\braket{\psi|\psi}},
\end{equation}
can be written as a stochastic expectation value as in
\cref{eq:stochasticexpectation}. This way, physical properties of
variational wave functions $\ket{\psi}$ can be evaluated in a
computationally efficient way. We will discuss this
\textit{Variational Monte Carlo}, short VMC, method in
\cref{sec:variationalmc}. For applying the VMC method we need to
compute coefficients of variational wave functions in a given
basis. The coefficients of Gutzwiller projected wave functions in the
local $S^z$ basis are presented in \cref{sec:gpwfvmc}.

\subsection{Markov Chain Monte Carlo}
\label{sec:mcmc}
Here, we summarize the basic principles of Markov Chain Monte Carlo
methods, short MCMC. For a detailed exposition see e.g.
Ref.~\cite{gilks1995markov}. The basic idea of MCMC methods is to
construct a Markov Chain $( X_i )_{i=1}^{\infty}$ with equilibrium
distribution $p(x)$. The stochastic average
\cref{eq:stochasticexpectation} is then approximated by the estimator
\begin{equation}
  \label{eq:meanestimatormcmc}
  \overline{\braket{f}} = \frac{1}{n}\sum\limits_{i=1}^n f(X_i),
\end{equation}
where $X_i$ are now samples from the evolution of the Markov chain
$( X_i )_{i=1}^{\infty}$. The samples $X_i$ in the estimator
\cref{eq:meanestimatormcmc} are taken once the distribution of the
Markov samples $X_i$ has converged to the equilibrium distribution
$p(x)$. The balance condition,
\begin{equation}
  \label{eq:balancecondition}
  \sum_{y}p(x)T(x\rightarrow y) = \sum_{y}p(y)T(y\rightarrow x),
\end{equation}
together with the ergodicity of the Markov chain are necessary and
sufficient conditions for the transition kernel $T(x \rightarrow y)$
to yield $p(x)$ as the equilibrium distribution. Each solution to this
equation yields a Markov chain suitable for computing the estimator
\cref{eq:meanestimatormcmc}. The most popular transition kernel
fulfilling this condition is the \textit{Metropolis}
kernel~\cite{Metropolis1953},
\begin{equation}
  \label{eq:metropoliskernel}
  T(x\rightarrow y) = G(x\rightarrow y) A(x\rightarrow y),
\end{equation}
where $G(x\rightarrow y)$ is called the \textit{proposal distribution}
and
\begin{equation}
  \label{eq:metropolisacceptancerate}
  A(x\rightarrow y) = \min\left(1, \frac{p(y)}{p(x)}\frac{G(y\rightarrow x)}{G(x\rightarrow y)}\right)
\end{equation}
is called the \textit{acceptance rate}. $G(x\rightarrow y)$ encodes
the strategy for choosing updates and $A(x\rightarrow y)$ defines a
probability, whether or not a proposed update is accepted. This kernel
additionally fulfills the more restrictive detailed balance condition,
\begin{equation}
  \label{eq:detailedbalancecondition}
  p(x)T(x\rightarrow y) = p(y)T(y\rightarrow x),
\end{equation}
for all microscopic $x$ and $y$. Although this condition is sufficient
to ensure convergence to the equilibrium distribution, it is not a
necessary condition. Transition kernels not fulfilling the detailed
balance condition have also been proposed \cite{Suwa2010} and can be
used for minimizing rejection rates. Since subsequent values $X_i$ of
a Markov chain are in general not independent, the estimator
$\sigma_{f}^2$ in \cref{eq:errorestimatoriid} cannot be used to
estimate the error of the mean value in \cref{eq:meanestimator}. The
correct general error estimator is given by
\begin{equation}
  \label{eq:errorestimatorgeneric}
  \sigma_{f,\text{dep}}^2 = \sigma_{f}^2
  (1 + 2\tau_f) =
  \frac{1}{n-1}\left[\sum\limits_{i=1}^n \left(f(X_i) - \overline{\braket{f}} \right)^2\right](1 + 2\tau_f),
\end{equation}
where the autocorrelation time $\tau_f$ of the quantity $f$ is given
by
\begin{equation}
  \label{eq:autocorrelationtime}
  \tau_f \equiv \frac{1}{ \braket{f^2} - \braket{f}^2}
  \sum\limits_{t=1}^\infty \left( \braket{f_{1}f_{1+t}} - \braket{f}^2\right).
\end{equation}
Hence, in order to estimate the true error
\cref{eq:errorestimatorgeneric} one can either directly estimate the
autocorrelation time \cite{Wolff2004} or use so-called \textit{binning
  analysis}.

\subsubsection{Binning analysis}
\begin{figure}[t]
  \centering
  \begin{minipage}[l]{0.45\linewidth}
    \caption{Typical error estimator $\sigma_{f,B}^2$ of a MCMC
      simulation for increasing bin size $B$. The error estimator
      finally reaches a plateau approximating the true error estimator
      $ \sigma_{f,\text{dep}}^2$.}  \label{fig:errorconvergence}
  \end{minipage}
  \quad\quad
  \begin{minipage}[r]{0.45\linewidth}
    \includegraphics[width=\textwidth]{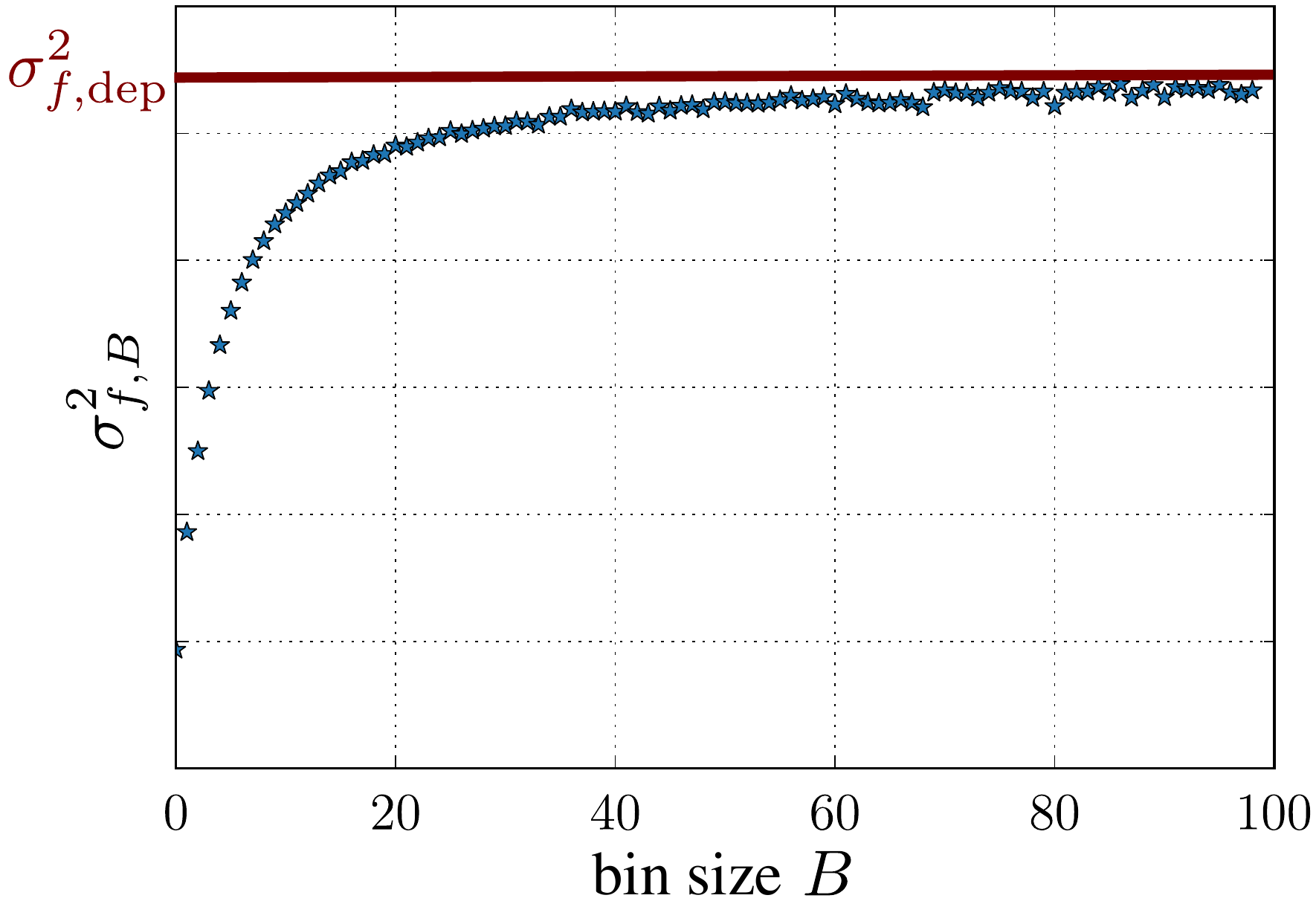}
  \end{minipage}
\end{figure}

We collect the measurements $X_i$ into $n_B$ bins of size $B$ such
that $n_B \cdot B = n$, where $n$ is the total number of measurements.
In each bin, we compute the bin average
\begin{equation}
  \label{eq:binaverage}
  f_k^{(B)} = \frac{1}{B}\sum\limits_{i=kB}^{k(B+1)} f(X_i) .
\end{equation}
In terms of these bin averages the mean estimator
\cref{eq:meanestimator} can be written as
\begin{equation}
  \label{eq:binmeanestimator}
  \overline{\braket{f}} = \frac{1}{n_B}\sum\limits_{k=1}^{n_B} f_k^{(B)}.
\end{equation}
The main point of binning analysis is to define new error estimators,
\begin{equation}
  \label{eq:binerrorestimator}
  \sigma_{f,B}^2 = \frac{1}{n_B-1}\sum\limits_{k=1}^{n_B}
  \left(f_k^{(B)} - \overline{\braket{f}} \right)^2,
\end{equation}
that converge to the generic error estimator
\cref{eq:errorestimatorgeneric},
\begin{equation}
  \label{eq:binningestimatorconvergence}
  \sigma_{f,\text{dep}}^2 = \lim\limits_{B\rightarrow \infty}  \sigma_{f,B}^2.
\end{equation}
Convergence is reached for bin sizes $B \gg \tau_f$. Thus, one
typically investigates the evolution of the error estimator
$\sigma_{f,B}^2$ for increasing $B$, as shown exemplarily in
\cref{fig:errorconvergence}.  The binning analysis can also be
performed by only storing $\mathcal{O}(\log n)$
mesurements~\cite{Ambegaokar2010}.

% Instead of using averages in the bins as in eq. \cref{eq:binaverage}
% we can also use so called \textit{jackknife} averages which are the
% averages over all values except the bin, i.e.
% \begin{equation}
%   \label{eq:binaverage}
%   \tilde{f}_k^{(B)} = \frac{1}{n-B}\sum\limits_{i\notin \{ kB,\ldots, k(B+1)\}} f(X_i) 
% \end{equation}

%%% Local Variables:
%%% mode: latex
%%% TeX-master: "../../thesis"
%%% End:

  %auto-ignore
\subsection{Stochastic sampling of quantum wave functions}
\label{sec:variationalmc}

Evaluating physical quantities of a pure quantum state $\ket{\psi}$
amounts to computing expectation values of the form
\begin{equation}
  \label{eq:physexpval}
  \braket{\mathcal{O}} = \frac{\braket{\psi|\mathcal{O}|\psi}}{\braket{\psi|\psi}}.
\end{equation}
Once the coefficients $\braket{x|\psi}$ of the wave function
$\ket{\psi}$ are known in a certain basis $\ket{x}$ of the Hilbert
space, \cref{eq:physexpval} can in principle be evaluated exactly by
computing
\begin{equation}
  \label{eq:physexpvalnumericalexact}
  \braket{\mathcal{O}} =  \frac{1}{\mathcal{N}}\sum\limits_{x,y} \braket{\psi|x}
  \braket{x|\mathcal{O}|y}\braket{y|\psi}, 
\end{equation}
where
\begin{equation}
  \label{eq:wfnormalization}
  \mathcal{N} =  \sum\limits_{x} \left|\braket{x|\psi}\right|^2 .
\end{equation}
The computational effort scales with the dimension of the Hilbert
space. Consequently, these calculations become infeasible for large
system sizes. This limitation can be overcome by rewriting
\cref{eq:physexpvalnumericalexact} to a form that allows for
stochastic Monte Carlo sampling,
\begin{align}
  \label{eq:wfmcreformulation}
  \braket{\mathcal{O}} =  &\frac{1}{\mathcal{N}}\sum\limits_{x,y} \braket{\psi|x}
                            \braket{x|\mathcal{O}|y}\braket{y|\psi} = \\
                          &= \sum\limits_{y}  \left(\sum\limits_{x}
                            \frac{\braket{\psi|x}}{\braket{\psi|y}}
                            \braket{x|\mathcal{O}|y}\right) \left|\braket{y|\psi}\right|^2/\mathcal{N}\\ 
                          &\equiv \sum\limits_{y} f(y) p(y).
                            \label{eq:operatormontecarloaverage}
\end{align}
\Cref{eq:operatormontecarloaverage} now corresponds to a stochastic
expectation value of the function
\begin{equation}
  f(y) = \sum\limits_{x} \frac{\braket{\psi|x}}{\braket{\psi|y}}
  \braket{x|\mathcal{O}|y},
\end{equation}
with respect to the probability measure
\begin{equation}
  p(y) = \left|\braket{y|\psi}\right|^2/\mathcal{N}.
\end{equation}
This can now be evaluated by the Metropolis Monte Carlo algorithm as
presented in \cref{sec:mcmc}. When using the Metropolis transition
kernel, only ratios ${\braket{x|\psi}}/{\braket{y|\psi}}$ of the
coefficients are needed to evaluate
\cref{eq:operatormontecarloaverage}. Sometimes, these ratios can be
computed more efficiently than the coefficients $\braket{x|\psi}$
itself, as is the case for Gutzwiller projected wave functions.
% This is actually the case for Gutzwiller projected wave functions
% when using a simple local spin flip updates for the Metropolis
% algorithm.

%%% Local Variables:
%%% mode: latex
%%% TeX-master: "../../thesis"
%%% End:

  %auto-ignore
\subsection{Coefficients of Gutzwiller projected wave functions}
\label{sec:gpwfvmc}

Gutzwiller projected wave functions, short GPWF, are correlated many
electron states. They were introduced by Martin Gutzwiller
\cite{Gutzwiller1963,Gutzwiller1965} as variational wave functions for
Hubbard models appearing in the study of high-temperature
superconductivity. We already discussed them in \cref{sec:rvbstate}
and \cref{sec:partonconstruction} in the context of the parton
construction of spin liquids. Here, we want to discuss how to compute
coefficients of these wave functions and how to perform quick Monte
Carlo updates via evaluating only ratios of coefficients.

The Gutzwiller projection operator is given by
\cite{Gutzwiller1963,Gros1989}
\begin{equation}
  \label{eq:gutzwillerprojector}
  P_{\text{GW}} = \prod\limits_i \left( 1 - n_{i\uparrow}n_{i\downarrow}\right),
\end{equation}
where $n_{i\sigma} = c_{i\sigma}^\dagger c_{i\sigma}$ for fermionic
creation and annihilation operators $c_{i\sigma}^\dagger$ and
$c_{i\sigma}$, $\sigma = \uparrow\downarrow$. It sets the part of a
many-body wave function with double site occupancy to zero. This can
be thought of as enforcing a hard-core constraint on a fermionic wave
function. A GPWF is now given by applying this projector to an
uncorrelated product state wave function
\begin{equation}
  \label{eq:gpwfdefinition}
  \ket{\psi_{\text{GPWF}}} = P_{\text{GW}} \ket{\psi_0}.
\end{equation}
$\ket{\psi_0}$ can be chosen as the ground state of the parton
mean-field Hamiltonian in \cref{eq:partonmeanfieldhamiltonian}. For
simplicity, we consider a simple tight binding parton ansatz without
pairing terms,
\begin{align}
  H = \sum\limits_{\substack{i,j\\ \sigma}} t_{ij}c^{\dagger}_{i\sigma}c_{j\sigma} + \textnormal{H.c.}, 
\end{align}
which can be diagonalized
\begin{align}
  a^{\dagger}_{i\sigma} = \sum\limits_{j} u^i{}_j c^{\dagger}_{j\sigma}\\
  H = \sum\limits_{i\sigma} \epsilon_i a^{\dagger}_{i\sigma}a_{i\sigma}, 
\end{align}
where $(u^i)_j$ denotes the $i$-th eigenvector of the hopping matrix
$(t_{ij})$ with eigenvalue $\epsilon_i$. Its many-body ground state is
given by
\begin{equation}
  \ket{\psi_0} = \prod\limits_{\substack{\epsilon_i < E_F\\ \sigma}} a^\dagger_{i\sigma}\ket{0},
\end{equation}
where energy levels are filled up to the Fermi energy $E_F$.  To
investigate GPWFs \cite{Gutzwiller1963,Gutzwiller1965} numerically, we
have to work out their coefficients in a computational basis. We are
interested in the coefficients of this wave function at half filling,
i.e.
\begin{equation}
  N_\text{up} = N_\text{down} = N/2.
\end{equation}
In this case, the coefficients of the ground state wave functions are
explicitly given by the following $N \times N$ Slater determinant,
\begin{equation}
  \label{eq:gpwfdeterminant}
  \braket{\sigma_1 \sigma_2\ldots\sigma_{N}| \psi_0} = \det
  \begin{vmatrix}
    w_1^\uparrow(1, \sigma_1) & w_1^\uparrow(2, \sigma_2) & \cdots&
    w_1^\uparrow({N}, \sigma_{N}) \\
    w_2^\uparrow(1, \sigma_1) & w_2^\uparrow(2, \sigma_2) & \cdots&
    w_2^\uparrow({N}, \sigma_{N}) \\
    \vdots & & & \vdots \\
    w_{N/2}^\uparrow(1, \sigma_1) & w_{N/2}^\uparrow(2, \sigma_2) &
    \cdots&
    w_{N/2}^\uparrow({N}, \sigma_{N}) \\

    w_1^\downarrow(1, \sigma_1) & w_1^\downarrow(2, \sigma_2) &
    \cdots&
    w_1^\downarrow({N}, \sigma_{N}) \\
    w_2^\downarrow(1, \sigma_1) & w_2^\downarrow(2, \sigma_2) &
    \cdots&
    w_2^\downarrow({N}, \sigma_{N}) \\
    \vdots & & & \vdots \\
    w_{N/2}^\downarrow(1, \sigma_1) & w_{N/2}^\downarrow(2, \sigma_2)
    & \cdots&
    w_{N/2}^\downarrow({N}, \sigma_{N}) \\
  \end{vmatrix},
\end{equation} 
where
\begin{equation}
  w_k^\uparrow(i,\sigma_i) = 
  \begin{cases}
    (u^k)_i  &\text{ if } \sigma_i = \uparrow \\
    0 &\text{ if } \sigma_i = \downarrow
  \end{cases}
  \quad
  \text{and}
  \quad
  w_k^\downarrow(i,\sigma_i) = 
  \begin{cases}
    0       &\text{ if } \sigma_i = \uparrow  \\
    (u^k)_i &\text{ if } \sigma_i = \downarrow
  \end{cases}.
\end{equation}
The coefficients of more general GPWFs with pairing terms as in the
general case in \cref{eq:partonmeanfieldhamiltonian} can also be
written in terms of Slater determinants, see
e.g.~\cite{Yanagisawa2003}.

The computational cost of evaluating the determinant in
\cref{eq:gpwfdeterminant} is of the order $\mathcal{O}(N^3)$. In a
typical Monte Carlo simulation, many of these coefficients have to be
evaluated which becomes the bottleneck of the computation.  A clever
method to speed up these computations was proposed
in~\cite{Ceperley1977}. The key observation is that in the Metropolis
Monte Carlo sampling only ratios of coefficients have to be
calculated. If two matrices differ only by one column the ratio of the
determinants can be evaluated with $\mathcal{O}(N)$ operations if the
inverse matrices are known.  If an update is accepted, the inverse
matrix has to be recomputed.  This can also be done more efficiently
with $\mathcal{O}(N^2)$ operations if the inverse of a matrix with
only one different column is known. For details on this update
procedure see~\cite{Ceperley1977}.

  \newpage
\part{Research projects}
\label{part:research}
\chapter[Large-Scale Exact Diagonalization]{Symmetries and Parallelization for Large-Scale Exact Diagonalization}
\epigraph{Computer, tea, Earl Grey, hot.}{Jean-Luc Picard}
\label{sec:largescaleed}
%auto-ignore

% \section{Large scale ED}
\subsubsection{Abstract}
We present novel algorithms for fast and memory-efficient use of
discrete symmetries in Exact Diagonalization computations of quantum
many-body systems. These techniques allow us to work flexibly in the
reduced basis of symmetry-adapted wave functions. Moreover, a
parallelization scheme for the Hamiltonian-vector multiplication in
the Lanczos procedure for distributed memory machines avoiding load
balancing problems is proposed. We show that using these methods
systems of up to $50$ spin-$1/2$ particles can be successfully
diagonalized.

\section{Introduction}
Exact Diagonalization, short ED, studies have in the past been a
reliable source of numerical insight into various problems in quantum
many-body physics. The method is versatile, unbiased and capable of
simulating systems with a sign problem. The main limitation of ED is
the typical exponential scaling of computational effort and memory
requirements in the system size. Nevertheless, the number of particles
feasible for simulation has steadily increased since the early
beginnings~\cite{Oitmaa1978,Lin1990,Weisse2013}. Not only does
increasing the number of particles yield better approximations to the
thermodynamic limit, but also several interesting simulation clusters
with many symmetries become available if more particles can be
simulated. Having access to such clusters becomes important if several
competing phases ought to be realized on the same finite size
sample. In this work, we present algorithms and strategies for the
implementation of a state-of-the-art large-scale ED code and prove
that applying these methods systems of up to 50 spin-$1/2$ particles
can be simulated on present day supercomputers. There are two key
ingredients making these computations possible:
\begin{enumerate}
\item \textbf{Efficient use of symmetries}. In
  \cref{sec:sublatticecoding} we present algorithms to work with
  symmetry-adapted wave functions in a fast and memory efficient
  way. These so-called \textit{sublattice coding techniques} allow us
  to diagonalize the Hamiltonian in every irreducible representation
  of a discrete symmetry group. The basic idea behind these algorithms
  goes back to H.Q. Lin~\cite{Lin1990}.
\item \textbf{Parallelization of the matrix-vector multiplications} in
  the Lanczos algorithm for distributed memory machines. We propose a
  method avoiding load-balancing problems in message-passing and
  present a computationally fast way of storing the Hilbert space
  basis in \cref{sec:edparallelization}.
\end{enumerate}
These ideas have been implemented and tested on various
supercomputers.  We present results and benchmarks of our
implementation in \cref{sec:edbanchmarks}.

We refer the reader to \cref{sec:ed} for a basic introduction to the
ED method. The Lanczos method is reviewed in \cref{sec:lanczosmethod}
and symmetry-adapted wave functions are discussed in
\cref{sec:symsnadaptesdwaves}. In this chapter, we also only consider
one-dimensional representations of the symmetry group. We recall some
important notions. Consider a generic spin configuration on $N$
lattice sites with local dimension $d$,
\begin{equation}
  \label{eq:spinconf_lsed}
  \ket{\bm{\sigma}} = \ket{\sigma_1,\ldots, \sigma_N}, \quad
  \sigma_i \in\{ 1, \ldots, d\}.
\end{equation}
The \textit{symmetry-adapted basis states} $\ket{\bm{\sigma}^\rho}$
are defined as (cf. \cref{eq:symmetryadaptedwavefunction1d})
\begin{align}
  \label{eq:symmetryadaptedwavefunction1d_lsed}
  \ket{\bm{\sigma}^\rho} \equiv \frac{1}{N_{\rho,\bm{\sigma}}}
  \sum\limits_{g \in \mathcal{G}} \chi^\rho(g)^* {g}\ket{\bm{\sigma}},
\end{align}
where $\mathcal{G}$ denotes a discrete symmetry group, $\rho$ a
one-dimensional representation of this group, $\chi^\rho(g)$ the
character of this representation evaluated at group element $g$, and
$N_{\rho,\bm{\sigma}}$ denotes the normalization constant of the state
$\ket{\bm{\sigma}^\rho}$. The set of basis state spin configurations
$\ket{\bm{\sigma}}$ is divided into \textit{orbits}
(cf. \cref{eq:hilbertspaceorbits}),
\begin{align}
  \label{eq:hilbertspaceorbits_lsed}
  \text{Orbit}(\ket{\bm{\sigma}}) = \{ {g} \ket{\bm{\sigma}}
  | g \in \mathcal{G}\}.
\end{align}
The \textit{representative} $\ket{\tilde{\bm{\sigma}}}$ within each
orbit is given by as the element with smallest integer value coding
(cf. \cref{eq:representativecondition}),
\begin{align}
  \label{eq:representativecondition_lsed}
  \ket{\tilde{\bm{\sigma}}} = g_{\bm{\sigma}}\ket{\bm{\sigma}}, \quad
  \text{where} \quad g_{\bm{\sigma}} = \argmin_{g \in \mathcal{G}}
  \,\text{int}( g\ket{\bm{\sigma}}).
\end{align}
The matrix element
$\braket{\tilde{\bm{\tau}}^\rho | H_k | \tilde{\bm{\sigma}}^\rho}$ for
non-branching terms $ H_k $ for two symmetry-adapted basis states with
representation $\rho$ is given by
(cf. \cref{eq:matrixelementrepresentatives})
\begin{align}
  \label{eq:matrixelementrepresentatives_lsed}
  \braket{\tilde{\bm{\tau}}^\rho | H_k | \tilde{\bm{\sigma}}^\rho} =
  \chi^\rho(g_{\bm{\tau}})  \frac{N_{\rho,\bm{\tau}}}{N_{\rho,\bm{\sigma}}}
  \braket{\bm{\tau} | H_k | \tilde{\bm{\sigma}}}.
\end{align}
In the following we define
\begin{equation}
  \ket{\bm{\sigma}} < \ket{\bm{\tau}} :\Leftrightarrow
  \integer(\ket{\bm{\sigma}}) < \integer(\ket{\bm{\tau}}).
\end{equation}

%%% Local Variables:
%%% mode: latex
%%% TeX-master: "../../thesis"
%%% End:

%auto-ignore

\section{Sublattice Coding techniques}
\label{sec:sublatticecoding}
Evaluating the matrix elements
$\braket{\tilde{\bm{\tau}}^\rho | H_k | \tilde{\bm{\sigma}}^\rho}$ in
\cref{eq:matrixelementrepresentatives_lsed} for all basis states
$\ket{\tilde{\bm{\sigma}}^\rho}$ and $\ket{\tilde{\bm{\tau}}^\rho}$
efficiently is the gist of employing symmetries in ED computations.
In an actual implementation on the computer we need to perform the
following steps:
\begin{itemize}
\item Apply the non-branching term $H_k$ on the representative state
  $\ket{\tilde{\bm{\sigma}}}$. This yields a possibly
  non-representative state $\ket{\bm{\tau}}$. From this, we can compute
  the factor $\braket{\bm{\tau} | H_k | \tilde{\bm{\sigma}}}$.
\item Find the representative $\ket{\tilde{\bm{\tau}}}$ of
  $\ket{\bm{\tau}}$ and determine the group element $g_{\bm{\tau}}$
  such that
  $\ket{\tilde{\bm{\tau}}} = {g}_{\bm{\tau}}\ket{\bm{\tau}}$. This
  yields the factor $\chi^\rho(g_{\bm{\tau}})$.
\item Know the normalization constants $N_{\rho,\bm{\sigma}}$ and
  $N_{\rho,\bm{\sigma}}$. These are usually computed when creating a
  list of all representatives and stored in a separate list.
\end{itemize}
The problem of finding the representative $\ket{\tilde{\bm{\sigma}}}$
of a given state $\ket{\bm{\sigma}}$ and its corresponding symmetry
${g}_{\bm{\sigma}}$ turns out to be the computational bottleneck of ED
in a symmetrized basis. It is thus desirable to solve this problem
fast and memory efficient. There are two straightforward approaches to
solving this problem:
\begin{itemize}
\item Apply all symmetries directly to $\ket{\bm{\tau}}$ to find the
  minimizing group element $g_{\bm{\tau}}$,
  \begin{equation}
    g_{\bm{\tau}} = \argmin_{g \in \mathcal{G}} \,
    \text{int}(g\ket{\bm{\tau}}).
  \end{equation}
  This method does not have any memory overhead but is
  computationally slow since all symmetries have to be applied to the
  given state $\ket{\bm{\tau}}$.
\item For every state $\ket{\bm{\sigma}}$ we store
  $\ket{\tilde{\bm{\sigma}}}$ and $g_{\bm{\sigma}}$ in a lookup
  table. While this is very fast computationally, the lookup table for
  storing all representatives grows exponentially in the system size.
\end{itemize}
The key to solving the representative search problem adequately is to
have an algorithm that is almost as fast as a lookup table, where
memory requirements are within reasonable bounds. This problem has
already been addressed by several authors \cite{Lin1990,
  Weisse2013}. The central idea in these so-called \textit{sublattice
  coding techniques} is to have a lookup table for the representatives
on a sublattice of the original lattice and combine the information of
the sublattice representatives to compute the total
representative. These ideas were first introduced in \cite{Lin1990,
  Weisse2013, Schulz1996}.  In the following paragraphs, we explain the
basic idea behind these algorithms and propose a flexible extension to
arbitrary geometries and number of sublattices.

\subsubsection{Sublattice coding on two sublattices}
\begin{figure}[t!]
  \centering \includegraphics[width=\textwidth]{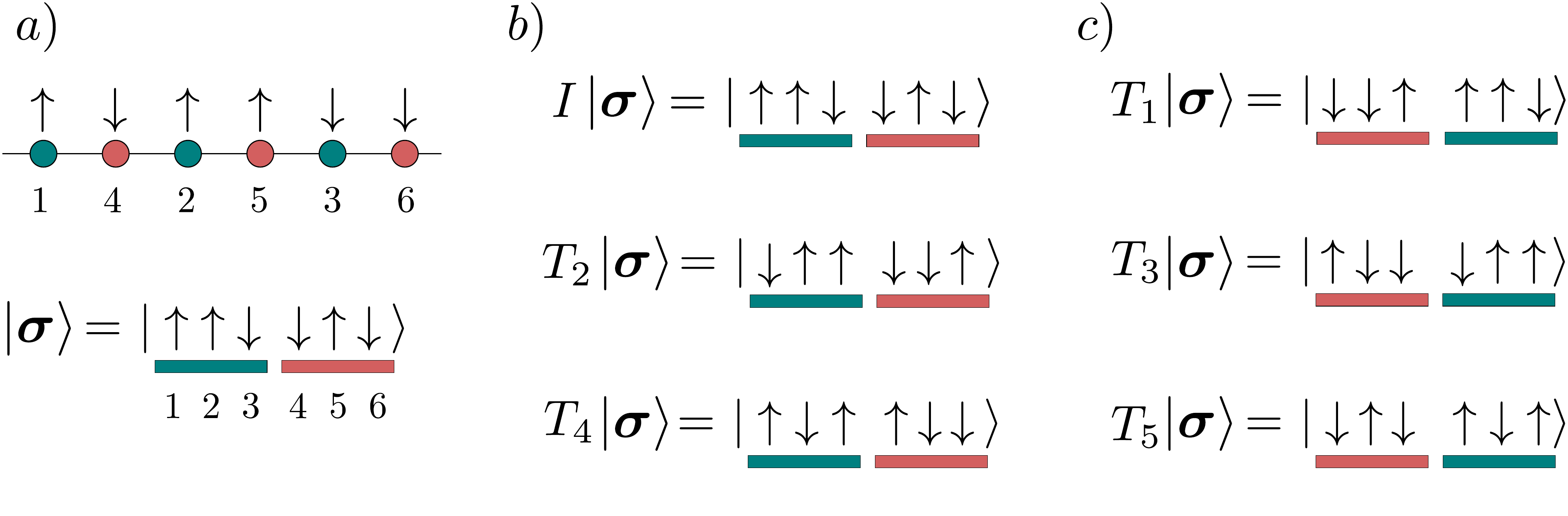}
  \caption{Two sublattice coding of the spin state $\ket{\bm{\sigma}}$
    on a six-site chain lattice and action of translational
    symmetries. The sites are enumerated such that site $1$-$3$ are on
    the blue sublattice $A$, $4$-$6$ on the red sublattice $B$. The
    representative state with this enumeration of sites is given by
    $\ket{\tilde{\bm{\sigma}}} = T_1\ket{\bm{\sigma}} =
    \ket{\downarrow \downarrow \uparrow \uparrow \uparrow\downarrow}$.
    Notice, that the symmetries act on real space and thus the
    transformation of the basis states also depends on the numbering
    of sites.}
  \label{fig:sublatticetransformations}
\end{figure}

For demonstration purposes, we consider a simple translationally
invariant spin-$1/2$ system on a six-site chain lattice with periodic
boundary conditions.  The lattice is divided into two sublattices as
in \cref{fig:sublatticetransformations}. The even sites form the
sublattice $A$ and the odd sites form the sublattice $B$. We enumerate
the sites such that the sites $1$ to $3$ are in sublattice $A$ and the
sites $4$ to $6$ are in sublattice $B$. We choose the integer
representation of a state $\ket{\bm{\sigma}}$ such that the most
significant bits are formed by the spins in sublattice $A$.  The
symmetry group we consider consists of the six translations on the
chain
\begin{equation}
  \mathcal{G} = \{ \text{Id}, T, T_2, T_3, T_4, T_5 \},
\end{equation}
where $T_n$ denotes the translation by $n$ lattice sites.  The
splitting of the lattice into two sublattices is stable in the sense
that every symmetry element $g \in \mathcal{G} $ either maps the $A$
sublattice to $A$ and the $B$ sublattice to $B$ or the $A$ sublattice
to $B$ and the $B$ sublattice to $A$.  We call this property
\textit{sublattice stability}. It is both a property of the partition
of our lattice into sublattices and the symmetry group. Hence, the
symmetry group is composed of two kinds of symmetries
\begin{align}
  \begin{split}
    \mathcal{G}_A &\equiv \{ g \in \mathcal{G} \, ; \quad
    g \text{ maps sublattice } A \text{ onto } A \},
    \\
    \mathcal{G}_B &\equiv \{ g \in \mathcal{G} \, ;
    \quad g \text{ maps sublattice } B \text{ onto } A \}. 
  \end{split}  
\end{align}
We denote by $\ket{\bm{\sigma}}_A$ (resp. $\ket{\bm{\sigma}}_B$) the
state restricted to sublattice $A$ (resp. $B$) and define the
\textit{sublattice representatives},
\begin{align}
  \label{eq:sublatreps2sl}
  \begin{split}
    \texttt{Rep}_A(\ket{\bm{\sigma}}_A) &\equiv h_A\ket{\bm{\sigma}}_A,
    \quad \text{where} \quad h_A = \argmin_{g \in \mathcal{G}_A} \,
    \text{int}(g \ket{\bm{\sigma}}_A),
    \\
    \texttt{Rep}_B(\ket{\bm{\sigma}}_B) &\equiv h_B\ket{\bm{\sigma}}_B,
    \quad \text{where} \quad h_B = \argmin_{g \in \mathcal{G}_B} \,
    \text{int}(g \ket{\bm{\sigma}}_B),
  \end{split}
\end{align}
and the \textit{representative symmetries},
\begin{align}
  \label{eq:repsyms2sl}
  \begin{split}
    \texttt{Sym}_A(\ket{\bm{\sigma}}_A) &\equiv \{ g \in \mathcal{G}_A
    \, ; \quad {g}\ket{\bm{\sigma}}_A =
    \texttt{Rep}_A(\ket{\bm{\sigma}}_A) \},
    \\
    \texttt{Sym}_B(\ket{\bm{\sigma}}_B) &\equiv \{ g \in \mathcal{G}_B
    \, ; \quad {g}\ket{\bm{\sigma}}_B =
    \texttt{Rep}_B(\ket{\bm{\sigma}}_B) \}.
  \end{split}
\end{align}
\bigbreak\noindent Let again
$\ket{\tilde{\bm{\sigma}}} = g_{\bm{\sigma}}\ket{\bm{\sigma}}$, where
$\ket{\tilde{\bm{\sigma}}}$ is the representative of
$\ket{\bm{\sigma}}$.  The minimizing symmetry $g_{\bm{\sigma}}$ can
only be an element of $\texttt{Sym}_A(\ket{\bm{\sigma}}_A)$ if
$\texttt{Rep}_A(\ket{\bm{\sigma}}_A) \leq
\texttt{Rep}_B(\ket{\bm{\sigma}}_B)$, or vice versa. Put differently,
\begin{equation}
  \label{eq:sublatticerepslogic}
  \texttt{Rep}_B(\ket{\bm{\sigma}}_B) < \texttt{Rep}_A(\ket{\bm{\sigma}}_A)
  \quad\Rightarrow \quad
  g_{\bm{\sigma}} \notin \texttt{Sym}_A(\ket{\bm{\sigma}}_A).  
\end{equation}
Otherwise, any symmetry element in $\texttt{Rep}_B(\ket{\bm{\sigma}}_B)$
would yield a smaller integer value than $g_{\bm{\sigma}}$. This is
the core idea behind the sublattice coding technique. We store
$\texttt{Rep}_{A,B}(\ket{\bm{\sigma}}_{A,B})$ for every substate
$\ket{\bm{\sigma}}_{A,B}$ in a lookup table together with
$\texttt{Sym}_{A,B}(\ket{\bm{\sigma}}_{A,B})$. In a first step, we
determine the sublattice representative with smallest most significant
bits. Then we apply the representative symmetries to
$\ket{\bm{\sigma}}$ in order to determine the true representative
$\ket{\tilde{\bm{\sigma}}}$. The number of representative symmetries
$|\texttt{Sym}_{A,B}(\ket{\bm{\sigma}}_{A,B})|$ is typically much
smaller than the total number of symmetries $|\mathcal{G}|$. The
following example illustrates the idea and shows how to compute the
representative given the information about sublattice representatives
and representative symmetries.

\paragraph{Example} We consider the state
$\ket{\bm{\sigma}} = \ket{\uparrow \uparrow \downarrow \downarrow
  \uparrow \downarrow}$ on a six-site chain lattice as in
\cref{fig:sublatticetransformations}~a). Notice that the sites
are not enumerated from left to right but such that sites $1$ to
$3$ belong to the sublattice $A$ and sites $4$ to $6$ belong to
sublattice $B$. The states restricted on the sublattices are
$\ket{\bm{\sigma}}_A = \ket{\uparrow \uparrow \downarrow}$ and
$\ket{\bm{\sigma}}_B = \ket{\downarrow \uparrow \downarrow}$. The
action of the sublattice symmetries
\begin{align}
  \begin{split}
    \mathcal{G}_A &\equiv \{ \text{Id}, T_2, T_4 \}, \\
    \mathcal{G}_B &\equiv \{ T_1, T_3, T_5 \}, \\
  \end{split}
\end{align}
on $\ket{\bm{\sigma}}$ is shown in
\cref{fig:sublatticetransformations}~b)~and~c). From this, we compute
the sublattice representatives as in \cref{eq:sublatreps2sl},
\begin{align}
  \begin{split}
    \texttt{Rep}_A(\ket{\bm{\sigma}}_A) &=
    \ket{\downarrow\uparrow\uparrow},
    \\
    \texttt{Rep}_B(\ket{\bm{\sigma}}_B) &=
    \ket{\downarrow\downarrow\uparrow},
  \end{split}
\end{align}
whose integer values are given by
\begin{align}
  \begin{split}
    \integer(\texttt{Rep}_A(\ket{\bm{\sigma}}_A)) &= (011)_2 = 3, \\
    \integer(\texttt{Rep}_B(\ket{\bm{\sigma}}_B)) &= (001)_2 = 1.
  \end{split}
\end{align}
Since
$\texttt{Rep}_B(\ket{\bm{\sigma}}_B) <
\texttt{Rep}_A(\ket{\bm{\sigma}}_A)$ the symmetry $g_{\bm{\sigma}}$
yielding the total representative $\ket{\tilde{\bm{\sigma}}}$ must be
contained in
\begin{equation}
  \texttt{Sym}_B(\ket{\bm{\sigma}}_B) = \{ T_1 \},      
\end{equation}
which in this case just contains a single element, namely $T_1$.
Consequently, the representative $ \ket{\tilde{\bm{\sigma}}} $ is
given by
\begin{equation}
  \ket{\tilde{\bm{\sigma}}} =  T_1 \ket{\bm{\sigma}} =
  \ket{\downarrow \downarrow \uparrow  \uparrow \uparrow \downarrow}.      
\end{equation}
\bigbreak
\paragraph{Lookup tables} If the quantities
$\texttt{Rep}_{A,B}(\ket{\bm{\sigma}}_{A,B})$ and
$\texttt{Sym}_{A,B}(\ket{\bm{\sigma}}_{A,B})$ are now stored in a lookup
table, this computation can be done very efficiently. Notice that
instead of having to store $2^N$ entries in the lookup table for the
representative we only need four lookup tables of order
$\mathcal{O}(2^{N/2})$. On larger system sizes the difference between
memory requirements of order $\mathcal{O}(2^{N})$ and
$\mathcal{O}(2^{N/2})$ is substantial.

To further speed up computations we also create lookup tables to store
the action of each symmetry $g \in \mathcal{G}$ on a substate
$\ket{\bm{\sigma}}_A$,
\begin{align}
  \begin{split}
    \texttt{SymmetryAction}_A(g, \ket{\bm{\sigma}}_A) &=
    {g}\ket{\bm{\sigma}}_A,
    \\
    \texttt{SymmetryAction}_B(g, \ket{\bm{\sigma}}_B) &=
    {g}\ket{\bm{\sigma}}_B.
  \end{split}
\end{align}
With this information, we can efficiently apply symmetries to a given
spin configuration by looking up the action of $g$ on the respective
substate and combining the results. The memory requirement for these
lookup tables is $\mathcal{O}(N_{\text{sym}}2^{N/2})$, where
$N_{\text{sym}} = |\mathcal{G}|$. This can be reduced by generalizing
the sublattice coding algorithm to multiple sublattices. The memory
requirement then scales as
$\mathcal{O}(N_{\text{sym}}2^{N/N_{\text{sublat}}})$, where
$N_{\text{sublat}}$ denotes the number of sublattices.

\subsubsection{Generic sublattice coding algorithm}
We start by discussing how we subdivide a lattice $\Lambda$ into
$N_{\text{sublat}}$ sublattices. The basic requirement is that every
symmetry group element either only operates within the sublattices or
exchanges sublattices. We do not allow for symmetry elements that
split up a sublattice onto different sublattices. Therefore we make
the following definition:

\paragraph{Definition (Sublattice stability)}
A decomposition,
\begin{equation}
  \Lambda = \bigcupdot\limits_{X=1}^{N_{\text{sublat}}} \Lambda_X ,
\end{equation}
of a lattice $\Lambda$ with symmetry group $\mathcal{G}$ into
$N_{\text{sublat}}$ disjoint sublattices $\Lambda_X$ is called
\textit{sublattice stable} if every $g\in \mathcal{G}$ maps each
$\Lambda_X$ onto exactly one (possibly different) $\Lambda_Y$,
i.e. for all $g\in \mathcal{G}$ and all $\Lambda_X$ there exists a
$\Lambda_Y$ such that
\[
  g(\Lambda_X) = \Lambda_Y.
\]
The set $\Lambda_X$ is called the $X$-\textit{sublattice} of
$\Lambda$.

\begin{figure}[t!]
  \centering
  \begin{minipage}[c]{0.3\textwidth}
    \centering \includegraphics[width=0.7\textwidth]{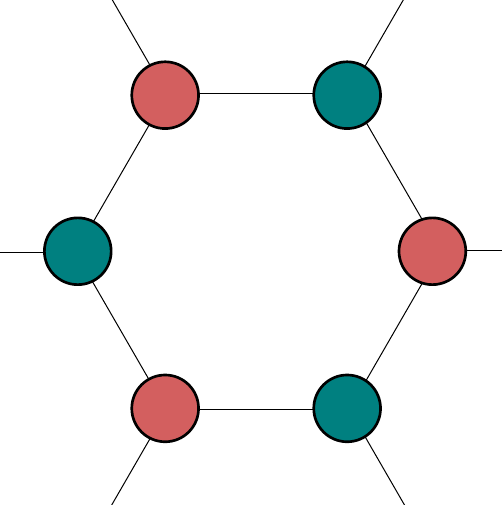}
  \end{minipage}%
  \quad
  \begin{minipage}[c]{0.3\textwidth}
    \includegraphics[width=\textwidth]{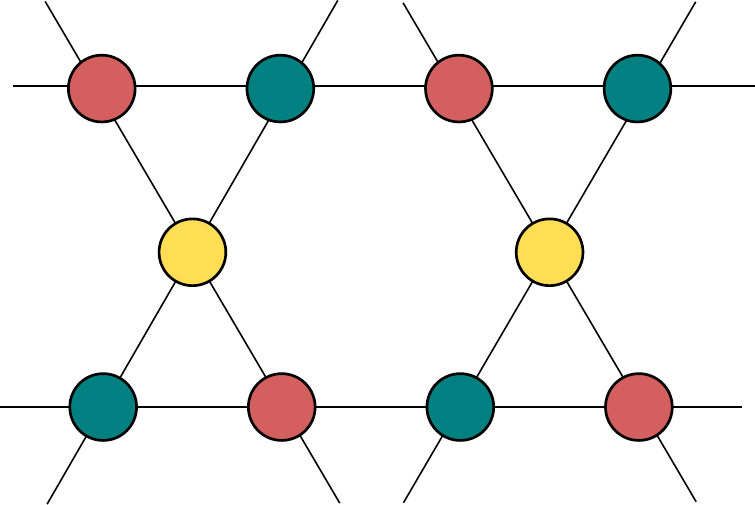}
  \end{minipage}
  \quad
  \begin{minipage}[c]{0.3\textwidth}
    \quad \includegraphics[width=0.9\textwidth]{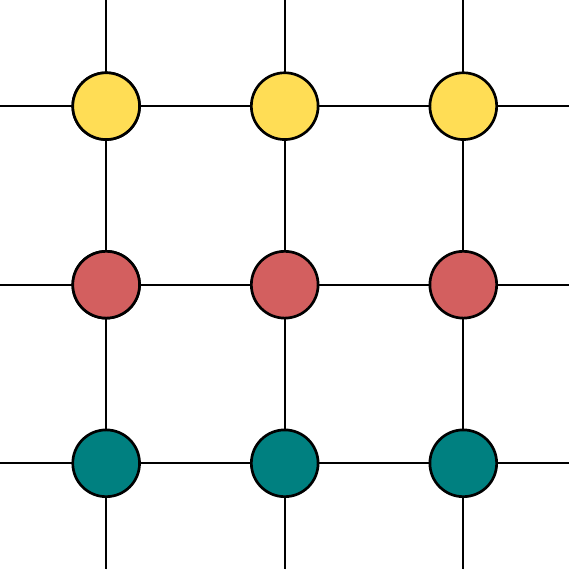}
  \end{minipage}
  \\
  \begin{minipage}[t]{0.3\linewidth}
    \subcaption{Two sublattice ordering in a honeycomb lattice.
      \label{fig:sublatticeorderings:honeycomb}}
  \end{minipage}
  \quad
  \begin{minipage}[t]{0.3\linewidth}
    \subcaption{Three sublattice ordering on a kagome lattice.
      \label{fig:sublatticeorderings:kagome}}
  \end{minipage}
  \quad
  \begin{minipage}[t]{0.3\linewidth}
    \subcaption{Three sublattice ordering on a square lattice.
      \label{fig:sublatticeorderings:square}}
  \end{minipage}

  \caption{Sublattice orderings for several common lattices. In
    figures \cref{fig:sublatticeorderings:honeycomb} and
    \cref{fig:sublatticeorderings:kagome} the sublattices are stable
    with respect to all spatial symmetries. In figure
    \cref{fig:sublatticeorderings:square} the sublattices are stable
    with respect to all translational symmetries, horizontal and
    vertical reflections, $180^\circ$ rotations but not with respect
    to $90^\circ$ rotations or diagonal reflections.}
  \label{fig:sublatticeorderings}
\end{figure}
\bigbreak The notion of \textit{sublattice stability} is illustrated
in \cref{fig:sublatticeorderings}. The sublattices $\Lambda_X$ are
drawn in different colors. A translation by one unit cell in
\cref{fig:sublatticeorderings:honeycomb} keeps the sublattices of the
honeycomb lattice invariant whereas a $60^\circ$ rotation exchanges
the sublattices.  For the kagome lattice in
\cref{fig:sublatticeorderings:kagome} a $60^\circ$ rotation around a
hexagon center for example cyclically permutes the three
sublattices. One checks that for both
~\cref{fig:sublatticeorderings:honeycomb,fig:sublatticeorderings:kagome}
all translational as well as all point group symmetries are sublattice
stable, so different color sublattices are mapped onto each
other. This is different for
\cref{fig:sublatticeorderings:square}. Still here all translational
symmetries just permute the sublattices, but a $90^\circ$ rotation
splits up a sublattice into different sublattices. Nevertheless, a
$180^\circ$ rotation keeps the sublattices stable, similarly a
vertical or horizontal reflection. Therefore, only the reduced point
group $\mathrm{D}2$ instead of the full $\mathrm{D}4$ point group for
the square lattice fulfills the sublattice stability condition in this
case. $\mathrm{D}2$ and $\mathrm{D}4$ denote the dihedral groups of
order $4$ and $8$ with two- and four-fold rotations and
reflections. Note, that for a square lattice a two or four sublattice
decomposition for which the full $\mathrm{D}4$ point group is
sublattice stable can be chosen instead. The choice of sublattice
decomposition in \cref{fig:sublatticeorderings:square} just serves
illustrational purposes.

From the definition of sublattice stability, it is clear that the
total number of sites $N$ has to be divisible by the number of
sublattices $N_{\text{sublat}}$.  The numbering of the lattice sites
is chosen such that the lattice sites from
$(X-1)N/N_{\text{sublat}} + 1$ to $XN/N_{\text{sublat}}$ belong to
sublattice $X$. We choose the most significant bits in the integer
representation to be the bits on sublattice $1$. Similar as in the
previous section we define the following quantities
\paragraph{Definition} \label{def:lookuptables} For every sublattice
$\Lambda_X$ we define the following notions:
\begin{itemize}
\item \textit{sublattice symmetries}:
  \begin{equation} \label{eq:sublatsymdef} \mathcal{G}_X \equiv \{ g
    \in \mathcal{G} \, | \, g \text{ maps sublattice } X \text{ onto
      sublattice } 1 \}.
  \end{equation}
\item \textit{sublattice representative}:
  \begin{equation} \label{eq:sublatrepdef}
    \texttt{Rep}_X(\ket{\bm{\sigma}}_X) \equiv h_X\ket{\bm{\sigma}}_X,
    \quad \text{where} \quad h_X = \argmin_{g \in \mathcal{G}_X} \,
    \text{int}(g \ket{\bm{\sigma}}_X),
  \end{equation}
  where $\ket{\bm{\sigma}}_X$ denotes the substate of
  $\ket{\bm{\sigma}}$ restricted on sublattice $\Lambda_X$.
\item \textit{representative symmetries}:
  \begin{equation} \label{eq:repsymdef}
    \texttt{Sym}_X(\ket{\bm{\sigma}}_X) \equiv \{ g \in \mathcal{G}_X
    \, | \, {g}\ket{\bm{\sigma}}_X =
    \texttt{Rep}_X(\ket{\bm{\sigma}}_X) \}.
  \end{equation}
\item \textit{sublattice symmetry action}:
  \begin{equation} \label{eq:symactiondef}
    \texttt{SymmetryAction}_X(g, \ket{\bm{\sigma}}_X) =
    {g}\ket{\bm{\sigma}}_X.
  \end{equation}
\end{itemize}
The symmetries in $\mathcal{G}_X$ map the sublattice $X$ onto the most
significant bits. Therefore, the symmetry that minimizes the integer
value in the orbit must be contained in the representative symmetries
of a minimal sublattice representative, i.e.
\begin{equation}
  g_{\bm{\sigma}} = \argmin\limits_{g \in \mathcal{G}} {g}\ket{\bm{\sigma}}
  \Rightarrow
  g_{\bm{\sigma}} \in \bigcup\limits_{\substack{Y\text{, \texttt{Rep}}_Y(\ket{\bm{\sigma}}_Y)
      \\ \text{minimal}}}{\texttt{Sym}}_Y(\ket{\bm{\sigma}}_Y). 
\end{equation}
To find the minimizing symmetry $g_{\bm{\sigma}}$, we only have to
check the symmetries yielding the minimal sublattice
representative. The quantities $\texttt{Rep}_X(\ket{\bm{\sigma}}_X)$
and $\texttt{Sym}_X(\ket{\bm{\sigma}}_X)$ are stored in lookup tables,
whose size scales as $\mathcal{O}(2^{N/N_{\text{sublat}}})$. In order
to quickly apply the symmetries, we can additionally store
$\texttt{SymmetryAction}_X(g, \ket{\bm{\sigma}}_X)$ in another lookup
table. The memory cost of doing so scales as
$\mathcal{O}(N_{\text{sym}}2^{N/N_{\text{sublat}}})$ and thus requires
the most memory. The generic sublattice coding algorithm consists of
two parts. The preparation of the lookup tables is shown as pseudocode
in \cref{alg:sublatticecodingprep}. The pseudocode of the actual
algorithm for finding the representative using the lookup tables is
shown in \cref{alg:sublatticecodingfindrep}.

%% \begin{algorithm}[H]
%%   \textbf{Preparation: }\\
%%   \For{ sublattice $X$} \For{ substate $\ket{\bm{\sigma}}_X$}
%%   compute the sublattice representative \cref{eq:sublatrepdef}, store it to $\texttt{Rep}_X(\ket{\bm{\sigma}}_X)$ \\
%%   compute the representative symmetries \cref{eq:repsymdef}, store them to $\texttt{Sym}_X(\ket{\bm{\sigma}}_X)$ \\
%%   \EndFor \EndFor \For{ sublattice $X$}{ \For{ substate
%%   $\ket{\bm{\sigma}}_X$}{ \For{ symmetry $g \in \mathcal{G}$}{
%%   compute ${g}\ket{\bm{\sigma}}_X$ and store it to
%%   $\texttt{SymmetryAction}_X(g, \ket{\bm{\sigma}}_X)$ } \EndFor
%%   \EndFor \EndFor

%%   \caption{Preparation step for initializing the lookup tables. The
%%   size of $\texttt{Rep}_X(\ket{\bm{\sigma}}_X)$ and
%%   $\texttt{Sym}_X(\ket{\bm{\sigma}}_X)$ scales as
%%   $\mathcal{O}(2^{N/N_{\text{sublat}}})$ whereas the size of the
%%   lookup table $\texttt{SymmetryAction}_X(g, \ket{\bm{\sigma}}_X)$
%%   scales as $\mathcal{O}(N_{\text{sym}}2^{N/N_{\text{sublat}}})$.}
%% \end{algorithm}

\begin{algorithm}[ht!]
  \caption{Preparation of lookup tables for sublattice coding
    algorithm}
  \begin{algorithmic}[0]
    \ForEach{substate $\ket{\bm{\sigma}_X}$} \ForEach{sublattice $X$}
    \State compute the sublattice representative
    \cref{eq:sublatrepdef}, store it to
    $\texttt{Rep}_X(\ket{\bm{\sigma}}_X)$ \State compute the
    representative symmetries \cref{eq:repsymdef}, store them to
    $\texttt{Sym}_X(\ket{\bm{\sigma}}_X)$ \ForEach{ symmetry
      $g \in \mathcal{G}$} \State compute ${g}\ket{\bm{\sigma}}_X$ and
    store it to $\texttt{SymmetryAction}_X(g, \ket{\bm{\sigma}}_X)$
    \EndForEach \EndForEach \EndForEach
  \end{algorithmic}
  \label{alg:sublatticecodingprep}
\end{algorithm}

\begin{algorithm}[ht!]
  \caption{Sublattice coding algorithm for finding the
    representative.}
  \begin{algorithmic}[0]
    \Require state $\ket{\bm{\sigma}}$ \Ensure representative
    $\ket{\tilde{\bm{\sigma}}}$ and $g_{\bm{\sigma}}$ \State Determine
    $\texttt{MinRep} = \min\limits_{X} \left\{
      \texttt{Rep}_X(\ket{\bm{\sigma}}_X)\right\}$ \State Set
    $\ket{\tilde{\bm{\sigma}}} = +\infty$ \ForEach{ sublattice $Y$
      with $\texttt{Rep}_Y(\ket{\bm{\sigma}}_Y) = \texttt{MinRep}$}
    \ForEach{ symmetry $ g \in \texttt{Sym}_Y(\ket{\bm{\sigma}}_Y)$}
    \State compute ${g} \ket{\bm{\sigma}}$ by using the lookup tables
    $\texttt{SymmetryAction}_X(g, \ket{\bm{\sigma}}_X)$
    \If{${g} \ket{\bm{\sigma}} < \ket{\tilde{\bm{\sigma}}}$} \State
    $\ket{\tilde{\bm{\sigma}}} \leftarrow {g} \ket{\bm{\sigma}} $
    \State $g_{\bm{\sigma}} \leftarrow {g}$ \EndIf \EndForEach
    \EndForEach \State \Return $\ket{\tilde{\bm{\sigma}}}$,
    $g_{\bm{\sigma}}$
  \end{algorithmic}
  \label{alg:sublatticecodingfindrep}
\end{algorithm}

\begin{figure}[t!]
  \centering \includegraphics[width=\textwidth]{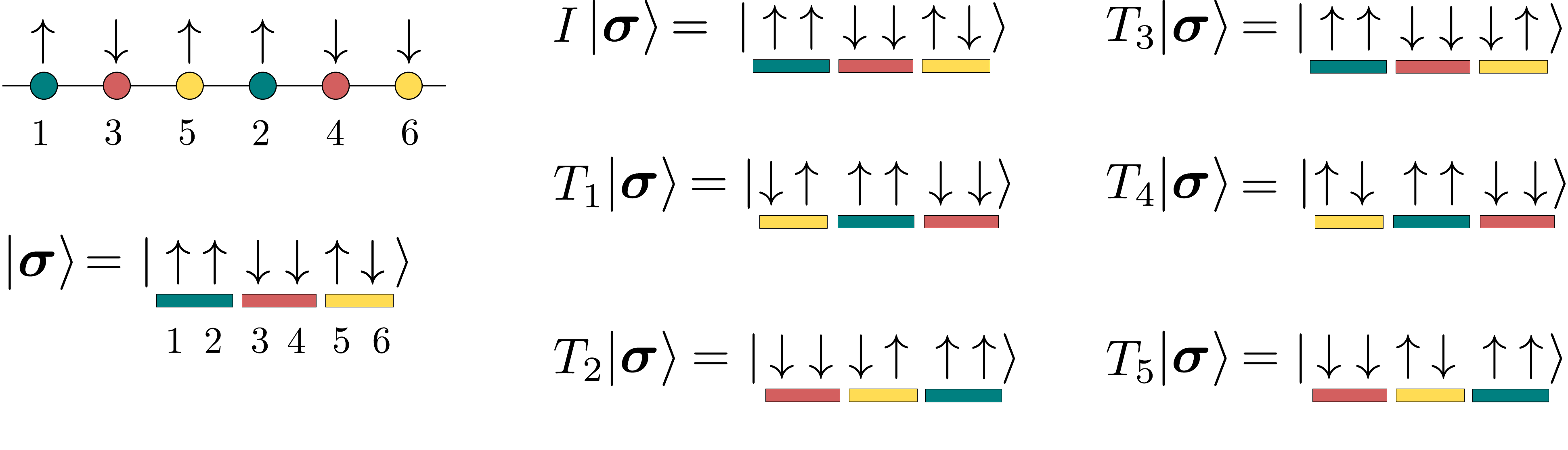}
  \caption{Three sublattice coding of the spin state
    $\ket{\bm{\sigma}}$ on a six-site chain lattice and action of
    translation symmetries. The sites are enumerated such that site
    $1$ and $2$ are on sublattice $A$, $3$,$4$ on $B$ and $5$,$6$ on
    $C$. The representative state with this enumeration of sites is
    given by
    $\ket{\tilde{\bm{\sigma}}} = T_2\ket{\bm{\sigma}} =
    \ket{\downarrow \downarrow \downarrow \uparrow \uparrow
      \uparrow}$}
  \label{fig:threesublatticetransformations}
\end{figure}

\paragraph{Example} We consider the same state on a six-site chain
lattice as in \cref{fig:sublatticetransformations}, but now using a
three sublattice decomposition in
\cref{fig:threesublatticetransformations}. We call the blue sublattice
the $A$ sublattice, the red $B$ and the yellow $C$. Notice, that due
to different sublattice structure the labeling of the real space sites
is different from the two sublattice case. In the three sublattice
case, we are now given the state
\begin{equation}
  \ket{\bm{\sigma}} = \ket{ \uparrow \uparrow \downarrow \downarrow \uparrow \downarrow}.
\end{equation}
Its substates are
\begin{equation}
  \ket{\bm{\sigma}}_A = \ket{ \uparrow \uparrow}, \quad
  \ket{\bm{\sigma}}_B = \ket{ \downarrow \downarrow}, \quad
  \ket{\bm{\sigma}}_C = \ket{ \uparrow \downarrow},
\end{equation}
with corresponding sublattice representatives
\begin{equation}
  \texttt{Rep}_A(\ket{\bm{\sigma}}_A) = \ket{ \uparrow \uparrow}, \quad
  \texttt{Rep}_B(\ket{\bm{\sigma}}_B) = \ket{ \downarrow \downarrow}, \quad
  \texttt{Rep}_C(\ket{\bm{\sigma}}_C) = \ket{ \downarrow \uparrow},
\end{equation}
and representative symmetries
\begin{equation}
  \texttt{Sym}_A(\ket{\bm{\sigma}}_A) = \{ I, T_3\}, \quad
  \texttt{Sym}_B(\ket{\bm{\sigma}}_B) = \{ T_2, T_5\}, \quad
  \texttt{Sym}_C(\ket{\bm{\sigma}}_C) = \{ T_1\}.
\end{equation}
The minimal sublattice representative $\texttt{MinRep}$ as in
\cref{alg:sublatticecodingfindrep} is given by
\begin{equation}
  \texttt{MinRep} = \texttt{Rep}_B(\ket{\bm{\sigma}}_B) =
  \ket{ \downarrow \downarrow}.
\end{equation}
The minimizing symmetry must now be in
$\texttt{Sym}_B(\ket{\bm{\sigma}}_B) = \{ T_2, T_5\}$.  We see that
\begin{equation}
  T_2\ket{\bm{\sigma}} =
  \ket{\downarrow \downarrow \downarrow \uparrow \uparrow \uparrow} <
  T_5\ket{\bm{\sigma}} =
  \ket{\downarrow \downarrow \uparrow \downarrow \uparrow \uparrow}. 
\end{equation}
Therefore, the representative $\ket{\tilde{\bm{\sigma}}}$ is given by
\begin{equation}
  \ket{\tilde{\bm{\sigma}}} = \ket{\downarrow \downarrow \downarrow \uparrow \uparrow \uparrow},
\end{equation}
with the minimizing symmetry $g_{\bm{\sigma}} = T_2$.  Notice, that
this state differs from the one found in the two sublattice example
since the labeling of the sites changes the integer representation of
a state and thus the definition of the representative. Once a given
labeling of sites is fixed the representative is of course unique.

%%% Local Variables:
%%% mode: latex
%%% TeX-master: "../../thesis"
%%% End:

%auto-ignore
\section{Distributed and hybrid memory parallelization}
\label{sec:edparallelization}

For reaching larger system sizes in ED computations a proper balance
between memory requirements and computational costs has to be
found. There are two major approaches when applying the Lanczos
algorithm. The Hamiltonian matrix can either be stored in memory in
some sparse-matrix format or generated on-the-fly every time a
matrix-vector multiplication is performed. Storing the matrix is
usually faster, yet memory requirements are higher.  This approach is
for example pursued by the software package \texttt{SPINPACK}
\cite{Spinpack}.  A matrix-free implementation of the Lanczos
algorithm usually needs more computational time since the matrix
generation, especially in a symmetrized basis can be expensive. Of
course, the memory cost is drastically reduced since only a few vectors
of the size of the Hilbert space have to be stored. It turns out that
on current supercomputing infrastructures the main limitation in going
to larger system sizes is indeed the memory requirements of the
computation. It is thus often favorable to use a slower matrix-free
implementation, as done by the software package ${\mathcal H}\Phi$
\cite{Kawamura2017}, for example. Due to this reasons, we also choose
the matrix-free approach.

The most computational time in the Lanczos algorithm is used in the
matrix-vector multiplication. The remaining types of operations are
scalar multiplications, dot products of Lanczos vectors or the
diagonalization of the $T$-matrix which are usually of negligible
computational cost. Today's largest supercomputers are typically
distributed memory machines, where every process only has direct
access to a small part of the total memory. It is thus a nontrivial
task to distribute data onto several processes and implement
communication amongst them once remote memory has to be
accessed. Also, when scaling the software to a larger amount of
processes load balancing becomes important. The computational work
should be evenly distributed amongst the individual processes in order
to avoid waiting times in communication. In the following, we explain
how we achieve this goal in our implementation using the Message
Passing Protocol (MPI) \cite{Forum:1994:MMI:898758}.

\paragraph{Matrix-vector multiplication}
The Hamiltonian can be written a sum of non-branching terms,
\begin{align}
  H = \sum \limits_k H_k,
\end{align}
as in \cref{eq:nonbranchingcondition}. To perform the full
matrix-vector multiplication we compute the matrix-vector
multiplication for the non-branching terms $H_k$ and add up the
results,
\begin{equation}
  \label{eq:branchingsum}
  H\ket{\psi} = \sum  \limits_k H_k \ket{\psi}.
\end{equation}
We denote by
\begin{equation}
  \label{eq:sigmabasis}
  \{ \ket{\bm{\sigma}_i} \}, \quad i=1,\ldots,D \quad ,
\end{equation}
a (possibly symmetry-adapted) basis of the Hilbert space.  A wave
function $\ket{\psi}$ is represented on the computer by storing its
coefficients $\braket{\bm{\sigma}_i| \psi}$.  Given an input vector,
\begin{equation}
  \label{eq:inputvector}
  \ket{\psi_{\text{in}}} = \sum\limits_{i=1}^D
  \braket{\bm{\sigma}_i| \psi_{\text{in}}} \ket{\bm{\sigma}_i},
\end{equation}
we want to compute the coefficients
$\braket{\bm{\sigma}_i| \psi_{\text{out}}}$ in
\begin{equation}
  H_k\ket{\psi_{\text{in}}} = \ket{\psi_{\text{out}}}.
\end{equation}
The resulting output vector $\ket{\psi_{\text{out}}}$ is given by
\begin{align}
  \begin{split}
    \ket{\psi_{\text{out}}} &= \sum\limits_{i=1}^D
    \braket{\bm{\sigma}_i| \psi_{\text{out}}} \ket{\bm{\sigma}_i}=
    \sum\limits_{i=1}^D
    \braket{\bm{\sigma}_i| H_k|\psi_{\text{in}}} \ket{\bm{\sigma}_i} \\
    & = \sum\limits_{i,j=1}^D c_k(\bm{\sigma}_j)\braket{\bm{\sigma}_j|
      \psi_{\text{in}}} \braket{\bm{\sigma}_i|\bm{\sigma}_j^\prime}
    \ket{\bm{\sigma}_i},
  \end{split}
      \label{eq:outputvector}  
\end{align}
where $ c_k(\bm{\sigma}_j)$ and $\ket{\bm{\sigma}_j^\prime}$ are given
by
\begin{equation}
  \label{eq:outpusonbasis}
  H_k\ket{\bm{\sigma}_j} = c_k(\bm{\sigma}_j)\ket{\bm{\sigma}_j^\prime}.
\end{equation}
Notice, that in a symmetry-adapted basis, evaluating
$c_k(\bm{\sigma}_j)$ requires the evaluation of
\cref{eq:matrixelementrepresentatives_lsed}, where the sublattice
coding technique of \cref{sec:sublatticecoding} can be
applied. Clearly, we have
\begin{equation}
  \label{eq:matmulkron}
  \braket{\bm{\sigma}_i|\bm{\sigma}_j^\prime} =
  \begin{cases}
    1 \text{ if } \ket{\bm{\sigma}_i} = \ket{\bm{\sigma}_j^\prime}, \\
    0 \text{ else.}
  \end{cases}
\end{equation}
For parallelizing the multiplication \cref{eq:outputvector}, we
distribute the coefficients in the basis $\{ \ket{\bm{\sigma}_i} \}$
onto the different MPI processes. This means we have a mapping,
\begin{equation}
  \label{eq:defprocdistro}
  \texttt{proc}: \ket{\bm{\sigma}_i} \rightarrow \{ 1, \dots, n_{\text{procs}} \},
\end{equation}
that assigns to every basis state of the Hilbert space its MPI process
number. Here, $n_{\text{procs}}$ denotes the number of MPI processes.
In general, $\ket{\bm{\sigma}_j}$ and $\ket{\bm{\sigma}_j^\prime}$ are
not stored in the same process. Hence, the coefficient
$c_k(\bm{\sigma}_j)\braket{\bm{\sigma}_j| \psi_{\text{in}}}$ has to be
sent from the process no. $\texttt{proc}(\ket{\bm{\sigma}_j})$ to
process no. $\texttt{proc}(\ket{\bm{\sigma}_j^\prime})$. This makes
communication between the processes necessary. This communication is
buffered in our implementation, i.e. for every basis state
$\ket{\bm{\sigma}_j}$ we first store the target basis state
$\ket{\bm{\sigma}_j^\prime}$ and the coefficient
$c_k(\bm{\sigma}_j)\braket{\bm{\sigma}_j| \psi_{\text{in}}}$
locally. Once every local basis state has been evaluated, we perform
the communication and exchange the information amongst all
processes. This corresponds to an \texttt{MPI\_Alltoallv} call in the
MPI standard.

After this communication step, every process has to add the received
coefficient to the locally stored coefficient
$\braket{\bm{\sigma}_j^\prime| \psi_{\text{out}}}$. For this, we have
to search, where the now locally stored coefficient of the basis state
$\ket{\bm{\sigma}_j^\prime}$ is located in memory. Typically, we keep
a list of all locally stored basis states defining the position of the
coefficients. This list is then searched for the entry
$\ket{\bm{\sigma}_j^\prime}$, which can also be time-consuming and
needs to be done efficiently. We are thus facing the following
challenges when distributing the basis states of the Hilbert space
amongst the MPI processes:
\begin{itemize}
\item Every process has to know which process any basis state
  $\ket{\bm{\sigma}_i}$ belongs to.
\item The storage of the information about the distribution should be
  memory efficient.
\item The distribution of basis states has to be fair, in the sense
  that every process has a comparable workload in every matrix-vector
  multiplication.
\item The search for a basis state within a process should be done
  efficiently.
\end{itemize}
We will now propose a method to address these issues in a satisfactory
way.

\begin{figure}[t!]
  \centering \includegraphics[width=0.8\textwidth]{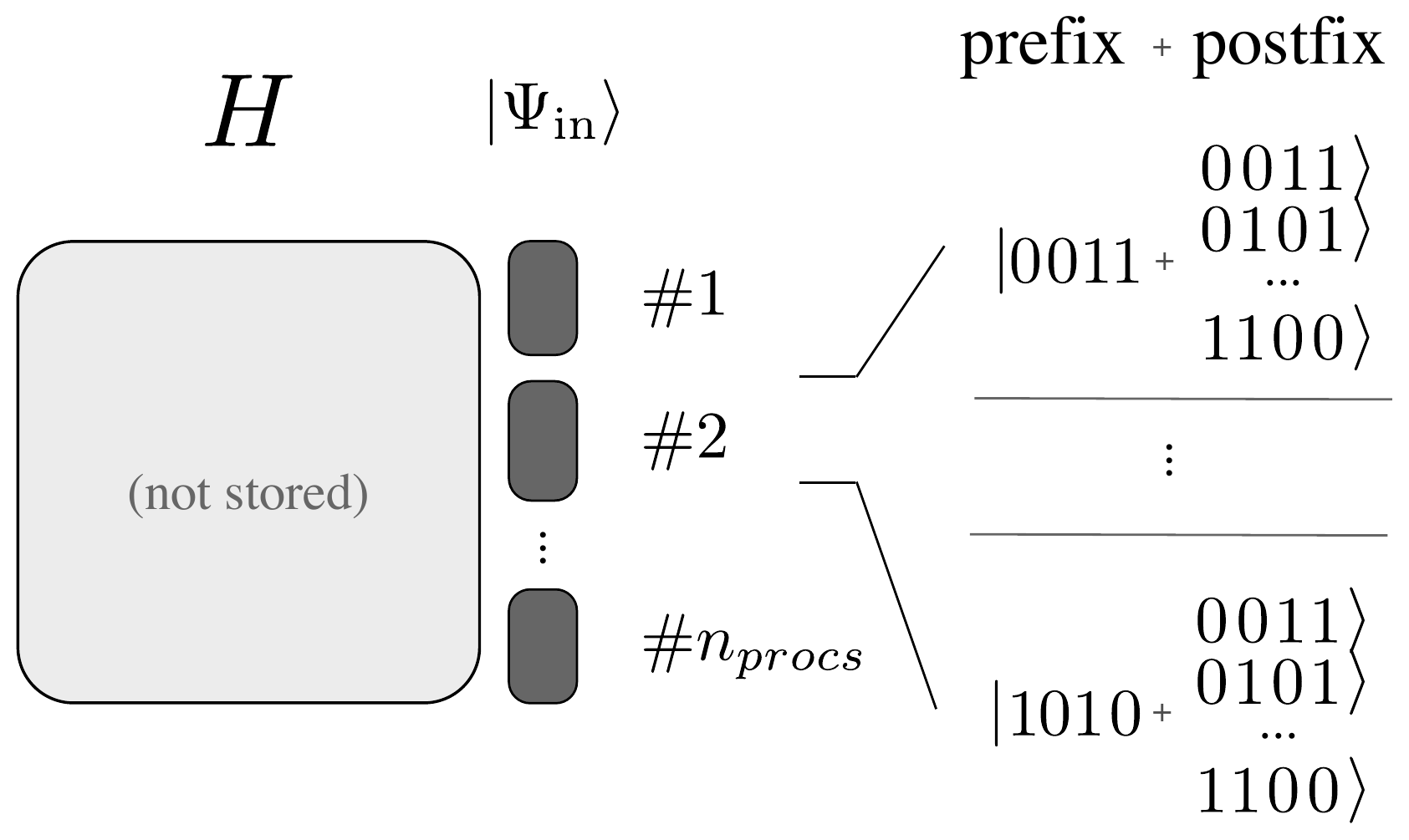}
  \caption{Storage layout of the distributed Hilbert space. The
    prefixes are randomly distributed amongst the MPI processes using
    a hash function. States with same prefixes are mapped to the same
    process. Within a process, the states are ordered
    lexicographically. The Hamiltonian matrix is not stored.}
  \label{fig:edparallelization}
\end{figure}

\paragraph{Distribution of basis states}
The central point of our parallelization strategy is the proper choice
of the distribution function $\texttt{proc}(\bm{\sigma})$ for the
basis states in \cref{eq:defprocdistro}. We split up every basis state
into prefix and postfix sites,
\begin{equation}
  \ket{\bm{\sigma}} =
  \ket{\underbrace{\sigma_1 \cdots \sigma_{n_{\text{prefix}}}}_{\text{prefix sites}}
    \quad
    \underbrace{\sigma_{n_{\text{prefix}}+1}\cdots\sigma_{n_{\text{prefix}}+ n_{\text{postfix}}}}_{\text{postfix sites}}},
\end{equation}
where $n_{\text{prefix}}$ and $n_{\text{postfix}}$ denote the number
of prefix and postfix sites. We decide that states with the same
prefix are stored in the same MPI process. The prefixes are randomly
distributed amongst all the processes. We do this by using a hash
function that maps the prefix bits onto a random but deterministic MPI
process. This hash function can be chosen such that every process has
a comparable amount of states stored locally. Moreover, a random
distribution of states reduces load balance problems significantly
since the communication structure is randomized. This is in stark
contrast to distributing the basis states in a linear
fashion. Thereby, single processes can often have a multiple of the
workload than other processes, thus causing idle time in other
processes.

\begin{algorithm}[p]
  \caption{Preparation of the distributed and symmetrized Hilbert
    space}
  \begin{algorithmic}[0]
    \State Perform the following steps on every process in parallel
    (no communication necessary) \State \texttt{myid} denotes the
    number of the current MPI process \State prepares data structures
    \texttt{Basis}, \texttt{Limits}
    on each process\\
    \ForEach{prefix spin configuration
      $\ket{\bm{\sigma}_{\text{prefix}}} = \ket{\sigma_1 \cdots
        \sigma_{n_{\text{prefix}}}}$}
    \If{$\texttt{proc}(\ket{\bm{\sigma}_{\text{prefix}}}) \neq $
      \texttt{myid}} \State continue \Else \State \texttt{begin} =
    length(\texttt{Basis}) \ForEach{spin configuration
      $\ket{\bm{\sigma}}$ with prefix
      $\ket{\bm{\sigma}_{\text{prefix}}}$} \State compute
    representative $\ket{\tilde{\bm{\sigma}}}$ of $\ket{\bm{\sigma}}$
    \If{$\ket{\bm{\sigma}} = \ket{\tilde{\bm{\sigma}}}$} \State append
    $\ket{\bm{\sigma}}$ to \texttt{Basis} \EndIf \EndForEach
    \texttt{end} = length(\texttt{Basis}) \If{end $\neq$ begin} \State
    insert ($\ket{\bm{\sigma}_{\text{prefix}}}$, begin, end) to
    \texttt{Limits} \EndIf \EndIf \EndForEach
  \end{algorithmic}
  \label{alg:hilbertspaceprep}
\end{algorithm}

\begin{algorithm}[p]
  \caption{Parallel matrix-vector multiply for a non-branching term
    $H_k$}
  \begin{algorithmic}[0]
    \Require input wave function $\ket{\psi_{\text{in}}}$
    \Ensure matrix-vector product $\ket{\psi_{\text{out}}} = H_k\ket{\psi_{\text{in}}}$\\
    %% \State \texttt{myid} denotes the number of the current MPI
    %% process
    \State $\rhd$ Preparation and sending step (communication may be
    buffered) \ForEach{basis state $\ket{\bm{\sigma}_j}$ stored
      locally in \texttt{Basis}} \State $\cdot$ apply non-branching
    term $H_k$ and apply sublattice coding technique \State $ $ to
    compute $ c_k(\bm{\sigma}_j)$ and $\ket{\bm{\sigma}_j^\prime}$,
    $$
    H_k\ket{\bm{\sigma}_j} =
    c_k(\bm{\sigma}_j)\ket{\bm{\sigma}_j^\prime}.
    $$
    \State $\cdot$ compute
    $c = c_k(\bm{\sigma}_j)\braket{\bm{\sigma}_j| \psi_{\text{in}}}$
    \State $\cdot$ send the pair $(\ket{{\bm{\sigma}}_j^\prime}, c)$
    to process no. $\texttt{proc}(\ket{{\bm{\sigma}}_j^\prime})$
    \EndForEach\\
    \State $\rhd$ Receiving and search step \ForEach{pair
      $(\ket{{\bm{\sigma}}_j^\prime}, c)$ received} \State $\cdot$
    determine indices (\texttt{begin}, \texttt{end}) from
    $\texttt{Limits}(\ket{{\bm{\sigma}}_j^\prime})$ \State $\cdot$
    determine index $i$ of $\ket{\bm{\sigma}^\prime}$ by binary search
    in array \State $ $ \texttt{Basis} between (\texttt{begin},
    \texttt{end}) \State $\cdot$ Set
    $\braket{\bm{\sigma}_j^\prime|\psi_{\text{out}}}[i] \leftarrow c$
    \EndForEach

  \end{algorithmic}
  \label{alg:matrixvector}
\end{algorithm}

By choosing this kind of random distribution of basis states, we also
don't have to store any information about their distribution. This
information is all encoded in the hash function.  Nevertheless, we
store the basis states belonging to a process locally in an
array. Finding the index of a given basis state also requires some
computational effort. Here, we use the separation between prefix and
postfix sites.  We store the basis states in an ordered way. This way,
states belonging to the same prefix are aligned in memory as shown in
\cref{fig:edparallelization}. We can store the index of the first and
the last states that belong to a given prefix. To find the index of a
given state we can now lookup the first and last index of the prefix
of this state and perform a binary search for the state between these
two indices. This reduces the length of the array we have to perform
the binary search on and, hence, reduces the computational effort in
finding the index. For implementing this procedure we need two data
structures locally stored on each process.
\begin{enumerate}
\item An array $\texttt{Basis}(i)$ storing all the basis states,
  \begin{equation}
    \texttt{Basis}(i) = \ket{\bm{\sigma}_i}, \quad i=1,\ldots,D.
  \end{equation}
\item An associative array
  $\texttt{Limits}(\ket{\bm{\sigma}_{\text{prefix}}})$ storing the map
  \begin{equation}
    \texttt{Limits}(\ket{\bm{\sigma}_{\text{prefix}}}) =
    \left[\texttt{begin}(\ket{\bm{\sigma}_{\text{prefix}}}),
      \texttt{end}(\ket{\bm{\sigma}_{\text{prefix}}})\right]
  \end{equation}
  where $\texttt{begin}(\ket{\bm{\sigma}_{\text{prefix}}})$ denotes
  the index of the first state with prefix
  $\ket{\bm{\sigma}_{\text{prefix}}}$ and
  $\texttt{end}(\ket{\bm{\sigma}_{\text{prefix}}})$ denotes the index
  of the last state with this prefix in the array $\texttt{Basis}(i)$,
  $\ket{\bm{\sigma}_{\text{prefix}}} = \ket{\sigma_1 \cdots
    \sigma_{n_{\text{prefix}}}}$.
\end{enumerate}
In \cref{alg:hilbertspaceprep} we summarize how to prepare these data
structures. The parallel matrix-vector multiplication in pseudocode is
shown in \cref{alg:matrixvector}. When working in the symmetry-adapted
basis, the lookup tables of the sublattice coding method need to be
accessible to every MPI process. One way to achieve this is of course,
that every process generates its own lookup tables.  However, in
present-day supercomputers, several processes will be assigned to the
same physical machine sharing the same physical memory.  To save
memory, the lookup tables are stored only once on a computing
node. Its processes can then access the lookup tables via shared
memory access. In our code, we use POSIX shared memory
functions~\cite{posixshm} to implement this hybrid parallelization.

%%% Local Variables:
%%% mode: latex
%%% TeX-master: "../../thesis"
%%% End:

%auto-ignore
\section{Benchmarks}
\label{sec:edbanchmarks}

\begin{figure}[t]
  \centering
  \begin{minipage}[c]{0.4\textwidth}
    \centering
    \includegraphics[width=0.7\textwidth]{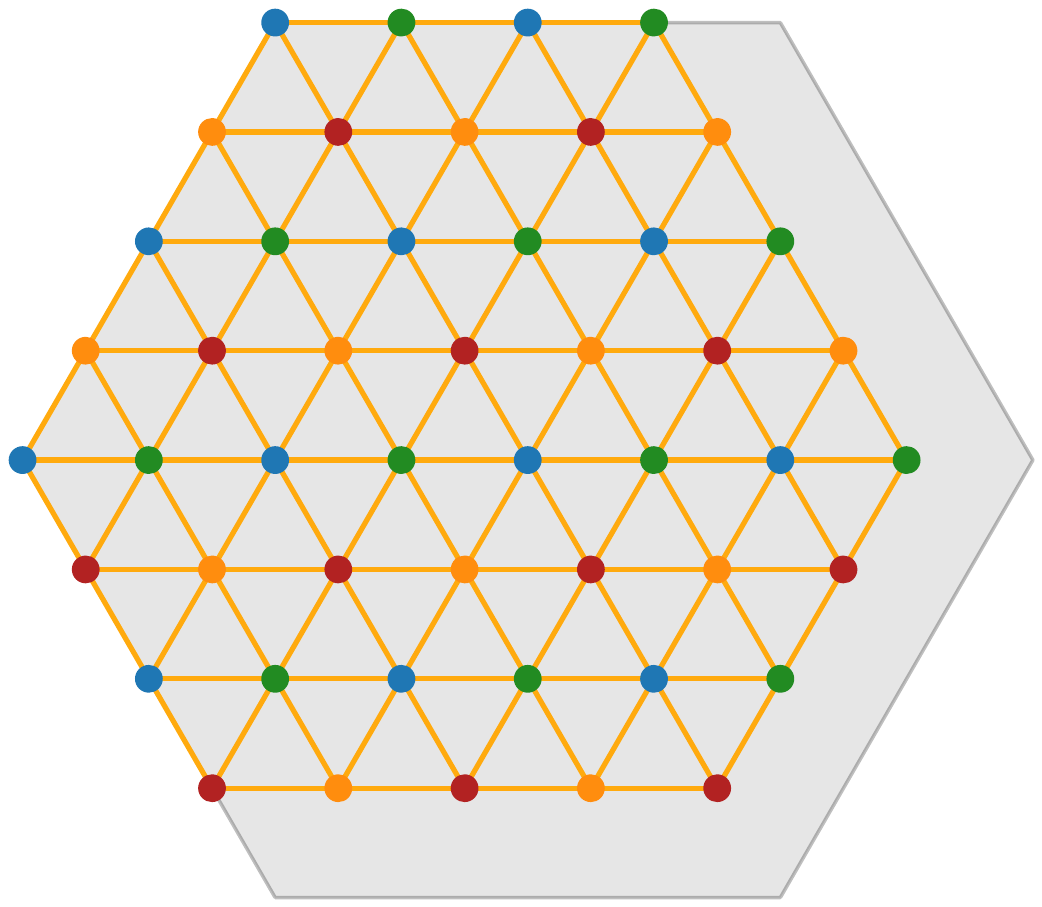}
  \end{minipage}\quad
  \begin{minipage}[c]{0.4\textwidth}
    \centering \includegraphics[width=0.8\textwidth]{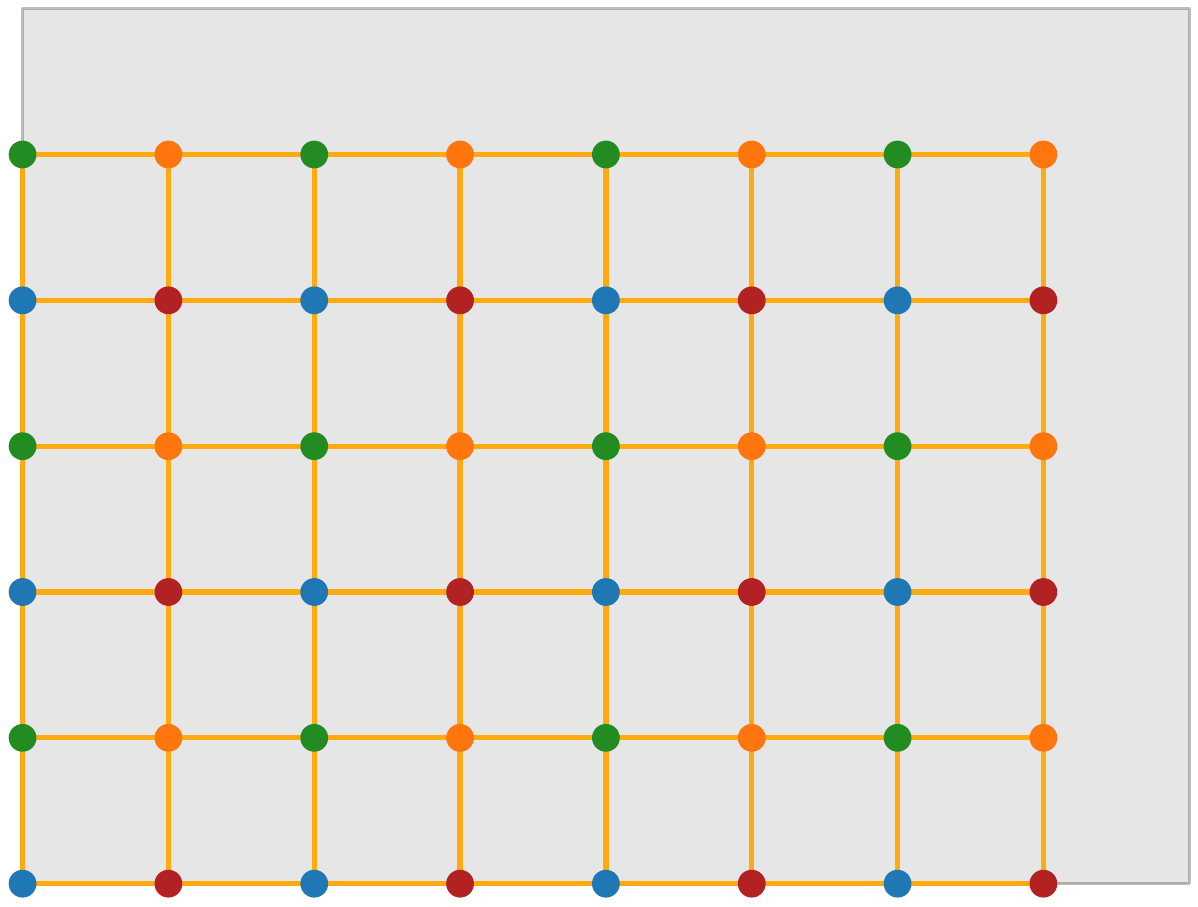}
  \end{minipage}\\
  \begin{minipage}[c]{0.4\textwidth}
    \subcaption{\label{fig:benchgeo:triangle48} Triangular lattice,
      $48$ sites, \\Four sublattice structure.}
  \end{minipage}\quad
  \begin{minipage}[c]{0.4\textwidth}
    \subcaption{\label{fig:benchgeo:square48} Square lattice, $48$
      sites,\\ Four sublattice structure.}
  \end{minipage}\\
  \begin{minipage}[c]{0.4\textwidth}
    \centering \includegraphics[width=0.65\textwidth]{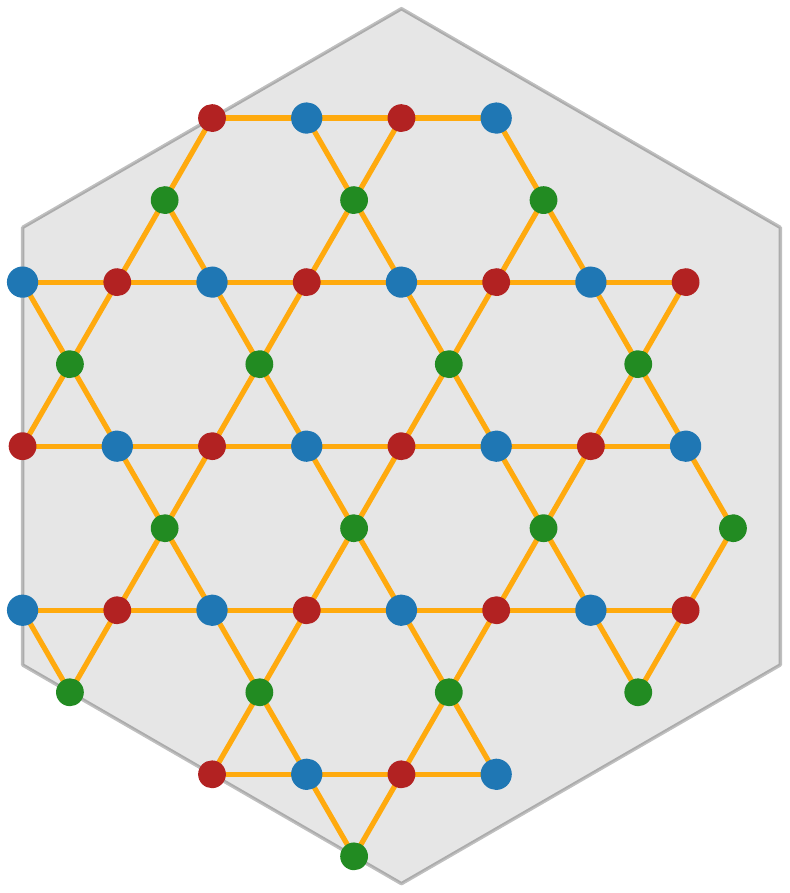}
  \end{minipage}\quad
  \begin{minipage}[c]{0.4\textwidth}
    \centering \includegraphics[width=0.7\textwidth]{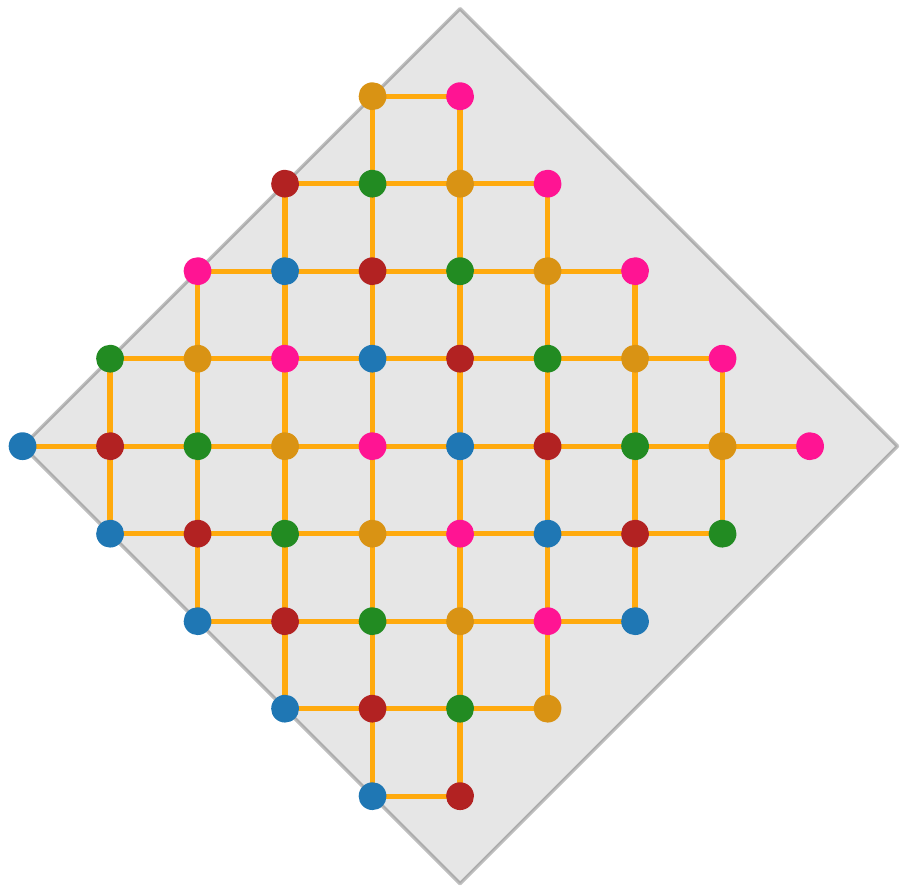}
  \end{minipage}
  \\
  \begin{minipage}[c]{0.4\textwidth}
    \subcaption{\label{fig:benchgeo:kagome48} Kagome lattice, $48$
      sites, \\Three sublattice structure.}
  \end{minipage}\quad
  \begin{minipage}[c]{0.4\textwidth}
    \subcaption{\label{fig:benchgeo:square50} Square lattice, $50$
      sites, \\Five sublattice structure.}
  \end{minipage}
 
  \caption{Geometries of Heisenberg spin-$1/2$ model
    benchmarks. Different colors show the sublattice structure used
    for the sublattice coding technique. Grey background shows the
    Wigner-Seitz cell defining the periodicity of the lattice.  }
  \label{fig:benchmarkgeometry}
\end{figure}

In order to assess the power of the methods proposed in the previous
sections, we performed test runs to compute ground state energies. We
considered the Heisenberg antiferromagnetic spin-$1/2$ nearest
neighbor model on four different lattice geometries: square ($48$
sites), triangular ($48$ sites), kagome ($48$ sites) and square ($50$
sites). \Cref{fig:benchmarkgeometry} shows the simulation clusters and
the sublattice structure we used. The benchmarks were performed on
three different supercomputers. The Vienna Scientific Cluster VSC3 is
built up from over 2020 nodes with two Intel Xeon E5-2650v2, 2.6 GHz,
8 core processors, the supercomputer Hydra at the Max Planck
Supercomputing \& Data Facility in Garching with over 3500 nodes with
20 core Intel Ivy Bridge 2.8 GHz processors and the System B Sekirei
at the Institute for Solid State Physics of the University of Tokyo
with over 1584 nodes with two Intel Xeon E5-2680v3 12 core 2.5GHz
processors. Both the Hydra and Sekirei use InfiniBand FDR
interconnect, whereas the VSC3 uses Intel TrueScale Infiniband for
network communication.

The benchmarks are summarized in \cref{tab:edbenchmarks}. We make use
of all translational, certain point group symmetries and spin-flip
symmetry. We show the memory occupied by a single lookup table for the
symmetries. Since we use a single buffered and blocking all-to-all
communication in the implementation it is straightforward to measure
the percentage of time spent for MPI communication by taking the time
before the communication call and afterward. In order to validate the
results of our computation, we compared the results of the
unfrustrated square case to Quantum Monte Carlo computations of the
ground state energy. We used a continuous time world-line Monte Carlo
Code \cite{Todo2001a} with $10^5$ thermalization and $10^6$
measurements at temperature $T=0.01$.  The computed energies per site
are $E/N = -0.676013 \pm 2\cdot 10^{-5}$ for the $48$ site square
cluster and $E/N = -0.67512 \pm 2\cdot 10^{-5}$ for the $50$ site
cluster. The actual values computed with ED are within the error bars.
% \begin{itemize}
% \item 48 QMC -0.676013 +- 2.02829e-05 100000 therm 1000000
%   production
% \item 50 QMC -0.67512 +- 1.98685e-05 100000 therm 1000000 production
% \end{itemize}
The ground state energy of the kagome Heisenberg antiferromagnet on
$48$ sites has been previously computed \cite{Lauchli2016} with a
specialized code and agrees with our results. We see that the amount
of time spent for communication is different for the three
supercomputers. On Sekirei, a parallel efficiency of $61\%$ on 3456
cores has been achieved.

\begin{table}[t]
  \centering
  \begin{tabular}{|c | c | c | c | c |}
    \hline
    Geometry & Triangular 48 & Square 48 & Kagome 48 & Square 50 \\
    computer & Sekirei & VSC3 & Hydra & Sekirei \\
    \hline
    point group & D6 & D2 & D6 & D2 \\
    \# symmetries & 1152 & 384 & 384 & 400 \\
    % \# sublattices & 4 & 4 & 3 & 5 \\
    dimension & $2.8 \cdot 10 ^{10}$ &  $8.3 \cdot 10^{10}$ & $8.4 \cdot 10^{10}$ & $3.2 \cdot 10^{11}$ \\
    \# cores & 3456 & 8192 & 10240 & 3456 \\
    total memory &  2.5 TB & n.A. & n.A. & 15.5 TB \\
    memory lookup & 151 MB & 50 MB & 604 MB & 17 MB \\
    Time / MVM & 399 s & 1241 s & 258 s & 3304 s \\
    \% comm. time & 39\% & 77\% & 48\% & 39\% \\ \hline
    g.s. sector & $\Gamma$.A1.even & $\Gamma$.A1.even & $\Gamma$.A1.even & M.A1.odd\\
    g.s. energy & -26.8129452715 & -32.4473598728 & -21.0577870635 & -33.7551019315\\
    \hline
  \end{tabular}
  \caption{Benchmark results for various problems on three different
    supercomputer systems described in the main text. The employed
    symmetries include translational, point group and spinflip
    symmetry. We show the total memory used by all MPI processes and
    the memory used by the lookup tables for the sublattice coding
    technique. We also show the amount of time spent for communication. }
  \label{tab:edbenchmarks}
\end{table}

%% \begin{table}[ht!]
%%   \centering
%%   \begin{tabular}{|c | c | c | c | c |}
%%       \hline
%%       Geometry & dimension & \# of cores & total memory & Time / MVM \\
%%       \hline
%%       Triangular 48 & $2.8 \cdot 10 ^{10}$ & 3456 & 2.5 TB & 399 s \\
%%       Square 48 & $8.4 \cdot 10^{10}$ & 8192 & n.A. & 1241 s \\
%%       Kagome 48 & $8.4 \cdot 10^{10}$ & 10240 & n.A. & 258 s \\
%%       Square 50 & $3.2 \cdot 10^{11}$ & 3456 & 15.5 TB & 3304 s \\
%%       \hline
%%     \end{tabular}
%%     \caption{Benchmark results for various problems on VSC3 and RZG
%%     Hydra.}
%%   \end{table}

%%% Local Variables:
%%% mode: latex
%%% TeX-master: "../../thesis"
%%% End:

%auto-ignore
\section{Conclusion}
\label{sec:edconclusion}
We proposed the generic sublattice coding algorithm for making
efficient use of discrete symmetries in large-scale ED
computations. The method can be used flexibly on most lattice
geometries and only requires a reasonable amount of memory for storing
the lookup tables. The parallelization strategy for distributed memory
architectures we discussed includes a random distribution of the
Hilbert space amongst the parallel processes. Lookup tables of the
sublattice coding technique are stored only once per node and are
accessed via shared memory. Using these techniques, we showed that
computations of spin-$1/2$ models of up to 50 spins have now become
feasible.

\subsubsection{Acknowledgements}
We thank Synge Todo for making the simulations on Sekirei at the ISSP
at the University of Tokyo possible. Further computations for this
chapter have been carried out on VSC3 of the Vienna Scientific
Cluster, the supercomputer Hydra at the Max Planck Supercomputing \&
Data Facility in Garching.

%%% Local Variables:
%%% mode: latex
%%% TeX-master: "../../thesis"
%%% End:

\newpage
\chapter{Nature of chiral spin liquids on the kagome lattice}
\label{sec:paperkagome}
%auto-ignore

% \documentclass[aps,prl,reprint]{revtex4-1}
% \usepackage{times}
% \usepackage[colorlinks=true, urlcolor=blue, linkcolor=red, citecolor=blue, pdftex]{hyperref}
% \usepackage{graphicx}
% \usepackage{amsmath}
% \usepackage{amssymb}
% \usepackage{color}

% \def\ket#1{\left|#1 \right\rangle}
% \def\bra#1{\left\langle #1 \right|}
% \def\braket#1#2{\left\langle #1 | #2 \right\rangle}
% \def\matrix22#1#2#3#4{\left(\begin{array}{cc}#1&#2\\#3&#4\end{array}\right)}
% \def\indentit{\mbox{\hspace{0.2em}l\hspace{-0.48em}1}}  
% \def\aml{\color{red}\bf}

% \begin{document}
% \title{Nature of chiral spin liquids on the kagome lattice}
% \author{Alexander Wietek}
% \email{alexander.wietek@uibk.ac.at}
% \author{Antoine Sterdyniak}
% \author{Andreas M. L\"auchli}
% \affiliation{Institut f\"ur Theoretische Physik, Universit\"at Innsbruck, A-6020 Innsbruck, Austria}
This chapter has been published as:
\begin{inlinecitation}[h]
  \fullcite{Wietek2015}
\end{inlinecitation}

\noindent
The numerical simulations and the development of simulation software has
been performed by the author of this thesis. He also wrote substantial parts
of the paper.

\subsubsection{Abstract}
We investigate the stability and the nature of the chiral spin liquids
which were recently uncovered in extended Heisenberg models on the
kagome lattice. Using a Gutzwiller projected wave function approach --
i.e.~a parton construction -- we obtain large overlaps with ground
states of these extended Heisenberg models. We further suggest that
the appearance of the chiral spin liquid in the time-reversal
invariant case is linked to a classical transition line between two
magnetically ordered phases.

% \maketitle
\section{Introduction}
The quest for quantum spin liquids~\cite{balents2010spin} is currently
a very active endeavour in condensed matter physics. This elusive
state of quantum matter comes in various forms and is theoretically
intensely studied, however was difficult to pin down in computational
studies of {\em realistic} quantum spin Hamiltonians and hard to
characterise unambigously in experiments on quantum magnets.

The $S=1/2$ Heisenberg antiferromagnet on the kagome lattice has
emerged as one of the paradigmatic systems where quantum spin liquid
phases are expected. A plethora of theoretical proposals have been put
forward, ranging from valence bond
crystals~\cite{Marston1991,Nikolic2003,Singh2007,Poilblanc2011,Capponi2013b},
algebraic spin
liquids~\cite{Hastings2000,Ran2007,Hermele2008,Iqbal2013},
$\mathbb{Z}_2$ spin
liquids~\cite{Sachdev1992,Moessner2001,Balents2002,Misguich2002,Wang2006,Lu2011,Iqbal2011},
to chiral spin
liquids~\cite{Marston1991,Messio2012,Messio2013,Yang1993}. Despite
tremendous theoretical and computational
progress~\cite{Leung1993,Lecheminant1997,Waldtmann1998,Capponi2004,Jiang2008,Schwandt2010,Yan2011,Laeuchli2011,Jiang2012,Depenbrock2012,Nishimoto2013,Clark2013,Capponi2013},
the true nature of the ground state and the low-lying excited states
of the nearest neighbour Heisenberg model on the kagome lattice is
still not settled completely.

Chiral spin liquids (CSL) are a particular family of spin liquids in
which time-reversal symmetry (TRS) and parity symmetry are
(spontaneously or explicitly) broken~\cite{Wen1989a,Wen1989b}. The
scalar chirality
$\langle\bm{S}_i\cdot(\bm{S}_j\times \bm{S}_k) \rangle$ is non-zero
and uniform and manifests the breaking of time-reversal and parity
symmetries, analogous to the presence of an orbital magnetic field. In
a favorable situation the breaking of these symmetries could
conceivably lead to a spin analogue of the Fractional Quantum Hall
Effect, although other types of ground states are possible as
well~\cite{Momoi1997,Lauchli2003}. Historically Kalmeyer and Laughlin
envisioned such a scenario by considering lattice versions of the
bosonic $\nu=1/2$ Laughlin wave function as candidate ground state
wave functions for the triangular lattice Heisenberg
model~\cite{Kalmeyer1987,Kalmeyer1989}.

In two recent papers~\cite{Gong2014,Bauer2014}, two forms of chiral
spin liquids have been discovered, which are stabilised away from the
nearest neighbour Heisenberg model upon adding further neighbour
Heisenberg interactions or scalar chirality terms to the
Hamiltonian. Both studies numerically demonstrate the required ground
state degeneracy and characterize the underlying topological order by
computing the modular matrices. For different models CSLs have also
been found in Refs.~\cite{Schroeter2007,He2014,He2015}.

This breakthrough lays the foundation for further investigations of
chiral spin liquids. Several pressing, important questions arise: i)
are the two chiral spin liquids phases distinct or are they related ?
ii) is there a simple physical (lattice-based) picture or a
variational wave function that describes the chiral spin liquid ?
iii) what is the "raison d'\^etre" of these chiral spin liquids,
i.e. why are the chiral spin liquids stabilized for the two reported
Hamiltonians ? Can we come up with some guiding principle which will
allow to stabilise CSL on other lattices ? In the following we will
address each of these questions. In short we find that the two chiral
spin liquids are indeed connected. We then demonstrate that
appropriate Gutzwiller projected parton wave functions can have large
overlaps with the numerically exact ground states of the studied
microscopic models. And finally we show that one location of the
chiral spin liquids in parameter space coincides largely with a
transition line in the phase diagram of the corresponding classical
model. The classical transition line lies between coplanar
$\mathbf{q}=0$ magnetic order and a chiral, non-coplanar magnetically
ordered phase ({\em cuboc1}~\cite{Messio2012}).

%%%%%%%%%%%%%%%%%%%%%%%%%%
\begin{figure}[b!]
  \centerline{\includegraphics[width=0.3\linewidth]{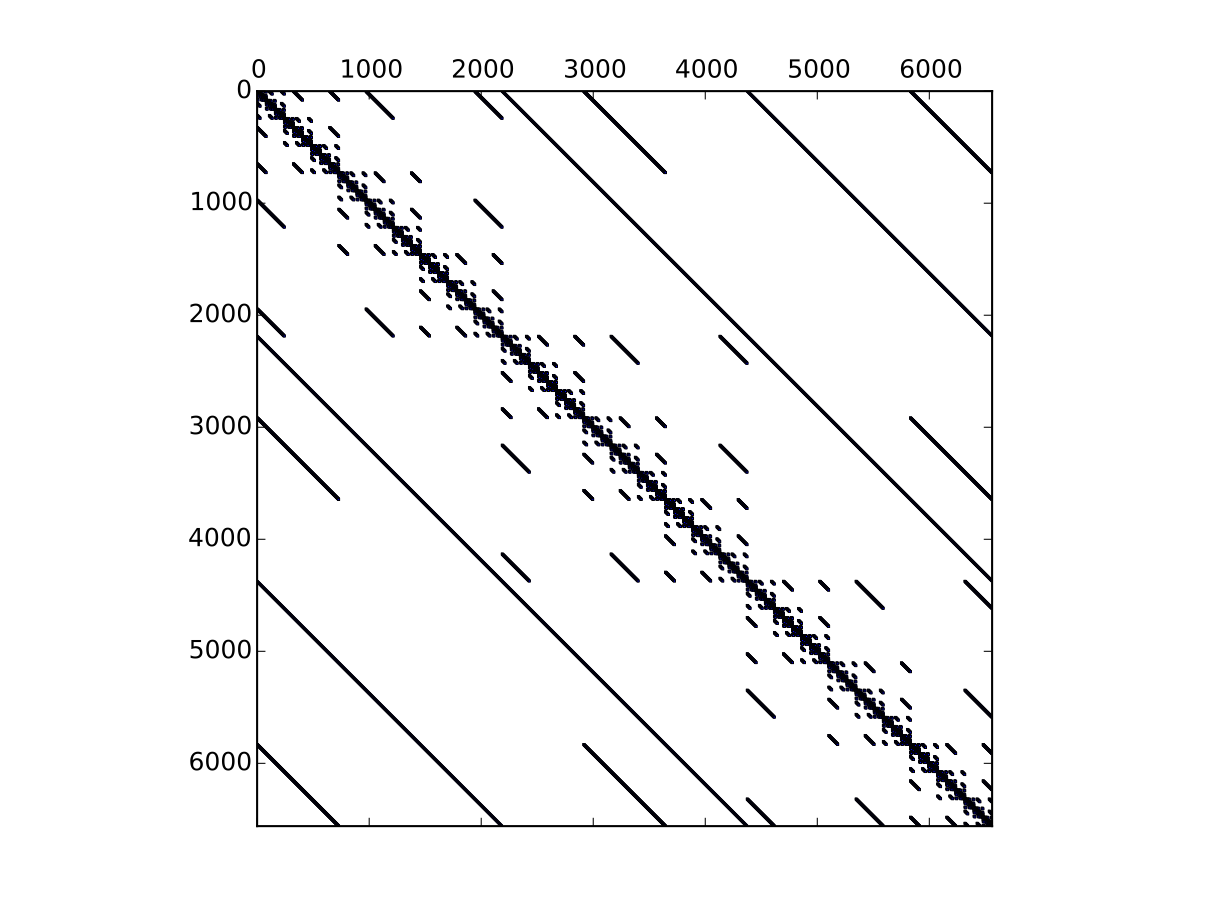}}
  \caption{ Sketch of the kagome lattice and of the different
    interaction terms of the Hamiltonian
    \eqref{eq:chiralhamiltonian}. Heisenberg interactions between
    first, second and third nearest neighbour are considered. The
    third nearest neighbour Heisenberg interactions are only
    considered across the hexagons. Three-spin scalar chirality
    interactions, breaking time-reversal and parity symmetries, are
    also considered on grey shaded triangles.  }
  \label{fig:Hamiltonian}
\end{figure}
%%%%%%%%%%%%%%%%%%%%%%%%%%

\section{Model}
We will consider the following Hamiltonian which unifies the two
models studied in Refs.~\cite{Gong2014,Bauer2014}:
\begin{equation}
  \label{eq:chiralhamiltonian}
  \begin{aligned}
    H = &J_1 \sum\limits_{\left<i,j\right>} \bm{S}_i \cdot \bm{S}_j +
    J_2 \sum\limits_{\left<\left<i,j\right>\right>} \bm{S}_i \cdot \bm{S}_j +  \\
    & J_3 \sum\limits_{\left<\left<\left<i,j\right>\right>\right>}
    \bm{S}_i \cdot \bm{S}_j + J_{\chi}\sum\limits_{i,j,k \in
      \bigtriangleup,\bigtriangledown} \bm{S}_i\cdot(\bm{S}_j\times
    \bm{S}_k).
  \end{aligned}
\end{equation}

This model includes first, second and third nearest neighbour
Heisenberg interactions with coupling constants $J_1$, $J_2$, $J_3$ as
sketched in \cref{fig:Hamiltonian}. The third nearest neighbour
Heisenberg interactions are only considered across the hexagons. While
these interactions preserve TRS and all the discrete lattice
symmetries of the kagome lattice, the additional three-spin scalar
chirality interactions on the triangles parametrized by $J_\chi$ break
explicitly TRS and spatial parity. Note that Hamiltonian
\cref{eq:chiralhamiltonian} features SU(2) invariance in spin
space. For simplicity we will set $J_1 = 1$ in the following.

In Ref.~\cite{Bauer2014} a CSL phase was found for
$0.05 \pi \lesssim \arctan |\frac{J_{\chi}}{J_1}| \lesssim \pi/2 $ and
$J_2=J_3=0$. In this case, TRS is explicitly broken. Interestingly a
two-fold degenerate ground state was found, which furthermore exhibits
the expected modular data and entanglement spectrum for a
topologically ordered chiral $\nu=1/2$ Laughlin state-like phase. On
the other hand in Ref.~\cite{Gong2014} a chiral spin liquid with {\em
  spontaneous} TRS breaking was discovered for $J_{\chi}=0$ and
$0.2 \lesssim (J_2 = J_3)/J_1 \lesssim 0.7$. Here the ground state
degeneracy is four, which can be understood as arising from two copies
of opposite chirality of a two-fold degenerate $\nu=1/2$ Laughlin
state. Unlike several topological phases as Toric code
\cite{Kitaev2003a} and double-semion \cite{Freedman2004428} phases
that also have a four-fold ground state degeneracy, we will show that
in this case time-reversal symmetry is spontaneously broken.

\section{Energy spectroscopy}

%%%%%%%%%%%%%%%%%%%%%%%%%%
\begin{figure}[t!]
  \centerline{\includegraphics[width=0.8\linewidth]{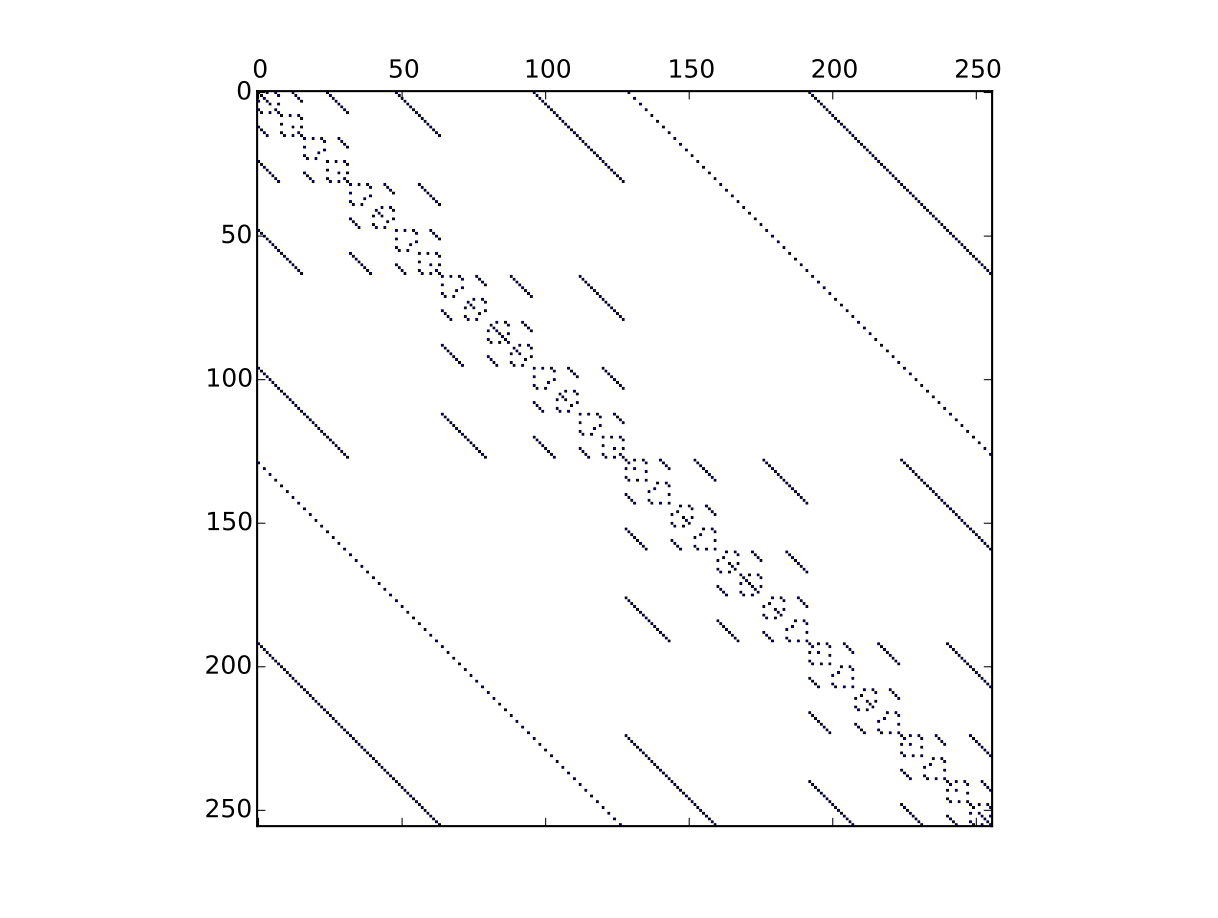}}
  \caption{ Excitation Spectra from Exact Diagonalization. Different
    symbols and colors correspond to different momentum/pointgroup
    symmetry sectors. We use the cluster geometries and notation
    explained in Ref.~\cite{Laeuchli2011} \textbf{(a)} Effect of
    $J_{\chi}$ term on spectrum on the 30 sites cluster. The four-fold
    degeneracy of the ground state is lifted to a two-fold degereracy
    which corresponds to one sign of the scalar
    chirality. \textbf{(b)} Scan across the classical transition line
    for $J_{\chi} = 0$ on the 36b sites cluster. The four-fold
    degereracy of the CSL is only present close to $J_2 = J_3$ (yellow
    shading).  \textbf{(c)} Energy spectra for $J_2=J_3=0.4$,
    $J_{\chi} = 0$ and various system sizes $N_s$ and geometries.
    \textit{Turquoise rectangle}: $(0,0)\ [\Gamma]$ momentum, even
    under $180^\circ$ rotation. \textit{Blue up triangle}:
    $(0,\pi)\ [M]$ momentum, odd under $180^\circ$
    rotation. \textit{Red down triangle}: $(0,0)\ [\Gamma]$ momentum,
    even under $180^\circ$ rotation, odd under reflection.}
  \label{fig:edspectra}
\end{figure}
%%%%%%%%%%%%%%%%%%%%%%%%%%

To investigate the persistence of this chiral spin liquid at the
thermodynamical limit, we studied the model for $J_2=J_3=0.4$ and
$J_{\chi}=0$ up to 42 sites. The low-energy spectra for different
system sizes are shown in \cref{fig:edspectra} c). While the energy
splitting between the four ground states has a non-monotonous
behaviour, the energy gap between the four lowest energy states and
the fifth one increases with the system size. Moreover, the ratio of
the energy splitting to the energy gap decreases with the system size,
this tends to indicate that this phase is indeed realized at the
thermodynamical limit. It is also important to notice that the
momentum sectors involved in the four-fold degenerate manifold depend
on the cluster shape and can be predicted in complete analogy to the
Fractional Quantum Hall and Fractional Chern insulator
states~\cite{Regnault2011,Laeuchli2013}.

In \cref{fig:edspectra} a) we investigate the energy splitting of the
four ground states as we switch on a finite $J_\chi$ coupling. At
$J_\chi=0$ the long-range order in the spin chirality is spectrally
encoded in the presence of two states per topological sector, where
the two states have to be at the same momentum, but differ in the
spatial reflection quantum number (if the sample allows this
symmetry). As is shown in \cref{fig:edspectra} a), the two states per
sector split very rapidly upon switching on $J_\chi \neq 0$. We can
understand the action of $J_\chi$ regarding the scalar chirality in
analogy to the effect of a longitudinal magnetic field on the two
degenerate ground states in a ferromagnetic Ising model in the ordered
phase, where the magnetic field immediately selects one of the two
ordered states. As we show later based on overlaps, the chiral spin
liquid thus selected by $J_\chi$ is of the same type as the one
stabilised in the $J_1-J_\chi$ model alone, and is connected to the
TRS symmetric situation in the absence of $J_\chi$.
%%%%%%%%%%%%%%%%%%%%%%%%%%
\begin{figure}[t!]
  \centerline{\includegraphics[width=0.7\linewidth]{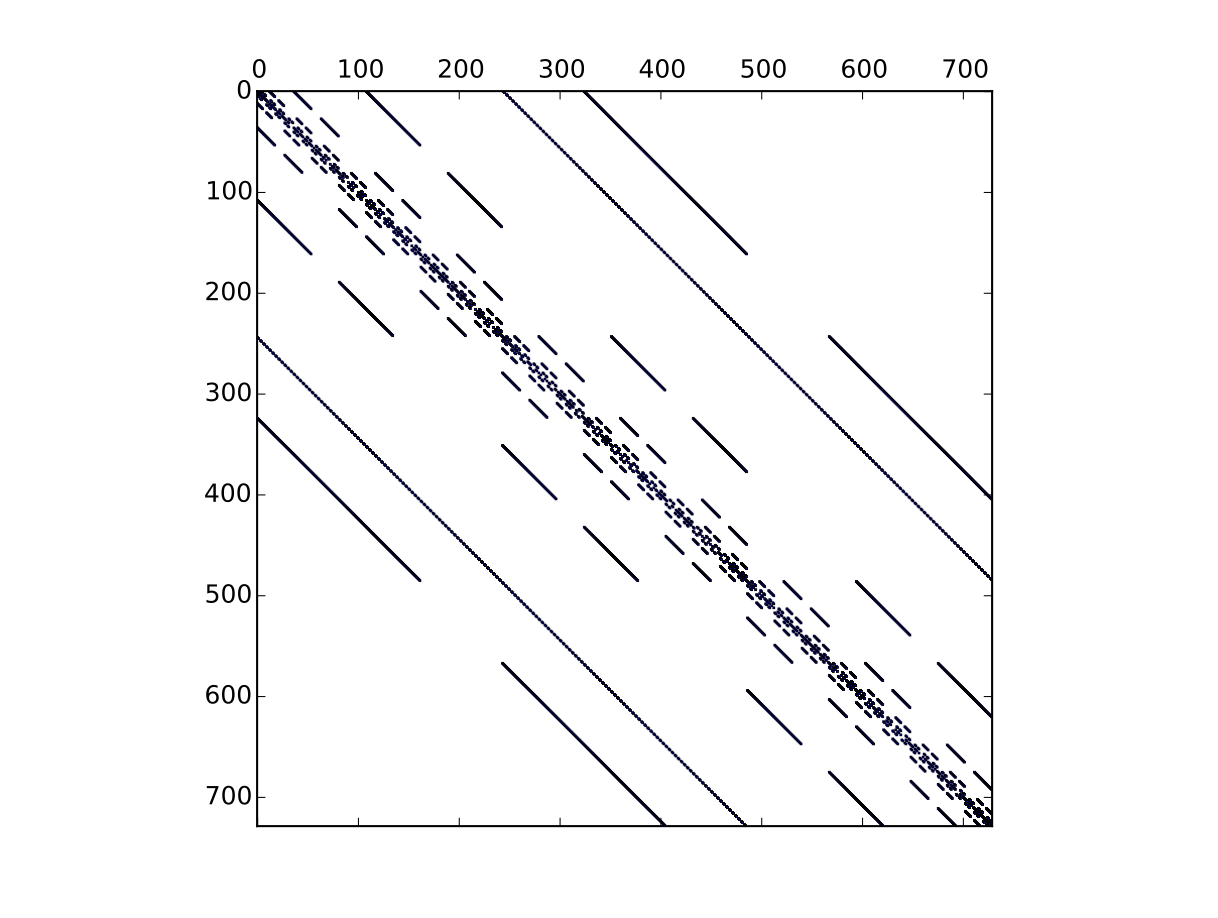}}
  \caption{Excitation Spectra for a path in phase space connecting the
    two CSLs from Refs.~\cite{Gong2014,Bauer2014} on a 30 sites
    lattice. The excitation spectra remain fully gapped over the whole
    path thus no phase transition takes place. The parameters in
    \cref{eq:chiralhamiltonian} are chosen
    $J_2 = J_3 = 0.5\cos(\frac{\pi}{2}\theta)$,
    $J_{\chi} = \frac{1}{\sqrt{2}} \sin(\frac{\pi}{2}\theta) $ and
    $J_1 = \cos(\frac{\pi}{2}\theta) +
    \frac{1}{\sqrt{2}}\sin(\frac{\pi}{2}\theta)$ \textit{Turquoise
      rectangle}: $(0,0)\ [\Gamma]$ momentum, even under $180^\circ$
    rotation.  \textit{Blue up triangle}: $(0,\pi)\ [M]$ momentum, odd
    under $180^\circ$ rotation.}
  \label{fig:connection_path}
\end{figure}
%%%%%%%%%%%%%%%%%%%%%%%%%%

In \cref{fig:edspectra} b) we investigate the effect of a deviation
from the $J_2=J_3$ condition (in the absence of $J_\chi$) by fixing
$J_2=0.5$ and varying $J_3$. One observes that the four-fold ground
state degeneracy is rapidly lifted when $J_3$ deviates more than about
$0.05\sim0.1$ from $0.5$. Interestingly the line $0< J_2=J_3 < 1$ is
the classical transition line between a magnetically ordered
$\mathbf{q}=0$ ground state for $J_3<J_2$ and the non-coplanar
magnetically ordered {\em cuboc1} phase for
$J_3>J_2$~\cite{Messio2012}. Below we will show that also the overlaps
with the variational wave functions are large only in the direct
vicinity of this classical transition line. A deeper understanding of
the classical ground state configurations on that line and of the
effect of quantum fluctuations on that manifold might thus lead to an
identification of the crucial ingredients required to predict and
uncover chiral spin liquids in TRS Hamiltonians on different lattices.
We note in passing that the explicitly TRS breaking Hamiltonian
\cref{eq:chiralhamiltonian} with $J_\chi\neq0$ can be considered as a
truncated version of a parent Hamiltonian for the CSL constructed in
Refs.~\cite{Nielsen2012,Greiter2014}, similar to the spin Hamiltonian
on the square lattice considered in Ref.~\cite{Nielsen2013}.

In order to finally prove that the two CSLs from
Refs.~\cite{Gong2014,Bauer2014} are in the same phase,
\cref{fig:connection_path} shows excitation spectra from Exact
Diagonalization for a path connecting the two CSLs. We find that no
gap is closing and thus the two phases are connected without a phase
transition.

\section{Parton construction and overlaps}
%%%%%%%%%%%%%%%%%%%%%%%%%%
\begin{figure}[t]
  \centerline{\includegraphics[width=0.7\linewidth]{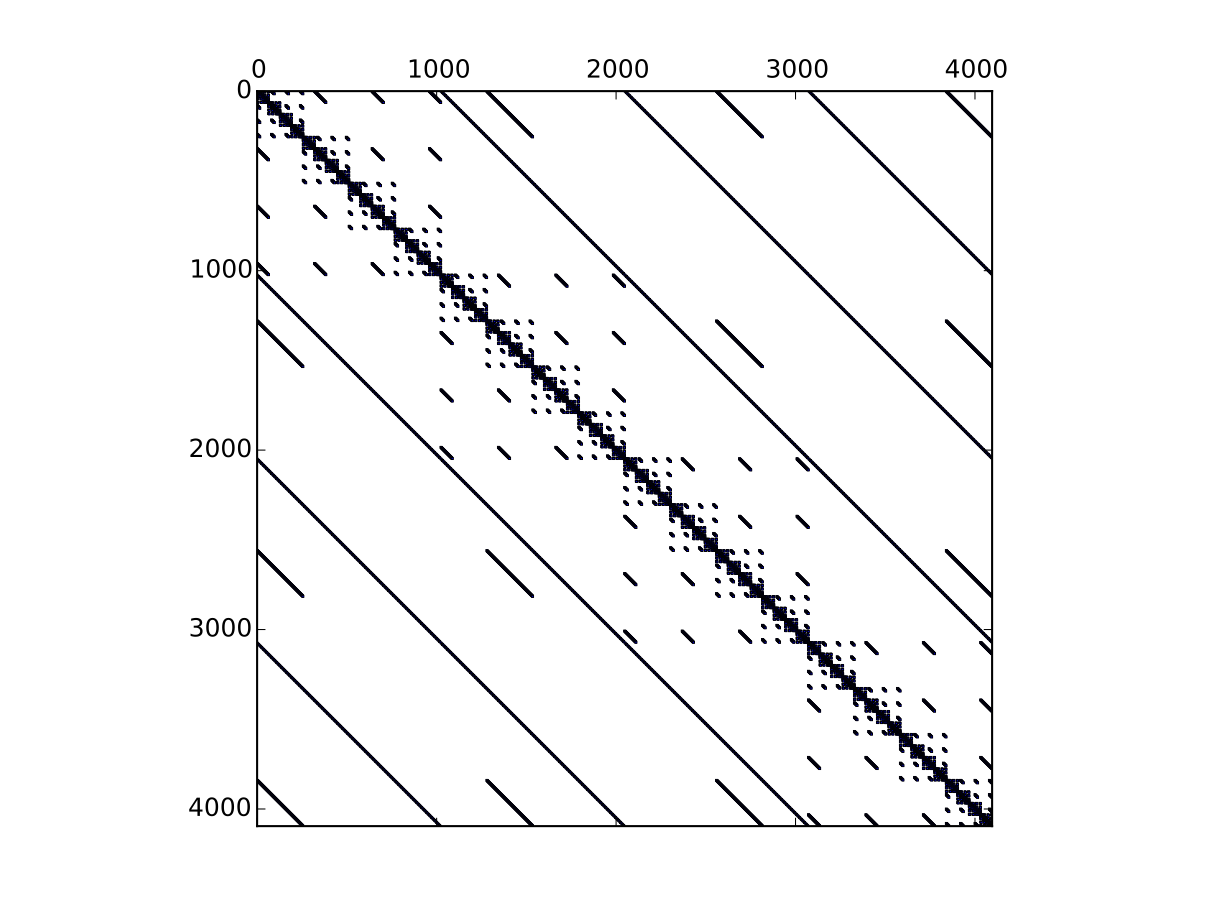}}
  \caption{ Band structure of the $[\pi/2, 0]$ - model used to
    generate the pair of model states with positive scalar chirality.
    {\textbf{a)}} geometry of the unit cell. The mean-field parameters
    $\chi_{ij}$ are chosen such that there are $\pi/2$ fluxes through
    the triangles and no flux through the hexagons.  {\textbf{b)}}
    Brillouin zone of the model (shaded) and conventional first
    Brillouin zone of the kagome lattice (hexagon).  {\textbf{c)}}
    Band structure of the model along the path between high symmetry
    points in the Brillouin zone as drawn in b).  The bands are
    separated by a finite gap. Each solid (dashed) band carries Chern
    number $-1$ ($+1$).}
  \label{fig:bandstructure}
\end{figure}
%%%%%%%%%%%%%%%%%%%%%%%%%%

As stated earlier on, the CSL can be considered as lattice analogues
of the bosonic $\nu=1/2$ Laughlin state. In recent years substantial
activity focused on realizing such states on fractionally filled Chern
insulators, so called Fractional Chern Insulators
(FCI)~\cite{Sheng2011,PhysRevLett.106.236804,Regnault2011}. It is thus
a natural question whether the CSLs under consideration might also
have such an interpretation. The natural bosonic $\nu=1/2$ FCI state
on the kagome lattice~\cite{2011arXiv1102.2406M,2014arXiv1409.2171K}
however does not have the correct magnetisation since it corresponds
to magnetisation $m/m_\mathrm{sat}=\pm2/3$ instead of the required
$m=0$~\footnote{An interesting idea for future study might be to
  combine a {\em featureless} Mott insulator wave function with
  bosonic density $n=1/3$ (i.e. magnetisation $m=2/3$) with a
  $\nu=1/2$ FCI state.}.

In the absence of a simple FCI candidate wave function we pursue an
alternative approach, based on a parton construction.  In order to
understand and classify the different spin liquids a generalized
construction scheme called parton construction has been introduced by
Refs.~\cite{Baskaran1987973,Baskaran1988,PhysRevB.38.745,PhysRevB.38.2926,PhysRevLett.76.503,senthil2000z}
- see \cite{Wen2004} for an introduction.  The main idea of this
technique is to split up each spin operator $\bm{S}_i$ at site $i$
into two fermionic parton operators $c_{i,\uparrow}$,
$c_{i,\downarrow}$ according to
\begin{equation}
  \begin{aligned}
    S^+_i = c_{i,\uparrow}^\dagger c_{i,\downarrow}\textnormal{, } S^-_i = c_{i,\downarrow}^\dagger c_{i,\uparrow}, \\
    S^z_i = \frac{1}{2}(c_{i,\uparrow}^\dagger c_{i,\uparrow} -
    c_{i,\downarrow}^\dagger c_{i,\downarrow}) .
  \end{aligned}
\end{equation}
Note that by introducing these operators the Hilbert space is enlarged
due to the possibility of doubly occupied or vacant
sites. Substituting the parton operators for the spin operators and
performing a mean-field approximation by introducing mean-field
parameters
$\chi_{ij} = \sum_{\sigma}\left < c_{i\sigma}^\dagger c_{j\sigma}
\right >$ yields (ignoring constants) a tight-binding model of type
\begin{equation}
  \label{eq:hbtypepartonmeanfield}
  H_{\textnormal{mean}} = \frac{1}{2} \sum\limits_{i,j,\sigma}  \left(\chi_{ij} c_{i\sigma}^\dagger c_{j\sigma}  +
    \textnormal{h.c.}\right ).  
\end{equation}

Several of these models have been investigated for the kagome
lattice~\cite{Marston1991,Hastings2000,Ran2007,Hermele2008,Iqbal2013}. Here
we focus on nearest neighbour $\chi_{ij}$ only and the norm is chosen
to be $|\chi_{ij}| = 1$. Physically different states can be created by
choosing $\chi_{ij}$ such that different magnetic fluxes thread the
triangles and the hexagons of the kagome lattice. Amongst these states
we consider states whose parent mean-field models have uniform
$\pm\pi/2$ flux through the triangles and zero flux through the
hexagons~\cite{Marston1991,Hastings2000,Ran2007}. To do so a magnetic
six sites unit cell is needed instead of the standard three sites unit
cell of the kagome lattice. In the following we will call these the
$[\pm\pi/2, 0]$ - models. On the parton level these wave functions
break time and parity symmetry. Thus the projected wave functions are
expected to break these symmtetries too. Moreover the states
constructed from the $[\pi/2, 0]$ - model are related to the states of
the $[-\pi/2, 0]$ - model by time-reversal symmetry.

The unit cell geometry, Brillouin zone and band structure of the
$[\pi/2, 0]$ - model are shown in \cref{fig:bandstructure}. All six
bands have non-zero Chern numbers as indicated in
\cref{fig:bandstructure}c). To obtain a $S_z = 0$ model state, the
three lowest bands are completely filled both for spins up and spins
down and an exact Gutzwiller projection is applied to project onto the
physical spin subspace.  As the filled bands are separated by a finite
gap from the empty ones, the spin-spin correlations after projection
are expected to decay exponentially with distance, and thus describe a
spin disordered state. The Chern number of the filled bands for the
$[\pi/2, 0]$ - model ($[-\pi/2, 0]$ - model) is $-1$ ($+1$). The
$[\pi/2, 0]$ - model ($[-\pi/2, 0]$ - model) yields a positive
(negative) scalar chirality expectation value for every basic
triangle. The average expectation value is the same for the two
topological partners within numercial precision and is given by
$\left<\bm{S}_i\cdot(\bm{S}_j\times\bm{S}_k)\right>_{i,j,k \in
  \bigtriangleup,\bigtriangledown} = \pm 0.2057 \pm 0.0005$. The
long-range chiral-chiral correlations are expected two be the square
of this value. Thus we get value of $0.042$ for the long-range
chiral-chiral correlation functions which is within the same order of
magnitude as the correlations computed in
Refs.~\cite{Hu2015,Gong2015}.

On the torus there are two independent non contractible loops. Some of
the gauge choices which leave the flux through the triangles and
hexagons invariant, correspond to different fluxes through these torus
loops. Threading flux through these loops corresponds to a Laughlin
flux insertion. Thereby different topological states can be
generated. These states cannot be distinguished by local observables
and therefore are degenerate for local Hamiltonians in the
thermodynamic limit. For the chiral spin liquid a two-fold topological
ground state degeneracy is expected. Thus, by threading different
fluxes through the torus we should only be able to create a
two-dimensional space.  We numerically computed the Gutzwiller
projected wave functions (GPWFs) of the $[\pm\pi/2, 0]$ - models with
a fixed gauge. In order to construct the topological partners of these
states we additionally thread fluxes through the torus as explained in
the previous section. We checked that for each of the $[\pm\pi/2, 0]$
- models, only two linearly independent states can be constructed as
expected for a CSL within a numerical accuracy of $10^{-3}$, similar
as in Ref.~\cite{2014arXiv1409.5427M}.

%%%%%%%%%%%%%%%%%%%%%%%%%%
\begin{figure*}[t!]
  \centerline{\includegraphics[width=\linewidth]{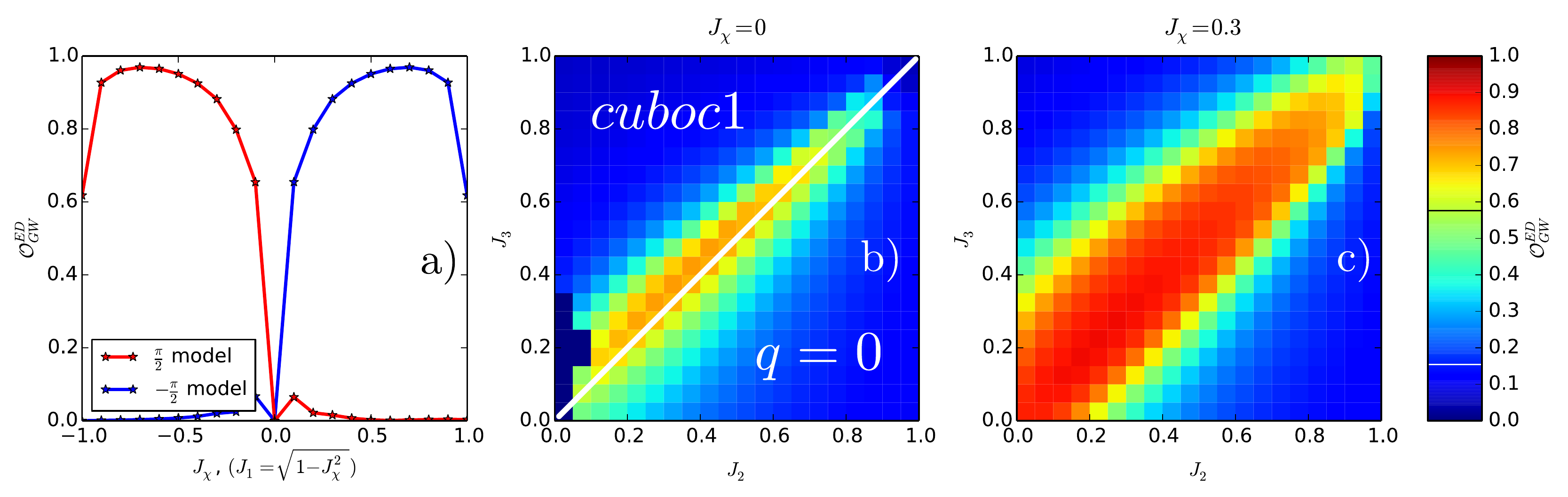}}
  \caption{Overlaps $\mathcal{O}_{\textnormal{GW}}^{\textnormal{ED}}$
    of GPWFs with ground states from Exact Diagonalization for
    $J_2 = J_3 = 0$ and $J_1 = \sqrt{1 - J_{\chi}^2}$ on a 30 sites
    lattice.  The overlaps of the GPWFs of the $\pm\pi/2$-models are
    symmetric under changing the sign of $J_\chi$.  The maximum
    overlap is equal to $0.97$ and is reached for $J_\chi = \pm
    0.7$. The CSL phase extends almost up to the Heisenberg point.  }
  \label{fig:overlaps_chiral}
\end{figure*}
%%%%%%%%%%%%%%%%%%%%%%%%%%

We compare now these four model states with the ground state
$\ket{\psi_{\textnormal{ED}}}$ of the Hamiltonian
(\ref{eq:chiralhamiltonian}) obtained using Exact Diagonalization. We
choose the overlap $\mathcal{O}_{\textnormal{GW}}^{\textnormal{ED}}$
of the ground state wave function with the four model states as our
figure of merit:
\begin{equation}
  \label{eq:overlap}
  \mathcal{O}_{\textnormal{GW}}^{\textnormal{ED}} \equiv \sqrt{\sum_\alpha \left| 
      \braket{\psi_{\textnormal{ED}}}{\psi_{GW}^\alpha} \right| ^2}
\end{equation}
Overlaps of the GPWFs with the ground state of the Hamiltonian
(\ref{eq:chiralhamiltonian}) for different parameters on a $N_s=30$
sites sample are shown in Fig.~\ref{fig:overlaps_chiral}. The overlaps
of our model state with the ground state wave functions of the model
of Ref.~\cite{Bauer2014} where $J_2 = J_3 = 0$, $J_1 = \cos \theta$
and $J_{\chi} = \sin \theta$ are shown in
Fig.~\ref{fig:overlaps_chiral}(a). We found that overlaps for $J_\chi$
between $0.1$ and $1$ range from $0.62$ to $0.97$. The overlap of the
two GPWFs of the $[-\pi/2, 0]$-model are by orders of magnitude larger
than those of the $[+\pi/2, 0]$-model. For $J_\chi$ between $-1$ and
$-0.1$ the overlaps are exactly the same within numerical precision as
for $J_\chi$ between $0.1$ and $1$ but the role of the GPWFs from the
$[+\pi/2, 0]$-model and $[-\pi/2, 0]$-model are exchanged. This is
expected since the model with negative $J_\chi$ should have a positive
scalar chirality and therefore only little overlap with the
variational states form the $[-\pi/2, 0]$-model with negative
chirality and vice versa.

For the time-reversal symmetric model with $J_{\chi} = 0$, our
variational wave functions have substantial overlap only close to the
line $J_2 = J_3$, in agreement with the energy spectroscopy results
discussed above [Fig.~\ref{fig:overlaps_chiral}(b)]. In this region
the overlaps reach up to $0.72$ for $N_s=30$.

As can be seen in Fig.~\ref{fig:overlaps_chiral} c) for
$J_{\chi} = 0.3$ and for $J_{\chi} = 0.6$ (not shown), the region of
the CSL broadens significantly when $J_{\chi}$ is increased from
zero. For $J_{\chi} = 0.3$ (resp. $J_{\chi} = 0.6$) the overlaps on
the classical transition line for $J_2=J_3$ between $0$ and $0.7$
range from $0.8$ to $0.9$ (resp. from $0.85$ to $0.95$).

\section{Conclusion}
We showed that the two recently found realizations of chiral spin
liquids on the kagome lattice~\cite{Gong2014,Bauer2014} are indeed
related and can be described by Gutzwiller projected parton wave
functions. This yields an intuitive microscopic picture of the CSL
phase stabilized in these models. The ansatz wave functions we chose
have been shown to describe a CSL on the kagome
lattice~\cite{Marston1991,Hastings2000,Ran2007}. We constructed a pair
of Gutzwiller projected parton CSL wave functions for each sign of the
scalar chirality. We suggested that these states describe the CSL
ground state found on the kagome lattice. To prove that indeed these
wave functions describe the novel CSL phases found in
Refs.~\cite{Gong2014,Bauer2014} we computed overlaps of these
variational wave functions with the ground state wave functions
computed by Exact Diagonalization. Substantial overlaps were found in
regions of the phase diagram where the CSL is expected. By further
investigation of excitation spectra, we showed that the CSL phase in
Ref.~\cite{Gong2014} is only present on the transition line between a
chiral \textit{cuboc1} and a coplanar $q=0$ phase of the classical
phase diagram~\cite{Messio2012}. This could serve as a guiding
principle for finding CSL phases in other models and on others
lattices. Being related to the Laughlin state, these states should
exhibit anyonic excitations. Their investigations will be pursued in a
future work.

\subsubsection{Note added} While completing the present manuscript we
became aware of parallel work reaching similar conclusions using
complementary methods~\cite{Hu2015,Gong2015}.

\subsubsection{Acknowledgments}
We acknowledge inspiring discussions with H.-H. Tu and A.B. Nielsen.
AW acknowledges support through the Austrian Science Fund project
I-1310-N27 (DFG FOR1807).  AS acknowledges support through the
Austrian Science Fund SFB FoQus (F-4018-N23). This work was supported
by the Austrian Ministry of Science BMWF as part of the
Konjunkturpaket II of the Focal Point Scientific Computing at the
University of Innsbruck.

\newpage
\chapter[Chiral Spin Liquid on the Triangular Lattice]{Chiral Spin
Liquid and Quantum Criticality in Extended $S=1/2$ Heisenberg Models
on the Triangular Lattice}
\label{sec:papertriangular}
%auto-ignore

% \documentclass[aps,prl,reprint,longbibliography]{revtex4-1}
% \usepackage{times}
% \usepackage[colorlinks=true, urlcolor=blue, linkcolor=red, citecolor=blue, pdftex]{hyperref}
% \usepackage{graphicx}
% \usepackage{amsmath}
% \usepackage{amssymb}
% \usepackage{color}

% \def\ket#1{\left|#1 \right\rangle}
% \def\bra#1{\left\langle #1 \right|}
% \def\braket#1#2{\left\langle #1 | #2 \right\rangle}
% \def\matrix22#1#2#3#4{\left(\begin{array}{cc}#1&#2\\#3&#4\end{array}\right)}
% \def\indentit{\mbox{\hspace{0.2em}l\hspace{-0.48em}1}}  
% \def\aw{\color{red}\bf}

% \begin{document}
% \title{{Chiral Spin Liquid and Quantum Criticality in Extended $S=1/2$ Heisenberg Models on the Triangular Lattice.}}
% \author{Alexander Wietek}
% \email{alexander.wietek@uibk.ac.at}
% \author{Andreas M. L\"auchli}
% \affiliation{Institut f\"ur Theoretische Physik, Universit\"at Innsbruck, A-6020 Innsbruck, Austria}
% \date{\today}

This chapter has been published as:
\begin{inlinecitation}[h]
  \fullcite{Wietek2017a}
\end{inlinecitation}

\noindent
The numerical simulations and the development of simulation software has
been performed by the author of this thesis. He also wrote substantial parts
of the paper.

\subsubsection{Abstract}
We investigate the $J_1$-$J_2$ Heisenberg model on the triangular
lattice with an additional scalar chirality term and show that a chiral
spin liquid is stabilized in a sizeable region of the phase
diagram. This topological phase is situated in between a
coplanar $120^\circ$ N\'{e}el ordered and a non-coplanar tetrahedrally
ordered phase. Furthermor we discuss the nature of the spin-disordered 
intermediate phase in the $J_1$-$J_2$ model. We compare the ground states from
Exact Diagonalization with a Dirac spin liquid wave function
and propose a scenario where this wave function describes the quantum
critical point between the $120^\circ$ magnetically ordered phase 
and a putative $\mathbb{Z}_2$ spin liquid.

\section{Introduction}
The emergence of quantum spin liquids in frustrated quantum 
magnetism is an exciting phenomenon in contemporary condensed matter
physics \cite{balents2010spin}. These novel states of matter
exhibit fascinating properties such as long-range ground state
entanglement \cite{Kitaev2006a,Levin2006} or
anyonic braiding statistics of quasiparticle excitations, relevant for
a potential implementation of topological quantum computation
\cite{Nayak2008}. Only very recently such phases have been found
to be stabilized in realistic local spin 
models~\cite{Gong2014,Bauer2014,Wietek2015,Hickey2015,He2014,Nataf2016,Kumar2015,
Gorohovsky2015, Thomale2009,Nielsen2012,Meng2015,Sachdev1992,Moessner2001,Misguich2002,Kitaev2006a}.

Triangular lattice Heisenberg models are a paradigm of frustrated
magnetism. Although the Heisenberg model with only nearest
neighbour interaction is known to stabilize a regular $120^\circ$
N\'{e}el order \cite{Jolicoeur1990,Bernu1992,Capriotti1999,White2007} adding further 
interaction terms may increase frustration and induce magnetic 
disorder to the system. 
Experimentally, several materials with triangular lattice geometry
do not exhibit any sign of magnetic ordering down to 
lowest temperatures \cite{Kurosaki2005, Shimizu2003,
  Yamashita2010,Itou2008}. These include 
for example the organic Mott insulators like
$\mathrm{\kappa-(BEDT-TTF)_2Cu_2(CN)_3}$ \cite{Kurosaki2005, Shimizu2003}
or $\mathrm{EtMe_3Sb[Pd(dmit)_2]_2}$ \cite{Yamashita2010,Itou2008}
and are thus candidates realizing spin liquid physics. 

Historically Kalmeyer and Laughlin \cite{Kalmeyer1987} introduced 
the \textit{chiral spin liquid} (CSL) state on the triangular lattice. This 
state closely related to the celebrated Laughlin wave function of
the fractional quantum Hall effect has recently been shown to 
be the ground state of several extended Heisenberg models
on the kagome lattice \cite{He2014,Gong2014,Bauer2014,Wietek2015}. 
The question arises whether a CSL
can indeed be realized on the triangular lattice as originally
proposed. In a recent study \cite{Nataf2016}
this was shown for SU($N$) models for $N\geq 3$.
In this letter we provide conclusive evidence that indeed the CSL is
stabilized in a spin-$1/2$ Heisenberg model upon adding a further
scalar chirality term $J_\chi\bm{S}_i\cdot(\bm{S}_j\times \bm{S}_k)$ 
similar as in Refs.~\cite{Bauer2014,Wietek2015,Hickey2015,Nataf2016}.
Such a term can be realized as a lowest order effective Heisenberg Hamiltonian
of the Hubbard model upon adding $\Phi$ flux through the elementary
plaquettes \cite{Sen1995,Motrunich2006, Bauer2014},
either via a magnetic field or by introducing artificial
gauge fields in possible cold atoms experiments 
\cite{Aidelsburger2013, Miyake2013}. The coupling 
constants then relate to the Hubbard model
parameters $t$ and $U$ as $J_1 \sim t^2/U$ and $J_\chi \sim \Phi
t^3/U^2$ where $J_1$ (resp. $J_\chi$) is the nearest neighbour
Heisenberg (resp. scalar chirality) coupling.

Another open question in frustrated magnetism of the triangular lattice 
is the nature of the intermediate phase in the phase diagram of the $S=1/2$
Heisenberg model with added next-nearest neighbour couplings around $J_2
/ J_1 \approx 1/8$.
Several authors \cite{Jolicoeur1990,Chubukov1992,Lecheminant1995} found
a spin disordered state. Recently several numerical studies~\cite{Kaneko2014,Zhu2015,Saadatmand2015,Hu2015a,Iqbal2016,Xu2013}
proposed that a topological spin liquid state of some kind might be realized in this regime.
The exact nature of this phase yet remains unclear. In this Letter
we advocate the presence of a $O(4)^*$ quantum critical point~\cite{Chubukov1994a,Chubukov1994b,Whitsitt2016,Thomson2016} 
separating the $120^\circ$ N\'eel order from a putative $\mathbb{Z}_2$ spin liquid. The diverging 
correlation length at this quantum critical point and the neighbouring first order phase 
transition into the stripy collinear magnetic ordered phase render the unambiguous identification of 
the intermediate spin liquid phase challenging however. 

%%%%%%%%%%%%%%%%%%%%%%%%%%
\begin{figure}[t!]
  \centering
  \includegraphics[width=0.8\linewidth]{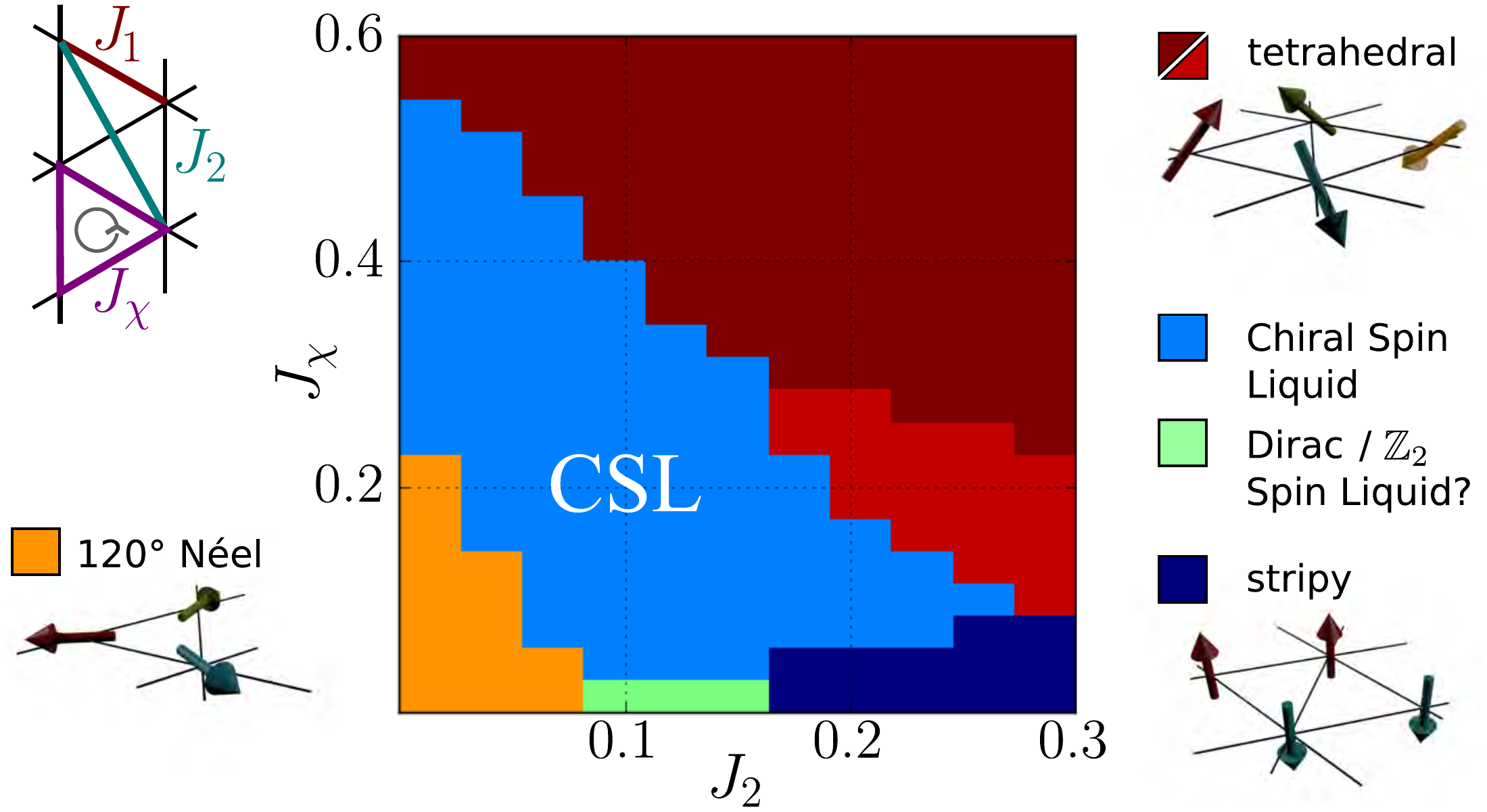}
  \caption{Approximate $T=0$ phase diagram of the $J_1$-$J_2$-$J_\chi$ model
   on the triangular lattice, c.f.~Eq.~\eqref{eq:hamiltonian}. The extent of phases is 
   inferred from excitation spectra from ED on a periodic $36$ sites 
   triangular simulation cluster, see main text for details.
  } 
  \label{fig:phasediagram}
\end{figure}
%%%%%%%%%%%%%%%%%%%%%%%%%%
\section{Model}
We investigate the Heisenberg model with nearest and next-nearest
neighbour interactions with an additional uniform scalar chirality
term on the triangular lattice
\begin{align}
  \begin{split}
    \label{eq:hamiltonian}
    \mathcal{H} \,=\, &J_1\sum\limits_{\left<
        i,j\right>}\bm{S}_i\cdot\bm{S}_j \ +
    J_2\sum\limits_{\left<\left<  i,j\right>\right>}\bm{S}_i\cdot\bm{S}_j + \\
    &J_{\chi}\sum\limits_{i,j,k \in \bigtriangleup}
    \bm{S}_i\cdot(\bm{S}_j\times \bm{S}_k)
  \end{split}
\end{align}
where we set $J_1 \equiv 1$ and consider $J_2,J_\chi \ge 0$. Amongst a
$120^\circ$ N\'{e}el order, a stripy and a tetrahedral magnetic order
we find a CSL being realized in an extended region of the phase
diagram in Fig.~\ref{fig:phasediagram}.  A first study of the
classical phase diagram for $J_\chi = 0$ \cite{Jolicoeur1990} found a
three sublattice $120^\circ$ N\'{e}el ordered ground state for
$J_2 < 1/8$ whereas for $1/8<J_2 < 1$ a two-parameter family of
magnetic ground states with a four-site unit cell was
found~\cite{Chubukov1992}. Two high-symmetry solutions within this
manifold are a two-sublattice collinear stripy magnetic order breaking
lattice rotation symmetry and a tetrahedral non-coplanar state with a
uniform scalar spin chirality on all triangles.  Taking into account
quantum fluctuations by applying spin-wave theory, large-$S$
perturbation theory and ED studies
\cite{Jolicoeur1990,Chubukov1992,Lecheminant1995} the degeneracy is
lifted by an \textit{order-by-disorder} mechanism. The true quantum
ground state for $ J_2 \gtrsim 0.18 $ exhibits stripy N\'{e}el order.
Yet the behaviour of the system close to the classical phase
transition point $J_2 = 1/8$ has not been fully understood.

\section{Phase diagram}
We performed ED calculations on a $N_s = 36$
sites simulation cluster with periodic boundary conditions to
investigate ground state properties and order parameters of the model
\eqref{eq:hamiltonian}. We have also checked selected results on smaller
clusters, but the $N_s=36$ cluster is particularly well suited because this 
single cluster can harbour all phases which we were able to detect.

\begin{figure}[t!]
  \centerline{\includegraphics[width=0.8\linewidth]{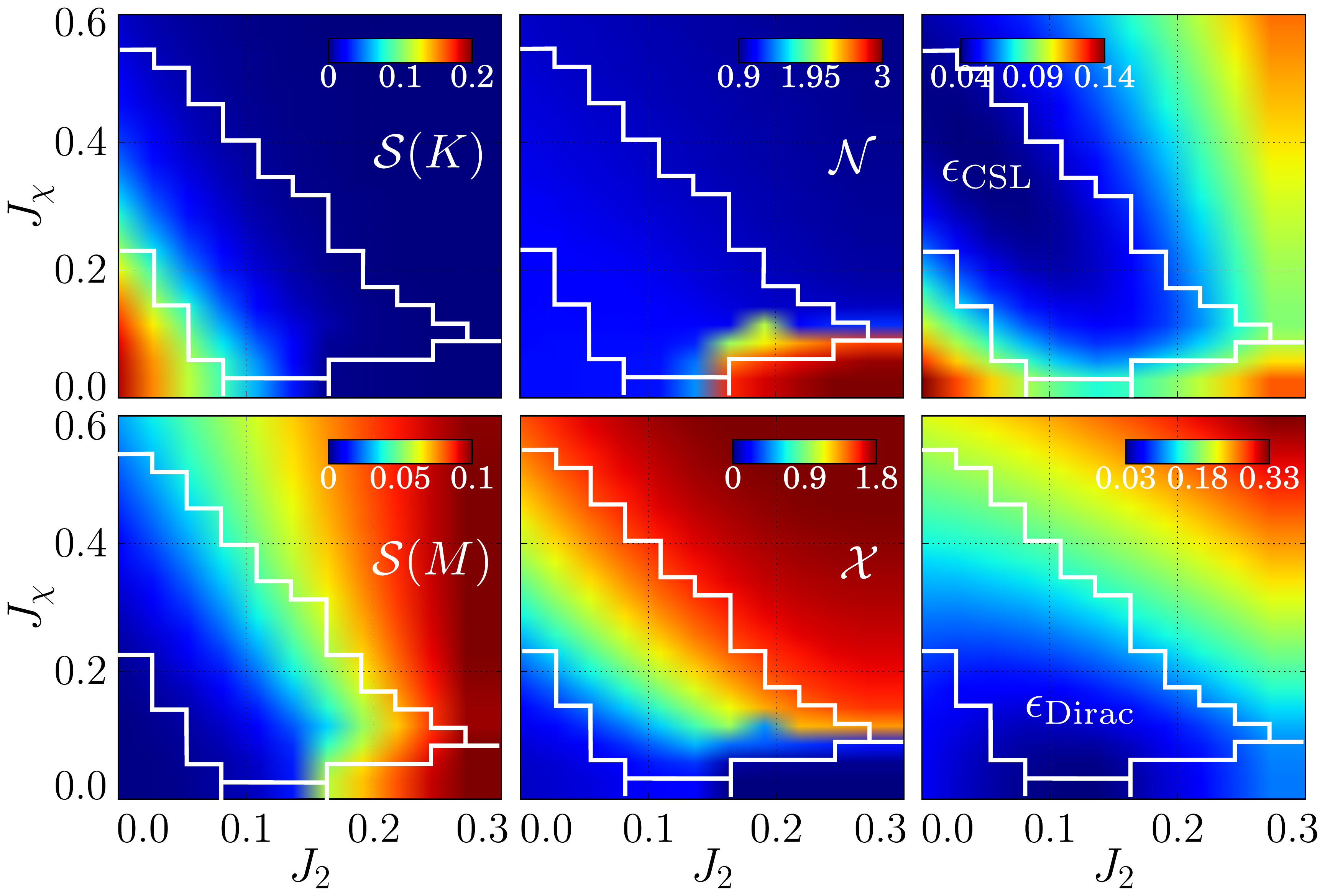}}
  \caption{Order parameters and variational energies of model
    wave functions. 
    \textit{Left:} static spin structure factor
    $\mathcal{S}(q)$. % as in Eq.~\eqref{eq:staticspincorrelation}
    evaluated at $K$ and $M$ point. 
    \textit{Middle:} nematic order parameter $\mathcal{N}$ as in
    Eq.~\eqref{eq:nematicorderparameter} and (disconnected) scalar chirality
    correlation $\mathcal{X}$ as in
    Eq.~\eqref{eq:chiralityorderparameter}
    \textit{Right:} Variational energies $\epsilon =
    (E_\textnormal{model} -E_\textnormal{ED} )/E_\textnormal{ED} $ for
    the chiral and Dirac spin liquid. }
  \label{fig:orderparameters}
\end{figure}

We present the approximate phase diagram in Fig.~\ref{fig:phasediagram} based on
the quantum numbers of the ground state level and the first excited state. The ground state 
is always in the $\Gamma$.A1 representation (except in the stripy phase
where $\Gamma$.A1 and the two $\Gamma$.E2
sectors are almost degenerate). The symmetry sector of the first
excited state determines the phase.
\textit{Orange}: $S=1$ $K$.A1 ($120^\circ$ N\'{e}el)
\textit{Light blue}: $S=0$ $\Gamma$.E2b (CSL), 
\textit{Green}: $S=0$ $\Gamma$.E2a,$\Gamma$.E2b degenerate (Dirac/$\mathbb{Z}_2$ spin liquid), 
\textit{Dark Blue}: $S=0$ $\Gamma$.A1, $\Gamma$.E2a, $\Gamma$.E2b 
degenerate (stripy magnetic order),
\textit{Dark red/Light red}: $S=1$ $M$.A / $S=0$ $\Gamma$.E2a (tetrahedral magnetic order)
For the magnetically ordered phases these quantum numbers follow from
a standard tower of states symmetry
analysis~\cite{Lhuillier2001,Rousochatzakis2008}, see supp. mat.~\cite{supmat} for details.
The spectral phase diagram is further corroborated by the analysis of
relevant order parameters and variational
energies of model wave functions, c.f.~Fig.~\ref{fig:orderparameters},
where the agreement is striking. 
% We now proceed to a detailed discussion of the phases and the corresponding order parameters.

%%%%%%%%%%%%%%%%%%%%%%%%%%
%%%%%%%%%%%%%%%%%%%%%%%%%%
% \textit{Magnetic order: } 
We find three magnetically ordered phases, 
a $120^\circ$ N\'{e}el order \cite{Bernu1992},
a stripy order \cite{Lecheminant1995,Jolicoeur1990,Chubukov1992},
and a non-coplanar tetrahedral order~\cite{Kubo1997,Messio2011,Hickey2015}.
The structure factor 
 $ \mathcal{S}(q) = |\sum_{j}
 e^{iq(\underline{r}_j-\underline{r}_0)}\langle \bm{S}_j\cdot\bm{S}_0
\rangle |^2/N_s$
is peaked at the Brillouin zone points $K$ for  $120^\circ$ N\'{e}el
and at $M$ for stripy and tetrahedral order~\cite{Messio2011}.
To distinguish between the latter we computed a nematic order parameter
\begin{equation}
  \label{eq:nematicorderparameter}
  \mathcal{N} = \sum\limits_{(i,j) \parallel (0,1)} 
  \left< \left(\bm{S}_0\cdot\bm{S}_1 \right)\left(\bm{S}_i\cdot\bm{S}_j \right)\right> _c
\end{equation}
indicative for the stripy phase and the summed scalar chirality correlations 
\begin{equation}
  \label{eq:chiralityorderparameter}
  \mathcal{X} = \sum\limits_{(i,j,k) \in \bigtriangleup}\left< \chi_{(0,1,2)}\cdot\chi_{(i,j,k)}\right> 
\end{equation}
where $\chi_{(i,j,k)} = \bm{S}_i\cdot(\bm{S}_j\times \bm{S}_k)$,
indicative for tetrahedral order. The regions where these quantities
are large in magnitude agree very well with the phase boundaries
derived from tower of states analysis, cf. Fig.~\ref{fig:orderparameters}. 
For the tetrahedral order the $S=1$ $M$.A level is lowest energy level
in the tower of states, depicted dark red in Fig
~\ref{fig:phasediagram}. Close to the stripy phase we observe 
that the first excited level is a $S=0$ $\Gamma$.E2a level, shown as the 
light red region in Fig.~\ref{fig:phasediagram}. We believe
that this level is an artifact of the finite size sample and is
related to the order by disorder mechanism. In neither of the ground state 
correlation functions we can see a difference between the light red region 
and the red region and thus conclude that also this 
region belongs to the same tetrahedral phase.

\section{Chiral Spin Liquids}
are spin disordered chiral topological states. 
Hallmark features of this phase are the topology dependent ground state
degeneracy, long-range entanglement, abelian anyonic excitations and 
gapless chiral edge modes. Several instances of this phase have recently 
been found in local spin models~\cite{He2014,Gong2014,Bauer2014,Wietek2015,Hickey2015,Nataf2016,Kumar2015,Gorohovsky2015, Thomale2009,Nielsen2012,Meng2015,Kamenev2015}.
It has been understood that a representative lattice model wave function for the CSL
is provided by Gutzwiller projected parton wave functions (GPWF) with a completely 
filled parton band with Chern number~$\pm1$~\cite{Wen2002,Wen2004,Wietek2015,Nataf2016,Messio2016}.
We observe no strong magnetic structure peak in between 
the $120^\circ$ N\'{e}el order and the tetrahedral, cf. Fig.~\ref{fig:orderparameters}.
 Therefore a spin disordered state is formed in a sizeable
 intermediate region. The summed scalar chirality correlations $\mathcal{X}$
in Fig.~\ref{fig:orderparameters} are relatively large in this regime compliant with the fact that here a CSL
with a uniform chirality is formed. We will now show conclusive evidence that this is indeed the case. 
We do so by constructing two GPWFs describing the two topological sectors of the
chiral spin liquid on the torus and by computing their overlaps with the two
lowest lying exact eigenstates from ED, similarly as in refs.~\cite{Nataf2016,Wietek2015}. 
%%%%%%%%%%%%%%%%%%%%%%%%%%
\begin{figure}[t!]
  \centering
    \includegraphics[width=0.7\linewidth]{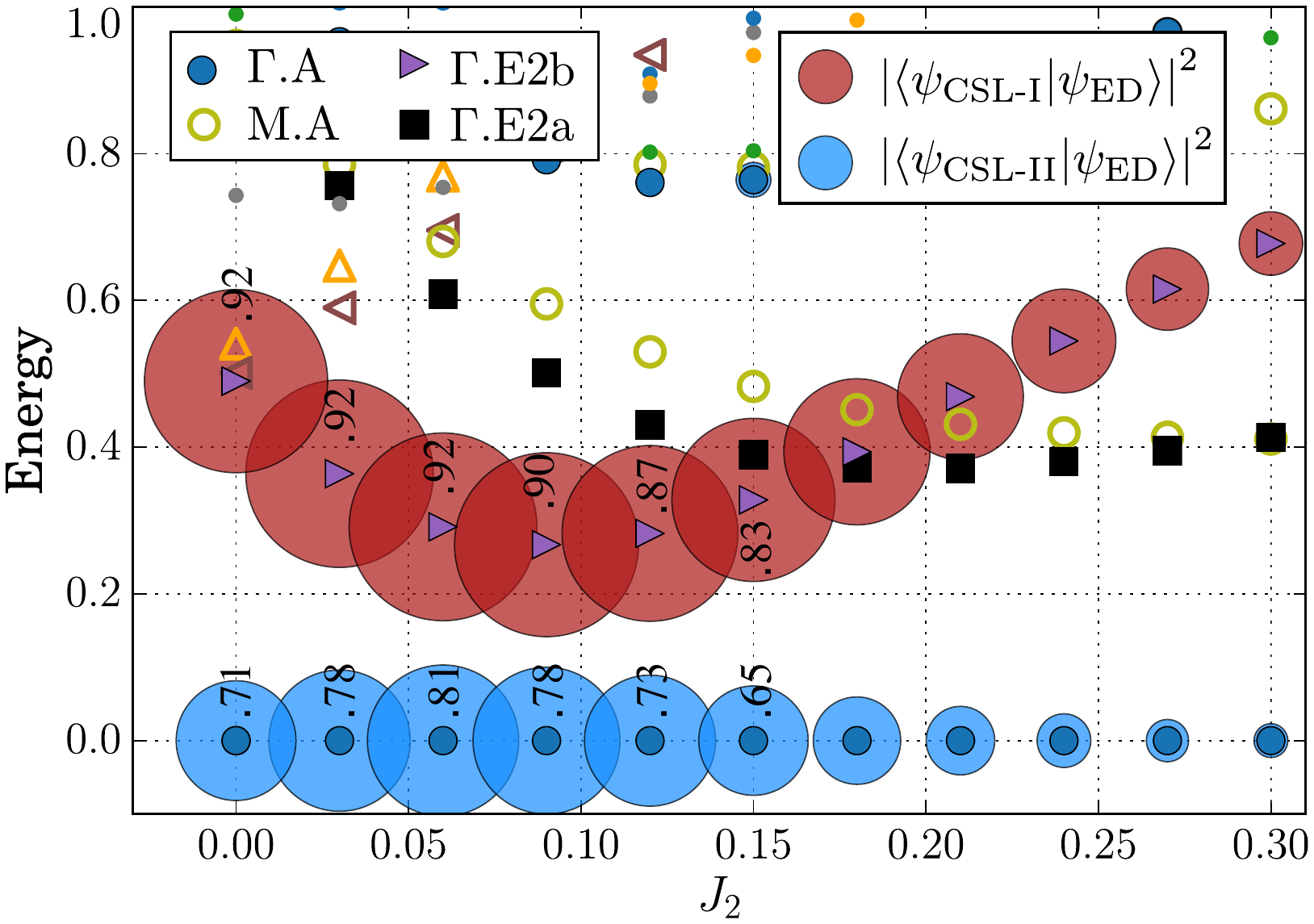}
  \caption{Excitation spectra of the model~\eqref{eq:hamiltonian}
    from ED for $J_\chi = 0.24$ and overlaps with the two CSL
    wave functions on a 36 site cluster. Full (empty) symbols denote even (odd)
    spin levels, different types of symbols denote different
    space-group representations. The numbers denote the summed overlaps
    $\mathcal{O}_{\textnormal{GW}-\textnormal{ED}}^\alpha$  as in
    Eq.~\eqref{eq:defoverlap}. We find overlaps up to $0.92$.
  }
  \label{fig:spectraoverlaps}
\end{figure}
%%%%%%%%%%%%%%%%%%%%%%%%

In Fig.~\ref{fig:spectraoverlaps} we show energy spectra for a
horizontal cut in the phase diagram at $J_\chi=0.24$. The first excited
level above the ground state for $J_2 \lesssim 0.16$ belongs to 
the irreducible representation $\Gamma$.E2b. The region where this 
representation is the first excited state is colored light blue in 
Fig.~\ref{fig:phasediagram}. The parton tight binding model for the 
GPWFs we choose has a two-site unit cell on the triangular lattice with
$\pi/2$ flux through the triangles. This yields a bandstructure with
two bands with Chern numbers $\pm 1$. The ground state
of this tight binding model at half filling is given by filling 
the orbitals of the lower band. After Gutzwiller projection such 
a state has been shown to yield a CSL wave function 
\cite{Mei2015,Wietek2015,Nataf2016,Hu2016}.
To construct the topological partner of the CSL wave function
the phases in the tight-binding model before projection can be tuned 
such that locally the flux through each triangle remains $\pi/2$
while the flux through incontractible loops around the torus changes.
The set of fluxes can be chosen arbitrarily, yet after Gutzwiller projection
these states only form a two dimensional space. This can be
verified by computing the overlap matrix for several GPWFs with
different fluxes through the torus. Indeed we find that thereby 
the rank of the overlap matrix is 2 with a numerical precision of 
$\sim10^{-3}$~\cite{Mei2015}. We chose two out of these wave 
functions spanning the CSL subspace and compute the overlaps with
the lowest two numerical eigenstates from ED.
We find that these two model wave functions
$\ket{\psi_{\text{GW}}^\alpha}$ yield very high overlaps 
\begin{equation}
    \mathcal{O}_{\textnormal{GW}-\textnormal{ED}}^\alpha \equiv
    \left|\left<\psi_{\textnormal{ED}}^0|\psi_{\textnormal{GW}}^\alpha\right>\right|^2
    + \left|\left<\psi_{\textnormal{ED}}^1|\psi_{\textnormal{GW}}^\alpha\right>\right|^2
  \label{eq:defoverlap}
\end{equation}
with the two lowest lying eigenstates of ED of up
to $0.92$~
\footnote{Note that both our model wave functions do not have a fixed (angular-)
momentum and thus overlap with both exact eigenstates. The fluxes of these two
wave functions have been chosen such that one state has mainly overlap
with the first excited state and the other mainly with the
ground state.}. In Fig.~\ref{fig:spectraoverlaps} we plot the square
overlap $|\braket{\psi_{\textnormal{ED}}^n}{\psi_{\textnormal{CSL}}^\alpha}|^2$
with the respective exact
eigenstate ($n$) as the diameter of the red ($\alpha=1$) and light
blue ($\alpha=2$) circles. The overlaps are large where the first
excited state is in the $\Gamma$.E2b representation and quickly decay 
afterwards. This region coincides approximately with the
region where the CSL model wave function has a low variational energy in 
the upper right panel of Fig.~\ref{fig:orderparameters}. We note that
the CSL phase in this phase diagram is located near a tetrahedral magnetic 
phase, reminiscent of a recent study of a frustrated honeycomb spin 
model~\cite{Hickey2015}. It would be interesting to investigate the nature of the phase 
transitions from the tetrahedral~\cite{Hickey2015} and the $120^\circ$ N\'eel phases into the CSL. Finally 
a recent purely variational study~\cite{Hu2016} also found evidence for a CSL in our 
model for selected values of $J_2$ and $J_\chi$.

%%%%%%%%%%%%%%%%%%%%%%%%%%
\begin{figure}[t!]
  \centering
  \includegraphics[width=0.6\linewidth]{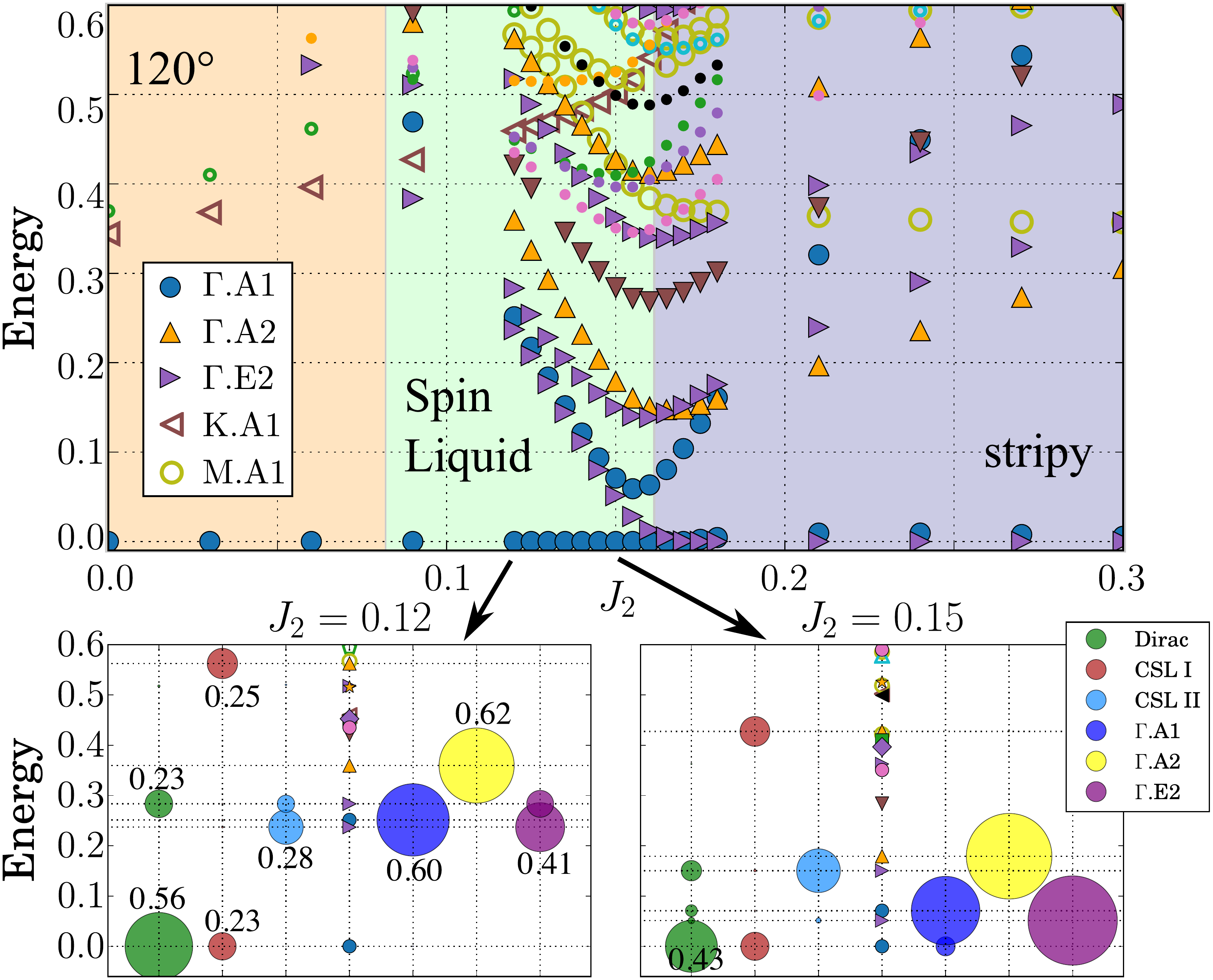}
  \caption{
    ED spectra for $J_\chi = 0$ and spectral decomposition of 
    several model wave functions for $J_2=0.12$ and $J_2=0.15$.
    Full (empty) symbols correnspond to even (odd) spin.
    The diameter of the poles  is proportional to the squareoverlap
    $|\braket{\psi_{\textnormal{ED}}}{\psi_{\textnormal{Model}}}|^2$. 
    Besides the CSL and Dirac spin liquid wave functions the three 
    wave functions denoted by $\Gamma$.A1, $\Gamma$.A2 and 
    $\Gamma$.E2b are the ground states in the respective symmetry 
    sectors at $J_2 = 0.3$.  }
  \label{fig:spectraldecomposition}
\end{figure}
%%%%%%%%%%%%%%%%%%%%%%%%%%
\section{Spin disordered state in the $J_1{-}J_2$ Heisenberg model} 
We now turn to the time-reversal invariant $J_1{-}J_2$ line with $J_\chi=0$. 
A number of recent numerical works~\cite{Kaneko2014,Zhu2015,Saadatmand2015,Hu2015a,Iqbal2016} involving flavors of variational 
Monte Carlo (VMC)~\cite{Kaneko2014,Iqbal2016} and Density Matrix Renormalization 
Group (DMRG) techniques~\cite{Zhu2015,Saadatmand2015,Hu2015a} found
a spin disordered region between the $120^\circ$ magnetic order region and the
stripy magnetic order at larger $J_2/J_1$. Multiple candidate 
phases for this intermediate parameter range have been
proposed, without a consensus so far. Whereas Ref.~\cite{Zhu2015} proposes a
gapped spin liquid phase, Ref.~\cite{Kaneko2014} proposes an extended gapless ASL
state. In Ref.~\cite{Hu2015a} it was argued that a CSL and a $\mathbb{Z}_2$ spin 
liquid are competing in the low energy sector in the intermediate region 
$0.07 \lesssim J_2 \lesssim 0.15$. Ref~\cite{Iqbal2016} compared
variational energies of several $\mathbb{Z}_2$ spin liquids based on 
Gutzwiller projected wave functions. Interestingly they find that among all
of these wave functions the lowest energy is not attained by a state with 
$\mathbb{Z}_2$ structure, but rather by a model whose band structure 
features gapless Dirac-like excitations before projection (see
supp. mat.~\cite{supmat} and Refs.~\cite{Iqbal2016,Lu}). 
After projection 
this state is called \textit{Dirac Spin Liquid} (DSL) and Ref.~\cite{Iqbal2016} finds
an extended gapless region described by a dressed wave function of the DSL kind.

In order to shed light on this open question we present the detailed energy spectrum
of the $N_s=36$ site cluster along the $J_\chi=0$ line in the top panel of Fig.~\ref{fig:spectraldecomposition}.
In the small $J_2$ region the first few levels are in agreement with the tower of state 
expectations for the $120^\circ$ N\'eel state~\cite{Bernu1992}, and similarly at the largest $J_2$ values shown
for the stripy collinear magnetic order~\cite{Lecheminant1995}~\footnote{Some additional levels are visible remnants
of the order by disorder mechanism~\cite{Lecheminant1995}}. 

Focusing on the intermediate region $ 0.08 \lesssim J_2  \lesssim 0.16$ we would expect to
see an approximate four-fold ground state degeneracy in either a non-chiral $\mathbb{Z}_2$
spin liquid or two time-reversal related copies of a CSL as in Refs.~\cite{Gong2014,Wietek2015}.
This is not the case for our system size. 
An additional complication
comes from the observation that some of the low-lying levels in the spin liquid region seem to be states which
become the ground state or low-lying levels in the stripy collinear region across the first order transition around $J_2\sim 0.16$.
 This illustrated by calculating overlaps of several low-lying eigenstates at $J_2=0.3$ with the eigenstates at $J_2=0.12$ ($J_2=0.15$) displayed in 
the lower left (right) panel of Fig.~\ref{fig:spectraldecomposition}.

Given the rather low variational energy of the DSL and to a lesser extent of the CSL model wave functions as shown in the 
right part of Fig.~\ref{fig:orderparameters} (and for the DSL in Refs.~\cite{Kaneko2014,Iqbal2016}) we
also compute the decomposition of these model wave functions onto the exact ED eigenstates for 
$J_2=0.12$ and $J_2=0.15$, as shown in the lower part of Fig.~\ref{fig:spectraldecomposition}. At $J_2=0.12$ 
the ground state has a sizeable overlap with the DSL model wave
function of $0.56$. Furthermore when going up to energies
of about $0.6$, we can also find four states which have non-vanishing overlap with the two different topological sectors
of the CSL model wave functions, although the integrated weight is lower than for the DSL state. This might explain the findings of~Ref.~\cite{Hu2015a}
and is due to the reported CSL stabilized at finite but small $J_\chi$.  In the future one should also 
explore overlaps with a $\mathbb{Z}_2$ spin liquid model wave function in order to address the propensity to this kind of spin liquid on an equal footing
the other model wave functions.

\begin{figure}[t!]
  \centering
  \includegraphics[width=0.9\linewidth]{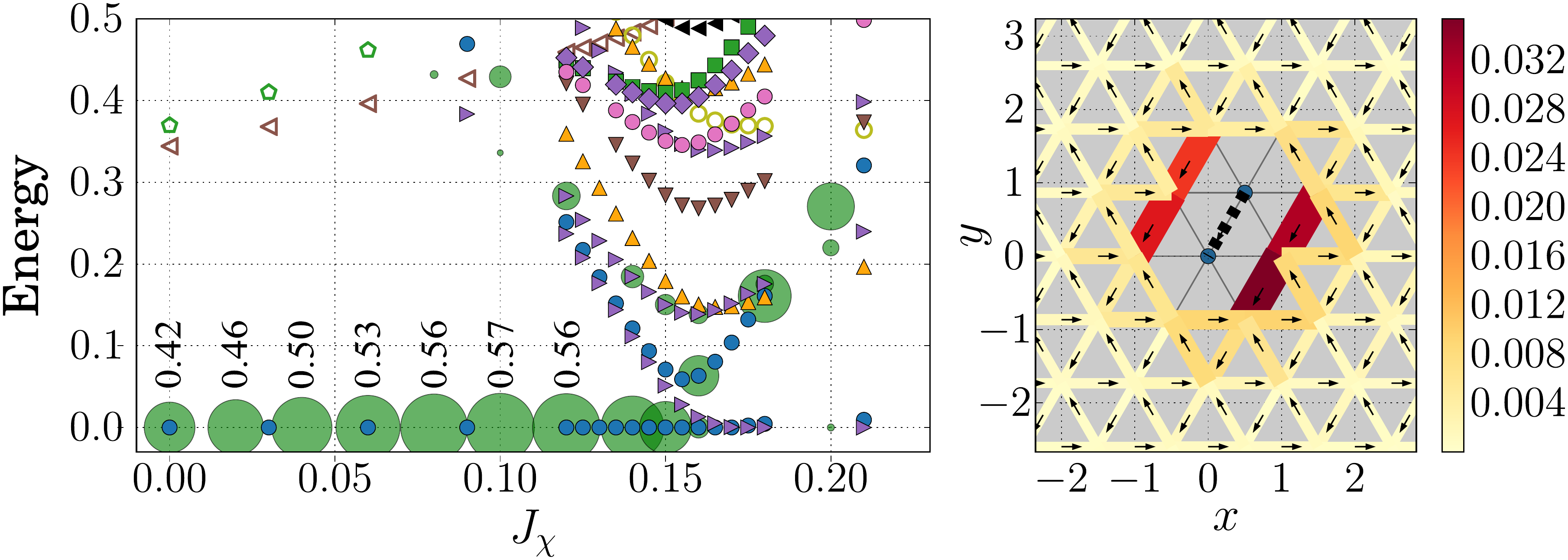}
  \caption{Overlaps of DSL wave function with ED eigenstates and decay
    of spin-spin and twist-twist corralation functions $\left< (\bm{S}_0\times \bm{S}_1)\cdot(\bm{S}_i\times \bm{S}_j)\right>$ of the DSL from VMC on a 144 sites lattice.
    The maximum ground state overlap is attained at $J_2=0.1$.
    %, close to
    %the phase transition point from the $120^\circ$ order to the spin
    %disordered phase. 
    The correlations decay algebraically over
    distance. 
}
  \label{fig:ovlpcorrdecay}
\end{figure}

We have then explored the overlap of the exact ED ground state with the DSL model wave function in a larger range of $J_2$ couplings and observe the
overlap to be maximal in the vicinity of the putative $120^\circ$
N\'eel to spin liquid quantum phase transition around $J_2\sim 0.08$
in Fig.~\ref{fig:ovlpcorrdecay}.
Motivated by this observation we have explored correlation functions in the DSL model wave function and we find likely power-law correlation functions 
which peak at the $K$ point in reciprocal space (consistent with Refs.~\cite{Kaneko2014,Iqbal2016}). We also investigated the spin vector chirality (twist) 
correlations and find them to exhibit likely power-law correlations with a real space pattern in agreement with the (ordered) pattern in the $120^\circ$ N\'eel 
ordered phase cf. Fig.~\ref{fig:ovlpcorrdecay}.

These nontrivial observations motivate us to conjecture that the DSL wave function should not be considered as a model wave function for an extended ASL region,
but instead as a lattice wave function correctly describing the long-distance properties the quantum critical point out of the $120^\circ$ N\'eel state into a spin liquid. The $O(4)^*$ theory~\cite{Chubukov1994a,Chubukov1994b,Whitsitt2016} is a strong contender describing this transition. Let us put this advocated picture
into a broader context: It is believed that Gutzwiller projected wave functions of partons with $SU(N)$ symmetry and a band structure with $n_D$ Dirac points 
capture some aspects (see e.g.~\cite{Tay2011}) of a lattice realization of QED$_3$: i.e. $N_f=N\times n_D$ two-component Dirac fermions coupled to a compact $U(1)$ gauge field in $2+1$~D.
 
It has been shown that in the limit of sufficiently large $N_f$ there are no relevant operators in the theory~\cite{Hermele2004,Hermele2005}, and therefore this 
wave function is representative for an extended ASL region at large $N_f$. For small $N_f<N_f^c$ on the other hand one expects QED$_3$ to become confining in general. The DSL wave function with its power-law decaying correlation functions could then describe a (multi)critical conformal field theory fixed point in between confining phases. The precise value for $N_f^c$ is not known, although recent work~\cite{Grover2014} bounds $N_f^c\lesssim 10$. In the particular case 
of the DSL on the triangular lattice we have $N=2$ and $n_D=2$ resulting in $N_f=4$, substantially lower than the presently known bound. There is also an earlier
observation in Ref.~\cite{Albuquerque2011} that a different $N_f=4$ DSL on the honeycomb lattice describes rather accurately the deconfined quantum critical point~\cite{Senthil1490} between collinear N\'eel order and a VBS phase, giving further evidence that $N_f=4$ DSLs should perhaps be seen as an approximate fixed point wave functions for exotic quantum critical points.

The quantum critical scenario naturally comes with divergent correlation lengths, which could be an explanation for the so far missing clear ground state 
degeneracy both in DMRG and ED. Using couplings frustrating both the $120^\circ$ and the stripy N\'eel orders, it might be 
possible to widen the spin-liquid region and to reduce the correlation lengths to numerically accessible scales, allowing to identify the spin liquid unambiguously. It would also be interesting to understand whether the CSL touches the $J_\chi=0$ line at the quantum critical point.

\section{Conclusion}
We established the phase diagram of an extended Heisenberg model
on the triangular lattice. Amongst several magnetic orderings 
we found a chiral spin liquid phase in an extended region. 
For the spin disordered region for $J_\chi=0$ we found that the 
DSL has sizeable overlap with ED ground states. We proposed a scenario
where this wave function is the quantum critical wave function at 
a transition from magnetic $120^\circ$ N\'{e}el order into a putative
spin liquid phase.

\subsubsection{Acknowledgments}
We thank F.~Becca, Y.~Iqbal, S.~Sachdev, M.~Schuler, P.~Strack and
S.~Whitsitt for discussions. We acknowledge support by the Austrian Science Fund 
through DFG-FOR1807 (I-1310-N27) and the SFB FoQus (F-4018-N23).
The computations for this manuscript have been carried out on
VSC3 of the Vienna Scientific Cluster and on the LEOIIIe cluster
of the Focal Point Scientific Computing at the University of Innsbruck.

%% \bibliography{triangular_j1j2jch}
% \input{triangularJ1J2Jchi_AML_4.bbl} %

% % \end{document}

\newpage
\chapter[Chiral Spin Liquids in $\mathrm{SU}(N)$ Fermionic Mott
Insulators]{Chiral Spin Liquids in Triangular-Lattice $\mathrm{SU}(N)$
  Fermionic Mott Insulators with Artificial Gauge Fields}
\label{sec:sunchiral}
%auto-ignore
% \documentclass{article}
% %\usepackage[ngerman]{babel}
% \usepackage{graphicx}
% \usepackage{hyperref}
 
% \title{Notes for the $SU(N)$-Chiral Spin Liquids on the triangular lattice}
% \author{Alexander Wietek, Andreas L\"auchli}

% \begin{document}

% \maketitle
% \tableofcontents
% \section{asdf}
This chapter has been published as:
\begin{inlinecitation}
  \fullcite{Nataf2016}
\end{inlinecitation}

\begin{figure}[ht!]
  \centering
  \begin{minipage}[]{0.55\linewidth}
    \centering
    \includegraphics[width=0.9\textwidth]{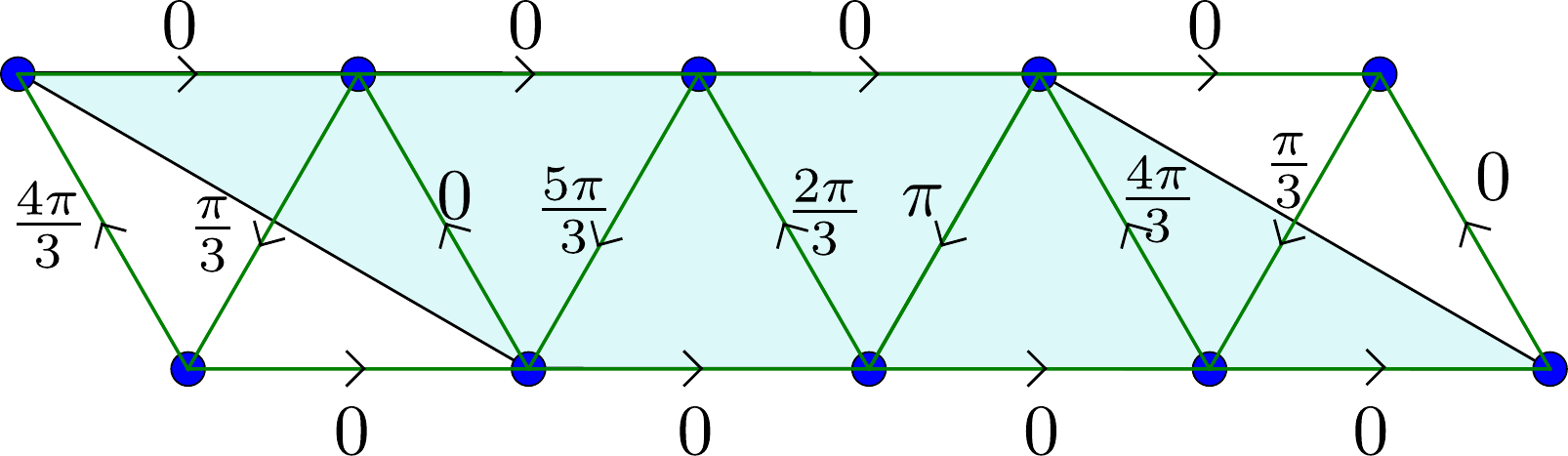}
  \end{minipage}
  \quad
  \begin{minipage}[]{0.4\linewidth}
    \centering
    \includegraphics[width=\textwidth]{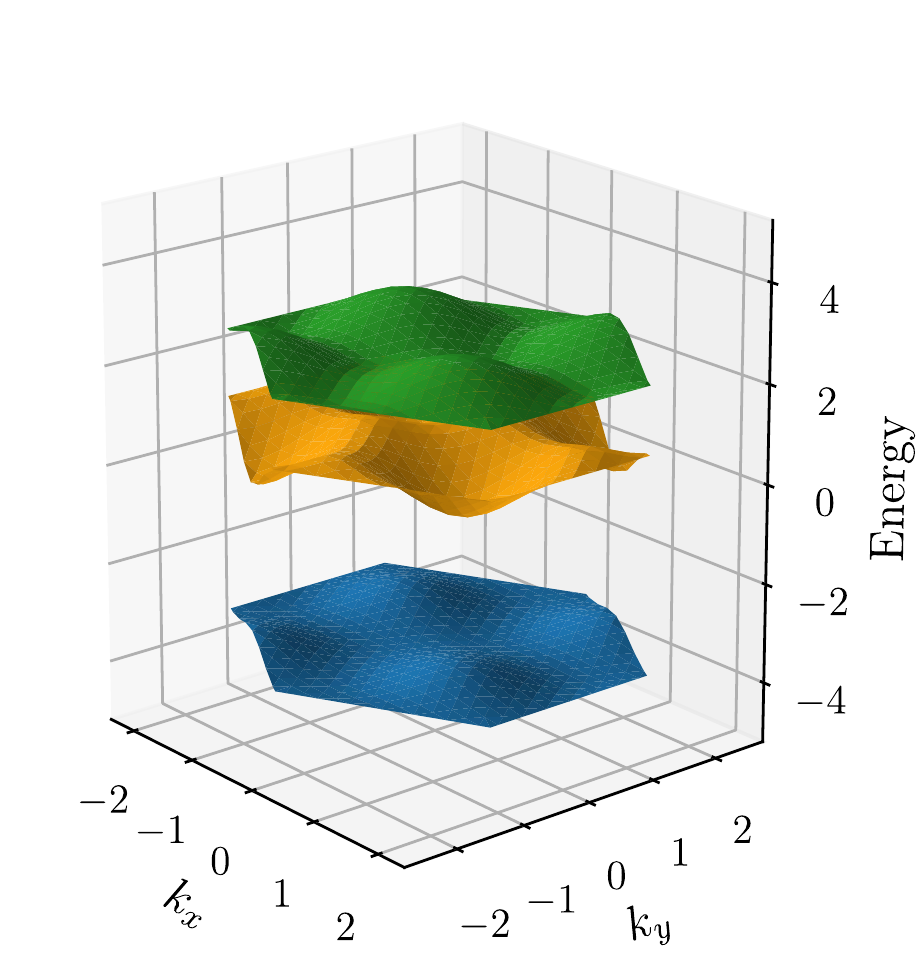}
  \end{minipage}
  \\
  \begin{minipage}[t]{0.55\linewidth}
    \subcaption{\label{fig:su3csl:ansatz}}
  \end{minipage}
  \quad
  \begin{minipage}[t]{0.4\linewidth}
    \subcaption{\label{fig:su3csl:bandstructure}}
  \end{minipage}
  \caption{Parton ansatz and band structure for the SU($3$) chiral spin
    liquid. (a) Geometry of the unit cell for the $\pi/3$ model and
    choice of Peierls phases. There is $\pi/3$ flux through each
    triangle. The periodicity of the unit cell has been chosen in order
    to fit this unit cell onto simulation clusters of Exact
    Diagonalization. (b) Band structure of the $\pi/3$ model. The
    three bands are gapped and carry Chern numbers -1, -1, 2 from
    bottom to top.}
  \label{fig:su3csl}
\end{figure}
\begin{figure}[ht!]
  \centering
  \begin{minipage}[t]{0.55\linewidth}
    \centering
    \includegraphics[width=0.7\textwidth]{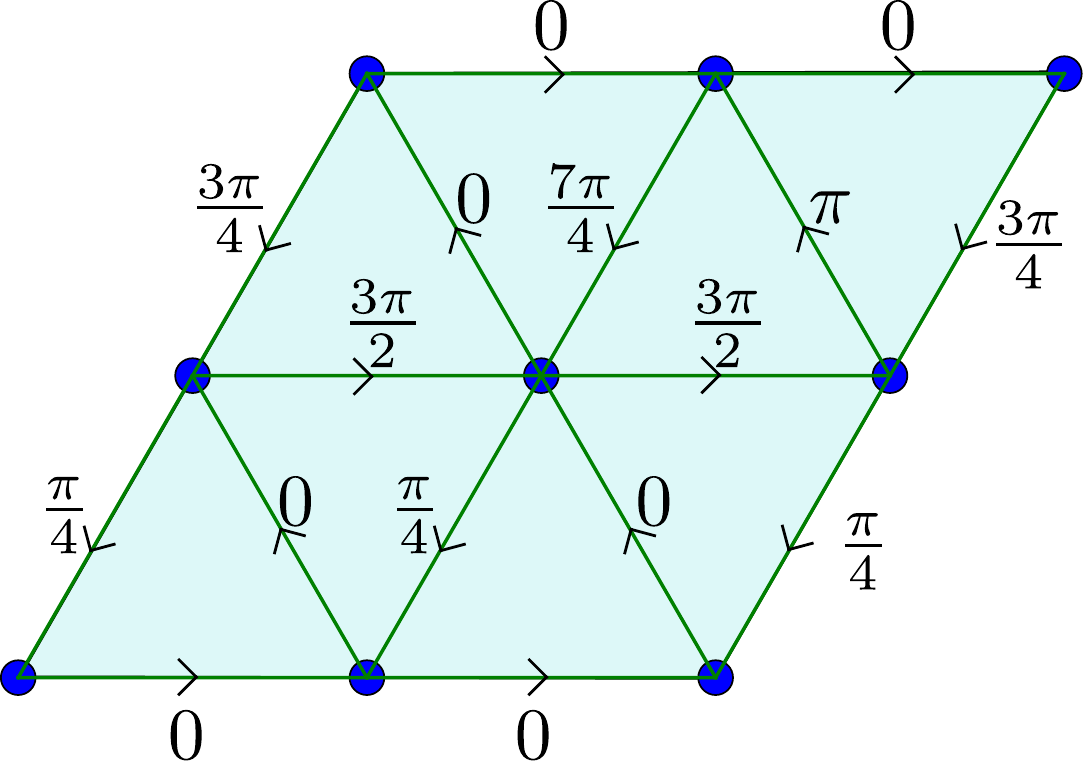}
  \end{minipage}
  \quad
  \begin{minipage}[t]{0.4\linewidth}
    \centering
    \includegraphics[width=\textwidth]{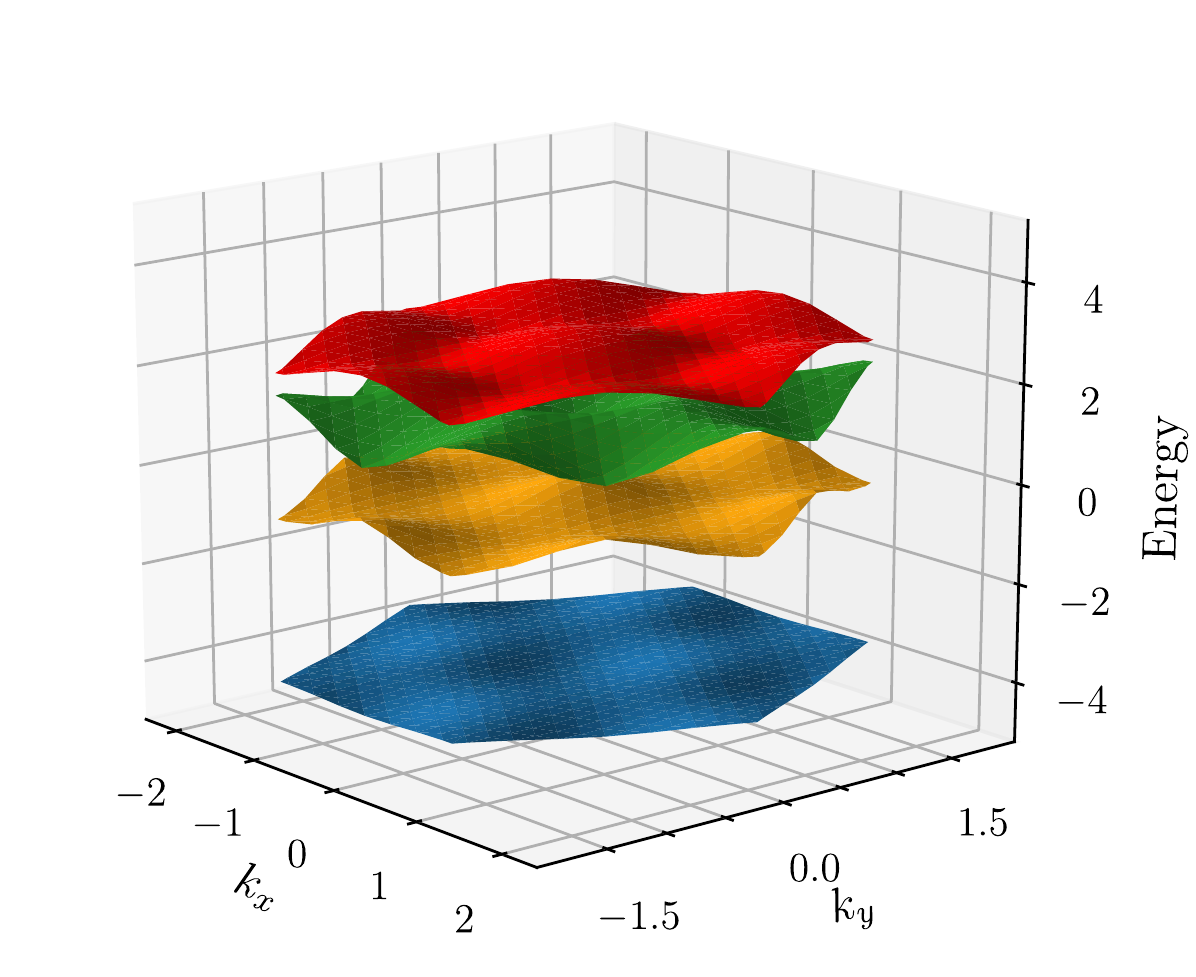}
  \end{minipage}
  \\
  \begin{minipage}[t]{0.55\linewidth}
    \subcaption{\label{fig:su4csl:ansatz}}
  \end{minipage}
  \quad
  \begin{minipage}[t]{0.4\linewidth}
    \subcaption{\label{fig:su4csl:bandstructure}}
  \end{minipage}
  \caption{Parton ansatz and band structure for the SU($4$) chiral spin
    liquid. (a) Geometry of the unit cell for the $\pi/4$ model and
    choice of Peierls phases. There is $\pi/4$ flux through each
    triangle. The periodicity of the unit cell has been chosen in order
    to fit this unit cell onto simulation clusters of Exact
    Diagonalization. (b) Band structure of the $\pi/4$ model. The
    three bands are gapped and carry Chern numbers -1, -1, -1, 3 from
    bottom to top.}
  \label{fig:su4csl}
\end{figure}
% \section{Author's contribution}
\noindent
We describe the author's contribution to the project and present the
published manuscript. We studied the SU($N$) Heisenberg model with an
additional imaginary ring exchange term on the triangular lattice,
\begin{equation}
  \label{eq:sunheisenbergringex}
  H =J\sum_{\langle i,j \rangle} P_{ij} + iK_3 \sum_{(i,j,k)} (P_{ijk}
  + \text{H.c.}),
\end{equation}
where $P_{ij}$ denotes the two-site exchange operator on $N$ colors,
\begin{equation}
  \label{eq:sunexchangeop}
  P_{ij}\ket{\ldots \alpha_i \ldots \alpha_j \ldots} = \ket{\ldots \alpha_j \ldots \alpha_i \ldots}, \quad \alpha_i \in 1, \ldots, N. 
\end{equation}
Analogously, $P_{ijk}$ denotes three-site cyclic exchange operator on
the elementary triangles of the triangular lattice. As explained in
the paper, such a model can be realized in the Mott insulating regime
of an SU($N$) Hubbard model with external magnetic
field. Experimentally such a system can be realized in ultracold atoms
experiments with earth alkaline
atoms~\cite{WuPRL2003,HonerkampHofstetter2004,Cazalilla2009,
  gorshkov2010,Scazza2014,Pagano2014,Zhang2014,CazalillaReyreview2014}.
In the SU($2$) case, $P_{ij}$ corresponds to the Heisenberg exchange
operator $(S_i^+S_j^- + S_i^-S_j^+ )$ and the ring exchange term
$i(P_{ijk} - P_{kji})$ to the scalar chirality interaction
$\bm{S}_i\cdot(\bm{S}_j\times\bm{S}_k)$. Hence, the model
\cref{eq:sunheisenbergringex} may be regarded as a generalization of
the model \cref{eq:hamiltonian} studied in \cref{sec:papertriangular}.
In our research manuscript, we conclude that a chiral spin liquid is
stabilized in an extended region of the phase diagram for
$N=3,\ldots,9$.

The contribution by the author of this thesis was to construct model
CSL wave functions for the SU($N$) chiral spin liquid and compute
their overlap with the numerical ground states. The coefficients of
the CSL have been evaluated in the conventional real space
computational basis. The model has been studied further using Exact
Diagonalization with a novel technique that allows for directly
working with irreducible representations of the SU($N$) symmetry
group~\cite{NatafMilaPRL2014} by Pierre Nataf and Fr\'ed\'eric
Mila. Variational energies and further properties of model CSL wave
functions have been studied by Mikl\'os Lajk\'o and Karlo Penc. The
project was conceived by Andreas M. Läuchli and the numerical edge
spectra together with their analytical prediction have been computed
by him.

\section{Construction of the $\mathrm{SU}(N)$ chiral spin liquid}
In order to construct the CSL model wave functions, we apply the
general parton construction explained in
\cref{sec:partonconstruction}.  We consider a fermionic tight binding
model on the triangular lattice. We choose an ansatz with $\pi/N$ flux
through each triangle. This corresponds to $2\pi/N$ flux per unit cell.
We will call this the $\pi/N$ model.  For defining this model a
$N$-site unit cell is used. The geometry of the unit cell for the
$N=3$ and $N=4$ CSL and the choice of the Peierls phases are shown in
\cref{fig:su3csl,fig:su4csl}.

The parton ansatz breaks time-reversal symmetry and yields $N$ bands,
which are well separated from each other. Moreover, the lowest $N-1$
bands carry Chern number $\pm1$ whereas the top band carries Chern
number $\mp (N-1)$. We fill the lowest band with fermions of $N$
colors to obtain the parton ground state $\ket{\psi_0}$ and perform a
Gutzwiller projection,
\begin{equation}
  \label{eq:gpwfdefinitionsun}
  \ket{\psi_{\text{GPWF}}} = P_{\text{GW}} \ket{\psi_0}.
\end{equation}
To compute the coefficients of this wave function, the formula
\cref{eq:gpwfdeterminant} is straightforwardly generalized to the case
of $N$ colors instead of the two up and down spins in the SU($2$)
case. By shifting the reciprocal space we can insert flux through the
torus. This way, we are able to construct a family of wave functions,
depending on the amount of flux inserted. The overlap matrix is of
rank 3 in the $\mathrm{SU}(3)$ case, and of rank 4 in the
$\mathrm{SU}(4)$ case, with a numerical precision of about
$0.01$. Consequently, we have been able to construct $N$ orthonormal
wave functions spanning the subspace of the CSL. These wave functions
are then compared to the ground state wave functions from Exact
Diagonalization in Fig.~3 of the paper.

\chapter{$\mathrm{SU}(N)$ J-Q model on the square lattice with
  multi-column representations}
\label{sec:sunmulticolumn}
\epigraph{ \parbox{5.4cm}{\flushright Topfpflanzen, bitte gehts
    spazieren.}}{Josef Hader, \textit{Topfpflanzen, Hey!}}
%auto-ignore
\subsubsection{Abstract}
In this preliminary Quantum Monte Carlo study, we consider the
antiferromagnetic $\mathrm{SU}(N)$ Heisenberg model with an additional
four-spin term, often called the $J$-$Q$ model~\cite{Sandvik2007}. We
allow for representations of $\mathrm{SU}(N)$ with a single row and
multiple columns, corresponding to higher spin-$S$ in the
$\mathrm{SU}(2)$ Heisenberg model. This model is expected to exhibit a
phase transition from a N{\'e}el state to a valence bond solid (VBS)
state in two dimensions, where the VBS state in the four-column
representation is a putative symmetry-protected topologically (SPT)
ordered state. We show results on the evaluation of the strange
correlator for the detection of SPT order in one dimension and
investigate the two-dimensional ``$Q$-only'' model with only four-spin
interactions. We give an estimate of the critical value
$N_{\text{crit}}$ for the transition to the VBS state in the $Q$-only
model.

\section{Introduction}
Exotic new states of matter, such as quantum spin liquids, may arise
in strongly interacting systems when fluctuations of local moments
become dominant and determine the behavior of the
system~\cite{balents2010spin}. Apart from geometrical frustration or
low-dimensionality, higher local symmetry may also increase quantum
fluctuations. This was proposed by Read and
Sachdev~\cite{Read1989,Read1990}, who suggested investigating
generalizations of the Heisenberg model with internal $\mathrm{SU}(2)$
symmetry to a larger symmetry group, $\mathrm{SU}(N)$. In the
antiferromagnetic $\mathrm{SU}(2)$ case, the ground state is N{\'e}el
ordered for bipartite lattices~\cite{Marshall1955a}. However, Read and
Sachdev~\cite{Read1989,Read1990} showed that in the
$N \rightarrow \infty$ limit a valence bond solid (VBS) state is
realized and gave an estimate at which value $N$ a phase transition
can be expected, depending on the representation of $\mathrm{SU}(N)$
chosen at every site. Numerically, these values have been determined
exactly by recent Quantum Monte Carlo (QMC) studies for the square
lattice~\cite{Harada2003, Kawashima2007, Okubo2015}.

Different local representations correspond to different values of the
spin-$S$ in the $\mathrm{SU}(2)$ case. Irreducible representations of
$\mathrm{SU}(N)$ are typically described by their Young
diagrams. Here, we only consider the symmetric representations with a
single row and multiple columns. The number of columns is denoted by
$n_{\text{c}}$. The exact form of this VBS in the
$N \rightarrow \infty$ limit is determined by the representation of
$\mathrm{SU}(N)$ chosen locally~\cite{Read1989,Read1990}. For the
two-dimensional square lattice, the value of $n_{\text{c}} \Mod{4}$
decides which kind of VBS is realized. If $n_{\text{c}}= 1,3 \Mod{4} $
a columnar VBS is realized, $n_{\text{c}}= 2 \Mod{4}$ yields a nematic
VBS. Interestingly, the case $n_{\text{c}}= 0 \Mod{4}$ admits a spin
disordered state without spontaneous symmetry breaking but with
possible symmetry-protected topological (SPT) order. This may be
considered as a two-dimensional analog of the famous Haldane phase in
the spin-$1$ Heisenberg chain~\cite{Haldane1983}. SPT order has
recently also attracted attention in the field of measurement-based
quantum computation, where several SPT states in two dimensions have
been proven to allow for universal quantum
computation~\cite{Else2012,Darmawan2012}. As of today, only few
examples of two-dimensional SPT phases emerging from local,
interacting models are known~\cite{Chen2011,You2016}, and often these
models feature complicated interactions. Hence, providing evidence for
the emergence of this kind of ordering in a simple local model would
be highly desirable.

For detecting SPT order in one and two dimensions, the so-called
\textit{strange correlator} has been proposed recently~\cite{You2014}.
Essentially, studying its behavior allows for differentiating between
trivial short-range entangled states and non-trivial SPT phases.  The
strange correlator has already been evaluated numerically for the
Haldane spin-$1$ chain~\cite{Wierschem2014}, where a saturation of the
strange correlator to a constant value is observed. Ultimately, we are
interested in computing the strange correlator for the two-dimensional
possibly SPT-ordered VBS state. In \cref{sec:strangecorr} we show
numerical results from QMC on the evaluation of the strange correlator
for the generalized $\mathrm{SU}(N)$ chain in the two-column
representation.  For $N=2$ this corresponds to the original Haldane
chain.

To study the nature of the phase transition between an ordered
N{\'e}el state and a VBS state a continuous parameter would be
desirable. The parameter $N$ is essentially discrete. Although a
generalization to continuous $N$ has been proposed~\cite{Beach2009},
another route of inducing a phase transition is to add further
interaction terms that favor VBS states. In this context, the
so-called $J$-$Q$ model with an additional four-site exchange term
has been proposed~\cite{Sandvik2007}.  Moreover, the phase transition
between the N{\'e}el state and the VBS state in the model is expected
to provide an example of \textit{deconfined
  criticality}~\cite{Senthil1490,Senthil2004}, a novel exotic kind of
quantum phase transition beyond Landau's theories.

We are thus interested in the $J$-$Q$ models for generalized
$\mathrm{SU}(N)$ interactions in two dimensions. In
\cref{sec:qonlymodel} we present results on the $Q$-only model on the
square lattice for the $\mathrm{SU}(N)$ representations with
$n_{\text{c}}= 1, 2, 3, 4$. We compute spin correlations using QMC to
differentiate between ordered and disordered states. This preliminary
study shows, for which values of $N$ a phase transition between the
N{\'e}el state and the VBS states may be observed using QMC
simulations.

\section{Model}
\begin{figure}[t]
  \begin{minipage}[l]{0.45\linewidth}
    \caption{Interactions of the $J$-$Q$ model. The $Q$-terms are
      defined on pairs of neighboring sites on the elementary square
      plaquettes of the lattice.}
  \label{fig:interactions_jq}
  \end{minipage}
  \quad\quad
  \begin{minipage}[r]{0.45\linewidth}
    \includegraphics[width=\textwidth]{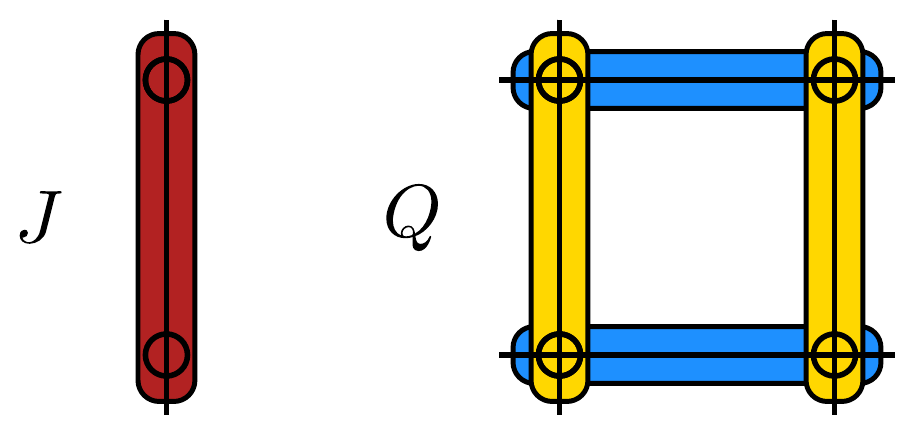}
  \end{minipage}
\end{figure}

We are interested in studying the $\mathrm{SU}(N)$ antiferromagnetic
Heisenberg model with an additional four-spin term,
\begin{equation}
  \label{eq:sun_jq_hamiltonian}
  H = -J\sum\limits_{\langle i j \rangle} B_{ij} -
  Q \sum\limits_{(ij)(kl)}B_{ij}B_{kl}.
\end{equation}
The second sum runs over pairs of nearest-neighbor bonds on an
elementary square plaquette of the lattice as shown in
\cref{fig:interactions_jq}.
The operator $B_{ij}$ is the generalization of the Heisenberg operator
for $\mathrm{SU}(2)$ and is defined by
\begin{equation}
  \label{eq:heisenbergsunop}
  B_{ij} = -\frac{1}{N} \sum\limits_{\alpha, \beta = 1}^{N}
  S_i^{\alpha\beta}\tilde{S}_j^{\beta\alpha}.
\end{equation}
Here, the operators $S_i^{\alpha\beta}$ denote generators of the
SU($N$) algebra, satisfying the commutation relations,
\begin{equation}
  \label{eq:suncommrels}
  [S_i^{\alpha\beta}, S_j^{\mu\nu}] =
  \delta_{ij}(\delta_{\beta\mu}S_i^{\alpha\nu} - \delta_{\alpha\nu}S_i^{\mu\beta}).
\end{equation}
The model is considered for different representations of the
generators $S_i^{\alpha\beta}$, corresponding to higher spin-$S$ in
Heisenberg $\mathrm{SU}(2)$ models. Moreover, for bipartite lattices
there are two choices for the Heisenberg operator $B_{ij}$. We can
either choose the same representation of generators on each sublattice
or choose the conjugate representation,
\begin{equation}
  \label{eq:conjugaterepresentation}
  \tilde{S}_i^{\alpha\beta} = -S_i^{\beta\alpha},
\end{equation}
on one of the two sublattices. The latter choice is then called the
\textit{antiferromagnetic} Heisenberg $\mathrm{SU}(N)$ operator, as
opposed to the ferromagnetic Heisenberg $\mathrm{SU}(N)$ operator with
the same representation on all sublattices~\cite{Auerbach}. The
antiferromagnetic model is sign-problem free and can, therefore, be
efficiently simulated with QMC techniques. We note, that the
Heisenberg $\mathrm{SU}(N)$ model considered in \cref{sec:sunchiral}
can be considered as the ferromagnetic Heisenberg model in the
fundamental represenation\footnote{The term \textit{ferromagnetic} can
  be confusing here. A ferromagnetic $\mathrm{SU}(N)$ Heisenberg bond
  with negative coupling constant actually induces antiferromagnetic
  correlations and vice versa.}. A convenient method of constructing
symmetric representations\footnote{By \textit{symmetric}
  representations we denote the single row, multiple column
  representations of $\mathrm{SU}(N)$.} of the $\mathrm{SU}(N)$
algebra is given by the Schwinger-boson
construction~\cite{Schwinger1952}. Let
\begin{equation}
  \label{eq:schwingerboson}
  a_{i\alpha}, a_{i\alpha}^\dagger, \quad \alpha= 1,\ldots, N,
\end{equation}
be bosonic operators fulfilling the canonical commutation relations % ,
% \begin{equation}
%   \label{eq:schwingerbosonccr}
%   [a_{i\alpha}, a_{j\beta}^\dagger] = \delta_{ij}\delta_{\alpha\beta},
% \end{equation}
and set
\begin{equation}
  \label{eq:sungeneratorsscwingerbosons}
  S_i^{\alpha\beta} = a_{i\alpha}^\dagger a_{i\beta}.
\end{equation}
The operators $S_i^{\alpha\beta}$ satisfy the $\mathrm{SU}(N)$
commutation relations \cref{eq:suncommrels} and conserve the total
number of bosons per site,
\begin{equation}
  \label{eq:schwingerbosonnumber}
  \sum\limits_{\alpha=1}^N a_{i\alpha}^\dagger a_{i\alpha} \equiv
  n_{\text{c}} = 1,2,\ldots.
\end{equation}
These operators are now used to derive the representation matrices for
the symmetric representations of $\mathrm{SU}(N)$ on the finite
dimensional spaces with fixed number of Schwinger-bosons. The
representation with $n_{\text{c}} = 1$ corresponds to the fundamental
representation. In general, the representation on the subspace with
$n_{\text{c}}$ bosons corresponds to the symmetric representation of
$\mathrm{SU}(N)$ with a Young diagram with a single row and
$n_{\text{c}}$ columns.

\section{Strange correlator of $\mathrm{SU}(N)$ Heisenberg chains}
\label{sec:strangecorr}
For detecting SPT phases in one and two dimensions the behavior of
so-called the \textit{strange correlator} can be investigated. It is
defined between two real-space coordinates $\bm{r}$ and
$\bm{r}^\prime$ by~\cite{You2014}
\begin{equation}
  \label{eq:strangecorrelator}
  C(\bm{r}, \bm{r}^\prime) =
  \frac{\braket{\Omega | \phi(\bm{r})\phi(\bm{r}^\prime) | \Psi_0} }
  {\braket{\Omega |\Psi_0}},
\end{equation}
where $\phi(\bm{r})$ is some local operator. Here, $\ket{\Psi_0}$
denotes the wave function under investigation, such as, for example, a
ground state wave function. $\ket{\Omega}$ denotes a SPT-trivial state
with short-range correlations, such as a product state of the form
\begin{equation}
  \label{eq:productstate}
  \ket{\Omega} = \bigotimes_{i=1}^{N_s} \ket{\sigma_i},
\end{equation}
where $N_s$ denotes the number of lattice sites and $\ket{\sigma_i}$ a
state for a local spin. According to Ref.~\cite{You2014}, for a
gapped, non-trivial short-range entangled state,
$C(\bm{r}, \bm{r}^\prime)$ will either converge to a constant or show
algebraic decay for $|\bm{r} - \bm{r}^\prime| \rightarrow \infty$ in
one and two dimensions. Gapped, SPT-trivial states are expected to
exhibit exponential decay of the strange correlator. The strange
correlator has been evaluated exactly for various relevant model
wavefunctions in Refs.~\cite{You2014, Wierschem2014, Wierschem2016}.

We consider the $\mathrm{SU}(N)$ antiferromagnetic Heisenberg model as
in \cref{eq:sun_jq_hamiltonian} without a $Q$-term on a
one-dimensional chain lattice. We choose periodic boundary
conditions. Although in one dimension also several other order
parameters exist for detecting non-trivial SPT order~\cite{Nijs1989,
  Nakamura2002, Duivenvoorden2012}, we compute the strange correlator
for the $\mathrm{SU}(N)$ generalization of the spin-$1$ Heisenberg
chain. Essentially, the string order parameters proposed by
Refs.~\cite{Nijs1989, Nakamura2002, Duivenvoorden2012} do not allow
for a straightforward generalization in two dimensions.
\begin{figure}[t]
  \centering
  \includegraphics[width=0.8\textwidth]{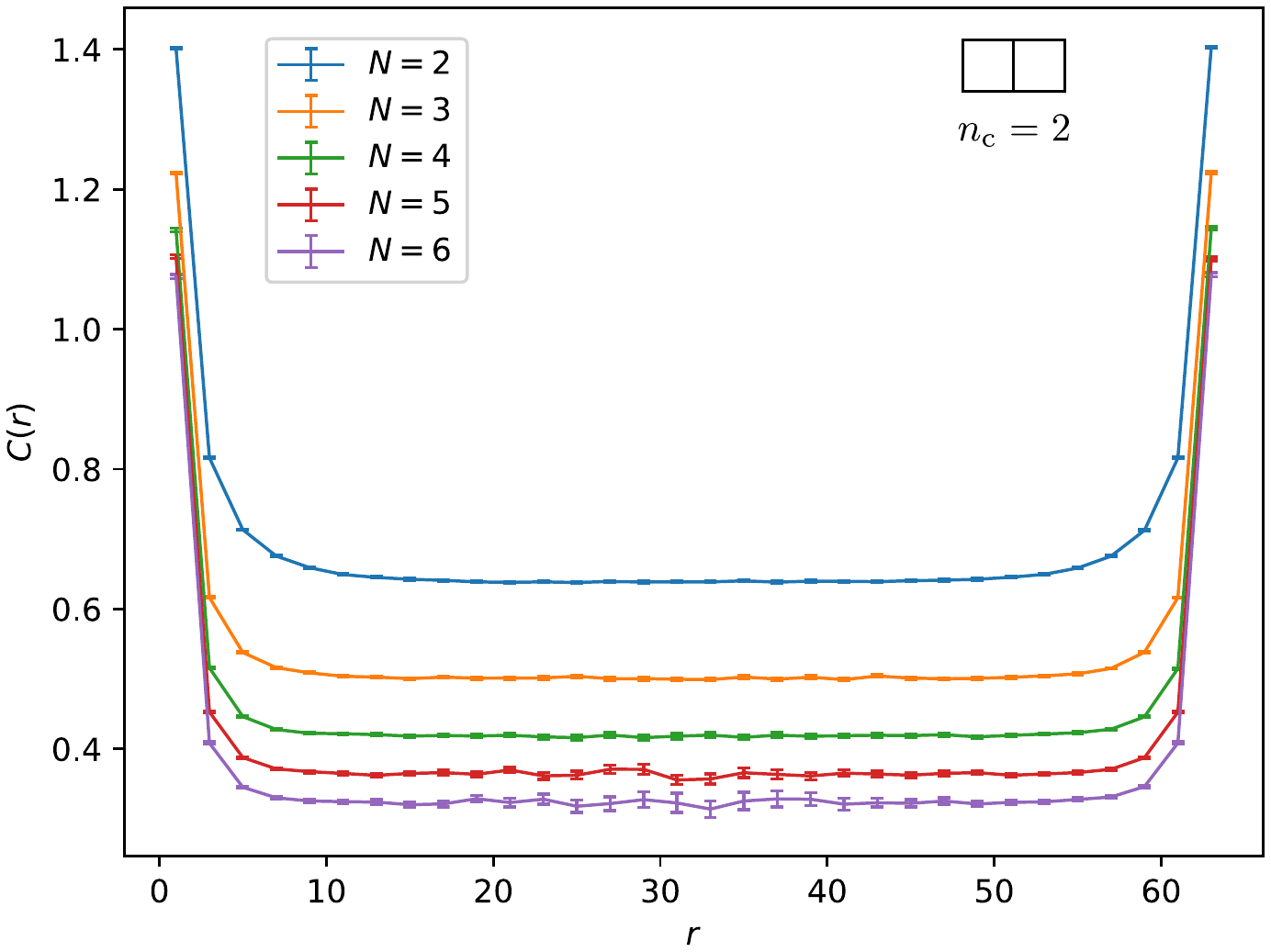}
  \caption{Strange correlator $C(\bm{r}) = C(\bm{r},\bm{0})$ as in
    \cref{eq:strangecorrelatorimagtimesun} for the $\mathrm{SU}(N)$
    antiferromagnetic Heisenberg chain on a $64$-site chain lattice in
    the $n_{\text{c}}=2$ representation. The case $N=2$ corresponds to
    the Heisenberg spin-$1$ chain. The imaginary time in
    \cref{eq:strangecorrelatorimagtimesun} is set to $\tau = 100$. For
    all values of $N$ considered, the strange correlator saturates to
    a constant, indicating SPT order. The error bars represent the
    standard errors.}
  \label{fig:strangecorrelator}
\end{figure}
The strange correlator for the ground state of a system can be
computed by QMC using a ground state projector technique, as
demonstrated by Ref.~\cite{Wierschem2014}. There, the strange
correlator for the $\mathrm{SU}(2)$ spin chain with spin $S=1$ has
successfully been computed. The local operator $\phi$ has been chosen
as the off-diagonal exchange operator,
\begin{equation}
  \label{eq:localopstrangewierschem}
  \phi(\bm{r})\phi(\bm{r}^\prime) = \frac{1}{2}(S^+_{\bm{r}} S^-_{\bm{r}^\prime} +
  S^-_{\bm{r}} S^+_{\bm{r}^\prime}),
\end{equation}
and the trivial product state as
\begin{equation}
  \label{eq:trivialstatewierschem}
  \ket{\Omega} = \bigotimes_{i=1}^{N_s} \ket{0}.
\end{equation}
For the generic $\mathrm{SU}(N)$ case, we consider the off-diagonal
local operators,
\begin{equation}
  \label{eq:heisenbergsunop}
  \phi(\bm{r})\phi(\bm{r}^\prime) = O_{_{\bm{r}\bm{r}^\prime}} = \frac{1}{N}
  \sum\limits_{\alpha\neq \beta }S_{\bm{r}}^{\alpha\beta}S_{\bm{r}^\prime}^{\beta\alpha},
\end{equation}
and the product states
\begin{equation}
  \label{eq:sunproductstates}
  \ket{\Omega} = \ket{\Omega_0} =
  \bigotimes_{i=1}^{N_s} \ket{\sigma_i},
  \text{ where } \ket{\sigma_i} = a_0^\dagger a_1^\dagger\ket{\emptyset}.
\end{equation}
$a_i^\dagger$ denote the Schwinger-boson operators as in
\cref{eq:sungeneratorsscwingerbosons}. Note, that the state
$a_0^\dagger a_1^\dagger\ket{\emptyset}$ in the Schwinger-boson
representation corresponds to the state $\ket{0}$ in the $S^z$ basis
for $\mathrm{SU}(2)$.

For an arbitrary wave function $\ket{\Omega_0}$, not orthogonal to the
true ground state $\braket{\Psi_0 | \Omega_0} \neq 0$, the imaginary
time evolved state,
\begin{equation}
  \label{eq:gsimagtime}
  \exp(-\tau H ) \ket{\Omega_0} \rightarrow \ket{\Psi_0} \text{ for }
  \tau \rightarrow \infty,
\end{equation}
converges to the ground state $\ket{\Psi_0}$ of $H$. $\tau$ denotes
the imaginary time variable. Hence, we consider the following strange
correlator:
\begin{equation}
  \label{eq:strangecorrelatorimagtimesun}
  C(\bm{r}) = \lim\limits_{\tau \rightarrow \infty}
  \braket{\Omega | O_{\bm{r}\bm{0}} \exp(-\tau H )| \Omega_0} /
  \braket{\Omega | \exp(-\tau H )| \Omega_0}.
\end{equation}
This quantity agrees with
\cref{eq:localopstrangewierschem,eq:trivialstatewierschem} for $N=2$
and can be evaluated for finite $\tau$ using the world-line QMC
technique. For an introduction and review see~\cite{Evertz2003,
  Todo2013}. We employ a non-reversible worm algorithm in the
continuous imaginary time integral representation~\cite{Syljuasen2002,
  Kawashima2004, Suwa2010}. We extended an existing implementation
available at~\cite{wormscode} to the generic $\mathrm{SU}(N)$ case.
In order to simulate representations with multiple columns, we apply a
variant of the algorithm proposed in Ref.~\cite{Todo2001a}. The
results of our computations are shown in
\cref{fig:strangecorrelator}. We thermalized the system with $10^7$
QMC sweeps and measured $10^8$ sweeps. Convergence in imaginary time
has been confirmed, also by investigating two different states
$\ket{\Omega_0}$ at $\tau=0$. On small lattices, we compared our
results to numerically exact data from Exact Diagonalization. We find,
that the strange correlator for higher $N$ also saturates to a
constant value. For gapped systems, this implies SPT order. We
computed conventional spin correlation functions which clearly exhibit
an exponential decay in the correlation function for small values of
$N$ \footnote{The spin correlation functions decay faster for higher
  values of $N$. Hence, differentiating between algebraic and
  exponential decay becomes challenging due to finite statistical
  errors.}. This indicates a gapped state, which together with the
long-ranged strange correlator implies SPT-order also for higher
values of $N$.

\section{Spin disordered states in $Q$-only model}
\label{sec:qonlymodel}
We now consider the two-dimensional square lattice case.  In order to
finally establish the full phase diagram of the model
\cref{eq:sun_jq_hamiltonian} we here consider the extremal points
$J=1, Q=0$ and $J=0, Q=1$. The pure Heisenberg case $J=1, Q=0$ has
already been studied for $n_{\text{c}}=1,2,3$ by
Refs.~\cite{Harada2003, Kawashima2007, Okubo2015}. Their findings
suggest that a disordered state is stablilized for $N=5$ if
$n_{\text{c}} = 1$, $N=10$ if $n_{\text{c}} = 2$ and likely $N=15$ if
$n_{\text{c}} = 3$.  This agrees well with the analytical estimate by
Read and Sachdev~\cite{Read1989,Read1990} who proposed a critical
value $n_{\text{c}}/N_{\text{crit}} = 0.19$ in the
$n_{\text{c}} \rightarrow \infty$ limit. This prediction yields a
critical value of $N \approx 20$ for $n_{\text{c}} = 4$.

\begin{figure}[t!]
  \centering
  \begin{minipage}{0.48\textwidth}
    \includegraphics[width=\textwidth]{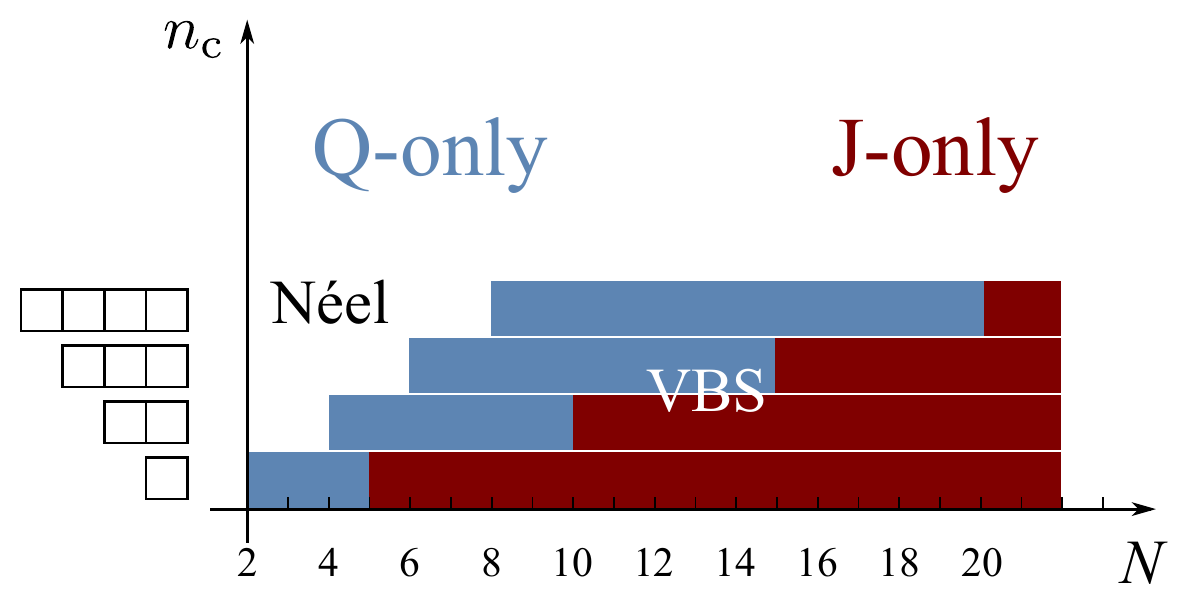}
  \end{minipage}
  \begin{minipage}{0.48\textwidth}
    \includegraphics[width=\textwidth]{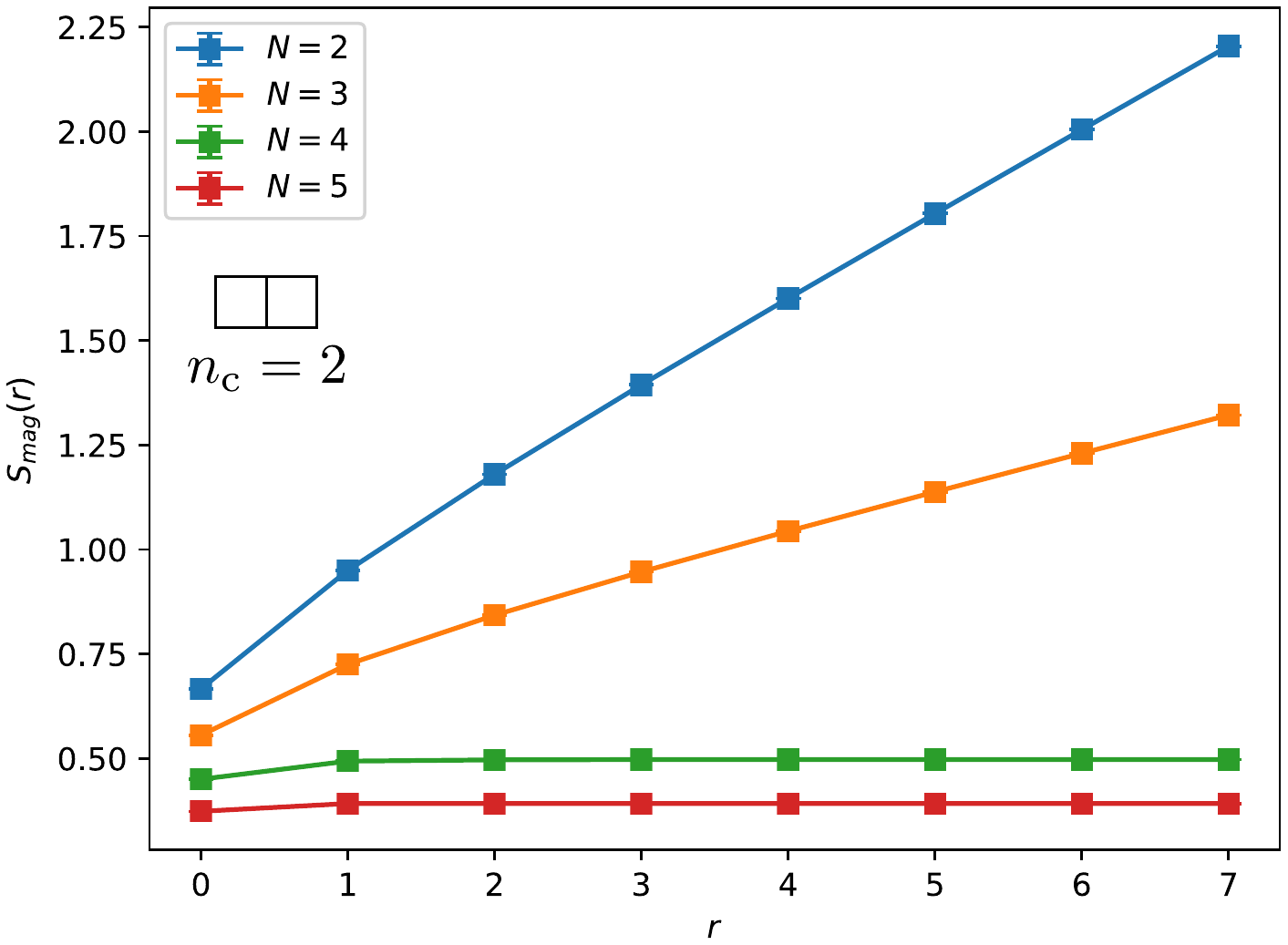}
  \end{minipage} \\
  \begin{minipage}{0.48\textwidth}
    \subcaption{}
    \label{fig:qonlyphasediagram:pt}
  \end{minipage}
  \begin{minipage}{0.48\textwidth}
    \subcaption{}\label{fig:qonlyphasediagram:n2}
  \end{minipage} \\
  % \vspace{1cm}
  % \begin{minipage}{0.48\textwidth}
  %   \includegraphics[width=\textwidth]{summed_corrs_3.pdf}
  % \end{minipage}
  % \begin{minipage}{0.48\textwidth}
  %   \includegraphics[width=\textwidth]{summed_corrs_4.pdf}
  % \end{minipage}\\
  % \begin{minipage}{0.48\textwidth}
  %   \subcaption{}\label{fig:qonlyphasediagram:overview}
  % \end{minipage}
  % \begin{minipage}{0.48\textwidth}
  %   \subcaption{}\label{fig:qonlyphasediagram:n2}
  % \end{minipage} \\
  \caption{ (a): Supposable phase diagram of the $Q$-only model and
    the $J$-only model. (b): Summed spin correlation functions
    $S_{\text{mag}}(\bm{r})$ of the $Q$-only model on a $16\times 16$
    square lattice as defined in \cref{eq:summedspincorrsun} for
    $n_{\text{c}}=2$ across the diagonal of a $16\times 16$ square
    lattice. Asymptotic linear behavior is expected for long-range
    order whereas convergence to a constant is due to exponential (or
    fast algebraic) decay. The error bars represent the standard
    errors.}
  \label{fig:qonlyphasediagram}
\end{figure}

The full $\mathrm{SU}(N)$ $J$-$Q$ model for $n_{\text{c}} = 1$ and
$N=2,3,4$ has been investigated in Ref.~\cite{Harada2013}. The points
of phase transition for the parameter
\begin{equation}
  \label{eq:jqredparm}
  q \equiv Q/(J+Q)
\end{equation}
between the N{\'e}el state and the VBS state have been determined
using QMC. The critical values found are $q_{\text{crit}}=0.9585$ for
$N=2$, $q_{\text{crit}}=0.3353$ for $N=3$ and $q_{\text{crit}}=0.0829$
for $N=4$.

In this preliminary study, we now considered the $Q$-only version of
\cref{eq:sun_jq_hamiltonian} with $J=0, Q=1$. We computed the diagonal
spin correlations,
\begin{equation}
  \label{eq:diagspincorrssun}
  C_{\text{mag}}(\bm{r}) = \frac{1}{N}\braket{  \sum\limits_{\alpha=1}^{N}
    S_{\bm{r}}^{\alpha\alpha}S_{\bm{0}}^{\alpha\alpha}},
\end{equation}
across the diagonal of a $L\times L = 16 \times 16$ square lattice for
$n_{\text{c}}=1,2,3,4$. To differentiate between long-range order and
exponential (or fast algebraic) decay we consider the summed
correlations,
\begin{equation}
  \label{eq:summedspincorrsun}
  S_{\text{mag}}(\bm{r}) = \sum_{\bm{r}^\prime= \bm{0}}^{\bm{r}}
  C_{\text{mag}}(\bm{r}^\prime).
\end{equation}
Our results for the case $n_{\text{c}} = 2$ are shown in
\cref{fig:qonlyphasediagram:n2}. For our simulation, we used $10^5$
thermalization sweeps and $10^6$ measurement sweeps. The temperature
has been set to $T=1/L = 1/16$. Long-range order implies an asymptotic
linear growth of $S_{\text{mag}}(\bm{r})$, whereas saturation to a
constant value suggests exponential (or fast algebraic) decay. We find
long-range order for $N \leq 3$ if $n_{\text{c}} = 2$. Further
preliminary computations suggest N{\'e}el order for $N \leq 5$ if
$n_{\text{c}} = 3$ and $N \leq 7 $ if $n_{\text{c}} = 4$. These
findings are summarized in
\cref{fig:qonlyphasediagram:pt}. Nevertheless, in order to determine
the exact critical value $N_{\text{crit}}$ a more refined analysis
should be performed. Especially, the system size dependence of the
spin correlations should be investigated in a further study. Our
results implicate, that a phase transition between a N{\'e}el ordered
state and a VBS state with the continuous parameter $q=Q/(J+Q)$ should
be observable in the blue region in \cref{fig:qonlyphasediagram:pt}.

\section{Conclusion and Outlook}
We considered the antiferromagnetic $J$-$Q$ model for generalized
$\mathrm{SU}(N)$ spins with multiple-column representations. As a
preparational step, we measured the strange correlator using world-line
QMC for the $\mathrm{SU}(N)$ generalization of the Heisenberg spin-$1$
chain, exhibiting SPT order in one dimension. We then showed first
results on the spin correlation functions of the $Q$-only model on the
two-dimensional square lattice. Our results yield an approximate phase
diagram for the number of columns $n_{\text{c}}$ of the
$\mathrm{SU}(N)$ representation and the value $N$.

In order to consolidate our results, we will compute the spin
correlations of the $Q$-only model also for different system sizes.
Furthermore, the phase transition from the N{\'e}el phase to the VBS
state for varying the parameter $q=Q/(J+Q)$ will be investigated in a
further study. Moreover, computing nematic and columnar order
parameters as in Ref.~\cite{Okubo2015} could determine the nature of
the VBS states for $n_{\text{c}}=1,2,3$. The case $n_{\text{c}}=4$ is
of particular interest since it supposedly exhibits SPT order in two
dimensions. To provide evidence for this kind of order, the strange
correlator can be evaluated, as performed for the one-dimensional case
in this chapter.

\subsubsection{Acknowledgements}
I am very grateful to Synge Todo for accepting me as an exchange
student in his group to work on this project. I greatly enjoyed the
hospitality of him and his group during my stay as well as the
scientific collaboration. I would also like to especially thank
Hidemaro Suwa for discussions and help with the QMC method. Moreover,
I am very grateful for the financial support by Andreas Läuchli and the
Marietta Blau Stipendium. The computations in this chapter have been
carried out on the LEOIIIe cluster of the Focal Point Scientific
Computing at the University of Innsbruck.

%%% Local Variables:
%%% mode: latex
%%% TeX-master: t
%%% End:

\chapter{Conclusion and Outlook}
\epigraph{\parbox{6cm}{\flushright
  Words are flowing out \\
  Like endless rain into a paper cup \\
  They slither while they pass \\
  They slip away across the universe.}}{The Beatles, \textit{Across The
    Universe}} %auto-ignore
% \section{Outlook}

The development and implementation of several numerical methods
allowed us to investigate interesting problems in frustrated quantum
magnetism. The Exact Diagonalization code employing sublattice coding
techniques and distributed memory parallelization presented in
\cref{sec:largescaleed} allowed for numerically exact evaluation of
ground state properties with system sizes up to 50 spin-$1/2$
particles, so far the largest feasible system size.

Using this software we were able to provide conclusive evidence for
the emergence of chiral spin liquids in several two-dimensional
frustrated quantum magnets in
\cref{sec:paperkagome,sec:papertriangular,sec:sunchiral}
(Refs.~\cite{Wietek2015,Nataf2016,Wietek2017a}).  Moreover, we
established approximate phase diagrams in these models and
investigated the nature of the phase transitions. Especially, we
proposed a scenario for the $J_1$-$J_2$ model on the triangular
lattice where a Dirac spin liquid might describe the quantum critical
point of the transition from the $120^\circ$ N{\'e}el ordered state to
a putative $\mathbb{Z}_2$ spin liquid. Our findings on the chiral spin
liquids on the triangular lattice have also recently been confirmed by
an independent DMRG study~\cite{Gong2017}.

We employed energy level spectroscopy (see \Cref{sec:towerofstates}
and Ref.~\cite{Wietek2017}) methods, directly computed ground state
properties from ED and compared with model spin liquid wave functions
derived from Gutzwiller projection, cf. \cref{sec:vmc}.  To
investigate properties of the Gutzwiller projected wave functions also
a Variational Monte Carlo simulation code has been developed.

The methods we developed are now readily applicable to various other
problems. As an extension of the projects in
\cref{sec:paperkagome,sec:papertriangular,sec:sunchiral}, where the
emergence of an Abelian $\mathrm{SU}(2)_1$ (resp. $\mathrm{SU}(N)_1$)
chiral spin liquid was shown, it would be interesting to investigate
systems possibly realizing non-Abelian topological order. Various
candidate systems have already been
proposed~\cite{Greiter2014,Grover2011,Huang2016,Hermele2011}. In the
course of these projects, it might be interesting also to compute
modular matrices of the ground state
manifold~\cite{Zhang2012,Kitaev2006b} or Wilson Loop observables (see
e.g.~\cite{He2015a}).

Dynamical properties of frustrated magnets are often directly
accessible in experiments. The dynamical spin structure factor of a
sample, for example, can be measured by Neutron scattering
experiments~\cite{Lacroix2011}. It would, therefore, be desirable to
compare measurements in the laboratory to numerical data derived from
a model Hamiltonian. For this purpose, the Exact Diagonalization
software can be extended by implementing the continuous fraction
expansion method~\cite{Gagliano1987}. Moreover, to study
non-equilibrium dynamics of quantum magnets, time evolution of wave
functions can also be implemented by Krylov subspace
methods~\cite{Park1986,Krimphoff2017}.

Another important direction of research is to investigate experimental
signatures of chiral spin liquids. Examples include the quantized spin
Hall conductivity measurements~\cite{Haldane1995} or thermal Hall
conductivity measurements~\cite{Katsura2010}. The spin Hall
conductivity can be obtained by computing the many-body Chern
number~\cite{Haldane1995}. A similar topological invariant for the
thermal Hall conductivity is not known so far. Yet, the behavior of
the system at low temperatures may be predicted by investigating
boundary modes~\cite{Nomura2012}. Hence, a comparison of thermal
conductivities derived numerically from the Kubo formula with these
analytical results would be highly interesting. Moreover, certain
dynamical signatures for topological order have been
proposed~\cite{Hazzard2014} and a numerical investigation of those
processes could yield new insight into the dynamical behavior of
quantum spin liquids.

Experimentally, several materials have been found as candidates for
realizing spin liquid behavior. First-principle calculations are
often capable of providing approximate low-energy effective
Hamiltonians for these systems, see e.g.~\cite{Yamaji2014}. Recently,
studies on the compound $\alpha$-\ch{RuCl_3} suggested a possible spin
liquid behavior at low temperatures~\cite{Plumb2014,Banerjee2016} and
effective Hamiltonians have been derived~\cite{Koitzsch2017}. Also,
effective Hamiltonians for Herbertsmithite including
Dzyaloshinskii-Moriya interactions have been proposed and
investigated~\cite{Rousochatzakis2009}. Applying energy level
spectroscopy and evaluation of ground state properties from Exact
Diagonalization might yield novel insights for these systems.

The preliminary results on the $\mathrm{SU}(N)$ Heisenberg models on
the square lattice with an additional four-site $Q$-term in
\cref{sec:sunmulticolumn} suggest a phase transition from a long-range
N{\'e}el ordered state to a paramagnetic state. In the case of the
fundamental representation of $\mathrm{SU}(N)$ it has already been
shown that the disordered state is a Valence Bond
solid~\cite{Lou2009}. For the representations with multiple columns, we
still need to evaluate observables to precisely determine their
nature. Also, the critical point separating those phases can be
evaluated by further computations. The four-column case is expected to
exhibit symmetry protected topological order~\cite{Okubo2015}. This
kind of order has rarely been observed in local models in two
dimensions.  We plan on establishing this order by computing the
strange correlator~\cite{You2014,Wierschem2014} in this system, as
already done in \cref{sec:sunmulticolumn} for the case of the
$\mathrm{SU}(N)$ Haldane chains.

During this thesis many tools have been developed in collaboration
with Michael Schuler for constructing lattice geometries and models,
computing their symmetries and visualizing the results of numerical
computations, such as energy spectra or correlation functions. Also,
parsers for several formats of data structures common in quantum
many-body computations have been written. This collection Python
scripts has been put into a form of a Python package and is by now
well documented. It also includes a flexible Exact Diagonalization
code as a C++ extension which can be used in a user-friendly and
didactic way. We think that these tools might be of broader interest
to the numerical community and plan on making the source code publicly
accessible as an open-source code called \textit{QuantiPy}.

%%% Local Variables:
%%% mode: latex
%%% TeX-master: "../../thesis"
%%% End:

\addtocontents{toc}{\protect\newpage}

\begin{appendices}
  \noindent
  \Cref{sec:spacegroupreptheory,sec:towerofstates} are part of lecture
  notes for the autumn school \textit{Quantum Materials: Experiments
    and Theory} held at Forschungszentrum Jülich form 12. -
  16. September 2016. The full manuscript is available as:
  \begin{inlinecitation}
    \fullcite{Wietek2017}
  \end{inlinecitation}
  
  %auto-ignore
\chapter{Representation theory for space groups}
\label[appendix]{sec:spacegroupreptheory}

For finite discrete groups such as the space group of a finite lattice
the full set of irreducible representations (irreps) can be worked
out.  Let us first discuss some basic groups.  Let's consider a
$n\times n$ square lattice with periodic boundary conditions and a
translationally invariant Hamiltonian like the Heisenberg model on
it. In the following we will set the lattice spacing to $a=1$. The
discrete symmetry group we consider is
$\mathcal{T} = \mathbb{Z}_n \times \mathbb{Z}_n$ corresponding to the
group of translations on this lattice.  This is an Abelian group of
order $n^2$. Its representations can be labeled by the momentum
vectors $\mathbf{k} = (\frac{2\pi i}{n}, \frac{2\pi j}{n})$,
$i,j \in \{ 0, \cdots, n-1\} $ which just correspond to the reciprocal
Bloch vectors defined on this lattice. Put differently, the vectors
$\mathbf{k}$ are the reciprocal lattice points of the lattice spanned
by the simulation torus of our $n\times n$ square lattice.  The
character $\chi_\mathbf{k}$ of the $\mathbf{k}$-representation is
given by
\begin{equation}
  \label{eq:transchars}
  \chi_\mathbf{k}(\mathbf{t}) = \text{e}^{i\mathbf{k}\cdot \mathbf{t}}
\end{equation}
where $\mathbf{t} \in \mathcal{T}$ is the vector of translation. This
is just the usual Bloch factor for translationally invariant systems.

Let us now consider a (symmorphic) space group of the form
$\mathcal{D} = \mathcal{T}\times \text{PG}$ as the discrete symmetry
group of the lattice where $\text{PG}$ is the pointgroup of the
lattice. For a model on a $n\times n $ square lattice this could for
example be the dihedral group of order 8, D$_4$, consisting of
fourfold rotations together with reflections. The representation
theory and the character tables of these point groups are
well-established.  An example for such character tables can be found
in Tab.~\ref{tab:d6chartable} for the dihedral group
$D_6$~\footnote{We follow the labeling scheme for point group
  representations according to Mulliken~\cite{Mulliken1955}.}.
\begin{table}[t]
  \begin{center}
    \begin{tabular}{c|c|c|c|c|c|c}
      $\text{D}_6$ & 1 & $2C_6$ & $2C_3$ & $C_2$ &
                                                   $3\sigma_d$ & $3\sigma_v$\\ \hline
      A1 & 1 & 1 & 1 & 1 & 1 & 1 \\
      A2 & 1 & 1 & 1 & 1 & -1 & -1 \\
      B1 & 1 & -1 & 1 & -1 & 1 & -1 \\
      B2 & 1 & -1 & 1 & -1 & -1 & 1 \\
      E1 & 2 & 1 & -1 & -2 & 0 & 0 \\
      E2 & 2 & -1 & -1 & 2 & 0 & 0 \\
    \end{tabular}
    \caption{Character table for pointgroup D$_6$.}
    \label{tab:d6chartable}
  \end{center}
\end{table}
Since $\mathcal{D}$ is now a product of the translation and the point
group we could think that the irreducible representations of
$\mathcal{D}$ are simply given by the product representations
$(\mathbf{k} \otimes \rho)$ where $\mathbf{k}$ labels a momentum
representation and $\rho$ an irrep of $\text{PG}$. But here is a small
yet important caveat. We have to be careful since $\mathcal{D}$ is
only a semidirect product of groups as translations and pointgroup
symmetries do not necessarily commute. This alters the representation
theory for this product of groups and the irreps of $\mathcal{D}$ are
not just simply the products of irreps of $\mathcal{T}$ and
$\text{PG}$. Instead the full set of irreps for this group is given by
$(\mathbf{k} \otimes \rho_\mathbf{k})$ where $\rho_\mathbf{k}$ is an
irrep of the so called \textit{little group} $L_\mathbf{k}$ of
$\mathbf{k}$ defined as \index{representation theory:little group}
\begin{equation}
  \label{eq:deflittlegroup}
  L_\mathbf{k} = \left\{ g \in \text{PG}; g(\mathbf{k}) = \mathbf{k} \right\}
\end{equation}
which is just the stabilizer of $\mathbf{k}$ in $\text{PG}$. For
example, all pointgroup elements leave $\mathbf{k}=(0,0)$ invariant,
thus the little group of $\mathbf{k}=(0,0)$ is the full pointgroup
PG. In general, this does not hold for other momenta and only a
subgroup of $\text{PG}$ will be the little group of $\mathbf{k}$. In
Fig.~\ref{fig:3sublatt_triangular} we show the $\mathbf{k}$-points of
a $6\times6 $ triangular lattice together with its little groups as an
example. The $K$ point in the Brillouin zone has a D$_3$ little group,
the $M$ point a D$_2$ little group. Having discussed the represenation
theory for (symmorphic) space groups we state that the characters of
these representations are simply given by
\begin{equation}
  \label{eq:spacechars}
  \chi_{(\mathbf{k},\rho_\mathbf{k})}(\mathbf{t},p) = \text{e}^{i \mathbf{k}\cdot \mathbf{t}} \chi_{\rho_\mathbf{k}}(p)
\end{equation}
where $\mathbf{t} \in \mathcal{T}$, $p \in \text{PG}$ and
$\chi_{\rho_\mathbf{k}}$ denotes the character of the representation
$\rho_\mathbf{k}$ of the little group $L_\mathbf{k}$.

%%% Local Variables:
%%% mode: latex
%%% TeX-master: "../../thesis"
%%% End:

  \chapter{Tower of States analysis}
  \label[appendix]{sec:towerofstates}
  %auto-ignore
\subsubsection{Symmetry analysis}
In the analysis of excitation spectra from Exact Diagonalization on
finite size simulation clusters the \textit{tower of states} analysis,
short TOS, is a powerful tool to detect spontaneous symmetry
breaking. As we have seen in \cref{sec:magneticorder} explicitly for
the Heisenberg antiferromagnet, symmetry breaking implies degenerate
ground states in the thermodynamic limit. On finite size simulation
clusters this degeneracy is in general not an exact degeneracy.  We
rather expect a certain scaling of the energy differences in the
thermodynamic limit.  We distinguish two cases:

\begin{itemize}
\item \textbf{Discrete symmetry breaking:} In this case we have a
  degeneracy of finitely many states in the thermodynamic limit.  The
  ground state splitting $\Delta$ on finite size clusters scales as
  $\Delta \sim \exp(-N/\xi)$, where $N$ is the number of sites in the
  system.
\item \textbf{Continuous symmetry breaking:} The ground state in the
  thermodynamic limit is infinitely degenerate. The states belonging
  to this degenerate manifold collapse as \mbox{$\Delta \sim 1/N$} on
  finite size clusters as we have seen in
  section~\ref{sec:magneticorder}.  It is important to understand that
  these states are not the Goldstone modes of continuous symmetry
  breaking.  Both the degenerate ground state and the Goldstone modes
  appear as low energy levels on finite size clusters but have
  different scaling behaviours.
\end{itemize}

The scaling of these low energy states can now be investigated on
finite size clusters. More importantly also their quantum numbers such
as momentum, pointgroup representation or total spin can be predicted
\cite{Lhuillier2001,Misguich2007,Rousochatzakis2008}.  The detection
of correct scaling behaviour together with correctly predicted quantum
numbers yields very strong evidence that the system spontaneously
breaks symmetry in the way that has been anticipated. This is the TOS
method. In the following we will discuss how to predict the quantum
numbers for discrete as well as continuous symmetry breaking.  The
main mathematical tool we use is the character-formula from basic
group representation theory.

Lattice Hamiltonians like a Heisenberg model often have a discrete
symmetry group arising from translational invariance, pointgroup
invariance or some discrete local symmetry, like a spinflip
symmetry. In this chapter we will first discuss the representation
theory and the characters of the representations of space groups on
finite lattices. We will then see how this helps us to predict the
representations of the degenerate ground states in discrete as well as
continuous symmetry breaking.

\subsubsection[Predicting irreducible representations in spontaneous
symmetry breaking]{Predicting irreducible representations in
  spontaneous symmetry breaking \subsectionmark{Predicting irreducible
    representations} }
\subsectionmark{Predicting irreducible representations} \index{tower
  of states!quantum numbers} \index{representation theory:irreducible
  representations} Spontaneous symmetry breaking at $T=0$ occurs when
the ground state $\ket{\psi_{\textnormal{GS}}}$ of $H$ in the
thermodynamic limit is not invariant under the full symmetry group
$\mathcal{G}$ of $H$. We will call a specific ground state
$\ket{\psi_{\textnormal{GS}}}$ a \textit{prototypical state} and the
\textit{ground state manifold} is defined by
\begin{equation}
  \label{eq:defdegenerategs}
  V_{\textnormal{GS}} = \spn\left\{ \ket{\psi_{\textnormal{GS}}^i }\right\}
\end{equation}
where $\ket{\psi_{\textnormal{GS}}^i }$ is the set of degenerate
ground states in the thermodynamic limit.  This ground state manifold
space can be finite or infinite dimensional depending on the
situation. For breaking a discrete finite symmetry this ground state
manifold will be finite dimensional, for breaking continuous SO($3$)
spin rotational symmetry\footnote{The actual symmetry group of
  Heisenberg antiferromagnets is usually SU($2$). For simplicity we
  only consider the subgroup SO($3$) in these notes which yields the
  same predictions for the case of sublattices with even number of
  sites (corresponding to integer total sublattice spin).}  as in
\cref{sec:magneticorder} it is infinite dimensional in the
thermodynamic limit. For every symmetry $g \in \mathcal{G}$ we denote
by $O_g$ the symmetry operator acting on the Hilbert space.  The
ground state manifold becomes degenerate in the thermodynamic limit and
we want to calculate the quantum numbers of the ground states in this
manifold. Another way of saying this is that we want to compute the
irreducible representations of $\mathcal{G}$ to which the ground states
belong to.  For this we look at the action $\Gamma$ of the symmetry
group $\mathcal{G}$ on $V_{\textnormal{GS}}$ defined by
\begin{align}
  \label{eq:representationongroundstates}
  \Gamma: &\mathcal{G} \rightarrow \aut(V_{\textnormal{GS}})\\
          & g \mapsto \left(  \braket{\psi_{\textnormal{GS}}^i| O_g |\psi_{\textnormal{GS}}^j}\right)_{i,j}
\end{align}
This is a representation of $\mathcal{G}$ on $V_{\textnormal{GS}}$, so
every group element $g \in \mathcal{G}$ is mapped to an invertible
matrix on $V_{\textnormal{GS}}$. In general this representation is
reducible and can be decomposed into a direct sum of irreducible
representations
\begin{equation}
  \label{eq:ssbirrepdecomp}
  \Gamma = \bigoplus_\rho n_\rho \rho
\end{equation}
These irreducible representations $\rho$ are the quantum numbers of
the eigenstates in the ground state manifold and $n_\rho$ are its
respective multiplicities (or degeneracies).  Therefore these irreps
constitute the TOS for spontaneous symmetry breaking
\cite{Lhuillier2001}. To compute the multiplicities we can use a
central result from representation theory, the \textit{character
  formula}
\begin{equation}
  \label{eq:characterformula}
  n_\rho = \frac{1}{|\mathcal{G}|}\sum\limits_{g\in \mathcal{G}}\overline{\chi_\rho(g)}\tr(\Gamma(g))
\end{equation}
where $\chi_\rho(g)$ is the character of the representation $\rho$ and
$\tr(\Gamma(g))$ denotes the trace over the representation matrix
$\Gamma(g)$ as defined in Eq.~\eqref{eq:representationongroundstates}.
Often we have the case that
\begin{equation}
  \label{eq:vanishingoverlapthermolim}
  \braket{\psi_{\textnormal{GS}} | O_g | \psi_{\textnormal{GS}}'} = 
  \begin{cases}
    1 \text{ if } O_g\ket{\psi_{\textnormal{GS}}'} = \ket{\psi_{\textnormal{GS}}} \\
    0 \text{ otherwise}
  \end{cases}
\end{equation}
With this we can simplify Eq.~\eqref{eq:characterformula} to what we
call the \textit{character-stabilizer formula}
\begin{equation}
  \label{eq:stabilizerformula}
  n_{\rho} = \frac{1}{|\text{Stab}(\ket{\psi_{\textnormal{GS}}})|}\sum
  \limits_{g \in \text{Stab}(\ket{\psi_{\textnormal{GS}}})}\chi_\rho(g)
\end{equation}
where
\begin{equation}
  \label{eq:stabilizerdefinition}
  \text{Stab}(\ket{\psi_{\textnormal{GS}}}) \equiv \{ g \in \mathcal{G} :
  \; O_g
  \ket{\psi_{\textnormal{GS}}} = \ket{\psi_{\textnormal{GS}}}\}
\end{equation}
is the stabilizer of a prototypical state
$\ket{\psi_{\textnormal{GS}}}$ ~\footnote{In some cases, the orbit of
  the prototypical state
  {$G.\ket{\psi_{GS}} = \{g \in G: \; O_g \ket{\psi_{GS}}\}$} does not
  span the full set of degenerate ground states {$\ket{\psi^i_{GS}}$}.
  In this case, we have to find a set of prototypical states with
  different orbits, such that the union of these orbits spans the full
  ground state manifold. Then,
  Eq.~\unexpanded{\eqref{eq:stabilizerformula}} has to be applied to
  each prototypical state, individually, and the final multiplicity is
  the sum of the individual results.  }.  We see that for applying the
character-stabilizer formula in Eq.~\eqref{eq:stabilizerformula} only
two ingredients are needed:
\begin{itemize}
\item the stabilizer $\text{Stab}(\ket{\psi_{\textnormal{GS}}})$ of a
  prototypical state $\ket{\psi_{\textnormal{GS}}}$ in the ground state
  manifold
\item the characters of the irreducible representations of the
  symmetry group $\mathcal{G}$
\end{itemize}
We want to remark that in the case of
$\mathcal{G} = \mathcal{D} \times \mathcal{C}$ where $\mathcal{D}$ is
a discrete symmetry group such as the spacegroup of a lattice and
$\mathcal{C}$ is a continuous symmetry group such as SO($3$) rotations
for Heisenberg spins the Eqs.~\eqref{eq:characterformula} and
\eqref{eq:stabilizerformula} include integrals over Lie groups
additionally to the sum over the elements of the discrete symmetry
group $\mathcal{D}$.  Furthermore also the characters for Lie groups
like SO($3$) are well-known. For an element $R \in $ SO($3$) the
irreducible representations are labeled by the spin $S$ and its
characters are given by
\begin{equation}
  \chi_S(R) = \frac{\sin\left[(S+\frac{1}{2})\phi\right]}{\sin\frac{\phi}{2}}
\end{equation}
where $\phi \in [0, 2\pi]$ is the angle of rotation of the spin
rotation $R$.
\subsubsection{SU($2$) symmetry breaking in square Heisenberg model}
\label{sec:cont_symm_breaking}
We now give a first example how the TOS method can be applied to
predict the structure of the tower of states for magnetically ordered
phases. We look at the N\'{e}el state of the antiferromagnet on the
bipartite square lattice with sublattices $A$ and $B$.  A prototypical
state in the ground state manifold is given by
\begin{equation}
  \label{eq:neel}
  \ket{\psi} =\ket{\uparrow\downarrow\uparrow\downarrow \cdots}
\end{equation}
where all spins point up on sublattice $A$ and down on sublattice $B$.
The symmetry group $\mathcal{G} = \mathcal{D} \times \mathcal{C}$ of
the model we consider is a product between discrete translational
symmetry
$\mathcal{D} =\mathbb{Z}_2\times \mathbb{Z}_2 = \left\{ 1, t_x, t_y,
  t_{xy}\right\}$ and spin rotational symmetry
$\mathcal{C} = \text{SO(3)}$.  We remark that we restrict our
translational symmetry group to
$\mathcal{D} = \mathbb{Z}_2\times \mathbb{Z}_2$ instead of
$\mathcal{D'} = \mathbb{Z}\times \mathbb{Z}$ because the N\'{e}el
state \index{N\'{e}el antiferromagnet} transforms trivially under
two-site translations $(t_x)^2,(t_y)^2$.  Thus, only the
representations of $\mathcal{D'}$ trivial under two-site translations
are relevant; these are exactly the representations of $\mathcal{D}$.
Put differently we only have to consider the translations in the
unitcell of the magnetic structure which in the present case can be
chosen as a $2$-by-$2$ cell. Furthermore, we will for now neglect
pointgroup symmetries like rotations and reflections of the lattice to
simplify calculations.  At the end of this section we give results
where also these symmetry elements are incorporated.

The ground state manifold $V_{\text{GS}}$ we consider are the states
related to $\ket{\psi}$ by an element of the symmetry group
$\mathcal{G}$, i.e.
\begin{equation}
  \label{eq:neelmanifold}
  V_{\text{GS}} = \left\{O_g\ket{\psi} ; g \in \mathcal{G}\right\}
\end{equation}
The symmetry elements in $\mathcal{G}$ that leave our prototypical
state $\ket{\psi}$ invariant are given by two sets of elements:
\begin{itemize}
\item No translation in real space or a diagonal $t_{xy}$ translation
  together with a spin rotation $R_z(\alpha)$ around the $z$-axis with
  an arbitrary angle $\alpha$.
\item Translation by one site, $t_x$ or $t_y$, followed by a rotation
  $R_a(\pi)$ of $180^\circ$ around an axis $a \perp z$ perpendicular
  to the $z$-axis.
\end{itemize}
So the stabilizer of our prototype state $\ket{\psi}$ is given by
\begin{equation}
  \label{eq:neelstab}
  \text{Stab}(\ket{\psi}) = \left\{1\times R_z(\alpha)\right\}\cup \left\{t_{xy}\times R_z(\alpha)\right\}\cup\left\{t_x\times R_a(\pi)\right\}\cup\left\{t_y\times R_a(\pi)\right\}
\end{equation}
The representations of the discrete symmetry group can be labeled by
four momenta
$\mathbf{k} \in \left\{ (0,0) ,\; (0,\pi) ,\; (\pi, 0) ,\; (\pi,\pi)
\right\}$ with corresponding characters
\begin{equation*}
  \chi_{\mathbf{k}}(t) = \text{e}^{i\mathbf{k}\cdot \mathbf{t}}
\end{equation*}
where $\mathbf{t}$ denotes the translation vector corresponding to
$t$.  The continuous symmetry group we consider is the Lie group
SO($3$). Its representations are labeled by the total spin $S$. The
character of the spin-$S$ representation is given by
\begin{equation*}
  \chi_S(R) = \frac{\sin\left[(S+\frac{1}{2})\phi\right]}{\sin\frac{\phi}{2}}
\end{equation*}
where $\phi \in [0, 2\pi]$ is the angle of rotation of the element
$R \in \text{SO(}3\text)$. We see that spin rotations with different
axes but same rotational angle give rise to the same character.  The
representations of the total symmetry group
$\mathcal{G} = \mathcal{D} \times \mathcal{C}$ are now just the
product representations of $ \mathcal{D}$ and $\mathcal{C}$. Therefore
also the characters of representations of $\mathcal{G}$ are the
product of characters of $\mathcal{D}$ and $\mathcal{C}$. We label
these representations by $(\mathbf{k},S)$ where $\mathbf{k}$ denotes
the lattice momentum and $S$ the total spin.  To derive the
multiplicities of the representations $(\mathbf{k},S)$ in the
ground state manifold, we now apply the character-stabilizer formula,
Eq.~\eqref{eq:stabilizerformula}.  In the case of the square
antiferromagnet this yields
\begin{align}
  \label{eq:neelcharacter}
  n_{(\mathbf{k},S)}\, =\quad& e^{i\mathbf{k}\cdot 0}
                               \frac{1}{4\left| R_z(\alpha)\right|}
                               \int\limits_0^{2\pi}d\alpha \chi_S(R_z(\alpha)) +
                               e^{i\mathbf{k}\cdot (\mathbf{e}_x + \mathbf{e}_y)}
                               \frac{1}{4\left|R_z(\alpha) \right|}
                               \int\limits_0^{2\pi}d\alpha \chi_S(R_z(\alpha)) \\
                             &+  e^{i\mathbf{k}\cdot \mathbf{e}_x}
                               \frac{1}{4\left| R_a(\pi)\right|}
                               \int\limits_0^{2\pi}da \chi_S(R_a(\pi)) +
                               e^{i\mathbf{k}\cdot \mathbf{e}_y}
                               \frac{1}{4\left| R_a(\pi)\right|}
                               \int\limits_0^{2\pi}da \chi_S(R_a(\pi))
\end{align}

We compute
\begin{equation*}
  \left| R_z(\alpha) \right| =  \left| R_a(\pi) \right|= \int\limits_0^{2\pi}d\phi = 2\pi
\end{equation*}
\begin{equation}
  \frac{1}{2\pi} \int\limits_0^{2\pi}d\alpha \chi_S(R_z(\alpha)) =
  \frac{1}{2\pi}\int\limits_0^{2\pi}d\alpha
  \frac{\sin\left[(S+\frac{1}{2})\alpha\right]}{\sin\frac{\alpha}{2}}  
  =  \frac{1}{2\pi}\int\limits_0^{2\pi}d\alpha \sum\limits_{l=-S}^S e^{il\alpha} = 1
  \label{eq:char_int1}
\end{equation}
and
\begin{equation}
  \frac{1}{2\pi} \int\limits_0^{2\pi}da \chi_S(R_a(\pi)) =
  \frac{1}{2\pi}\int\limits_0^{2\pi}da
  \frac{\sin\left[(S+\frac{1}{2})\pi\right]}{\sin\frac{\pi}{2}} = (-1)^S
  \label{eq:char_int2}
\end{equation}
Putting this together gives the final result for the multiplicities of
the representations in the tower of states
\begin{align}
  n_{\left((0,0),S\right)}& = \frac{1}{4}\left(1\cdot 1 + 1\cdot1 +
                            1\cdot (-1)^S + 1\cdot (-1)^S\right) =  
                            \left\{ \begin{array}{l} 1 \text{ if } S
                                      \text{ even} \\ 
                                      0 \text{ if } S \text{ odd}
                                    \end{array} \right. \\
  n_{\left ((\pi,\pi),S\right)} &= \frac{1}{4}\left(1\cdot 1 + 1\cdot1
                                  - 1\cdot (-1)^S - 1\cdot (-1)^S\right)  =
                                  \left\{ \begin{array}{l} 0 
                                            \text{ if } S 
                                            \text{ even} \\ 1 \text{ if } S \text{ odd}
                                          \end{array} \right. \\
  n_{\left ((0,\pi),S\right)} &= \frac{1}{4}\left(1\cdot 1 - 1\cdot1
                                + 1\cdot (-1)^S - 1\cdot
                                (-1)^S\right) = 0\\
  n_{\left ((\pi,0),S\right)} &= \frac{1}{4}\left(1\cdot 1 - 1\cdot1
                                - 1\cdot (-1)^S + 1\cdot
                                (-1)^S\right) = 0
\end{align}
Tab.~\ref{tab:TOS_Neel} lists the computed multiplicities of the
irreducible representations where additionally the $\text{D}_4$ point
group was considered in the symmetry analysis. These irreps and their
multiplicities exactly agree with the irreps and multiplicities in the
TOS of the square lattice Heisenberg model from ED in
Fig.~\ref{fig:andersontowersquare32}. The spectroscopic predictions
together with the numerical data thus constitute a firm and solid
evidence of N\'{e}el order.

%% Comparing this to Fig.~\ref{fig:TOS_Neel} we observe that these are
%% exactly the irreducible representations (momenta and point group
%% irreps) and multiplicities observed in the tower of states for the
%% Heisenberg model on the square lattice obtained from ED.

\begin{table}[ht]
  \centering
  \begin{tabular}{c|cc}
    $S$ & $\Gamma$.A1 & $M$.A1\\ \hline
    0 & 1 & 0\\
    1 & 0 & 1\\
    2 & 1 & 0\\
    3 & 0 & 1\\
  \end{tabular}
  \caption{Multiplicities of irreducible representations in the TOS
    for the N\'eel Antiferromagnet on a square lattice.}
  \label{tab:TOS_Neel}
\end{table}

  %auto-ignore
\subsubsection{Magnetic order in triangular lattice geometries}
On the triangular lattice several magnetic orders can be
stabilized. The Heisenberg nearest neighbour model has been shown to
have a $120^\circ$ N\'{e}el \index{triangular antiferromagnet} ordered
ground state where spins on neighbouring sites align in an angle of
$120^\circ$~\cite{Jolicoeur1990,Chubukov1992}.  Upon adding further
second nearest neighbour interactions $J_2$ to the Heisenberg nearest
neighbour model with interaction strength $J_1$ it was shown that the
ground state exhibits \textit{stripy order} for
$J_2/J_1 \gtrsim 0.18$~\cite{Lecheminant1995}.  Here spins are aligned
ferromagnetically along one direction of the triangular lattice and
antiferromagnetically along the other two. Interestingly, it was shown
that a phase exists between these two magnetic orders whose exact
nature is unclear until today. Several articles propose that in this
region an exotic \textit{quantum spin liquid} is stabilized
\cite{Iqbal2016,Kaneko2014,Hu2015a,Zhu2015}.  In a recent proposal two
of the authors established an approximate phase diagram of an extended
Heisenberg model with further scalar chirality interactions
$J_{\chi}\mathbf{S}_i\cdot(\mathbf{S}_j\times
\mathbf{S}_k)$~\cite{Wietek2017a} on elementary triangles. The
Hamiltonian of this model is given by
\begin{equation}
  \label{eq:hamiltonianj1j2jch}
  H \,=\, J_1\sum\limits_{\left<
      i,j\right>}\mathbf{S}_i\cdot\mathbf{S}_j \ + 
  J_2\sum\limits_{\left<\left<
        i,j\right>\right>}\mathbf{S}_i\cdot\mathbf{S}_j +
  J_{\chi}\sum\limits_{i,j,k \in \bigtriangleup}
  \mathbf{S}_i\cdot(\mathbf{S}_j\times \mathbf{S}_k)
\end{equation}
Amongst the already known $120^\circ$ N\'{e}el and stripy phases an
exotic \textit{Chiral Spin Liquid} and a magnetic
\textit{tetrahedrally ordered} phase were found. \index{tetrahedral
  order} Here we will only discuss the magnetic orders appearing in
this model.  The non-coplanar tetrahedral order has a four-site
unitcell where four spins align such that they span a regular
tetrahedron. In this chapter we discuss the tower of states for the
three magnetic phases in this model.
 
\begin{figure}[ht]
  \centering
  \begin{subfigure}[c]{.5\textwidth}
    \includegraphics[width=\textwidth]{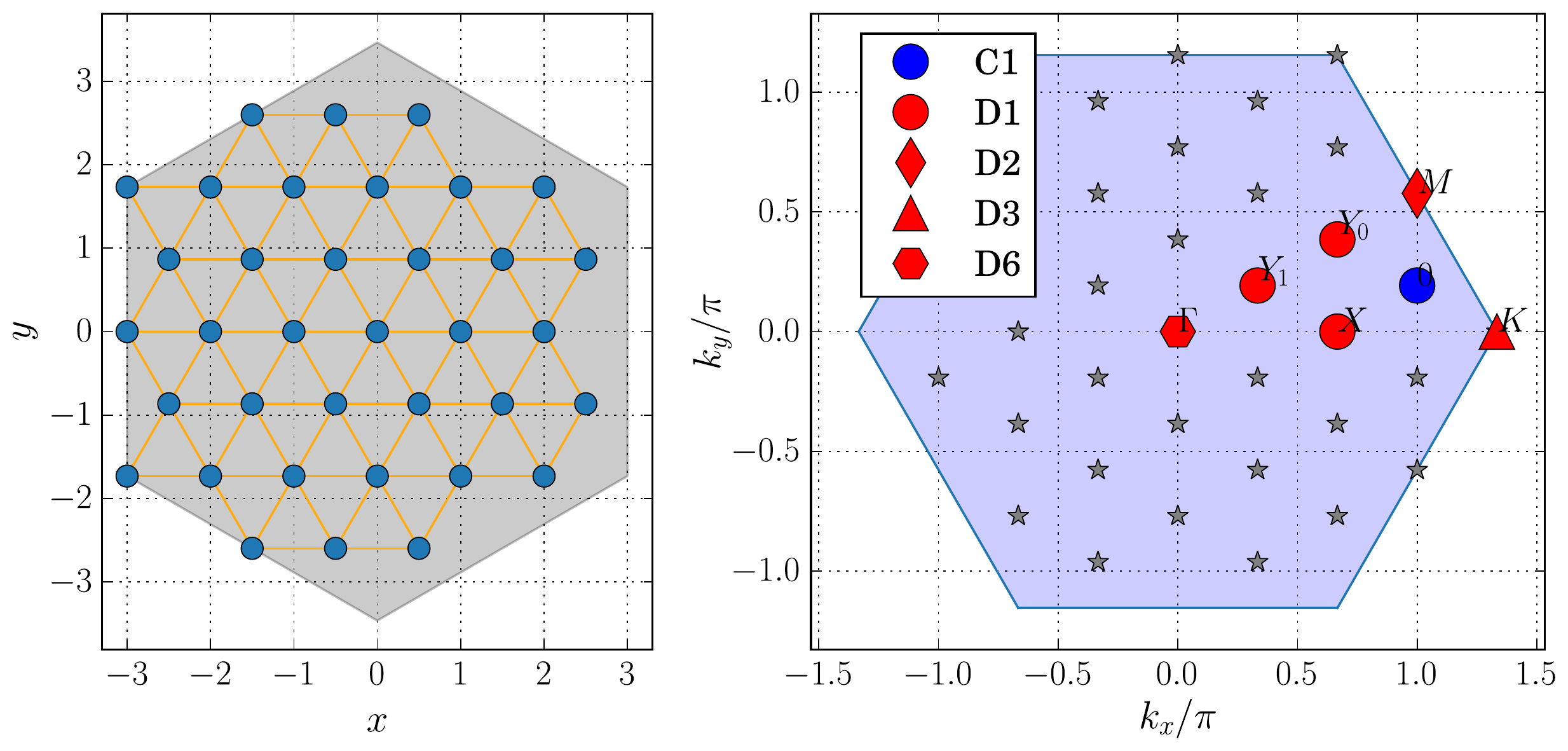}
  \end{subfigure}
  \hfill
  \begin{subfigure}[c]{.45\textwidth}
    \includegraphics[width=\textwidth]{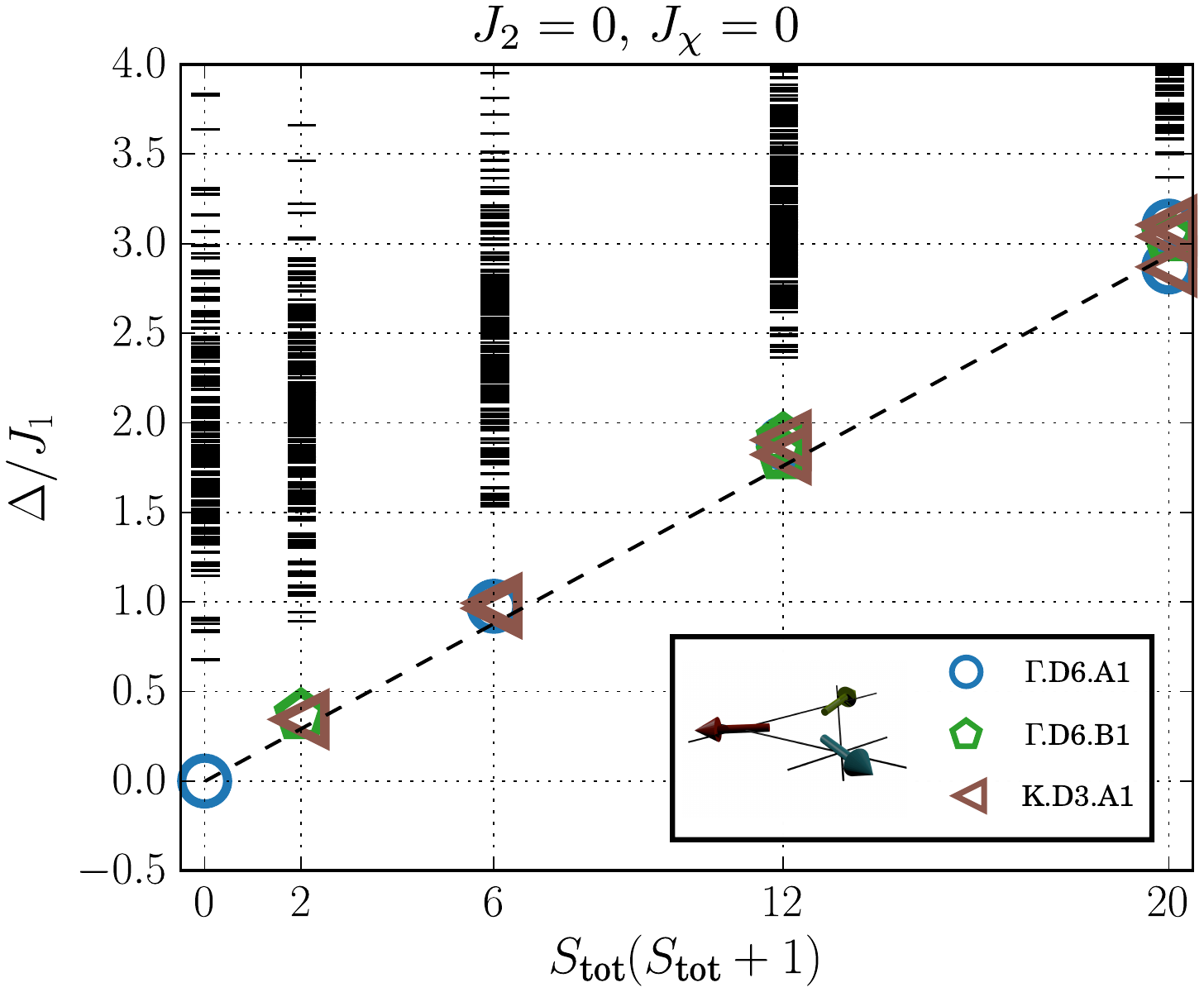}
  \end{subfigure}
  \caption{(Left): Simulation cluster for the Exact Diagonalization
    calculations. (Center): Brillouin zone of the triangular lattice
    with the momenta which can be resolved with this choice of the
    simulation cluster.  Different symbols denote the little groups of
    the corresponding momentum.  (Right): TOS for the $120^{\circ}$
    N\'eel order on the triangular lattice. The symmetry sectors and
    multiplicities fulfill the predictions from the symmetry analysis
    (See Tab.~\ref{tab:andersontowermagtri}).  One should note, that
    the multiplicities grow with $S_{\text{tot}}$ for non-collinear
    states.}
  \label{fig:3sublatt_triangular}
\end{figure}

\begin{figure}[ht]
  \centering
  \begin{subfigure}[c]{.45\textwidth}
    \includegraphics[width=\textwidth]{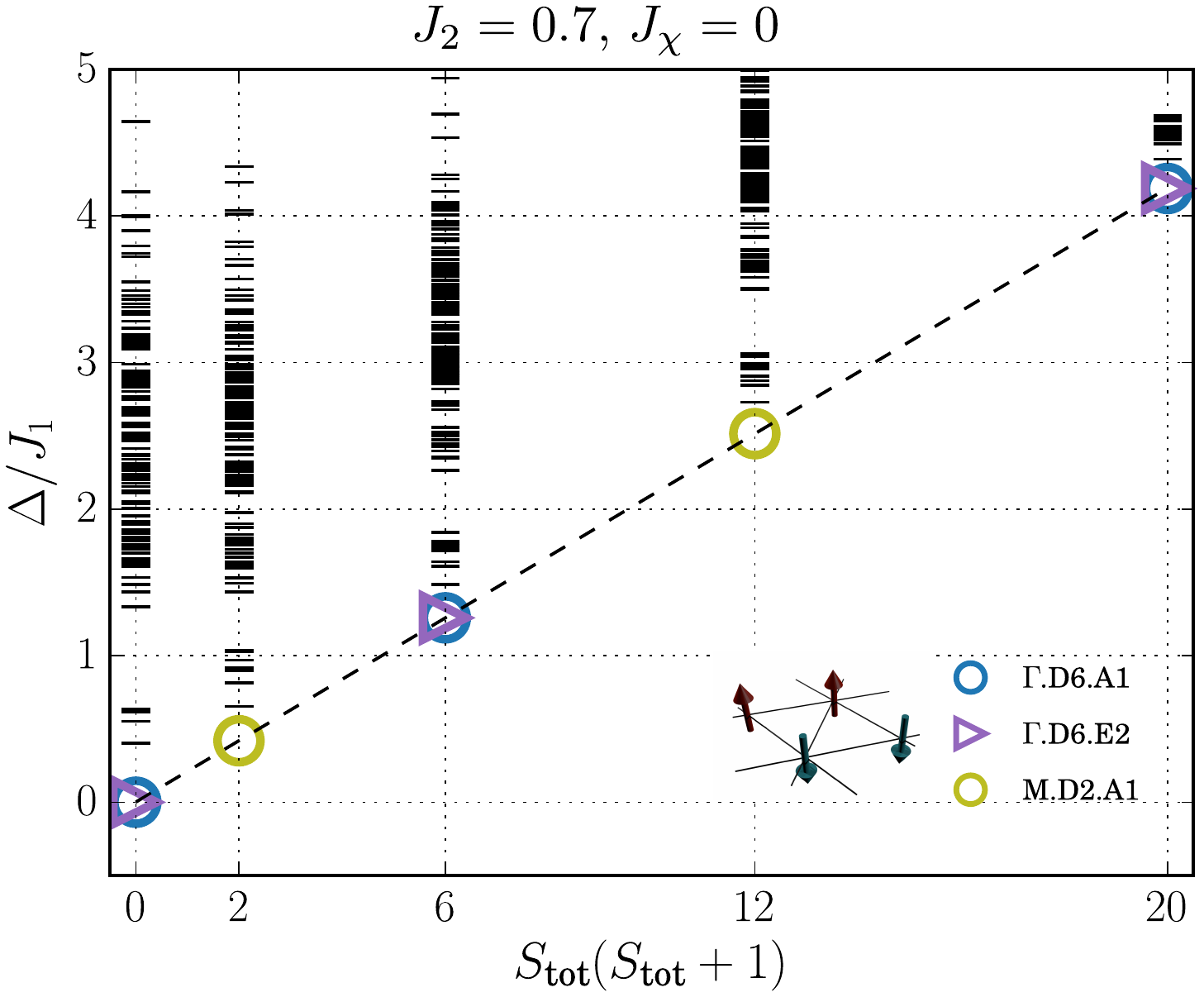}
  \end{subfigure}
  \begin{subfigure}[c]{.45\textwidth}
    \includegraphics[width=\textwidth]{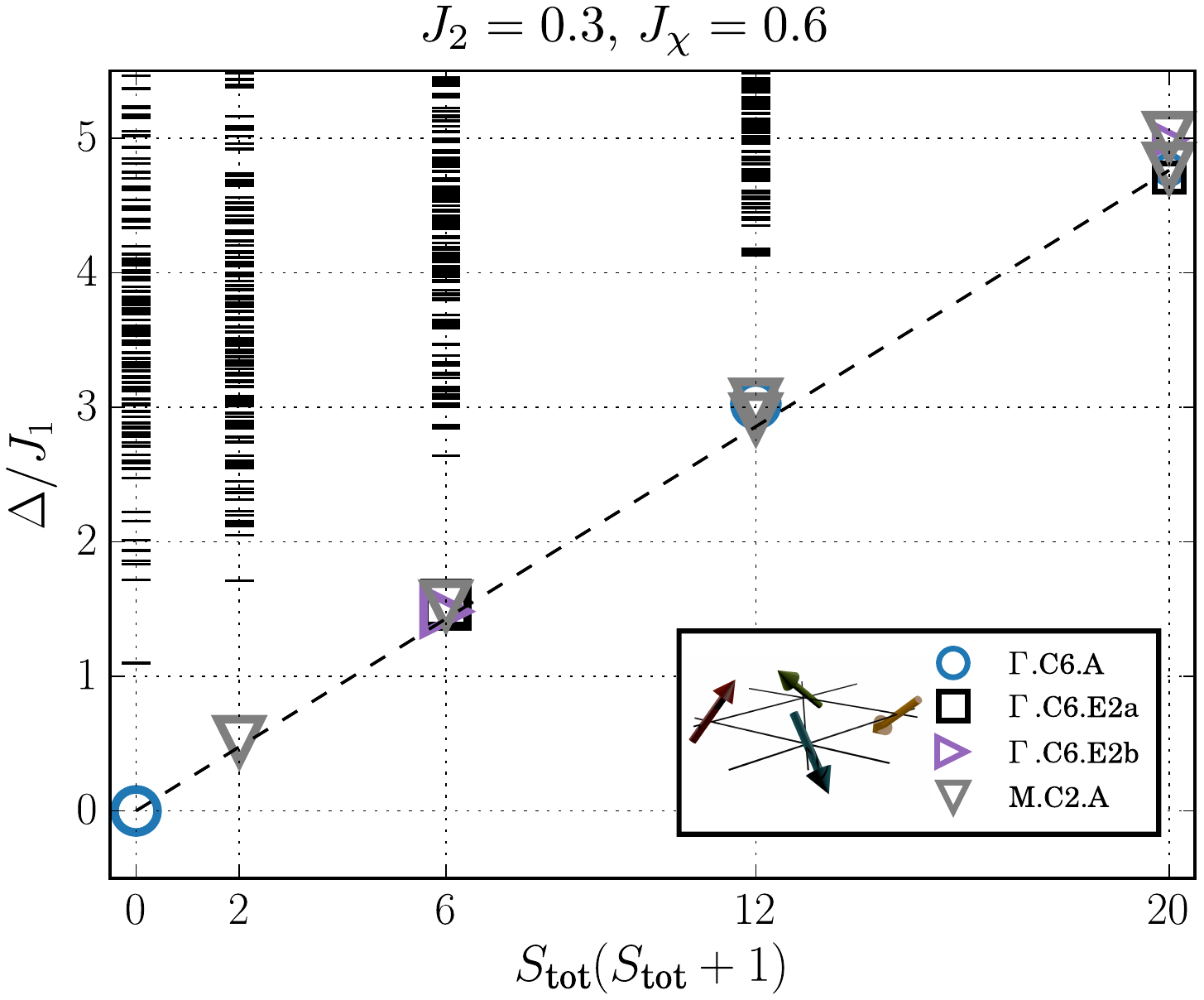}
  \end{subfigure}
  \caption{(Left): TOS for the stripy phase on the triangular
    lattice. The multiplicities for each even/odd $S_{\text{tot}}$ are
    constant for collinear phases. (Right): TOS for the tetrahedral
    order on the triangular lattice.}
  \label{fig:stripy_tetra_triangular}
\end{figure}

First of all Fig.~\ref{fig:3sublatt_triangular} shows the simulation
cluster used for the Exact Diagonalization calculations in
\cite{Wietek2017a}. We chose a $N=36=6\times 6$ sample with periodic
boundary conditions. This sample allows to resolve the momenta
$\Gamma$, $K$ and $M$, amongst several others in the Brillouin
zone. The $K$ and $M$ momenta are the ordering vectors for the
$120^\circ$, stripy and tetrahedral order. Furthermore, this sample
features full sixfold rotational as well as reflection symmetries (the
latter only in the absence of the chiral term, i.e. $J_{\chi} =
0$). Its pointgroup is therefore given by the dihedral group of order
12, D$_6$.  The little groups of the individual $\mathbf{k}$ vectors
are also shown in Fig.~\ref{fig:3sublatt_triangular}.  For our tower
of states analysis we now want to consider the discrete symmetry group
\begin{equation}
  \mathcal{D} = \mathcal{T} \times \text{D}_6
\end{equation}
where $\mathcal{T}$ is the translational group of the magnetic
unitcell. The full set of irreducible representations of this symmetry
group is given by the set $(\mathbf{k}\otimes\rho_{\mathbf{k}})$ where
$\mathbf{k}$ denotes the momentum and $\rho_{\mathbf{k}}$ is an irrep
of the little group associated to $\mathbf{k}$. The points $\Gamma$,
$K$ and $M$ give rise to the little groups D$_6$, D$_3$ and D$_2$ (the
dihedral groups of order $12$, $8$, and $4$), respectively.  For the
stripy and tetrahedral order we can choose a $2\times2$ magnetic
unitcell, and a $3\times3$ unitcell for the $120^\circ$ N\'{e}el
order. The spin rotational symmetry gives rise to the continuous
symmetry group
\begin{equation}
  \mathcal{C} = \text{SO(}3\text{)}
\end{equation}
We can therefore label the full set of irreps as
$(\mathbf{k}, \rho_{\mathbf{k}}, S)$ where $S$ denotes the total spin
$S$ representation of SO($3$).  Similarly to the previous chapter we
now want to apply the character-stabilizer formula,
Eq.~\eqref{eq:stabilizerformula}, to determine the multiplicities of
the representations forming the tower of states. The characters of the
irreps $(\mathbf{k},\rho_{\mathbf{k}},S)$ are given by
\begin{equation}
  \chi_{(\mathbf{k},\rho_{\mathbf{k}},S)}(t, p, R) =
  \text{e}^{i\mathbf{k}\cdot \mathbf{t}}
  \chi_{\rho_{\mathbf{k}}}(p)
  \frac{\sin\left[(S+\frac{1}{2})\phi\right]}{\sin\frac{\phi}{2}}
\end{equation}
where again $\phi \in [0, 2\pi]$ is the angle of rotation of the spin
rotation $R$.
% \begin{table}[htb]
%   \begin{center}
%     \begin{tabular}{c|c|c|c|c|c|c}
%       $\text{D}_6$ & 1 & $2C_6$ & $2C_3$ & $C_2$ &
%                                                               $3\sigma_d$ & $3\sigma_v$\\ \hline
%       A1 & 1 & 1 & 1 & 1 & 1 & 1 \\
%       A2 & 1 & 1 & 1 & 1 & -1 & -1 \\
%       B1 & 1 & -1 & 1 & -1 & 1 & -1 \\
%       B2 & 1 & -1 & 1 & -1 & -1 & 1 \\
%       E1 & 2 & 1 & -1 & -2 & 0 & 0 \\
%       E2 & 2 & -1 & -1 & 2 & 0 & 0 \\
%     \end{tabular}
%     \caption{Character table for pointgroup D$_6$.}
%     \label{tab:d6chartable}
%   \end{center}
% \end{table}
The characters of the pointgroup D$_6$ are given in
Tab.~\ref{tab:d6chartable}.  We skip the exact calculations which
follow closely the calculations performed in the previous chapter,
although now pointgroup symmetries are additionally taken into
account. The results are summarized in
Tab.~\ref{tab:andersontowermagtri}.
\begin{table}[h]
  \centering
  % |l|*{3}{c}|c|l|*{4}{c}|c|
  \begin{tabular}{|l|*{3}{c}c|*{3}{c}c|*{4}{c}|}
    \hline
    &\multicolumn{4}{l|}{$120^\circ$ N\'{e}el} 
    &\multicolumn{4}{l|}{stripy order} 
    & \multicolumn{4}{l|}{tetrahedral order}\\
    \hline\hline
    $S$ & $\Gamma$.A1 & $\Gamma$.B1 & K.A1  & \quad 
                      & $\Gamma$.A1 & $\Gamma$.E2 & M.A   & \quad
                                    & $\Gamma$.A & $\Gamma$.E2a & $\Gamma$.E2b & M.A  \\
    \hline
    0   &  1  & 0 & 0 & \quad & 1 & 1 & 0 & \quad & 1&0&0&0\\
    1   &  0  & 1 & 1 & \quad & 0 & 0 & 1 & \quad & 0&0&0&1\\
    2   &  1  & 0 & 2 & \quad & 1 & 1 & 0 & \quad & 0&1&1&1\\
    3   &  1  & 2 & 2 & \quad & 0 & 0 & 1 & \quad & 1&0&0&2\\
    \hline
  \end{tabular}
  \caption{Multiplicities of irreducible representations in the
    Anderson tower of states for the three magnetic orders on the
    triangular lattice defined in the main text. }
  \label{tab:andersontowermagtri}
\end{table}
We remark that the tetrahedral order is stabilized only for
$J_\chi \neq 0$ where the model in Eq.~\eqref{eq:hamiltonianj1j2jch}
does not have reflection symmetry any more since the term
$\mathbf{S}_i\cdot(\mathbf{S}_j\times \mathbf{S}_k)$ does not preserve
this symmetry. Therefore we used only the pointgroup C$_6$ of sixfold
rotation in the calculations of the tower of states for this phase.

If we compare these results to
Figs.~\ref{fig:3sublatt_triangular},\ref{fig:stripy_tetra_triangular}
we see that these are exactly the representations appearing in the TOS
from Exact Diagonalization for certain parameter values $J_2$ and
$J_\chi$. This is a strong evidence that indeed SO($3$) symmetry is
broken in these models in a way described by the $120^\circ$ N\'{e}el,
stripy and tetrahedral magnetic prototype states.

It is worth noting, that the sum of the multiplicities is constant
with $S_{\text{tot}}$ for collinear phases, e.g. the stripy order
shown here, whereas it is increasing for non-collinear orders.

\end{appendices}

\cleardoublepage\phantomsection\addcontentsline{toc}{chapter}{References}
\printbibliography

\chapter*{List of Publications}
\phantomsection\addcontentsline{toc}{chapter}{List of Publications}
\begin{inlinecitation}[h]
  \begin{minipage}[h]{1.0\linewidth}
    \fullcite{Wietek2015}
  \end{minipage}
  
  \vspace{1cm}

  \begin{minipage}[h]{1.0\linewidth}
    \fullcite{Nataf2016}
  \end{minipage}

  \vspace{1cm}

  \begin{minipage}[h]{1.0\linewidth}
    \fullcite{Wietek2017a}  
  \end{minipage}

  \vspace{1cm}

  \begin{minipage}[h]{1.0\linewidth}
    \fullcite{Wietek2017}  
  \end{minipage}
\end{inlinecitation}

\chapter*{Eidesstattliche Erkl{\"a}rung}
\phantomsection\addcontentsline{toc}{chapter}{Eidesstattliche
  Erkl{\"a}rung} Ich erkl{\"a}re hiermit an Eides statt durch meine
eigenh{\"a}ndige Unterschrift, dass ich die vorliegende Arbeit
selbst{\"a}ndig verfasst und keine anderen als die angegebenen Quellen
und Hilfsmittel verwendet habe. Alle Stellen, die w{\"o}rtlich oder
inhaltlich den angegebenen Quellen entnommen wurden, sind als solche kenntlich gemacht.\\
\noindent
\\
Die vorliegende Arbeit wurde bisher in gleicher oder {\"a}hnlicher
Form noch nicht als Dissertation eingereicht.  \vspace*{3cm}

\noindent Innsbruck, den  \hspace*{6.0cm} Alexander Wietek, M.Sc.,
M.Sc.

\end{document}